\documentclass{aa}  

\usepackage{graphicx}
\usepackage{float}

\usepackage{txfonts}
\begin{document}

\title{The VANDELS survey: Discovery of massive overdensities of galaxies at $z>2$} % in the first half of the VANDELS survey}
\subtitle{Location of Ly$\alpha$-emitting galaxies with respect to environment}

\author{L. Guaita \inst{1,2,3}
\and E. Pompei \inst{2} 
\and M. Castellano \inst{4} 
\and L. Pentericci \inst{4} 
\and O. Cucciati \inst{5} 
\and G. Zamorani \inst{5}
\and A. Zoldan \inst{6}
\and F. Fontanot \inst{6}
\and F. E. Bauer \inst{1,12,13,14} 
\and R. Amorin \inst{7,15}
\and M. Bolzonella \inst{5}
\and G. de Lucia \inst{6}
\and A. Gargiulo \inst{8}
\and N. P. Hathi \inst{10} 
\and P. Hibon \inst{2}
\and M. Hirschmann \inst{9}
\and A. M. Koekemoer \inst{10} 
\and R. McLure \inst{16}
\and L. Pozzetti \inst{5}
\and M. Talia \inst{11,5}
\and R. Thomas \inst{2}
\and L. Xie \inst{17}
\fnmsep \thanks{Based on data obtained with the European Southern Observatory Very Large Telescope, Paranal, Chile, under Large Program ID 194.A-2003}
}

\offprints{Lucia Guaita, \email{lguaita@astro.puc.cl}}

\institute {Instituto de Astrof\'isica, Universidad Cat\'olica de Chile, Vicu\~na Mackenna 4860, Santiago, Chile %1
\and ESO-Chile, Alonso de Cordova 3107, Vitacura, Chile %2
\and N\'ucleo de Astronom\'ia, Facultad de Ingenier\'ia, Universidad Diego Portales, Av. Ej\'ercito 441, Santiago, Chile %3
\and INAF-Osservatorio Astronomico di Roma, Via Frascati 33, 00078 Monteporzio, Roma, Italy %4
\and INAF-Osservatorio di Astrofisica e Scienza dello Spazio di Bologna, via Gobetti 93/3, I-40129, Bologna, Italy %5
\and INAF-Osservatorio Astronomico di Trieste, via G.B. Tiepolo 11, I-34143 Trieste, Italy 0000-0003-4744-0188 %6
%\email{lucia.guaita@mail.udp.cl}
\and Departamento de Astronom\'ia, Universidad de La Serena, Av. Juan Cisternas 1200 Norte, La Serena, Chile %7
\and INAF-Istituto di Astrofisica Spaziale e Fisica Cosmica Milano, via Bassini 15, 20133, Milano, Italy %8
\and Sorbonne Universit\'es, UPMC-CNRS, UMR7095, Institut d'Astrophysique de Paris, F-75014, Paris, France %9
\and Space Telescope Science Institute, 3700 San Martin Drive, Baltimore, MD 21218, USA %10
\and Dipartimento di Fisica e Astronomia, Universit\'a di Bologna, Via Gobetti 93/2, I-40129, Bologna, Italy %11
\and Centro de Astroingenier{\'{\i}}a, Facultad de F{\'{i}}sica, Pontificia Universidad Cat{\'{o}}lica de Chile, Casilla 306, Santiago 22, Chile %12
\and Millennium Institute of Astrophysics (MAS), Nuncio Monse{\~{n}}or S{\'{o}}tero Sanz 100, Providencia, Santiago, Chile %13
\and Space Science Institute, 4750 Walnut Street, Suite 205, Boulder, Colorado 80301 %14
\and Instituto de Investigaci\'on Multidisciplinar en Ciencia y Tecnolog\'ia, Universidad de La Serena, Ra\'ul Bitr\'an 1305, La Serena, Chile %15
\and Institute  for  Astronomy,  University  of  Edinburgh,  RoyalObservatory, Edinburgh, EH9 3HJ, UK %16
\and Tianjin Normal Univeristy, Binshuixilu 393, 300387, Tianjin, China %17
}

\date{accepted for publication}

% \abstract{}{}{}{}{} 
% 5 {} token are mandatory
 
\abstract
   {%Clusters of galaxies are the largest virialized structures in the Universe. %; as such their formation and evolution are interesting both for the influence they have on their member galaxies and their immediate surroundings, and to place important constraints on cosmological models.
      %Going beyond $z=1$ however a systematic study of clusters is increasingly difficult, becoming really challenging beyond $z=2$.
      The advent of deep, multi-wavelength surveys, %, such as CANDELS and VANDELS, 
together with the availability of extensive numerical
      simulations, %of protoclusters 
now allow us for the systematic search and study of (proto)clusters and their surrounding
      environment as a function of redshift.}
   {We aim to define the environment and to identify overdensities in the VANDELS {\it{Chandra}} Deep Field-South (CDFS) and UKIDSS Ultra Deep Survey (UDS) fields. %We want to study the galaxy populations in the identified overdensities, in order to gain a better understanding on their evolution from galaxy overdensity to mature cluster and on how the dominant galaxy population evolves from active star forming to passive. 
%Also, w
We want to investigate whether we can use Ly$\alpha$ emission to obtain additional information of the environment properties and whether Ly$\alpha$ emitters show different characteristics as a function of their environment.}
   {We estimated local densities using a three-dimensional algorithm which works in the RA-dec-redshift space. %; % by \citet{Trevese2007}; 
%the appropriate input parameters have been validated through comparison with the GAEA models.  %\citet{Hirschmann2016}. 
We took advantage of the physical parameters of all the sources in the VANDELS fields to study their properties as a function of environment. In particular, we focused on the rest-frame $U-V$ color to evaluate the stage of evolution of the galaxies located in the overdensities and in the field.
Then we selected a sample of 131 Ly$\alpha$-emitting galaxies (EW(Ly$\alpha)>0${\AA}), unbiased with respect to environmental density, from the first two seasons of the VANDELS survey to study their location with respect to the over- or under-dense environment and infer whether they are useful tracers of overdense regions. 
%
%more???
%MAss sSFR
%
%The resulting catalogue has been further cleaned by applying a spectroscopic selection criteria, i.e. only those structures with 3 to 5 spectroscopically confirmed members have been retained.
}
   {We identify 13 (proto)cluster candidates in the CDFS and nine in the UDS at $2<z<4$, based on photometric and spectroscopic redshifts from VANDELS and from all the available literature. %, which could be high-z protoclusters. 
%
%Among all the overdensity members 50????\% are spectroscopically confirmed. 
%In this redshift bin, we do no
%redeer/bluer than the field
%
%Our final catalogue has been cross-checked with other identified high-redhsift structures: we got positive identifications in all cases where a structure was already known.
No significant difference is observed in the rest-frame $U-V$ color between field and galaxies located within the identified overdensities, but the star-forming galaxies in overdense regions tend to be more massive and to have low specific SFRs than in the field.
%
%J-K simulations observations
%
We study the distribution of the VANDELS Ly$\alpha$ emitters (LAEVs) %with that of all the other galaxies with redshifts from VANDELS 
and we find that Ly$\alpha$ emitters lie preferentially outside of overdense regions as the majority of the galaxies with spectroscopic redshifts from VANDELS.
%In all cases the Ly$\alpha$ emitters were located at the outskirt of the overdensity.
The LAEVs in overdense regions tend to have low Ly$\alpha$ equivalent widths and low specific SFRs, and they also tend to be more massive than the LAEVs in the field. %Also, some protocluster LAEVs show signs of non-zero dust content. 
Their stacked Ly$\alpha$ profile shows a dominant red peak and a hint of a blue peak. There is evidence that their Ly$\alpha$ emission is more extended and offset with respect to the UV continuum.
}
   {%Even if the LAEVs seem not to trace the environment, they 
LAEVs are likely to be influenced by the environment. In fact, our results favour a scenario that implies outflows of low expansion velocities and high HI column densities for galaxies in overdense regions. 
%Geometrical combinations of interstellar media with low expansion velocities and high HI column densities seem to be favoured to interpret our results as characteristics of the galaxies in overdense regions. 
An outflow with low expansion velocity could be related to the way galaxies are forming stars in overdense regions; the high HI column density can be a consequence of the gravitational potential of the overdensity. Therefore, Ly$\alpha$-emitting galaxies can provide useful insights on the environment in which they %{\bf{and also other galaxies}} 
reside.
%The final catalog contains XX protoclusters, ranging from z$\sim$ to z$\sim$, making this the first systematic protocluster catalogue with spectroscopically confirmed members.
%
%The Ly$\alpha$ emitters in overdense regions are characterized by masses of sSFR of , 1/2/3x larger than the average of the field galaxies. 
%
%They show a blue bump which could be related to the HI
%The blue bump observed in Ly$\alpha$ emitters in overdensities may be related to a lower HI fraction, which may have been stripped by the i%nteraction with the
%ther galaxies or by the forming potential well of the larger scale structure.
%
%Combined to the low specific star-formation rate property, this could tell us that in certain overdensities the gas has been consumed in forming stars and/or some
%form of interaction,
%and left with low column density??????, i.e. a clear signpost that the environmental influence is already at work.
%The galaxies inside these overdensities are more evolved with respect to the field, more massive, and currently characterized by low specific star-formation rates. 
%Therefore, Lyman alpha emitters could be used to infer environment properties and infer the properties of galaxies in general in dense environments.
}

\keywords{Galaxies: high-redshift, Galaxies: evolution, Galaxies: protoclusters, Galaxies: kinematics and dynamics, Galaxies: interactions
}

\titlerunning{Overdensities at $z>2$}
\authorrunning{Guaita L.}

\maketitle
%
%-------------------------------------------------------------------

\section{Introduction}

%stage of evolution of virialized clusters
%
%proto-cluster and inside out scenario
%
%radio galaxy studied masses of clusters??????
%
Clusters of galaxies are interesting laboratories for the study of the evolution of galaxies as a function of their environment and they are important cosmological tracers to constrain models of the Universe. The maximum mass which can collapse and virialize at any epoch
depends on $\Omega_M$, the power spectrum $\sigma_8$, and the dark energy equation of state. Hence, the study of a statistically significant sample of clusters, whose mass can be reliably measured, is a powerful tool for deriving observational constraints on cosmological models.

A mature cluster can be recognized for the presence of a red sequence \citep[significant number of galaxies with similar colors redder than the field galaxies, for example][]{GladdersYee2000}. 
However, the formation and the evolution of clusters of galaxies with a developed red sequence is still under investigation. Relatively recent results at low redshift ($z\le$ 0.5) have revealed
that the influence of the potential well of the cluster extends much farther than previously thought, up to three times the effective radius (a size that contains about 40\% of the total mass of the cluster), % (hereafer R$_e$????????),
implying a larger environmental influence on the galaxies and groups in the vicinity of the cluster \citep[see, e.g.,][]{haines15}. Groups interacting with
clusters have clearly shown the importance of lower mass structures in stripping the gas from galaxies before they fall into the larger potential well
of the cluster, the so-called pre-processing \citep[][]{cortese06}.

Going back in time, evolved galaxy overdensities %and extended X-ray emission 
have been observed up to redshift $z\sim2$ \citep{Kodama2007, willis13, andre14}. %(see Andreon et al., 2014;  Willis et al, 2013)These observations indicate that %have pushed back the assembly era
These %large
structures \citep[more commonly referred to as protoclusters\footnote{This name is a simplification since we do not know for sure the fate of a $protocluster$ and if it will really evolve into a cluster},][]{Overzier2016} assembled at $z\sim4$ %and beyond, with recognizable cluster cores observed at least as far back as $z\simeq4.3$ 
\citep[e.g.,][]{miller18,Lemaux2018}. However, there are also observations of dense structures at $z\sim2$ in which the galaxies are more massive and older than the field galaxies, but do not show a red sequence \citep[e.g.,][]{Steidel2005}. 
%(see Miller et al., 2018). 

Different methods have been used to identify protoclusters at $z>2$, %high redshift (z$\ge$) 2 clusters: 
including the detection of overdensities of Lyman Break Galaxies \citep[LBGs;][]{Steidel1998} and of galaxies detected in surveys covering large areas \citep{Castellano2007, Salimbeni2009, KangIm2015,Franck2016}, the search for overdensities around radio galaxies \citep[][and references therein]{Pentericci2000,Venemans2007} and around submillimeter galaxies \citep{miller18}, the detection of overdensities of Lyman alpha (Ly$\alpha$) emitters \citep[LAEs;][]{Kubo2013,Zheng2016}. %, see Kubo et al. %, or , to Ly$\alpha$ emission surrounding the central galaxy (Prescott et al., 2008). 
In the majority of the cases, the galaxies observed in the most dense regions show some enhanced mass assembly and evolution \citep[e.g.,][]{Hatch2011, Zirm2012,Lemaux2014}, can be distributed in more than one main density peak, %\citep[e.g.,][]{Cucciati2018}, 
and are typically surrounded by starburst and active sources \citep[e.g.,][]{Koyama2013, Shimakawa2015}. There is also evidence to support that actively star-forming galaxies could coexist with more evolved ones in the cores of dense structures at $z\sim2$ \citep[e.g.,][]{Strazzullo2013}. %However, all the aforementioned methods suffer from both incompleteness of the photometric and spectroscopic catalogs, and from the diversity of the samples used for the characterization of the overdensities. 
%, and in most cases the work concentrates on the study of a specific target; despite this, some interesting results have been obtained. All the observed argets show and excess of star-forming galaxies; in some AGN activity and extreme starbursts have been observed;  galaxies in protoclusters

%{\bf{I think it is important to talk about Lya, because the reason of the paper}}
Ly$\alpha$ emission is a powerful tool to detect high-redshift galaxies, because Ly$\alpha$ is the strongest recombination line of neutral hydrogen (HI) and it is produced in star-forming regions. For this reason, many of the protocluster searches are based on Ly$\alpha$-emitting galaxies, especially at the highest redshifts. It is, therefore, important to understand how the properties of the Ly$\alpha$ line are related to the environmental densities, since it is only in this way that we can understand if and how protocluster searches based on Ly$\alpha$ emitters may bias the nature of the structures found.

The escape of Ly$\alpha$ photons out of a galaxy is, however, strongly regulated by the resonant scattering of HI in the interstellar and circumgalactic medium. The presence of a blue peak in addition to the main red peak of the Ly$\alpha$ emission line could be an indication of low HI column densities in the surrounding of the galaxy star-forming regions \citep[][]{Verhamme2017, Guaita2017}. %, Marchi2019}. Therefore, Ly$\alpha$ emission reflects the intrinsic properties of the host galaxy and of its environment. %, allowing a biased identification of overdensities. 
As shown for the protoclusters detected around radio-galaxies at $2<z<3$ \citep{Venemans2007,Shimakawa2015}, LAEs tend to mainly trace the outskirts of the structures, rather than the cores.  

%Lya vs environment
%
Different cosmological simulations have attempted to provide a broad view on the formation and evolution of protoclusters of galaxies \citep[e.g.,][]{Chiang2013,muldy15,contini2016}.
They typically show that the progenitors of the present day clusters extend over very large areas on the sky,
of the order of 10-20 comoving Mpc (cMpc) and are characterized by one or multiple overdense cores, in different evolutionary stages. %, which may or may not merge at lower redshift to form a cluster of galaxies.  
Therefore, any systematic search for these structure needs to encompass very large areas with uniform criteria. %, in order to avoid missing parts of the structure.
\citet{Muldrew2015} studied the size and the structure of protoclusters in the framework of the Millennium simulation \citep{Springel2005}. They found that protoclusters at $z>2$ have a variety of evolutionary stages, independent of the mass they are expected to gather by $z=0$, and can be very extended. The progenitors of $z=0$ clusters with $\sim$10$^{14}$ M$_{\odot}$ (>10$^{15}$ M$_{\odot}$) can have sizes of the order of 20 h$^{-1}$ cMpc (35 h$^{-1}$ cMpc) at $z\sim2$.  
%The protocluster structures can comprise many halos linked by filaments. The authors noticed that selecting galaxies by their star-formation rate biases against low-mass galaxies, while there are more massive and evolved galaxies in the cores of the protoclusters than the field. 

Observationally, \citet{Franck2016} inspected cylindrical volumes of 20 cMpc radii on the sky and redshift depths of $\pm20$cMpc, and looked for associations of at least four galaxies with a galaxy overdensity, $\delta_{gal}$ = (n$_{gal}$-n$_{field}$)/n$_{field}>0.25$. With this method, they compiled the Candidate Cluster and Protocluster Catalog (CCPC). The CCPC contains 216 spectroscopically-confirmed overdensities at $2<z<7$. These overdensities have a median $\delta_{gal}=2.9$, average $z=0$ collapsed mass of the order of 10$^{14}$ M$_{\odot}$, and average dispersion velocity of $\sim650$ km sec$^{-1}$. However, 30\% of the structures in the CCPC have masses larger than 10$^{15}$ M$_{\odot}$ and dispersion velocities as large as 900 km sec$^{-1}$, values which are difficult to explain with simulations \citep{Chiang2013}.
\citet{KangIm2015} identified massive structures of galaxies at $0.6<z<4.5$ as the regions where the projected number density is larger than 3.5$\sigma$ above the average value in circular top-hat filters with 1 physical Mpc (pMpc) diameter. They found that even if the 1 pMpc diameter is efficient in identifying massive structures, the entire structures are generally more extended than that value. They noted that the number density of the identified massive ($>$ 10$^{13}$ M$_{\odot}$) structures at $z>2$ was five times larger than the value expected by simulations. From the observation point of view, this discrepancy could be related to photometric redshift uncertainty and the difficulty in estimating structure masses. Therefore, large spectroscopic samples are needed to improve this issue. % {\bf{I am saying from the observational point of view so I leave like this}}. 
%They noticed that only 20\% of the identified structures can be spatially associated with known AGN/radio galaxies. The number density of the identified massive ($>$ 10$^{13}$ M$_{\odot}$) structures at $z>2$ is 5 times larger than the value expected by simulations. From the observation point of view, this discrepancy could be related to photometric redshift uncertainty and the difficulty in estimated structure masses. Therefore, large spectroscopic samples are needed to improve on this issue. 

These results raise some interesting questions about what a protocluster is exactly and how a cluster might evolve, how the cluster galaxies transform from very active star forming to passive, whether %Can we identify a {\it transition} protocluster? 
we can identify some evolutionary difference among protoclusters at the same redshift, and %and among them and the surrounding environment? 
whether we can find a unique observational tracer to identify protoclusters in a similar way as the red sequence is used to identify clusters.
%Can we identify a unique observational tracer for protoclusters in a similar way to the red sequence for nearby clusters?

To attempt to answer some of these questions, we need to define a set of objective criteria to identify a protocluster that rely on deep, spectroscopic surveys which cover relatively large areas on the sky.
%{\bf{or now I leave this here}}
%\citet{Muldrew2015} studied the size and the structure of protoclusters in the framework of the Millennium simulation \citep{Springel2005}. They found that protoclusters at $z>2$ have a variety of evolutionary stages, independently of the mass they will gathered up to $z=0$, and can be very extended. The progenitors of $z=0$ cluster with 10$^{14}$ M$_{\odot}$ (>10$^{15}$ M$_{\odot}$) can have sizes of the order of 20 h$^{-1}$ cMpc ($\sim$35 h$^{-1}$ cMpc) at $z\sim2$.  
%The protocluster structures can comprise many halos linked by filaments. The authors noticed that selecting galaxies by their star-formation rate biases against low-mass galaxies, while there are more massive and evolved galaxies in the cores of the protoclusters than the field. Therefore, large-scale observations of overdensities of comprehensive sets of galaxies are needed to trace the whole structures.

%, such as CANDELS \citep[][]{Koekemoer2011, Guo2011, Galametz2013}, and spectroscopic redshifts to confirm the protocluster candidates. 
Large-area spectroscopic surveys have already been very useful to characterize $z\sim2$ structures \citep[e.g.,][]{Lemaux2014,Cucciati2018} and we wish to push this further using as many as possible spectroscopic redhifts to build reliable overdensity catalogs.
We adopt the dataset from VANDELS \citep{McLure2018, Pentericci2018}, 
which is a deep VIMOS survey of the {\it{Chandra}} Deep Field-South (CDFS) and UKIDSS Ultra Deep Survey (UDS) fields in 
CANDELS \citep{Grogin2011,Koekemoer2011}, to systematically search for 
candidate protoclusters.
%We adopt the dataset from VANDELS \citep[a deep VIMOS survey of the CANDELS CDFS and UDS fields; ][]{McLure2018, Pentericci2018}, which is a deep VIMOS survey of the CDFS ({\it{Chandra}} Deep Field-South) and UDS (UKIDSS Ultra Deep Survey) fields in CANDELS \citep[Cosmic Assembly Near-infrared Deep Extragalactic Legacy Survey][]{Grogin2011, Koekemoer2011}, to systematically search for candidate protoclusters. %They are identified as overdensities of galaxies on the sky with a minumum number (3-5) of confirmed spectroscopic members in the central part of the overdensity, using the code by REFS. Following this, 
For the search, we use the method described by \citet{Trevese2007, Castellano2007, Salimbeni2009, Pentericci2013}. %, considering the results from the Millenium simulation by \citet{Chiang2013} as a starting point to define the search parameters (Sect. \ref{method}), and afterwards we check the robustness of the search against the GAEA models by \citet{Hirschmann2016}. %On the output of the search we further apply a spectroscopic membership criteria, as a function of the redshift, see for more details Sect. 3
Then we aim to understand how Ly$\alpha$-emitting galaxies are distributed in and outside our candidate protoclusters, to
see if they can be used as unique identifiers of high-redshift overdensities.

The paper is organized as follows. In Sect. \ref{vandels}, we summarize the dataset used in this work. In Sect. \ref{method}, we describe the method adopted to define local
densities in the VANDELS fields. In Sect. \ref{mock}, %we perform some tests with mock catalogs. 
we test the robustness of the method and the properties of our identified overdensities using mock galaxy catalogs.
In Sect. \ref{structures}, we present the properties of the most dense structures identified with the method outlined in
Sect. \ref{method}. In Sect. \ref{LAEs}, we present the sample of Ly$\alpha$-emitting galaxies and study their properties and location with respect to the environment. % used to \ref{discussion}, we discuss any general property of the detected overdensities. 
In Sect. \ref{summary}, we summarize our work and provide some considerations on the way Ly$\alpha$-emitting galaxies could trace the environment properties. In the appendices, we show the figures relative to the overdensity space distribution and to the physical and morphological properties of the members of the detected overdensities. Throughout the paper, we use AB magnitudes and we adopt a standard cosmology (H$_0$=70, $\Omega_{0}=0.3$).
%The paper is organized as follows. In Sec. \ref{vandels}, we summarize VANDELS survey. In Sec. \ref{method}, we describe the method adopted to define local densities in the VANDELS fields. In Sec. \ref{structures}, we present the properties of the most dense structures identified with the method outlined in Sec. \ref{method}. In Sec. \ref{summary}, we summarize our work. Throughout the paper, we use AB magnitudes and and we adopt a standard cosmology (H0=70, $\Omega_{0}=0.3$).
 
%--------------------------------------------------------------------
\section{Dataset}
\label{vandels}

The data used for this work are all part of the VANDELS survey \citep[][]{McLure2018,Pentericci2018}. %, either for the unique spectroscopy and for the photometry. 
VANDELS (a deep VIMOS survey of the CDFS and UDS fields) is an ESO public spectroscopic survey. It targets the CDFS and the UDS fields, over a total area of about 0.2 square degree. These fields have the {\it{HST}} multi-wavelength coverage from the CANDELS treasury survey \citep{Grogin2011, Koekemoer2011} 
in their central parts, as well as a wealth of ancillary data, including near-IR and far-IR wavelengths. % including ultra-deep IRAC photometry.

For the CANDELS regions, we consider the photometric redshift solutions by the CANDELS team \citep{Santini2015}. For the area outside CANDELS, new photometric redshifts are generated by the VANDELS team. 
%{\bf{The parameters of the spectral energy distribution (SED) templates used for determining the photometric redshifts are the ones also adopted for estimating the galaxy physical parameters (see Sect. \ref{specVAN}).}} 
As described in detail in \citet{McLure2018}, the photometric redshifts were estimated by 14 different members of the VANDELS team by using a variety of different  publicly available codes on state of the art multiwavelength catalogs. The codes include a wide variety of different spectral energy distribution (SED) templates, star-formation histories, metallicities, and emission-line prescriptions. The adopted photometric redshift for each galaxy is the median value of the 14 estimates. This median was used in a final run of SED fitting carried out to derive physical parameters and was also used for the selection of the spectroscopic targets \citep[passive galaxies, bright star-forming galaxies, SF\_ $2.4<z<5.5$, and Lyman break galaxies, LBG\_$3.0<z<5.5$, following the definition of the VANDELS galaxy populations in][]{Pentericci2018}. 

The final run of SED fitting was performed using \citet{Bruzual:2003} templates with solar metallicity, no nebular emission, exponentially-declining star-formation histories, \citet{Calzetti2000} dust attenuation law, and \citet{Madau:1995} prescription for the intergalactic-medium absorption \citep[the details are presented in the Section 4.4 of][]{McLure2018}. This parameter set allows us to recover the total star-formation rate of main-sequence galaxies and provides stellar-mass values in good agreement with those derived for the CANDELS CDFS and UDS photometric catalogs by \citet{Santini2015}.
%As described in \citet{McLure2018}, the physical parameters of all the galaxies in the VANDELS database are derived through SED fitting on state of the art multiwavelength catalog (see section). 
It has been shown \citep[e.g.,][]{Santini2015} that stellar mass is almost independent of model assumptions, while sSFR and Av are dependent, for instance, on the choice of the star-formation history assumed. Given that we have spectroscopic redshifts and state of the art multiwavelength photometry, the typical error on the rest-frame colors is at the $\pm$0.2 mags level and the errors on the stellar-mass measurements are typically at the level of $\pm$0.2 dex. 
The typical errors on the Av and sSFR determination are about $\pm$0.3 mags and $\pm$0.4 dex, respectively (see the upcoming work on SED fitting parameters). 
%The sSFR parameter is also affected by an enforced upper limit due to the limit on how young the stellar populations can be.
%
%The typical errors on the Av determination is about $\pm$0.3 mags. 
%
%The physical property with the largest uncertainty is the sSFR, which carries an uncertainty of about 0.4 dex (see the upcoming paper of SED fitting parameters). This is also affected by an enforced upper limit due to the limit on how young the stellar populations can be.

The galaxies selected for VANDELS observations have $i_{sel}<=27.5$  %$H_{sel}<=27$, $i_{sel}<=27.5$  %$H_{sel}$ and 
when they are located in the CANDELS areas and $1<z_{phot}<7$, with $i_{sel}$ being the magnitudes in the I814 {\it{HST}} filter. These limiting magnitudes are consistent with the 50\% completeness limits of the CANDELS catalogs %, which depend on the depth of the CANDELS images 
%In fact, the 50\% completeness limit is 26.6 in the deep area of the CDFS H160 image and is 27.05 in the UDS H160 image
 \citep{Guo2013,Galametz2013}. %For comparison the 5$\sigma$ detection limits were estimated to be 28.16 and 27.45 in the same images. 
About 90\% of the galaxies in the CANDELS areas are brighter than $i_{sel}<=27.5$.
In the extended areas of both fields, galaxies were selected based on $i_{sel}$ and photometric redshift as explained in \citet{McLure2018}. %$H_{sel}<=25$ at $1<z_{phot}<7$ and also $z_{sel}<=26$, $i_{sel}<=27$ at $3<z_{phot}<7$ in UDS;  $J_{sel}<=24.7$ at $1<z_{phot}<7$ and also $i_{sel}<=26.1$ at $3<z_{phot}<7$ in CDFS. 
About 95\% of the galaxies in the extended areas are brighter than $i_{sel}<=26.1$ independent of their redshift. 
Therefore, for our analysis, we consider all the sources with $i_{sel}<=27.5$ in the CANDELS areas and $i_{sel}<=26.1$ in the extended areas. 

As stated in \citet[][their Fig. 6 and Table 5]{Pentericci2018}, the photometric redshifts in VANDELS have a catastrophic outlier rate of about 2.0\%. The uncertainty on the photometric redshifts slightly depends on the source magnitude while the dependence on galaxy type and position is negligible.
%For the sources for which we have both photometric and spectroscopic redshifts (either from VANDELS or from the literature), we estimate the quantity q=$(z_{phot}-z_{spec})/(1+z_{spec})$, that is therefore related to the uncertainty on the photometric redshifts. In CDFS, we find that mean(q) $\pm$ stdev(q) = $-0.05\pm0.16$ ($-0.004\pm0.025$ excluding the outliers) for the galaxies at $z_{spec}>2$. In UDS, we find that mean(q) $\pm$ stdev(q) = $0.01\pm0.09$ ($-0.008\pm0.030$ excluding outliers) for sources at $z_{spec}<=4$.
%{\bf{Doing comparisons between zphot and zspec (Pentericci2018), they have found that the uncertainty on zphot depends on the magnitude cut. }}
The VANDELS observations were organized in three seasons, from 2015 to 2018. %This work encompasses only the data from the first two seasons.  
The second public data release is already available on the ESO phase 3 website \footnote{http://archive.eso.org/cms/eso-archive-news/new-data-release-of-spectra-and-catalogue-from-the-vandels-eso-public-spectroscopic-survey.html} \citep{Pentericci2018c}.

\subsection{Spectroscopic data from VANDELS}
\label{specVAN}

For this work, we consider VANDELS redshifts and spectra from season 1 and 2 \citep{Pentericci2018c,Pentericci2018}. The spectra were all obtained with the red medium-resolution grism of VIMOS, that covers a wavelength range of 4800-10000 {\AA}, with an average resolution of 580. The spectroscopic redshifts were estimated using the Pandora software \citep{Garilli2010} as described in \citet{Pentericci2018}. We choose the redshifts with quality flag 3 and 4 (95-100\% probability to be correct) and we prioritize them over other measurements of the same sources from the literature (see Sect. \ref{specLIT}). %, when there are duplicated spectroscopic redshifts in the literature.
%Details on VANDELS survey in data release paper (in prep). Status of observations. Data flags

Among the galaxies with spectroscopic redshifts from VANDELS, we define Ly$\alpha$-emitting galaxies (LAEVs) as the sources with EW(Ly$\alpha)>0$ {\AA} in the VANDELS spectra and $z>3$ due to the VIMOS grism wavelength coverage. 
Since SED fitting on multiwavelength photometry is the basis of the selection of the VANDELS spectroscopic targets, the LAEVs are Lyman break galaxies by selection and we expect they may present differences in the physical parameters with respect to typical narrow-band selected  Ly$\alpha$ emitters (see Sect. \ref{propLAEs}).
We choose the LAEVs only among the sources with VANDELS spectra to avoid space and redshift inhomogeneities, that could come from different survey coverage, and also to avoid survey depth inhomogeneities that could prevent the detection of the Ly$\alpha$ emission line in the galaxy spectra.

The physical parameters of all the sources in the VANDELS spectroscopic redshift catalog were obtained by using SED fitting with redshifts fixed to their spectroscopic (instead of photometric) values and SED templates as described above. This is also the case for Ly$\alpha$-emitting galaxies. Semi-analytical models \citep[e.g., ][]{Gurung-Lopez2019b} can reproduce observational properties of Ly$\alpha$-emitting galaxies assuming sub-solar metallicities. Therefore, in addition to the run at solar metallicity, a second run of SED fitting was also performed with sub-solar metallicities (m42 choice of the Bruzual and Charlot models) and we see that stellar mass, specific star-formation rate, rest-frame magnitudes, and dust reddening 
%do not significantly change with respect to the run with solar metallicity 
are consistent with the parameters obtained in the run with solar metallicities within the parameter uncertainties \citep[see the upcoming work on SED fitting parameters and][]{Carnall2019}. 
%n addition to this, the fitting was performed using \citet{Bruzual:2003} templates with solar metallicity, no nebular emission, exponentially-declining star-formation histories, \citet{Calzetti2000} dust attenuation law, and \citet{Madau:1995} prescription for the intergalactic-medium absorption \citep[the details are described in][]{McLure2018}.
%This SED fitting was performed using Bruzual \& Charlot (2003) templates with solar metallic- ity and no nebular emission. 
%Exponentially-declining star-formation histories were employed, with \tau in the range $0.3\leq \tau \leq20$ Gyr, and ages were constrained to lie between 50 Myr and the age of the Universe at the redshift of interest. Dust attenuation was described using the Calzetti et al. (2000) starburst attenuation law, with AV in the range $0.0\leq AV\leq2.5$, and IGM absorption was accounted for using the Madau (1995) prescription. These parameters were adopted following the results of Wuyts et al. (2011), who showed that this parameter set does a reasonable job of recovering the total star-formation rate of main-sequence galaxies, provided that they are not heavily obscured. We also note that this SED parameter set is very similar to that adopted by the 3D-HST survey team (Momcheva et al. 2016) and delivers stellar-mass estimates in good agreement with those derived for the CANDELS CDFS and UDS photometric catalogues by Santini et al. (2015). During the SED-fitting process the redshift was fixed at the median value derived from the multiple photometric-redshift runs described in Section 4.2.

\subsection{Spectroscopy from the literature}
\label{specLIT}

We augment the VANDELS redshifts with previous spectroscopy in the literature. 
In the CDFS, we consider the compilation made by Nimish Hathi (private communication), which includes
%The spectroscopic redshifts in CDFS include 
 the redshift surveys published up to November 2017\footnote{\citet{Grazian2006,Vanzella2008,Wuyts2008, Vanzella2009,Rhoads2009, Straughn2009,Wuyts2009,Balestra2010, Cooper2012, Kurk2013,LeFevre2013, Trump2013,LeFevre2015, Kriek2015, Morris2015, Momcheva2016}}. 
In the UDS, we consider the spectroscopic compilation from Maltby et al. (in prep). %{\bf{Hartley et al. or Almaini et al.; to be decided}}. 
It includes redshifts from the UDSz ESO large program\footnote{http://www.nottingham.ac.uk/astronomy/UDS/UDSz/} and the redshift lists from \citet{CurtisLake2012,Maltby2016}, from 3DHST\footnote{http://3dhst.research.yale.edu/Data.php}, VIPERS\footnote{http://vipers.inaf.it}, and archival redshifts\footnote{http://www.nottingham.ac.uk/astronomy/UDS/data/data.html}.

From the two compilations, we consider the highest-flag redshifts, corresponding to the quality flag 3 and 4 in VANDELS.
The properties of the sources in these catalogs are quite different, since the surveys were designed with different targets. Therefore, the whole data compilation is relatively inhomogeneous, but offer the currently most complete list of spectroscopic redshifts, and, for the purpose of detecting secure dense structures at $z>2$, it is important to rely on as many spectroscopic redshifts as possible. 
However, since the spectroscopic redshifts represent a small fraction of the total sample of sources considered in this analysis (i.e., mostly photometric), we do not expect that the lack of spectroscopic homogeneity would affect our results in a relevant way. %For sources with several redshift measurements, we prioritize those with the highest quality flag.
%There are 60 sources at $0.9<z<5.8$ observed by VANDELS and with previous redshift measurements from the literature. The quantity (z$_{VANDELS}$-z$_{literature}$)/(1+z) provides an estimation of the agreement between surveys. Its mean and standard deviation are 0.008 and 0.06, which become 0.001 and 0.002 if we exclude the outliers. {\bf{The outliers can be galaxies with measurements from field spectroscopic surveys, such as VVDS \ref{LeFevre2013} and 3D-HST \ref{Kriek2015}.}}

%In this field, there are 87 sources at $0.6<z<4.8$ observed by VANDELS and with previous redshift measurements from the literature. Among those sources, the mean and the standard deviation of (z$_{VANDELS}$-z$_{literature}$)/(1+z) are 0.02 and 0.17, which become 0.003 and 0.002 if we exclude the outliers. {\bf{The outliers can be galaxies with measurements from field spectroscopic surveys, such as 3D-HST \ref{Kriek2015}.}}
The agreement between the spectroscopic redshifts obtained within VANDELS and previous measurements from the literature is very good \citep[see ][]{Pentericci2018}. Sources with spectroscopic redshifts from the literature are usually brighter than the magnitude limits discussed in Sect. \ref{vandels}. In the VANDELS database, instead, we have spectra for sources as faint as $i_{sel}\sim27.5$.

%--------------------------------------------------------------------
\section{Algorithm for the estimation of local densities}
\label{method}

We estimate local densities by using the three-dimensional algorithm `3dv4'.  For a detailed description of the method, we refer the reader to %described in detail in %Trevese at al. (2007), Castellano et al. (2007), and Salimbeni et al. (2008). 
\citet{Trevese2007, Castellano2007, Salimbeni2009, Pentericci2013}. The algorithm was widely tested up to $z=2.5$ \citep{Salimbeni2009}, but it is used in this work at $2<z<4$. 

%For a detailed description of the method, we refer the reader to the cited papers; however w
We briefly summarize it here for clarity.
The algorithm receives an input catalog and a few configuration parameters. Some of the parameters were optimized in the works listed above and we confirm them to be valid when we use the code at $z>2.5$ (see also Sect.\ref{mock}).

\subsection{Input catalogs}
\label{catalog}

The input catalog is composed of the coordinates for each target galaxy and of either photometric ($z_{phot}$) %(because they are the majority) 
or spectroscopic ($z_{spec}$) %(to increase the redshift accuracy) 
redshifts. For each galaxy for which we have more than one spectroscopic redshift available (from VANDELS and from
the literature), we keep the spectroscopic redshift coming from the VANDELS survey and discard the others, so that at the end there are no duplicated spectroscopic redshifts in the input catalog.
% (Sect.\ref{catalog}). 

\begin{figure*}
 \centering
\includegraphics[width=8cm]{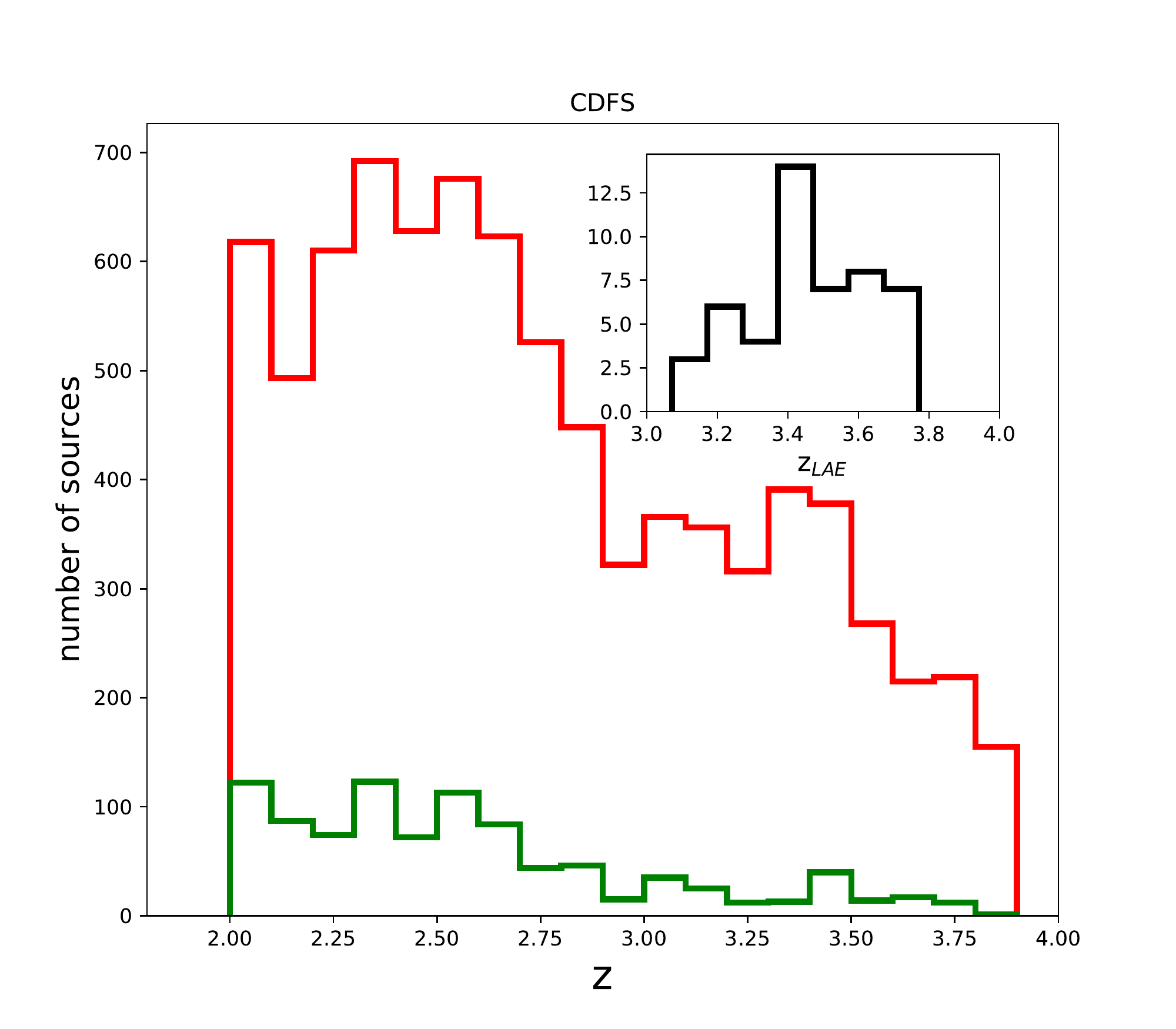}
\includegraphics[width=8cm]{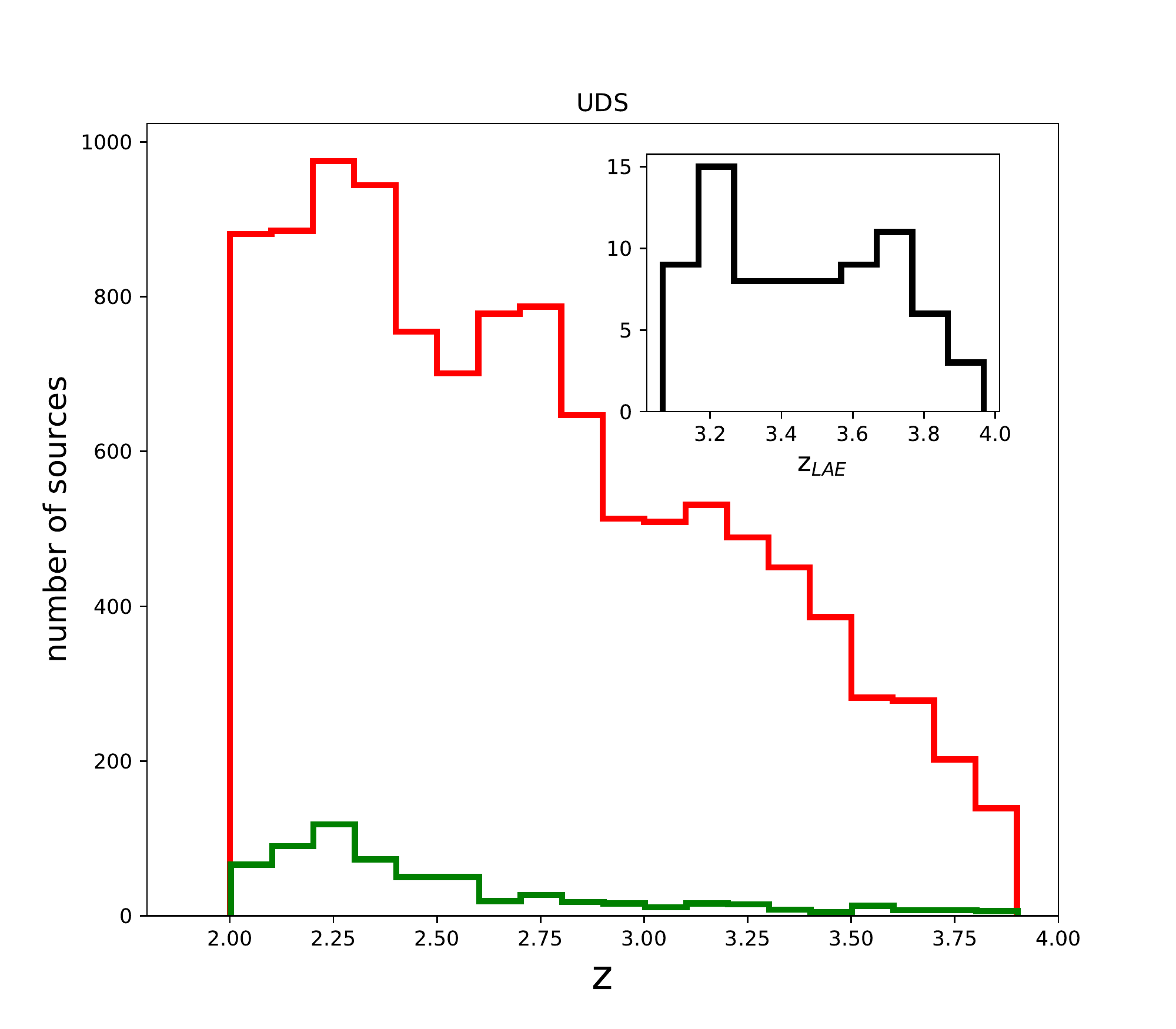}
\caption{Redshift distribution of the sources in the input catalog of the CDFS ($left~panel$) and of the UDS ($right ~panel$). The red histograms represent the sources with photometric and spectroscopic redshifts, the green histograms the sources with spectroscopic redshifts from the literature. The inserts show the redshift distributions of the Ly$\alpha$-emitting galaxies selected from the VANDELS database and used in this work.}
\label{inputcathist}%
\end{figure*}

\begin{figure*}
 \centering
\includegraphics[width=18cm]{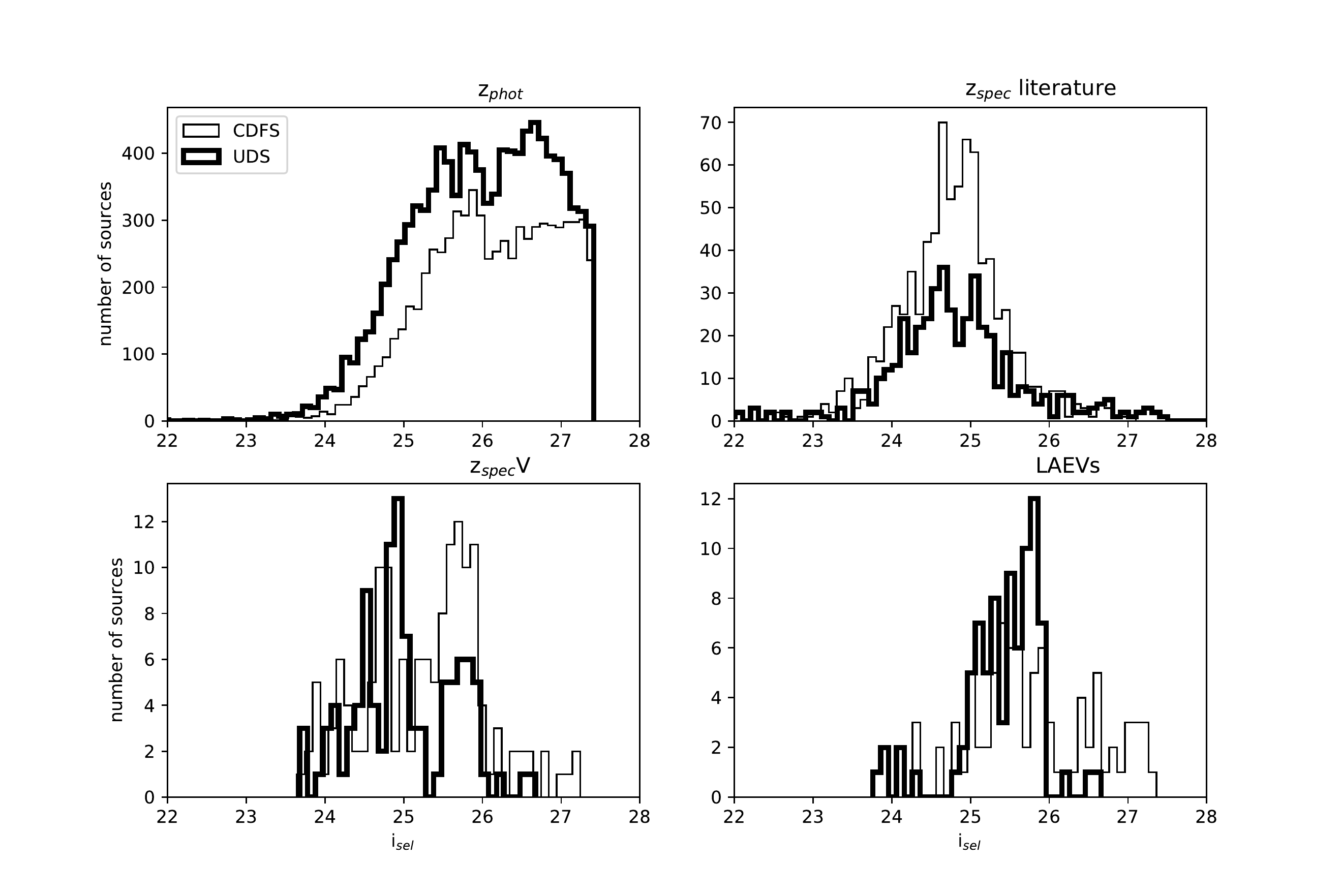}
\caption{Magnitude distribution of sources with photometric ($upper ~left$), spectroscopic redshift from the literature ($upper ~right$), from VANDELS ($lower ~left$), and of the Ly$\alpha$-emitting galaxies selected from the VANDELS database ($lower ~right$) in the CDFS (weak line) and in the UDS (strong line).}
\label{maghist}%
\end{figure*}

In Fig. \ref{inputcathist}, we show the redshift distribution of the sources in the input catalog. Between $2<z<4$, we have 1103 (733) $z_{spec}$ and 7342 (10581) $z_{phot}$ sources in the CDFS (UDS). Among the spectroscopic redshifts, we include 151 and 103 $z_{spec}$ from VANDELS in the CDFS and in the UDS, respecively.
%candels 2<z<4\\
%zphot CDFS 5681\\
%zphot UDS 8227\\
%
%extended 2<z<4\\
%zphot CDFS 1661\\
%zphot UDS 2353\\
%
%zspec 2<z<4\\
%zspec CDFS 1103\\
%zspec UDS 733\\
In Fig. \ref{maghist}, we show the magnitude distribution of the sources of the input catalog with photometric redshifts, spectroscopic redshifts from the literature, from VANDELS, and of the LAEVs in the CDFS and in the UDS.

\subsection{Running the algorithm}
\label{algo}

Once the input catalog is defined, the algorithm works as follows. The entire survey volume is first divided into a grid of cells. Each grid cell is characterized by a position in the right ascension (ra), declination (dec), and redshift (z) space, according to its location in the survey volume, and by the same size $\Delta$ra, $\Delta$dec, and $\Delta$z in the three directions. 
The number of galaxies from the input catalog in each cell depends on the density of galaxies in our field and on the cell dimension. Some cells may be empty, some can contain more than one galaxy. % {\bf{[we are talking only about the cells in the initial grid]}}. 

To measure local densities, the three dimensions of each cell of the initial grid are increased in size by steps of $\Delta$ra, $\Delta$dec, and $\Delta$z, 
%one cell at a time 
 and the code counts the number of galaxies within the increased volume of the cell.
%To measure the local density, associated to each cell of the initial regular grid, % and to all the galaxies contained in it,  each cell is then increased in size by steps of one initial cell along the three dimensions and the code counts the number of galaxies within the increased-cell volume. 
The density associated to that cell is then defined as $\rho_{N}=$N/V$_{N}$, where V$_{N}$ is the comoving volume which includes the N nearest neighbours. 
%When a certain number $N$ %[{\bf{this is descriptive, the value N as the other chosen parameters are defined in the next paragraph}}] 
%of galaxies is reached, the comoving volume (V$_{N}=4/3 \pi D_{N}^3$, where D$_{N}$ is the distance to the Nth nearest neighbour) containing the $N$ galaxies is estimated and a comoving local density is calculated as $\rho_{N}=$N/V$_{N}$.
The value $\rho_{N}$ is also the density associated to the galaxies contained in that cell in the initial regular grid.
%In the final step, 
Then the code studies the $\rho_{N}$ values in the field, detects and extracts overdense structures in the 3D space. This procedure is performed with a SExtractor \citep{bertin1996} approach. The code measures the mean and the standard deviation of the local densities in each redshift bin, applying a 2$\sigma$ clipping in an iterative way. %{\bf{there is afile which tells us how many times the clipping was performed in each redshift slice}} 
%Finally, 
It extracts as overdensities the regions with densities larger than a certain THRESHOLD (defined in standard deviations above the mean local density), comprising a minimum number of cells from the initial regular grid (MIN\_CELL), and a minimum number of galaxies (MIN$\_{obj}$). % ({\bf{in a structure = 10}}). 

We use two sets of configuration parameters. The first set is related to the calculation of local densities; the second one is related to the detection and extraction of overdensities. 
We start with the parameters described in the papers cited at the beginning of Sect. 2. %, further refined by the use of more complete input catalogs \citep{McLure2018}, CANDELS images, and photometry.
We define the first set as follows.
The volume of a cell in the starting grid is defined as $\Delta$ra$\times \Delta$dec$\times \Delta$z. 
We adopt $\Delta$A = $\Delta$ra = $\Delta$dec = 3 arcsec and $\Delta$z = 0.02. These values correspond to 75 cKpc (spatial direction) and 30 cMpc (redshift direction) at $z=2$ considering our adopted cosmology. The extension of the cells depends on the positional accuracy and is characterized by a parallelepiped volume, elongated towards the redshift direction. 
%$\Delta$A would depend on the number density of the sources in the input catalog. Our choice of $\Delta$A allows running feasibility of the algorithm[{\bf{I'd just say this}}]. %depends on the number density of objects in our field and allows running feasibility of the algorithm.
%The choice of is a trade off between image quality and . 
The choice of $\Delta$z is related to the photometric redshift accuracy, 0.02 $\times$ (1+z), which corresponds to $\sim0.1$ at $z>2$. % (Sect.\ref{catalog}). 
%
%are chosen to be equal to 3depends on the astrometry of the input catalog and $\Delta$z depends on the redshift accuracy. 
%
%The astrometry of CANDELS+VANDELS catalogs is based on HST and high-resolution ground based images (ref). %Therefore, we adopt $\Delta$ra and $\Delta$dec equal to 3 arcsec, which correspond to        comoving
%Given the fact we are combining zphot and zspec, we adopt a $\Delta$z=0.02, which correspond to        comoving
%This way the cell assumes the shape of a parallelepiped, elongate towards the redshift direction. 
We fix the maximum cell length in the spatial directions to be 15 arcsec (380 cKpc at $z\sim2$) and to be $2\times \Delta$z$\times$ (1+z) %(related to a 2$\sigma$ redshift uncertainty) 
in the redshift direction to avoid infinitely elongated cells with unphysically-low local densities \citep[details can be found in][]{Salimbeni2009}. %This maximum redshift dimension would correspond to about 0.1 at $z>2$. 
%The choice of adopting comoving quantities relies in the fact that we do not want our results to be redshift dependent. 
Also, we fix N=10 and we verify that we do not obtain significant differences when we vary N from 10 to 20. In the case that ten sources are not counted after reaching the maximum size of a cell, the density is calculated from the number of sources contained in the cell of maximum size.   %Smallest is the volume containing 15 neighbours, highest is the comoving density in that region of the ra,dec,z space.
%{\bf{mostrare qualche esempio a Gianni di CC di density map di sorgenti???}}
We define MIN\_CELL = 10 as previously optimized in \citet{Trevese2007}, THRESHOLD $=6$, and MIN$\_{obj}=5$ or 10 (at $3<z<4$ or $2<z<3$, respectively). These values are tested on our data and with mock galaxies (see Sec. \ref{mock}), and they allow us to recover the structures previously identified at $z \simeq2.3$ by \citet{Salimbeni2009}. % and to identify as parts of the identified structures the density peaks observed in the density map at $z > 2$. % --> why only z>2.5 and not also z<2.5 where peaks are seen in both fields?
%These values allow us to recover the structures previously identified at $z\sim2.3$ , to identify as parts of structures the density peaks found in the density map at $z>2.5$, and have been tested with mocks . 
The choice of a lower MIN$\_{obj}$ value at $z>3$ is justified by the fact that the number of sources in our input catalog decreases by a factor of 2 going from $z\sim2.5$ to $z\sim3.5$ (Fig. \ref{inputcathist}).
%We run the code from $z=1.5$ up to $z=4.5$ in bins of one redshift (i.e., bins of $z =1.5-2.5$, $2.0-3.0$, $2.5-3.5$, $3.0-4.0$, and $3.5-4.5$). {\bf{The choice of redshift bin as $\Delta z_{bin}=1$ is made to ensure feasibility in running the code.
%These bins, overlapping by 0.5 in redshift, allow the detection of overdensities at $2<z<4$ not affected by border effects. }} 

\subsection{Identification of overdensities}
\label{clusters}

\begin{figure}
 \centering
 \includegraphics[width=8cm]{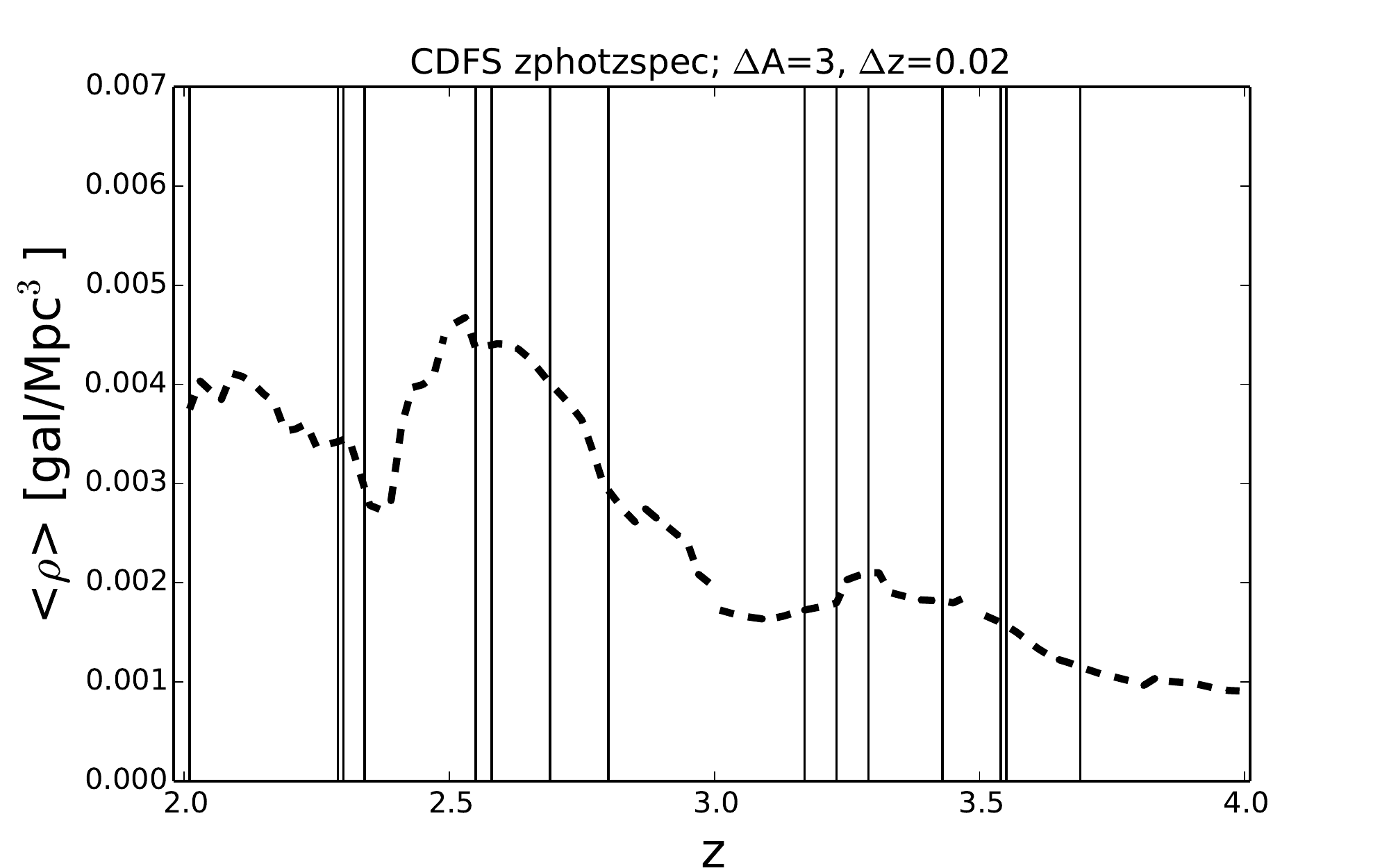} 
\includegraphics[width=8cm]{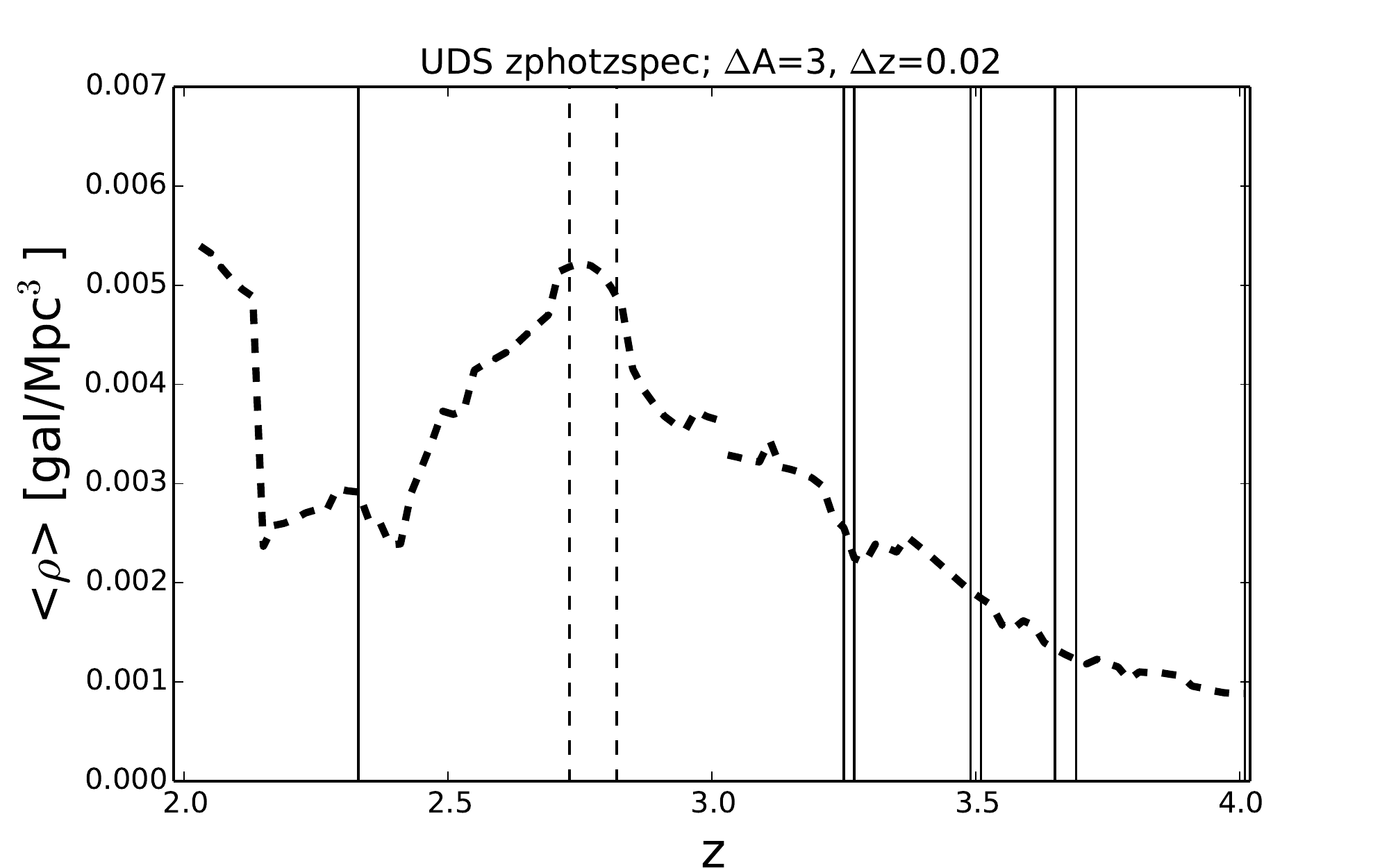}
\caption{ %. At z=2.7 in UDS peak????}}
Mean local density %(number of galaxies per cubic comoving Mpc) 
versus redshift for the CDFS ($upper ~panel$) and for the UDS ($lower ~panel$) shown as dashed solid curves. The mean corresponds to the average of the local density values in each bins of $\Delta z=0.02$, after the application of sigma clipping, as explained in the text. 
Vertical solid lines represent the redshifts of the detected overdensities. In the lower panel, vertical dashed lines correspond to two slightly lower-threshold overdensities at $z\sim2.8$ (THRESHOLD=4 instead of 6).
}
\label{rhoz}%
\end{figure}

In Fig. \ref{rhoz}, we show the average of the local densities associated to every galaxy of the input catalog ($<\rho>$) as a function of redshift. 
The density, averaged over the entire field, is expressed as the number of galaxies per Mpc$^3$ and decreases as the richness of sources in the input catalog declines. We are studying small-size areas, so we expect to see overdensities diffuse all over the field. The structures embedded in these diffuse overdensities are expected to have redshift peaks corresponding to the ones in the $<\rho>$ vs redshift function \citep[e.g.,][]{Trevese2007,Salimbeni2009}. The code outputs the list of extracted overdensities together with the position in right ascension, declination, and the redshift of their highest signal-to-noise density peak.

%CDFScont_zbin_zphotzspecNONE.py

%In the final step, the algorithm detects/extracts overdense structures with a SExtractor \citep{bertin1996} approach. It measures the mean and the standard deviation of the local densities in each redshift bin, applying a 2$\sigma$ clipping in an iterative way. %{\bf{there is afile which tells us how many times the clipping was performed in each redshift slice}} 
%Finally, it extracts as structures the regions with density larger than a certain number (THRESHOLD) of standard deviations above the mean local density, at least a minimum number of cells from the initial regular grid (MIN\_CELL), and a minimum number of galaxies (MIN$\_{obj}$). % ({\bf{in a structure = 10}}). 

Among the overdensities in the output list, we select a subsample of them characterized by dense cores. These are expected to be the most reliable structures (see also the test with the mock catalogs in Sect. \ref{mock}). We consider the local densities associated to all the galaxies of the input catalog, we focus on the sources in bins of 0.1 redshifts (for instance at $1.95\leq z\leq2.05$ or $2.05\leq z\leq2.15$, and so on), and we calculate the average and the standard deviation of their associated densities ($\rho_{m01}$ and $\sigma_{m01}$). 
We identify as dense cores any region inside an overdensity with at least a given number of galaxies with associated densities larger than $\rho_{m01} +2\times \sigma_{m01}$, and contained in a certain circular projected area, where the number of galaxies is five (three) at $z\sim2.5$ ($z\sim3.5$) according to the number of sources in the input catalog (Fig. \ref{inputcathist}) and %$\rho_{m01}$ is
%the spatial mean density of the cell and 
the radius of the circular area varies between 3.5-7 cMpc at $z\sim2$ and 4.5-8 cMpc at $z\sim3$. 
%We find that a region above a THRESHOLD of 6 would correspond to a core of galaxies with associated density larger than $\rho_{m01} + 2 \times \sigma_{m01}$. This is related to the fact that the spatial mean density of the cells is lower than the mean of the densities associated only to galaxies.
%
%The radii of the circular areas are chosen to be 3.5-7 cMpc at $z\sim2$ and 4.5-8 cMpc at $z\sim3$. 
The choice of these values of radii is supported by simulations. %{\bf{but that depends on the completeness and nature of tracers used: i.e. depends if you use all galaxies down to the faintest/least massive ones, and in principle also might depend on the subsample (selection function) observed.}}. 
\citet{Franck2016} % (and references therein) 
developed a method to identify the most massive protoclusters at $z>2$ and investigated which volumes must be inspected to find them. While the most massive clusters have radii of the order of a few Mpc at $z=0$, the protoclusters that would evolve into them are much more extended. %, with sizes larger than 50 cMpc. 
By studying the Millennium simulation \citep{Springel2005}, \citet{Chiang2013} found that the effective radius of the progenitors of local clusters with masses of the order of $2 \times 10^{14}$ M$_{\odot}$ is 3.5 (4.5) cMpc on average at $z\sim2$ ($z\sim3$), while more massive clusters with masses larger than $10^{15}$ M$_{\odot}$ can have effective radii up to 7 (8) cMpc at the same redshifts. 
With the aforementioned parameter choices, we expect to be able to identify the progenitors of $z=0$ clusters with the lowest, but also the highest expected effective sizes.
%{\bf{REMOVE?}}
%We are interested in studying protoclusters of a wide range of masses, because overdensities of Ly$\alpha$ emitters have been discovered even in massive clusters \citep[$2-9 \times 10^{14}$ M$_{\odot}$ at $z=2-5$][]{Venemans2007}, likely to be progenitors of massive clusters of galaxies at $z=0$ and the goal of our project is studying the distribution of Ly$\alpha$ emission versus environment 
%
%In Tab.\ref{tab:TabCDFSUDSstructures}), we present the 22 structures detected in CDFS and UDS at $2<z<4$. 

We identify 22 overdensities at $2<z<4$ (Sect. 5); %(Tab. \ref{tab:TabCDFSUDSstructures}); 
we recover the structures found by \citet{Salimbeni2009} at $z\sim2.3$ with our same algorithm (Fig. \ref{Auno}) and the dense structure detected by \citet{Forrest2017} at $z\sim3.5$ (Fig. \ref{Aquattro}). Our final list of identified overdensities does not contain some of the dense regions found by \citet{Franck2016} and \citet{KangIm2015}, which, however, are located in the positions of high-density peaks recognized by our algorithm (Fig. \ref{Auno}, \ref{Aquattro}, \ref{Acinque}). %and that are present in the initial list of structures found by our algorithm.

%At $2<z<4$, the total VANDELS volume is about 8.5E+6 cMpc$^3$. Therefore, in the two VANDELS fields we detected 2.6E-6 structures/cMpc$^3$. 
%
%As we will show in Sect. \ref{mock}, our choice of parameters allows us to compile a trustful list of detected structures, characterized by only about 20\% of contamination from fake detections at $z>3$. We will also show that the detected structures could be composed of more than one high density peak.

%--------------------------------------------------------------------
\section{Algorithm performance on simulated galaxies}
\label{mock}

\citet{Trevese2007} tested the reliability of our algorithm in detecting clusters of various types and redshifts. They simulated galaxies of reasonable magnitude and distribution, and clusters of various richness numbers. They showed that a richness-0 cluster \citep{Abell1958} was
detected with acceptable contamination up to $z=1$ at a magnitude limit of 25 %(close to the magnitude corresponding to the 50/% completeness of the input catalog used there) 
and it was still visible at $z=2$ in a survey with magnitude depth of 27. Also, they calculated that it was possible to separate aligned clusters when their difference in redshift was larger than 0.15. By increasing the detection threshold of the algorithm, they showed it was then possible to separate overdensities with initially unseparated multiple peaks. 
Furthermore, \citet{Salimbeni2008} showed that our algorithm allows us to preserve high purity even for the smallest structures at $z=2.5$. The comoving volume, probed by VANDELS is larger than that studied in the previous papers, so we could expect to detect structures even of the highest numbers of richness. Also, the images VANDELS is based on are at least one magnitude deeper than the ones studied in the previous papers. Therefore, we could expect reliable performance of our algorithm at redshift up to 4.
%{\bf{ON GOING ANALYSIS -- SECTION TO BE ORGANIZED}}
%{\bf{QUICK COMMENT: FIRST I COMPILED A LIST OF ALL THE STRUCTURES IN THE MOCK LIGHT CONE AT A SPECIFIC REDSHIFT. THEY HAVE A NUMBER N OF MEMBERS LARGER THAN 1. THEN I COMPARED THE STRUTURES IDENTIFIED WITH MY METHOD AND THAT LIST. THE STRUCTURES I RECOVERED ARE FOR SURE THE ONES WITH MORE THAN 10 MEMBERS A CERTAIN 'LARGE' MASS.}}
%{\bf{THERE ARE A FEW DIFFERENCES WHEN I CONSIDER A CUT AT isel=26.1 OR isel=27.5}}
%{\bf{TABLE WITH ALL THE STRUCTURES AND THE ONES RECOVERED WITH THEIR PROPERTIES (MVIR, MSTELLAR, MGASS, observed-dust mag) for isel<=27.5}}
%{\bf{TABLE WITH ALL THE STRUCTURES AND THE ONES RECOVERED WITH THEIR PROPERTIES (MVIR, MSTELLAR, MGASS, observed-dust mag) for isel<=26.1}}
%{\bf{FIGURE WITH THE RECOVERED STRUCTURES AT 3<z<4 AND FIGURE WITH STRUCTURE SHAPE}}
In this work, to check the performance of the algorithm in detecting overdensities in our current data, % relying mostly on photometric redshifts, 
we make use of mock catalogs of VANDELS observations. 

%As described in the previous section, our code is designed to detect $overdensities$ which are not necessarily either virialized or bound structures, as the structures identified in the light cone of the GAEA model by the friends-of-friends algorithm. We use an input catalog mainly composed of photometric redshifts and the uncertainty on the photometric redshifts is the most significant limitation in the search and detection of structures. We want to verify with mocks how this uncertainty affects the process. %serious is the uncertainty effect.
%Therefore, w
We first use the mock catalogs as an ideal Universe in which all the galaxies have secure redshifts and we identify overdensities. As described in the previous section, our code is designed to detect overdensities which are not necessarily either virialized or bound structures. To perform the identification of the overdensities, we generate an input catalog with the same magnitude cuts used for the real data and we run our code without considering photometric redshift uncertainties. We call this first run the fiducial-run and the identified overdensities are the fiducial-run overdensities. %Since the mocks provide galaxy physical and photometric properties, we study the characteristics of the galaxies inside the detected overdensities. 
Then we apply photometric uncertainties to a given fraction of redshifts in the mock input catalog and compare the properties of the overdensities characterized by photometric-redshift uncertainties with the fiducial-run overdensities.
%test which combination of parameters allows us to better recover the $fiducial$-run overdensities.

 \subsection{Mock of VANDELS observations}

The VANDELS mock catalog is created from the GAlaxy Evolution and Assembly (GAEA) semi-analytic model \citep{Hirschmann2016}, embedded in the dark matter Millennium simulation. %, a model embedded in a dark matter simulation. 
This model represents an evolution of the \citet{DeLuciaBlaizot2007} code. % and contains several important updates. Most notably, the reference model used to build VANDELS-like mock includes a sophisticated chemical enrichment scheme that follows the evolution of individual elements and accounts for mass dependent stellar lifetimes \citet{DeLucia2014}, and an updated stellar feedback scheme that is partly based on results from hydrodynamical simulations \citet{Hirschmann2016}. Also, it has been modified to follow more accurately processes on the scale of the Milky Way satellites as described in \citet{DeLuciaHelmi2008} and \citet{Li2010}. 
%
%It has been shown to reproduce a number of important observational constraints, such as the galaxy stellar mass function up to $z\sim3$, the observed correlation between gas/stellar metallicity and galaxy stellar mass in the local Universe, and the measured evolution of the mass-gaseous metallicity up to $z\sim2$ \citep{DeLuciaHelmi2008, Li2010}.
%%The model assumes that the re-ionisation of the Universe started at $z = 15$ and that the gas cooling is suppressed below $10^{4}$K. 
%The evolution of baryons is traced in four different reservoirs, the stellar component of galaxies, the cold gas in galaxy discs, the hot gas associated with the dark matter haloes, and the ejected gas component. The mass and energy transfer between these different reservoirs is modelled with physically and/or observationally motivated prescriptions for gas cooling, re-ionisation, star formation, stellar feedback and gas recycling, metal evolution, black hole growth, AGN feedback, disk instabilities, and hot gas stripping. 
%
The dark matter merger trees are extracted from the Millennium simulation \citep{Springel2005}. This is a cosmological N-body, dark-matter only simulation that follows the evolution of $2160^3$ particles of mass $8.6\times10^8$ h$^{-1}$ M$_{\odot}$ within a comoving box of 500 h$^{-1}$ Mpc on a side. % and that assumes the following cosmological parameters Omega\_M=0.25, Omega\_L=0.75, h=0.73. 
Dark matter haloes are identified using a standard friends-of-friends algorithm \citep[e.g.,][]{PressDavis1982, Davis1985} with a linking length of 0.2 in units of the mean particle separation. 
The most-massive self-bound subhalo in a friends-of-friends group is its main subhalo, which contains a `central galaxy'. 
%The centre of a friends-of-friends group is its potential minimum and its virial radius is the radius of the largest sphere with that centre and a mean overdensity exceeding 200 times the critical value. The total mass within the virial radius is defined as the virial mass of the group \citep{Guo2011}. 
%The main subhalo and the satellite subhalos compose a bound structure. Bound substructures have a minimum of 20 particles, which corresponds to a halo mass resolution limit of $1.7\times10^{10}$ M$_{\odot}$ h$^{-1}$. Given the resolution of the Millennium dark-matter particles, the model galaxies in GAEA are reasonably well-resolved for log(M$_{*}$/M$_{\odot})>9$.}}
%Type: 0-> central; 1-> Sateliite with Halo; 2-> Satellite without halo

From GAEA outputs, we generate light cones as described in \citet{Zoldan2017}. 
The light-cone galaxies have all the observational properties we consider for the VANDELS galaxies, such as dust-dimmed magnitudes in $U$, $B$, $V$, {\it{HST}} $ACS\_F775W$, $i$, $z$, $J$, {\it{HST}} $WFC3\_F160W$-band filters and observed redshift (including peculiar velocities), in addition to 
%galaxy type (either central or satellite), total stellar mass, cold gas mass, hot gas mass, 
virial mass, virial radius, virial velocity, and star-formation rate directly coming from the model.  %, scale radius of the disk, and mass of metals, directly coming from the model. 

%Fig. \ref{compmockdataRAdec} shows the distribution of the galaxies of CDFS, UDS, and of the mock light cone in right ascension and declination at $2<z<4$.  
%\begin{figure*}[h!]
 %\centering
%\includegraphics[width=20cm]{RAdecdistribution_CDFSUDScandels_mock.pdf}
%\caption{Right ascension versus declination for the $2<z<4$ galaxies in the CANDELS areas of the CDFS ($left$), UDS ($middle$), and mock light cone ($right$). The shade of the hexagons scales with the number of galaxies as we can see from the vertical color bar.}
%\label{compmockdataRAdec}
%\end{figure*}
%In Fig. \ref{compmockdata}, we compare the redshift versus magnitude plane for the galaxies in CDFS, UDS, and for the mock galaxies.  
%\begin{figure*}[h!]
% \centering
%\includegraphics[width=14cm]{iselvsz_CDFSUDScandels_mock.pdf}
%\includegraphics[width=14cm]{iselvsz_CDFSUDSextended_mock.pdf}
%\caption{$First ~ row$: redshift versus $i_{sel}$ magnitude for the CANDELS areas (blue) and for the mock (red). The redshifts include  photometric and spectroscopic redshifts. $Second ~ row$: redshift versus $i_{sel}$ magnitude for the extended areas of CDFS and UDS (blue) and for the mock (red). The mock catalog is limited to $i_{sel}\leq26.1$ like the photometric catalog in the extended areas.  In the $right ~panels$, the observed redshifts of a random 85\% of the mock galaxies are perturbed, as described in the text, to mimic photometric redshifts.}
%\label{compmockdata}
%\end{figure*}
To mimic the observations in the CANDELS (extended) area, we apply the magnitude cut of $i_{sel}=27.5$ ($i_{sel}=26.1$) to the mock catalog. The i-band distributions in the data and in the model are consistent down to the faintest considered flux and the mocks are complete down to the magnitude of the sample.

No treatment of Ly$\alpha$ radiation is included in the GAEA model. Therefore, it is just used to check the performance of our code in detecting overdensities of galaxies and the effect of photometric redshift uncertainties, but not to verify the observed physical properties of Ly$\alpha$ emitters. % Both distributions are qualitatively consistent in the data and in the mocks. 

\subsection{Fiducial run on mock galaxies}

After applying the same magnitude cuts as for the VANDELS input catalog, we build the fiducial run. %of our code to identify the overdensities of the mock lightcone (not necessarily isolated bound/virialized structures). 
%%%%%%%The friends-of-friends algorithm groups gravitationally connected galaxies independently of their number and magnitude. When we assume a spatial resolution as in the real data ($\Delta$A parameter), our code is not able to identify associations of galaxies composed of either too few bright or too many faint galaxies [{\bf{more details?}}]. 
%
%Considering the spatial and redshift distribution of the lightcone galaxies, we choose reasonable setting parameters to estimate the local densities. 
The redshifts of the mock galaxies include peculiar velocities (z$_{obs}$), but they are not affected by uncertainties as large as those from photometric redshifts. Therefore, to estimate the local densities we set a narrow width in the redshift direction ($\Delta$z = 0.005) %%%%%maybe not relevant%and $\Delta$A = 3 arcsec 
for the cells of the initial regular grid. Also, we require THRESHOLD $=7$ and MIN$\_{obj}=10-5$ (at $z\sim2.5$ and $z\sim3.5$, respectively) to be able to detect dense regions. %{\bf{This is true, in the sense of instrumental measurement, but you might add the comment that you work in redshift space, so peculiar velocities are included.}}
These detection parameters allow us to identify density peaks that contain massive central galaxies (with virial masses larger than 10$^{12}$ M$_{\odot}$) and to take into account the fact that the number of galaxies decreases by at least half from redshift 2 to 3. % (Fig. \ref{compmockdata}). %Examples of the overdensities detected in the fiducial run at $z\sim2.3$ and $z\sim3.4$ are shown in Fig. \ref{fidrunstructuresrecovering} (blue polygons).

\begin{figure*}
 \centering
\includegraphics[width=16cm]{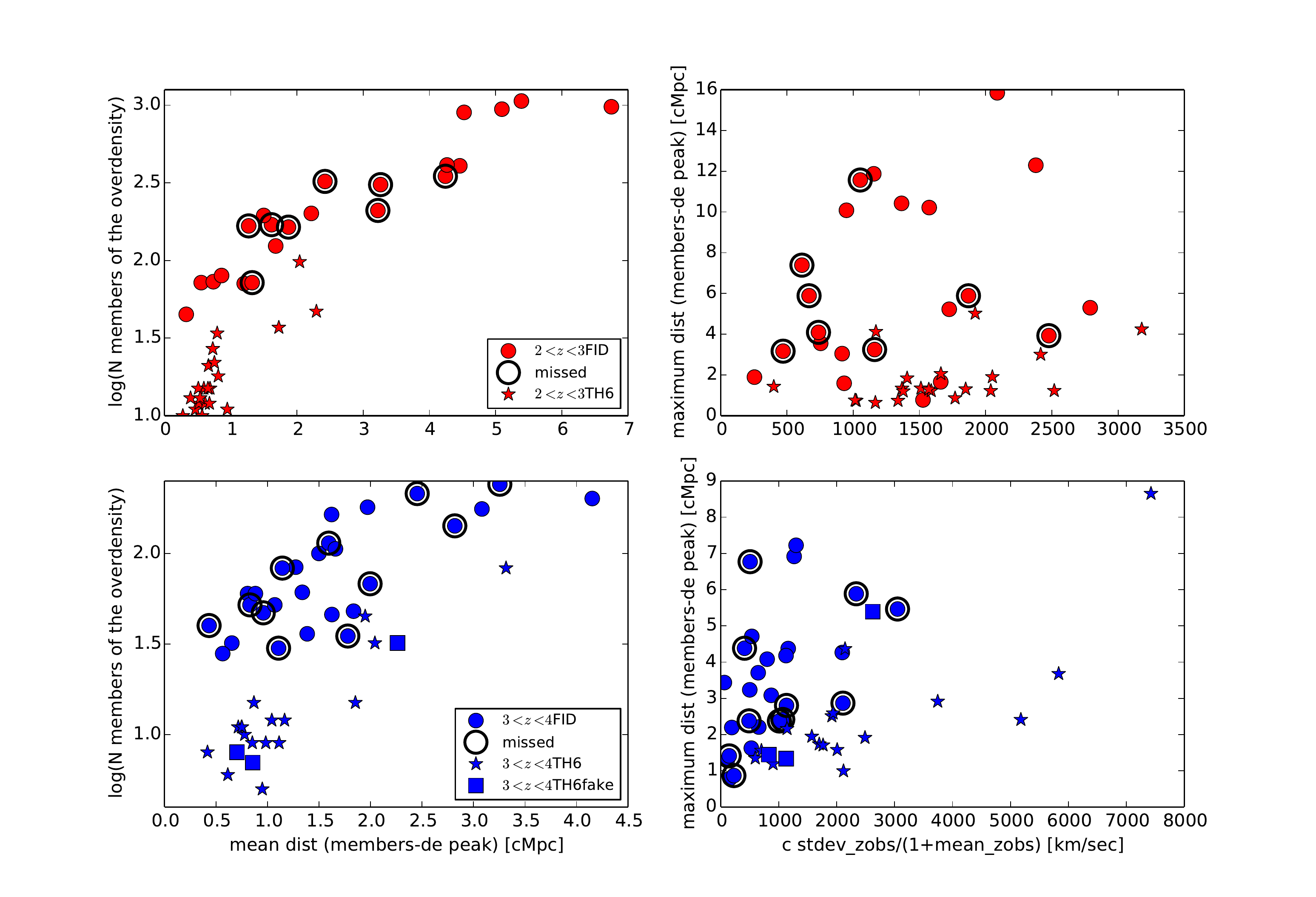}
\caption{ %{\bf{LARGER LEGEND,   ok zphot?}}
Properties of the overdensities detected among the mock galaxies with $i_{sel}\leq27.5$ at $2<z<3$ (red) and $3<z<4$ (blue). The properties of the fiducial run overdensities ($2<z<3$FID and $3<z<4$FID) are shown as dots and the properties of the overdensities detected after including z$_{phot}$ errors (mimicking the VANDELS data) and adopting THRESHOLD=6 ($2<z<3$TH6 and $3<z<4$TH6) are shown as stars. The $left ~panels$ show the total number of members versus the mean distance between the members and the highest-density peak, the $right ~panels$ show the maximum distance between the members and the highest-density peak and the standard deviation of the observed redshifts in velocity [c $\times$ stdev(z$_{obs}$)] / [1+mean(z$_{obs}$)].
%
%dispersion velocity in redshift ($lower ~ left$)
%maximum virial radius of all the galaxies contained in an overdensity versus their mean $i$ magnitude ($upper ~ left$), maximum distance towards the highest density peak versus maximum virial mass of all the galaxies contained in an overdensity ($upper ~ right$), total number of members versus dispersion velocity in redshift ($lower ~ left$), mean distance towards the highest density peak versus dispersion velocity in redshift ($lower ~right$).
%The dots correspond to the overdensities detected in the fiducial run. The stars to the overdensities recovered by setting parameters as for the real data ($\Delta$ z = 0.02, $\Delta$A = 3, THRESHOLD $=6$, MIN$\_{obj}=10-5$, scattering the observed redshifts). 
The fiducial-run overdensities missed after applying photometric uncertainties are indicated with black circles around the dots (missed). The `fake' overdensities at $z\sim3$ ($3<z<4$TH6fake), not detected in the fiducial run, are shown as squares. To read this figure, we remind that each star (recovered overdensity) corresponds to a dot (fiducial-run overdensity) without black circle. The overdensities detected among the mock galaxies with $i_{sel}\leq26.1$ share the same properties of the ones detected among the mock galaxies with $i_{sel}\leq27.5$.
}
\label{fidrecproperties}
\end{figure*}

Our algorithm detects overdensities by linking together more than one bound or virialized structure as identified by the friends-of-friends algorithm in the simulation and more than one massive central galaxy. We perform a merger tree analysis of the mock structures in the lightcones till $z=0$ and we find that 97\% of the detected overdensities are still identified as such at $z=0$.
To better understand the properties of the overdensities we identify in the fiducial run, we consider the total number of members, the standard deviation of the redshifts of the members, the mean and maximum value of the distance of the members with respect to the highest-density peak (Fig. \ref{fidrecproperties}).  
%\end{document}
%At $2<z<3$, the overdensities are characterized by brighter galaxies with larger virial radii than at $3<z<4$. 

The overdensities detected at $2<z<3$ tend to be larger than those detected at $3<z<4$. In fact, the maximum distance between the overdensity members and the location of the highest-density peak is on average 6.3 cMpc at $z\sim2$ and 3.5 cMpc at $z\sim3$. They also tend to be composed of a larger number of members.
%In Tab. \ref{table:mockFID275}, we show the average of the overdensity properties for the two redshift bins.
%\input{mock0FIDproperties.tex}
The standard deviation of the observed redshifts in velocity [c $\times$ stdev(z$_{obs}$)] / [1+mean(z$_{obs}$)] is of the order of 1000 km sec$^{-1}$ and it can be up to 3000 km sec$^{-1}$ both in the lower and in the higher redshift bins.
The overdensities that contain at least five massive central galaxies are the largest among all. %and the most massive 

%\subsection{Optimization of the observational parameters needed to recover the fiducial run structures}
\subsection{Effects of observational uncertainties}

To simulate the conditions of real observations, we perturb the observed redshifts of the mock galaxies. Less than one fifth of the galaxies in the VANDELS input catalog have reliable spectroscopic redshifts. About 85-90\% of the sources only have photometric redshifts. The number of photometric redshifts dominates that of spectroscopic redshifts and a $\sim$15\% (in the CDFS) rather than a $\sim$10\% (in the UDS) of spectroscopic redshifts equally contribute to the detection of overdensities. % with a typical uncertainty of the order of 0.1. %$\Delta$z $\times$ (1+z). 
Therefore, we perturb the observed redshifts of a random 85\% of the mock galaxies. The mock perturbed redshifts are obtained adding an error extracted from a Gaussian distribution centered on the unperturbed redshift and with sigma equal to the photometric redshift error. As we can see in the Table 5 of Pentericci et al. (2018a), the photometric redshift uncertainty has a weak dependence on magnitude and so the treatment of the photo-z errors is realistic enough for our purpose.
% of an amount proportional to the photometric redshift uncertainty ({\bf{?}}Fig. \ref{mockzdistr}). {\bf{
%
The availability of a robust spectroscopic redshift can be magnitude dependent. However, %, but the perturbed 85\% dominates over the unperturbed 15\% and 
the magnitude of the galaxies for which we have spectroscopic redshifts from VANDELS can be much fainter than that from other surveys and we verify that choosing the 15\% of unperturbed redshifts only among the brightest galaxies does not change our results.

We, then, set the parameters of our code to account for the inclusion of redshift uncertainties. We use cells wider in the redshift direction than for the fiducial run case ($\Delta$z = 0.02) and $\Delta$A = 3 arcsec as for the real data. We try three different values of THRESHOLD and check the fraction of recovered overdensities. The THRESHOLD values explored are lower than the value adopted in the fiducial run to account for the fact that the structures become more spread out due to the inclusion of photometric-redshift uncertainties. We also calculate the fraction of `fake' detections, i.e. the ones detected by setting the code parameters as for the real data, but not detected in the fiducial run. In Table \ref{table:fractions}, we report these fractions for THRESHOLD $=5$, $=6$, and $=7$. % in the case of $mag\_i<27.5$.
%{\bf{why not fiducial run with TH=6. So we see the effect of photom redshift uncertainty at the TH used in the real data? Should we report fractions also at TH=4, given Fig. 2?}}
%
To estimate these fractions, we assume that a recovered detection is real if it contains at least the highest-density peak of a fiducial-run overdensity. 
%?????n the majority of the cases, close-by density peaks are considered as one structure in the fiducial run and as disconnected peaks after perturbing the redshifts, even at low thresholds. However, a pair or group of close-by peaks can be interpreted as one structure after the inclusion of z$_{phot}$ errors. % (Fig. \ref{fidrunstructuresrecovering}).  
%
\begin{table*}
\centering
\caption{Fraction of recovered and fake overdensities}  
\label{table:fractions}
\scalebox{0.8}{
\begin{tabular}{|c|c|c|c|}
\hline  
number of & TH=7, MIN$\_{obj}=10-5$ & {\bf{TH=6, MIN$\_{obj}=$ 10-5}} & TH=5, MIN$\_{obj}=10-5$ \\
FID overdensities & recovered/fake & recovered/fake &  recovered/fake\\
\hline
i$_{sel}\leq27.5$ & & & \\
22 ($2<=z<3$) &41\%/ 0\% &  68\%/ 0\% & 73\%/ 11\%  \\
27 ($3<=z<=4$) &55\%/ 12\% &  63\%/ 15\% & 63\%/ 26\% \\ 
\hline
i$_{sel}\leq26.1$ & & & \\
6 ($2<=z<3$) &50\%/ 0\% &  83\%/ 0\% &  83\%/ 0\%\\
4 ($3<=z<=4$) &50\%/ 0\% &  75\%/ 25\% & 75\%/ 25\%\\   
\hline
\end{tabular}
}
\tablefoot{Fraction of recovered and fake overdensities, detected by setting the code parameters as for the real data. %We separate the calculation in two redshft bins. 
We report the results for three values of detection threshold (TH) and we use MIN$\_{obj}=10$(5) at $2<z<3$($3<z<4$). In the upper part of the table, we present the results after applying the magnitude cut typical of the CANDELS area (i$_{sel}\leq27.5$), in the lower part that of the extended area (i$_{sel}\leq26.1$). 
}
\end{table*}
With THRESHOLD $=6$, we obtain the best trade-off between recovered fiducial-run overdensities and low number of fake detections %are able to recover more than 60\% of the fiducial-run overdensities and at the same time keep an acceptable low number of fake detections 
(Table \ref{table:fractions}), even for the case with i$_{sel}\leq26.1$. 
With the same uncertainties and magnitude cut as for the real data, we detect 68\% and 63\% of the fiducial-run over densities at $2<z<3$ and $3<z<4$, respectively. %, which means 32 overdensities in total from the complete mock catalog. %As summarize in Sect 5, we detect only 30\% more overdensities in the mocks than in the real catalog, that is 50\% complete. 
%With THRESHOLD $=6$, we also obtain the best trade-off between recovered fiducial-run overdensities and low number of fake detection for the case with i$_{sel}\leq26.1$.
%In the case of i$_{sel}\leq26.1$, by using THRESHOLD $=6$ we can recover 80\% of the fiducial-run overdensities and keep an acceptable number of fake detections (0\% at $2<=z<3$ and 25\% at $3<=z<=4$) as well. 
%This test on the mocks tells us that to recover the fiducial-run overdensities and account for photometric-redshift uncertainty, we need to increase the cell size in the redshift direction, but also to decrease the detection THRESHOLD with respect to a run without photometric-redshift uncertainties. 
%Moreover, in the case of our data, a THRESHOLD $=6$ allowed us to recover the overdensities previously discovered at $z\sim2.3$ \citep{Salimbeni2009}

As we can see in Fig. \ref{fidrecproperties}, at THRESHOLD $=6$ the spatial size of a recovered overdensity (expressed as the maximum separation between the overdensity members and the overdensity highest-density peak) is smaller than that of the corresponding overdensity detected in the fiducial run. The difference in spatial size is more evident for the overdensities at $z\sim2$ (1.8 cMpc on average compared to 6.3 cMpc on average for the fiducial run) than at $z\sim3$ (2.5 cMpc on average versus 3.5 cMpc on average for the fiducial run), where the number of members is also much more comparable than at $z\sim2$. 
The difference in size along the redshift direction is larger for the overdensities at $z\sim3$ (2500 km sec$^{-1}$ on average versus 900 km sec$^{-1}$ on average for the fiducial run) than at $z\sim2$. %This is due to a larger photometric redshift uncertainty at higher redshift.
This could happen because when photometric redshift uncertainties are introduced, we see two effects: several peaks can be detected as disconnected structures and also very close pairs of peaks can be merged in a single, larger structure.
The first effect is likely caused by galaxy members moving in redshift bins due to the errors. The second is caused by the larger smoothing of the distribution due to the introduction of the errors.
%This could be because at $z\sim3$ we have about half of the sources that at $z\sim2$. % and that the magnitude distribution of the mock galaxies at $3<z<4$ is flatter than at $2<z<3$.  
The standard deviation of the observed redshifts of the galaxies of a recovered overdensity can be twice that of the fiducial-run overdensity 
 %, but are characterized by larger dispersion velocities in redshift 
(right panels of Fig. \ref{fidrecproperties}). This is an effect of the inclusion of the photometric redshift uncertainty. %%%This is something we do not observe in the VANDELS data when we combine photometric and spectroscopic redshifts. 

%{\bf{say something about Nmembers}}
In terms of spatial and redshift sizes, there are no clear differences between the fiducial-run structures that are missed and those that are recovered when setting the parameters to those used on the real data.
The three fake detections at $z\sim3$ have less than ten members and a standard deviation of the observed redshifts larger than 1000 km sec$^{-1}$. One of them (N$_{members}=33$) is located at the border of the field.

This test on the mock galaxies demontrates that the largest uncertainty in detecting overdensities is given by the %need to rely on a 
large number of photometric redshifts, rather than spectroscopic ones. To recover the fiducial-run overdensities and account for photometric-redshift uncertainty, we increase the cell size in the redshift direction. 
%need to increase the cell size in the redshift direction, but also to decrease the detection THRESHOLD with respect to a run without photometric-redshift uncertainties. 
After including photometric-redshift uncertainties, our algorithm is able to identify mainly the highest-density peaks of the fiducial-run overdensities. The maximum distance between the members and the highest-density peak is of the order of 2 cMpc both at $z\sim2$ and at $z\sim3$ on average. The spatial size of the overdensities detected after the inclusion of z$_{phot}$ errors is more similar to that of the fiducial run at $z\sim3$ than at $z\sim2$, but the size along the redshift direction changes much more at $z\sim3$ than at $z\sim2$. % due to z$_{phot}$ uncertainties.
%Moreover, in the case of our data, a THRESHOLD $=6$ allowed us to recover the overdensities previously discovered at $z\sim2.3$ \citep{Salimbeni2009}.

%The main conclusions we can draw from this exercise are the following: (i) most of the detected overdensities at high redshift do not show the central/satellites scheme used in many models {\bf{say better???}}; (ii) 

%--------------------------------------------------------------------
\section{Properties of the overdensities identified in the CDFS and in the UDS}
\label{structures}

%Typically, clusters are expected to be composed, on average, by galaxies with redder colors than the field ones. Due to the inside-out evolution of galaxies in clusters, we could also expect that the youngest and bluest galaxies are located in the outskirts, while the most massive and reddest galaxies in the densest regions \citep[e.g.,][]{Shimakawa2017}. If the evolution of the galaxies around a peak of density is not yet determined (it can be the case at $z>2$), we could expect to have galaxies in different stages of evolution either in the core and in the outskirts of the structure. This could also occur if an identified structure is a more complex composition of sub-structures either located close-by in redshift or merging together.
%
With the scope of studying the properties of star-forming galaxies, and in particular Ly$\alpha$ emitters, as they relate to their environment, we use the 3dv4 algorithm to detect overdensities. To run the algorithm, we chose the set of parameters reported in the previous sections ($\Delta$ z = 0.02, $\Delta$A = 3 arcsec, THRESHOLD = 6, MIN$\_{obj}=10$ and 5 at $2<z<3$ and at $3<z<4$, respectively) and we identify 22 overdensities, 13 in the CDFS and nine in the UDS.

In the following section, we describe the properties of the identified overdensities. We explain the observational properties we consider for the analysis in Sect. \ref{prop}, then we qualitatively describe the most interesting overdensities  (composed of more than one density peak, containing more than one spectroscopic redshifts), including the ones with Ly$\alpha$ emitters (Sect. \ref{LAEs}) as members.

%--------------------------------------------------------------------
\subsection{Observational properties of the identified overdensities}
\label{prop}

To characterize the observational properties of our overdensities, we determine the number of members, their redshift distribution, their physical properties, the dispersion velocity, and the overdensity total mass. %, the associated mass, and the galaxy  (Tab. \ref{tab:TabCDFSUDSstructures}). 
For the majority of the overdensities, at least two members have spectroscopic redshifts. However, one and four structures, respectively in the CDFS and in the UDS, are defined solely by photometric redshifts. %We will check the existent of any spectroscopic redshift in the VANDELS final release.

Even if the overdensities tend to be elongated in the redshift direction, to calculate their volumes we adopt the formula of a sphere with radius equal to the mean member-to-member distance (R$_{meandist}$),  4/3 $\pi$ R$_{meandist}^3$.
For each overdensity redshift range, we define a corresponding field, which is composed of galaxies outside the area occupied by the overdensity, with density within $\pm3\sigma$ around %the value associated to the overdensity members.
%For each overdensity we define the corresponding field. The galaxies in the field are those in the same redshift range of the overdensity, but local densities lower than 3 times 
the average local density ($<\rho >$, see Sect. 3.3). 
To estimate the number of galaxies of the field in a volume equal to the volume occupied by the overdensity, we count the number of field galaxies in a spherical volume centered at more than 2.5 cMpc\footnote{R$_{meandist}$ is always lower than 2.5 cMpc/2 to avoid overlaps between the overdensity and the field volumes.} away from the center of the overdensity and with the same R$_{meandist}$ radius of the overdensity. We repeat the calculation of the number of field galaxies in nine spherical volumes located at nine different centroids and we take the median value.

%For the redshifts of the structure members (spectroscopic if available, photometric otherwise), we show the redshift distribution. 
For the overdensities clearly composed of more than one peak in the spectroscopic-redshift distribution, we estimate the location and the scale of each peak, by following the formalism described in \citet{Beers1990}. We calculate the location of the center of each peak with the biweight estimator and their width with the biweight scale estimator and the gapper scale estimator \citep{WainerThissen1976}. The former mainly takes into account the difference between the redshifts of members and the median redshift of the structure, the latter the redshift difference between members. 
To derive the scale estimator via the gapper method, we place an upper limit on the maximum velocity difference with respect to the central one of $\pm$ 1500 km sec$^{-1}$ \citep[we remind the reader that one of the largest velocity dispersions measured for a cluster of galaxies is 1200 km sec$^{-1}$ for the Coma cluster,][]{Zabludoff1993}. %{\bf{are the disp velocties consistent with this limit of 1500 km/sec. OK because gapper looks at the difference between z$i$ and z$i+1$}}
 %estimate the redshift standard deviation with respect to the density-peak redshift and the structure dispersion velocity by following \citet{Beers1990}. We adopted the $gapper$ methods from this author to estimate the dispersion velocity of the members within 3000 km sec$^{-1}$ from the redshift of the highest-density peak \citep[see also][]{Cucciati2018}. 
We perform a bootstrap with replacement technique to estimate the uncertainties of the dispersion velocities. 
 
%According to this work, the line-of-sight component of the velocity of a galaxy with respect to the overdensity peak can be defined as,
%\begin{equation}
%v_{//}=\frac{V_{//}-\overline{V_{//}}}{1+\overline{V_{//}}/c}
%\end{equation}
%where $V_{//}=c \times z$ is the radial velocity, $\overline{V_{//}}=c \times <z>$, and <z> is the average redshift of the overdensity members, which, in some cases, corresponds to the redshift of the highest density peak of the ovrdensity. The v$_{//}$ of a member i can be expressed as,
%\begin{equation}
%v_{//}i=\frac{c \times (zi-<z>)} {1+<z>}.
%\end{equation}
%Therefore, the radial dispersion velocity becomes,
% \begin{equation}
%\sigma^2_{//}=\frac{\Sigma^n_i v^2_{//}i}{n-1}-\frac{\delta^2} {(1+<z>)^2}
%\label{raddispvel}
%\end{equation}
%where $\delta$ is related to the redshift uncertainty, and the error on the radial %dispersion velocity becomes, 
%\begin{equation}
%(\delta\sigma_{//})^2\sim \left({\frac{c \times \delta(<z>)}{(1+<z>)}}\right)^2 \frac{1+\left(\frac{c \times \delta(<z>)/(1+<z>)^2}{2 \sigma^2_{//}}\right)}{n} 
%\end{equation}
%where $\delta(<z>)=stdev(zi)/\sqrt{n}$ is the standard error of the mean of the redshifts of the structure members.

%For each structure, w
We estimate the total mass, M, associated with our identified overdensities as that proportional to the matter overdensity  \citep[see][]{Steidel1998, cuccia14,Lemaux2014}.  %We calculate the dynamical mass based on the velocity dispersion in equation \ref{raddispvel},  
%\begin{equation}
%M_{dyn,vir}=\frac{3 \sqrt{3} \sigma^3_{//}}   {11.4??? G H(z)}
%\label{dynmass}
%\end{equation}
%where G is the gravitational constant and H(z) is the Hubble constant at the redshift of the overdensity \citep[e.g.,][]{Lemaux2014}. This mass estimation assumes that the %overdensity is in a virialized state. Therefore, depending on the evolutionary stage of the structure, it can provide a lower or upper limit of the true value. 
%
%We calculate the mass, M,   
We list here some of the equations used to derive M, additional details can be found in the cited papers. M is defined as  
\begin{equation}
M = \rho_{0} V (1+\delta_{m}),
\label{Meq}
\end{equation}
where $\rho_{0}$ is the comoving mean density of the Universe ($= \Omega_{m} \rho_{0crit} = 4.079 \times 10^{10}$ M$_{\odot}$ cMpc$^{-3}$, given the adopted cosmology), $\delta_{m}$ is the matter overdensity in the structure, and V is the volume occupied by the overdensity in real space. %, taking into account peculiar velocities and the growth of perturbations. 
For each overdensity, we estimate n$_{overdensity}$ as the number of members divided by the observed overdensity volume and n$_{field}$ as the median number of the field galaxies in the redshift range of the overdensity and in a volume equal to the observed volume of the overdensity. 
Hence, the matter overdensity is proportional to the galaxy overdensity as
\begin{equation}
(1+b \delta_{m}) = C (1+\delta_{gal}),
\label{rhomgal}
\end{equation}
where $\delta_{gal} = $(n$_{overdensity}$-n$_{field}$)/n$_{field}$, C relates the volumes in real (V) and observed space (V$_{obs}$) and depends on the adopted cosmology through f(z) = $\Omega_{m}(z)^{4/7}$. 
\begin{equation}
C = 1+ f(z) - f(z) (1+\delta_{m})^{1/3},
\end{equation}
such that V =  V$_{obs}$/C (where V$_{obs}$ is in comoving coordinates), and b is a bias factor ranging from 2 to 4 at $2<z<4$ \citep{Durkalec2015}. 
%Then, we estimate the stellar mass of an overdensity as the sum of the stellar masses of its members. This estimation provides a lower limit of the total mass of the structures.
We do not report mass estimates for the overdensities composed of a large number of multiple peaks %, except for those with a large enough number of spectroscopic redshifts to securely identify the individual peaks (see Sect. 5.2), 
nor for the overdensities with zero spectroscopic redshifts. %In fact, for these overdensities not even the matter density would have a clear meaning.
% $b$ is the bias factor, and $\delta_{gal}/b=200$. We assume $b=2-4$ at $2<z<4$ \citep{Durkalec2015} and $\rho_{0} = \Omega_{m} \rho_{0crit} = 4.079E-10$ M$_{\odot}$ Mpc$^{-3}$, given the adopted cosmology.

%As described in \citet{Chiang2013}, there is a redshift at which a protocluster first contains a halo of 10$^{14}$ M$_{\odot}$. This is called cluster assembly redshift. Before this redshift, protoclusters are not virialized. They are composed of a main overdensity peak, surrounded by small halos which will assemble onto the main one by $z=0$ or even later. %More massive protoclusters occupy larger comoving volumes. 
%The progenitor of a cluster with mass larger than $10^{15}$ M$_{\odot}$ can reach the cluster assembly redshift at $z\sim2.3$, while low-mass clusters made the transition from protocluster to cluster more recently. According to the mass and the size of the overdensities we detect, we could therefore infer  their stage of virialization. 
Given the redshift of a structure and $\delta_{gal}$, \citet{Chiang2013} simulations provide a prediction of the kind of $z=0$ clusters the overdensities may evolve to, either a Fornax-type (with $1.3-3.0$ 10$^{14}$ M$\odot$), or a Virgo-type (with $3-10$ 10$^{14}$ M$\odot$), or a Coma-type (with $>$ 10$^{15}$ M$\odot$) cluster at $z=0$. 
%According to the $\delta_{gal}$ we estimate, we could predict that the overdensity studied here are progenitors of $z=0$ cluster with a certain mass. By following the simulations in \citet{Chiang2013}, we could expect that a protocluster at $z\sim2$ with $\delta_{gal}=2$ could be progenitor of a Fornax-type cluster at $z=0$. For this structure the assembly redshift would be $0.2<z<1$. %, therefore it is likely not to be virialized yet at $z\sim2$. 
%However, if its current mass is much lower than 10$^{13}$ M$\odot$, it is possible that the structures will not assemble in time to become a cluster at $z=0$. 
For example, we could expect that a protocluster at $z\sim2$ with $\delta_{gal}=4$(6) could be progenitor of a Virgo(Coma)-type cluster at $z=0$. It can also happen that high-z overdensities may not be able to assemble a virialized cluster by $z=0$.
We use these simulation results to determine the fate of our identified overdensities in the next section.
To understand the stage of evolution of an overdensity and its members, we study rest-frame $U-V$ color, stellar mass (M$_{*}$), and the specific star-formation rate (sSFR=SFR$_{SED}$/M$_{*}$) of the structure members. %These quantities are provided by SED fitting \citep{McLure2018}. 
The rest-frame $U$ and $V$ magnitudes correspond to the rest-frame absolute magnitudes at 3670{\AA} and 5500{\AA}, respectively, obtained assuming a 200 {\AA}-wide tophat filter on the best-fit SED template. The rest-frame $U-V$ color resembles the observed-frame $J-K$ color, frequently used at $z\sim2$ to define the protocluster red sequence \citep{Stott2012,willis13,Strazzullo2016} of evolved, possibly passive galaxies. %, we choose the $B$-band magnitude.
%restUV                      #(rest-frame absolute magnitude at 1500A, 200A-wide tophat filter)
%MAGabsV_top=scidata.field(18)       #restOPTICAL                  #(rest-frame absolute magnitude at 5500A, 200A-wide tophat filter)

We compare rest-frame $U-V$ color, M$_{*}$, and sSFR of the overdensity members with those corresponding to the galaxies in the field (Appendix \ref{appendix2}) and we estimate if the set of quantities of the members and the field galaxies are drawn from the same distribution by using a KS test.  % and in regions with density in between the field and the overdensity values. 
%{\bf{cut?}} 
For reference, the rest-frame $U-V$ colors for passive galaxies in the VANDELS photometric catalog is about 1.7, while it is $\sim0.62$ for star-forming galaxies at $z\sim2.8$, and $\sim0.63$ for LBGs at $z\sim3.4$ \citep[following the definition of the VANDELS galaxy populations in][]{Pentericci2018}. %{\bf{this is the reference for the definition of passive LBGs and SFGs}}

We also consider the rest-frame color-color diagram composed of the $U-V$ and $V-J$ colors to investigate any possible physical vs morphological property trend among the galaxies of each structure \citep{Strazzullo2013,Newman2014}. Given the availability of the morphological catalogs from CANDELS \citep{vanderWel2012}, we study the morphology of the overdensity members in terms of GALFIT Sersic index and axis ratio. The rest-frame $J$ magnitude corresponds to the rest-frame absolute magnitudes at 12500{\AA}, assuming a 200{\AA}-wide tophat filter. %In Appendix \ref{appendix3}, we show the color-color diagram as a function of sSFR and morphological quantities. 
In Appendix C and D, we show all these trends and we comment them in detail only for the most notable structures discussed in Sect. \ref{structurecdfsuds}. We do not report morphological considerations for the galaxies located in the extended areas of the CDFS and of the UDS.

\subsection{Overdensities identified in the CDFS and in the UDS}
\label{structurecdfsuds}

The main observational properties of the 22 identified overdensities, such as the number of members estimated by our algorithm and the number of members with spectroscopic redshifts, are presented in Table \ref{tab:TabCDFSUDSstructures}. % and shown in the figures of the Appendices. 
The redshift distributions %(zphot+zspec and only zspec) 
of each identified overdensity are shown in Fig. \ref{zdistrCDFS1}, \ref{zdistrCDFS2}, and Fig. \ref{zdistrUDS1}. In Appendix A, we show the regions occupied by our 22 identified overdensities, together with the location of structures identified by \citet{KangIm2015, Franck2016} in the CDFS. In Appendix B, we show the space distribution of each of the identified overdensity; in Appendix C and D, we show the physical properties of the most interesting overdensities described here below.
\begin{figure*}
 \centering
\includegraphics[width=5cm]{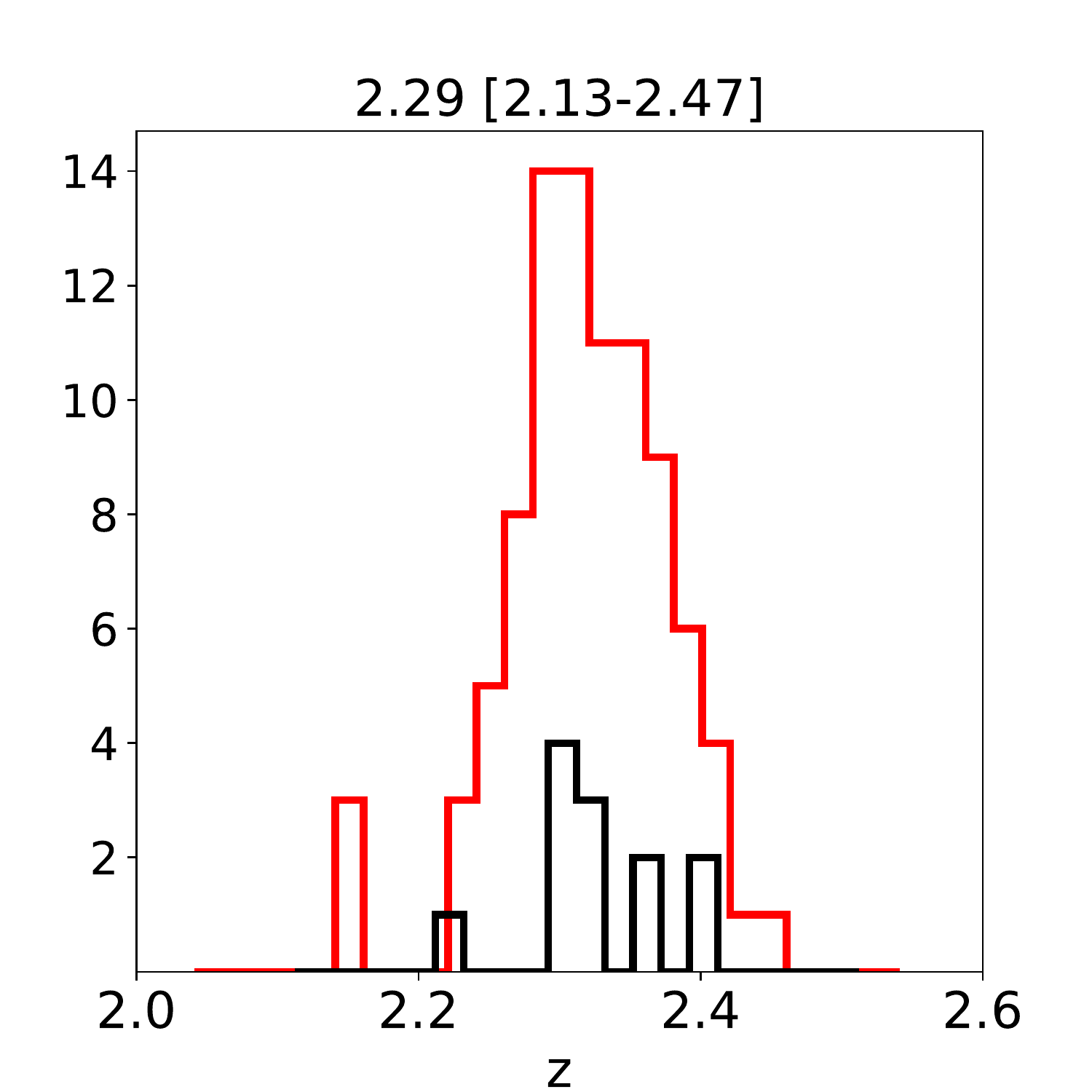}
\includegraphics[width=5cm]{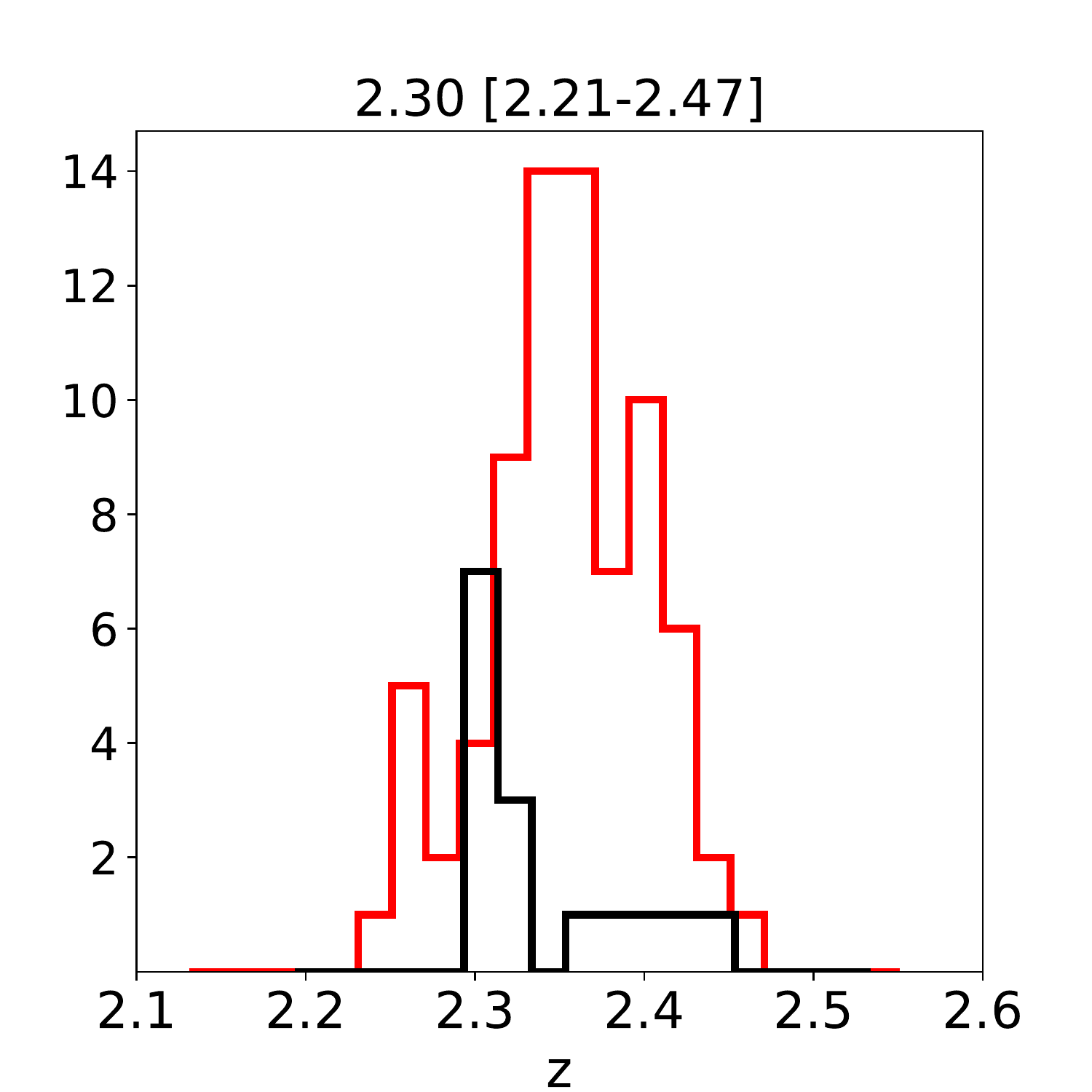}
\includegraphics[width=5cm]{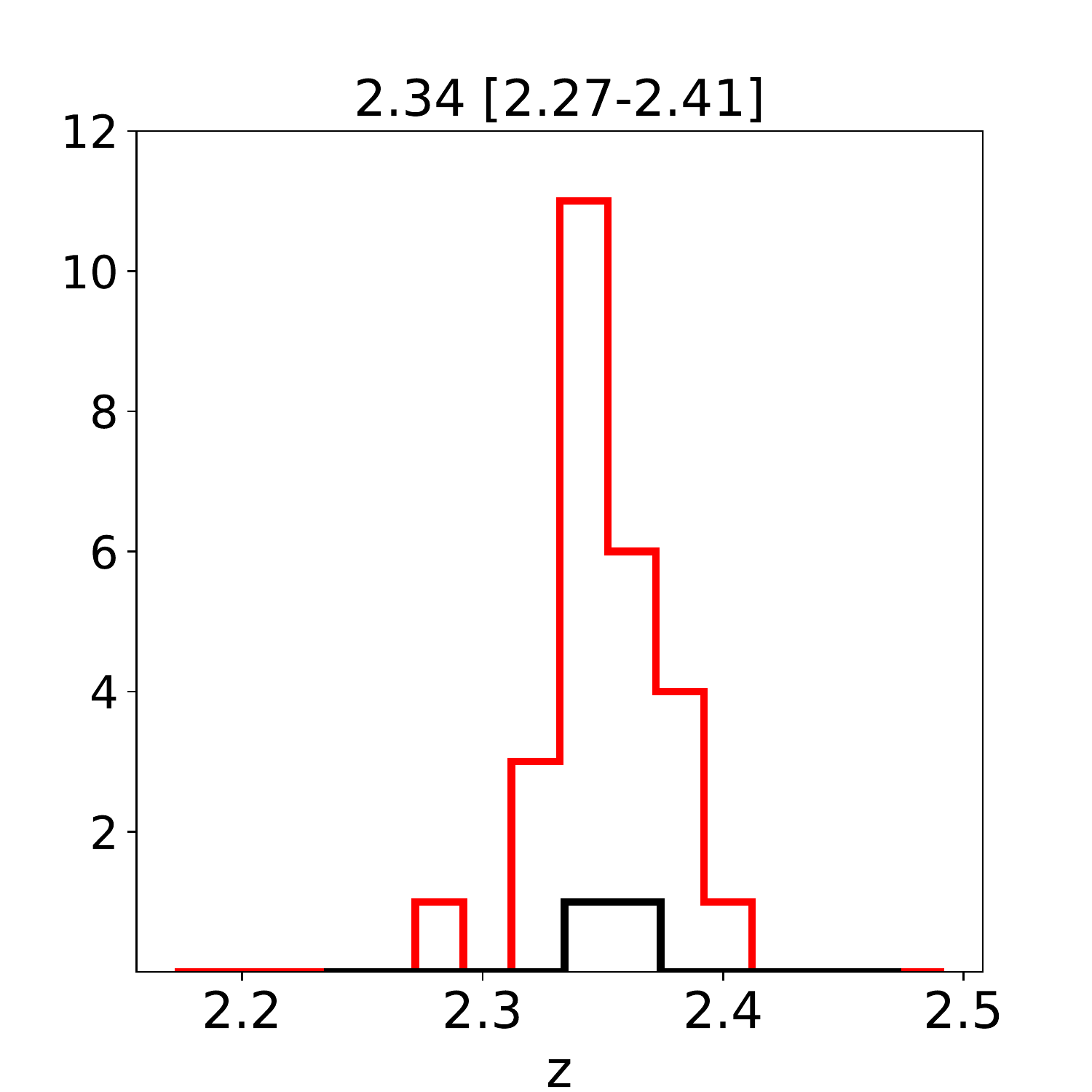}
\includegraphics[width=5cm]{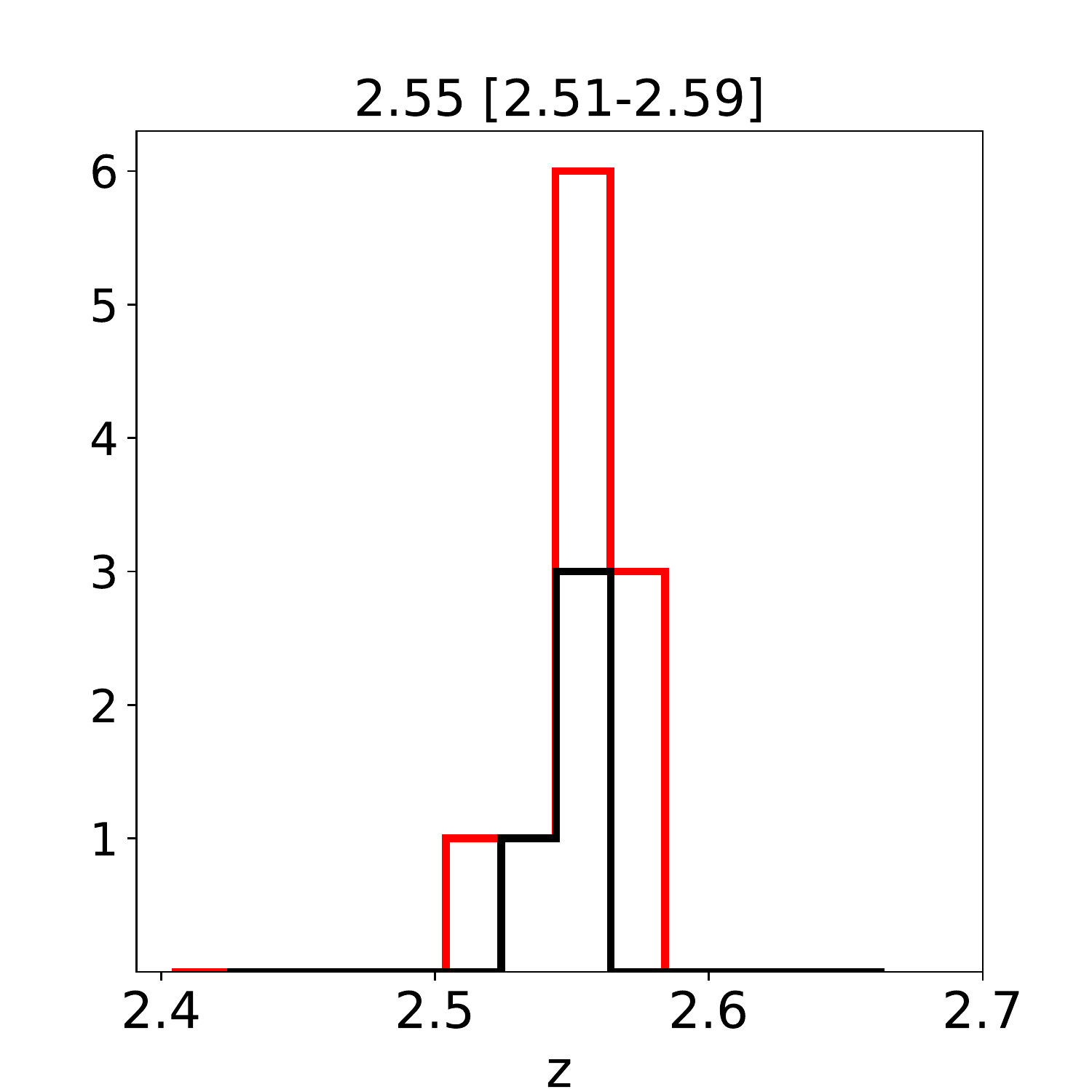}
\includegraphics[width=5cm]{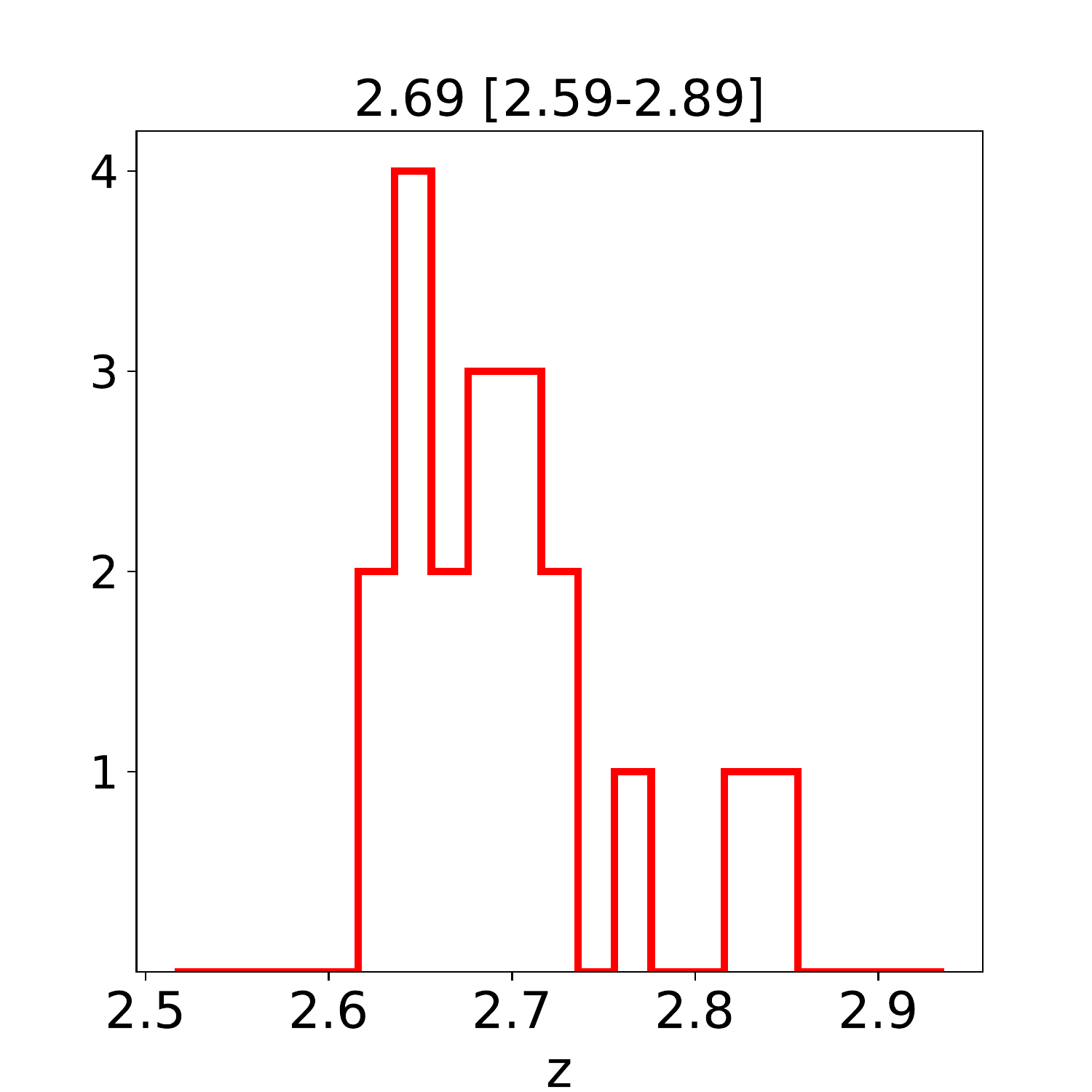}
\includegraphics[width=5cm]{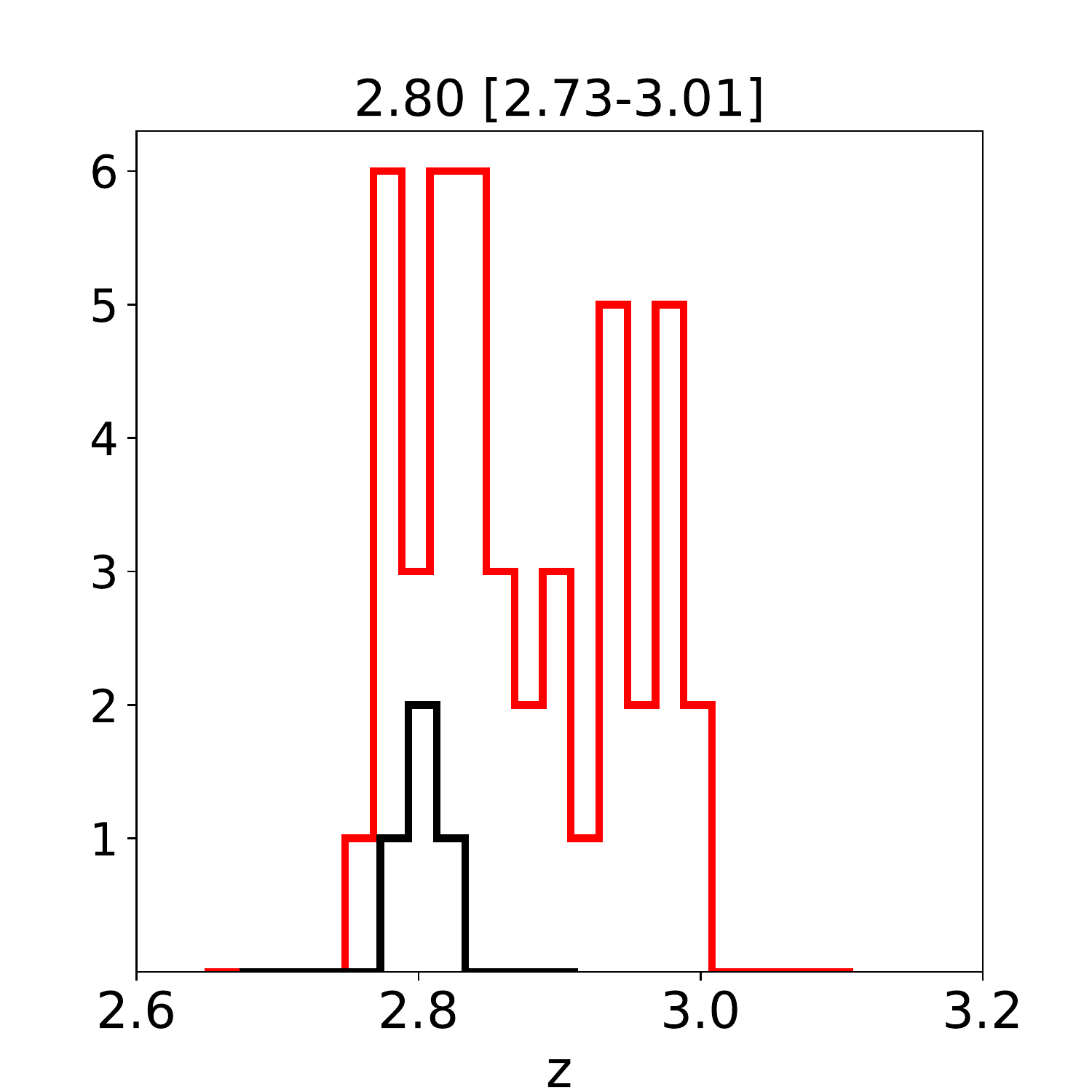}
\caption{%{\bf metterla con le strutture}
Redshift distribution of the galaxies in the overdensities identified in the CDFS. The red histograms include both spectroscopic and photometric redshifts, while black histograms only the spectroscopic ones. The histogram bin is 0.02 in redshift. Some of overdensities are likely composed of more than one density peak.} 
\label{zdistrCDFS1}%
\end{figure*}
\begin{figure*}
 \centering
\includegraphics[width=5cm]{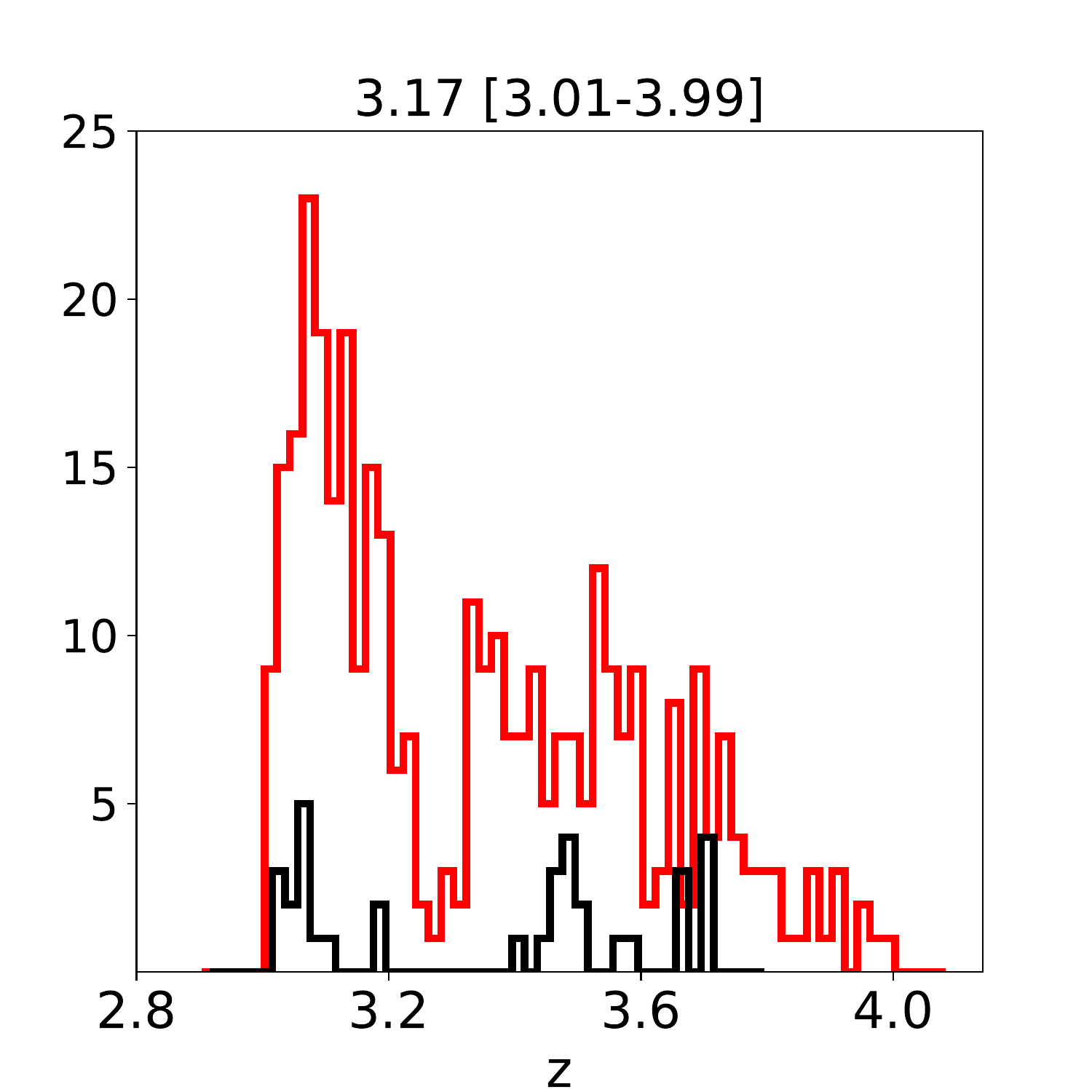}
\includegraphics[width=5cm]{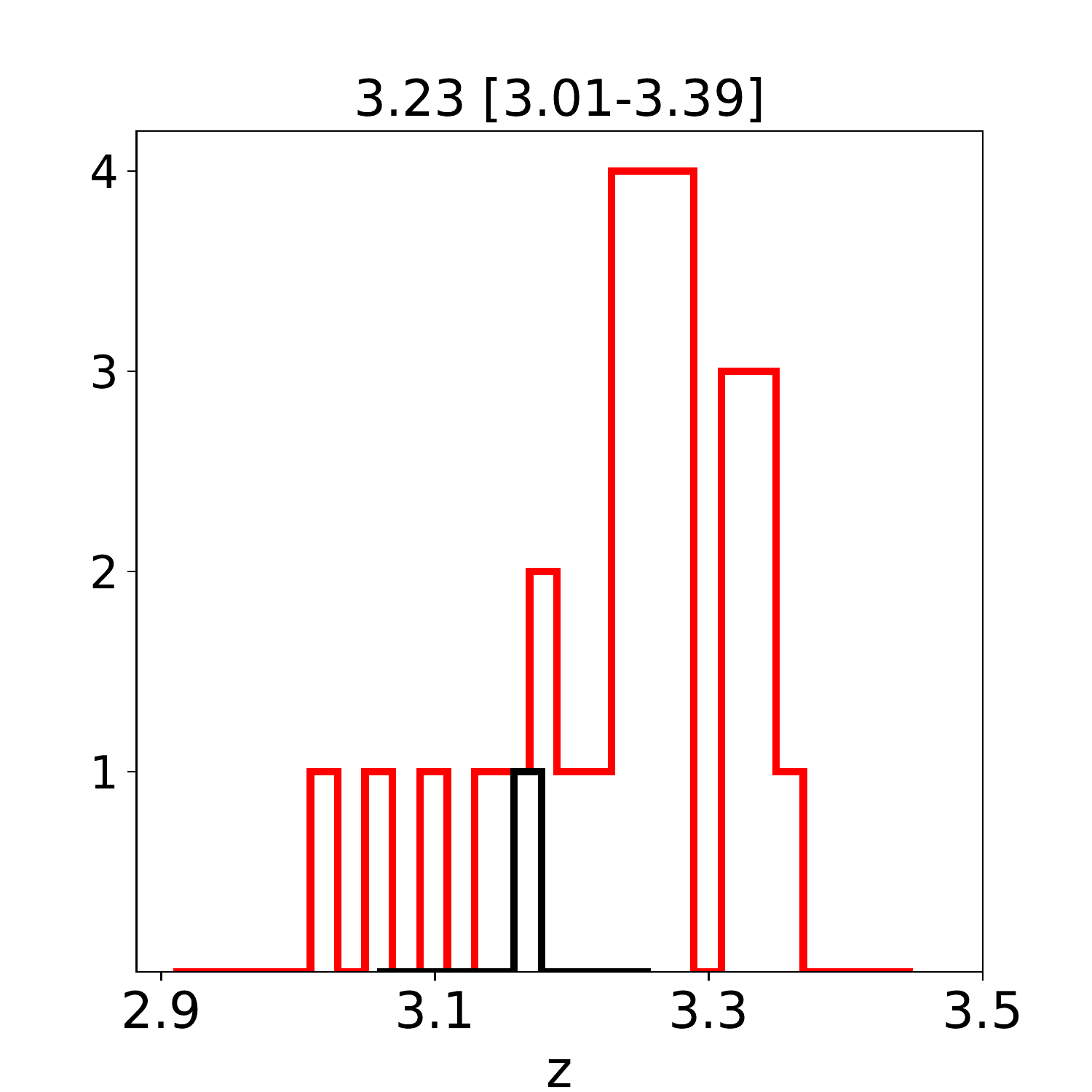}
\includegraphics[width=5cm]{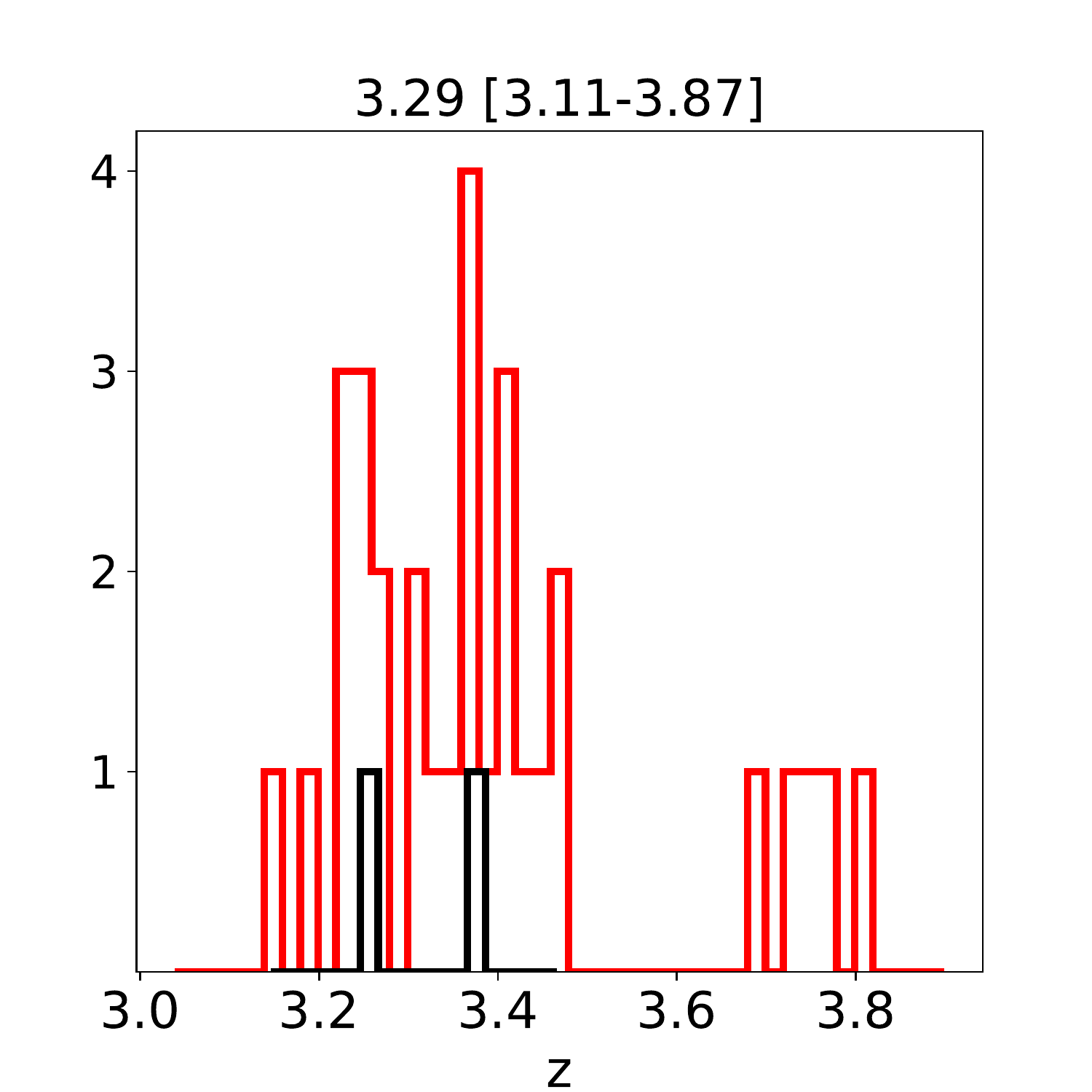}
\includegraphics[width=5cm]{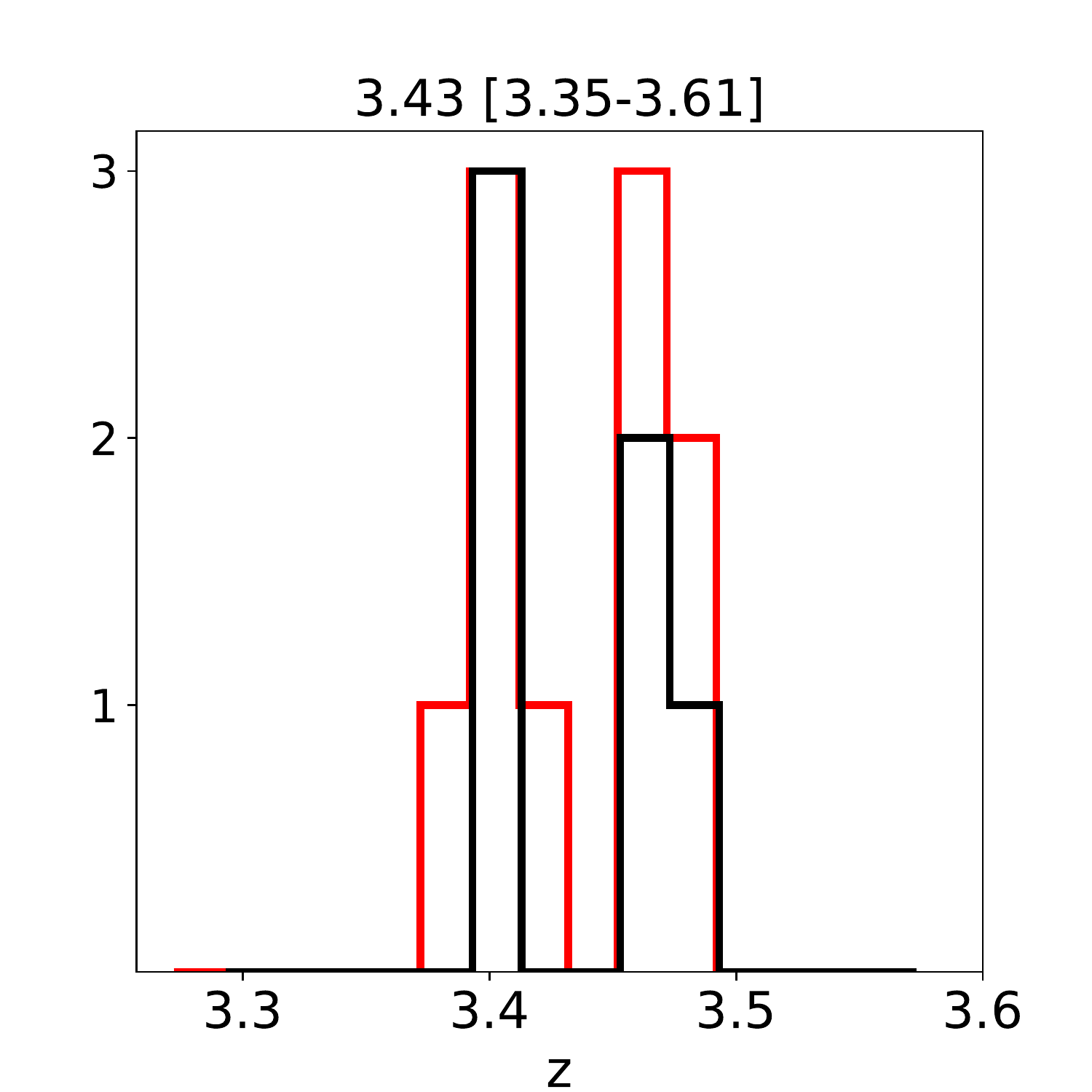}
\includegraphics[width=5cm]{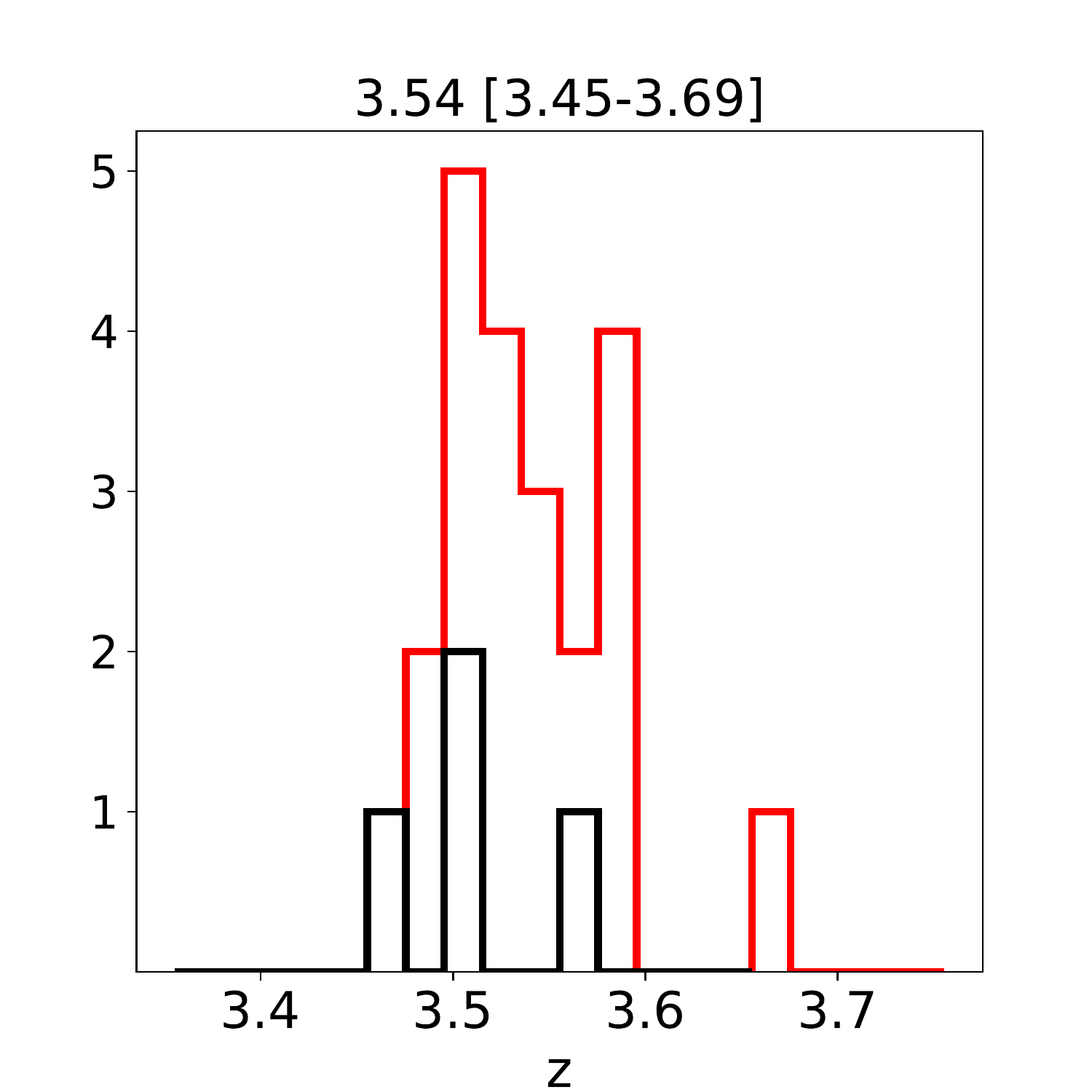}
\includegraphics[width=5cm]{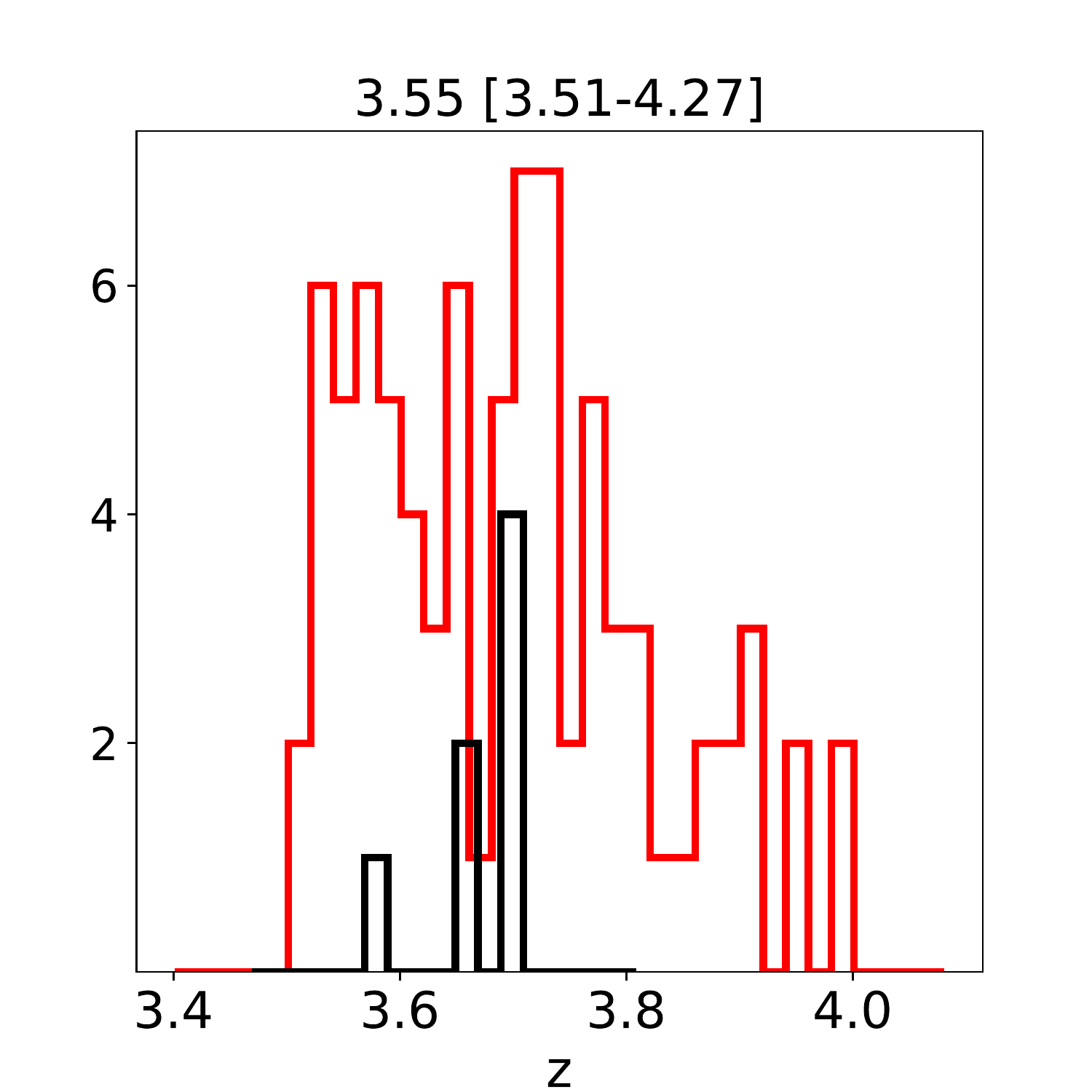}
\includegraphics[width=5cm]{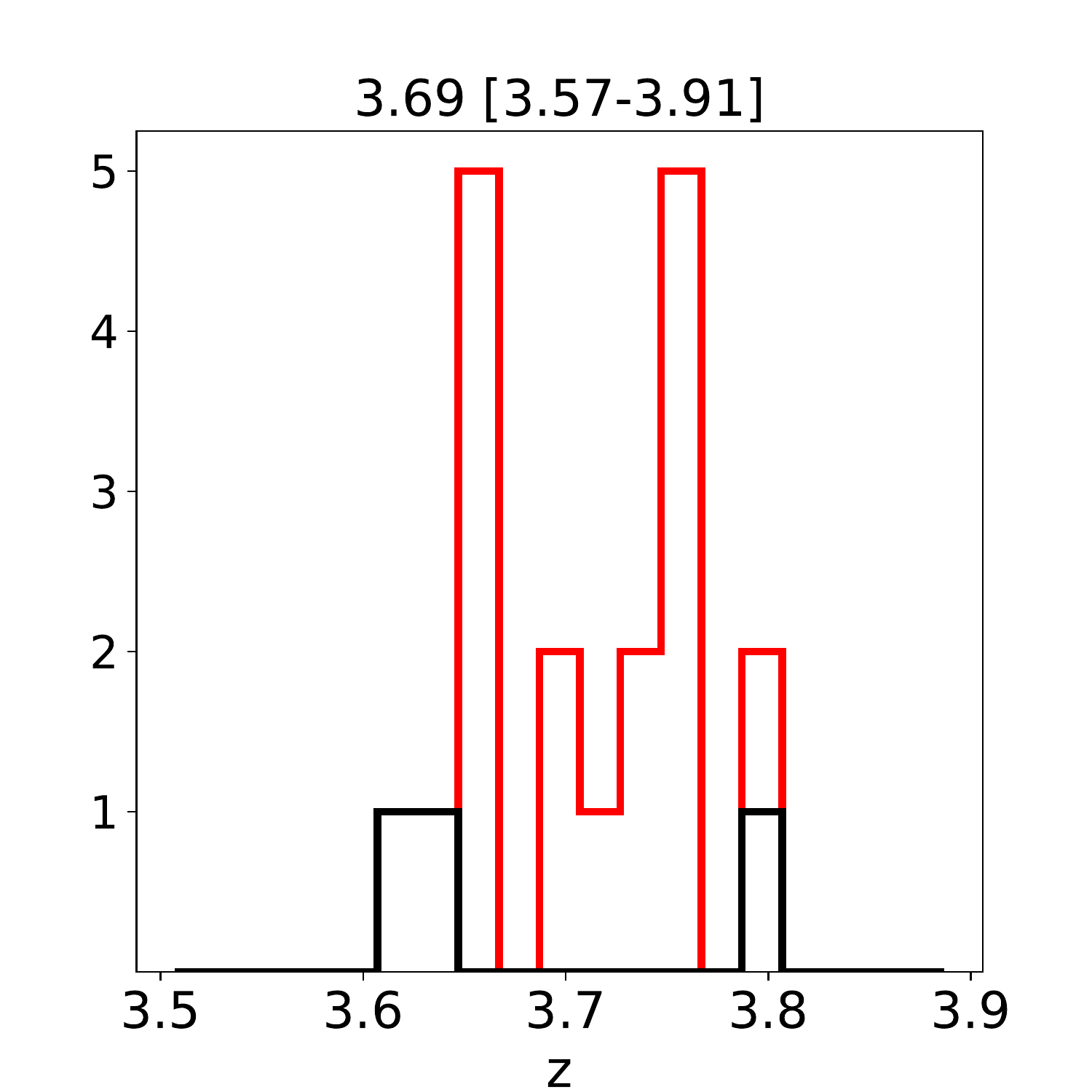}
\caption{%{\bf metterla con le strutture}
Continuation of Fig. \ref{zdistrCDFS1}} %, such as the structure at $z\simeq3.17$.}
\label{zdistrCDFS2}%
\end{figure*}

\begin{figure*}
 \centering
 \includegraphics[width=5cm]{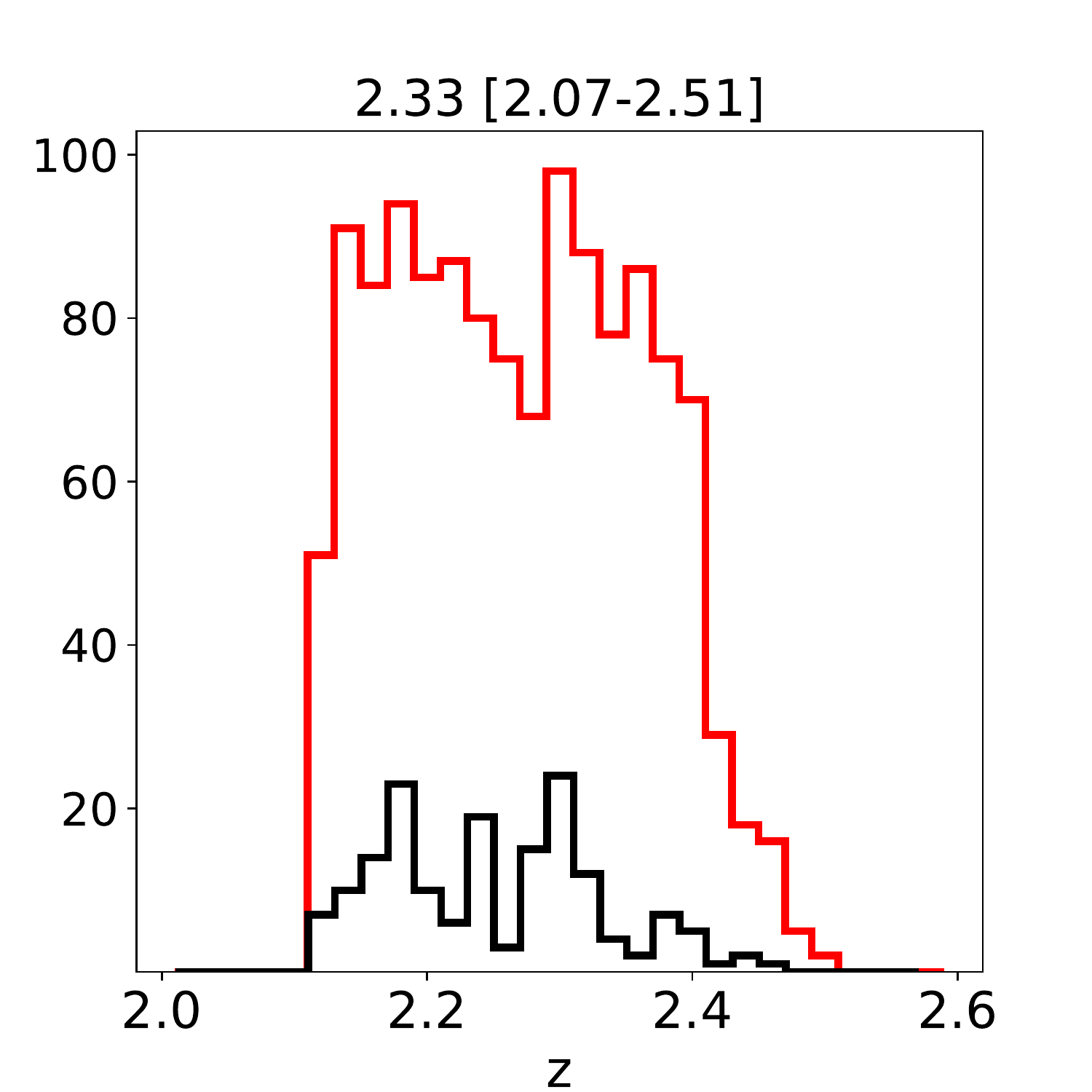}
 \includegraphics[width=5cm]{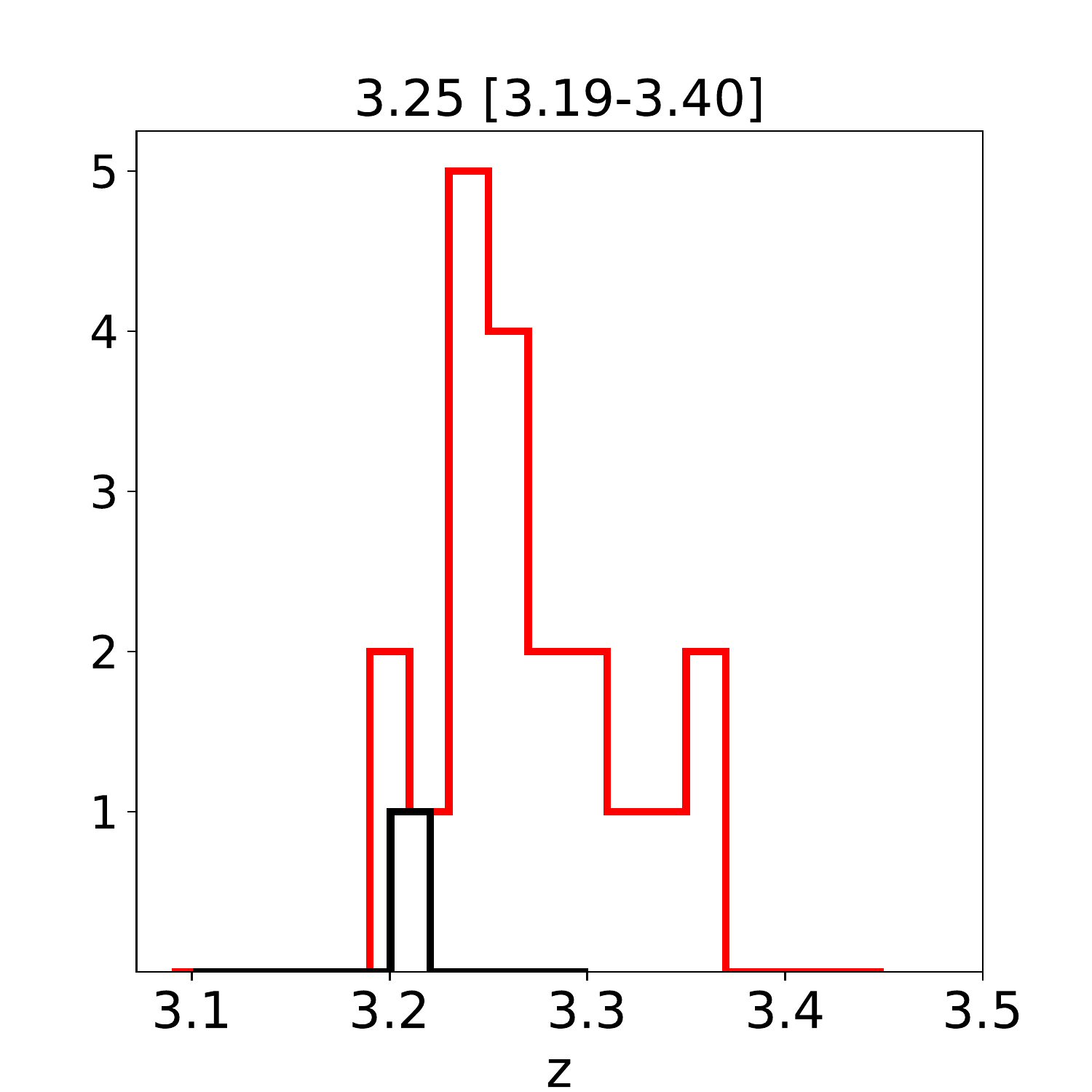}
 \includegraphics[width=5cm]{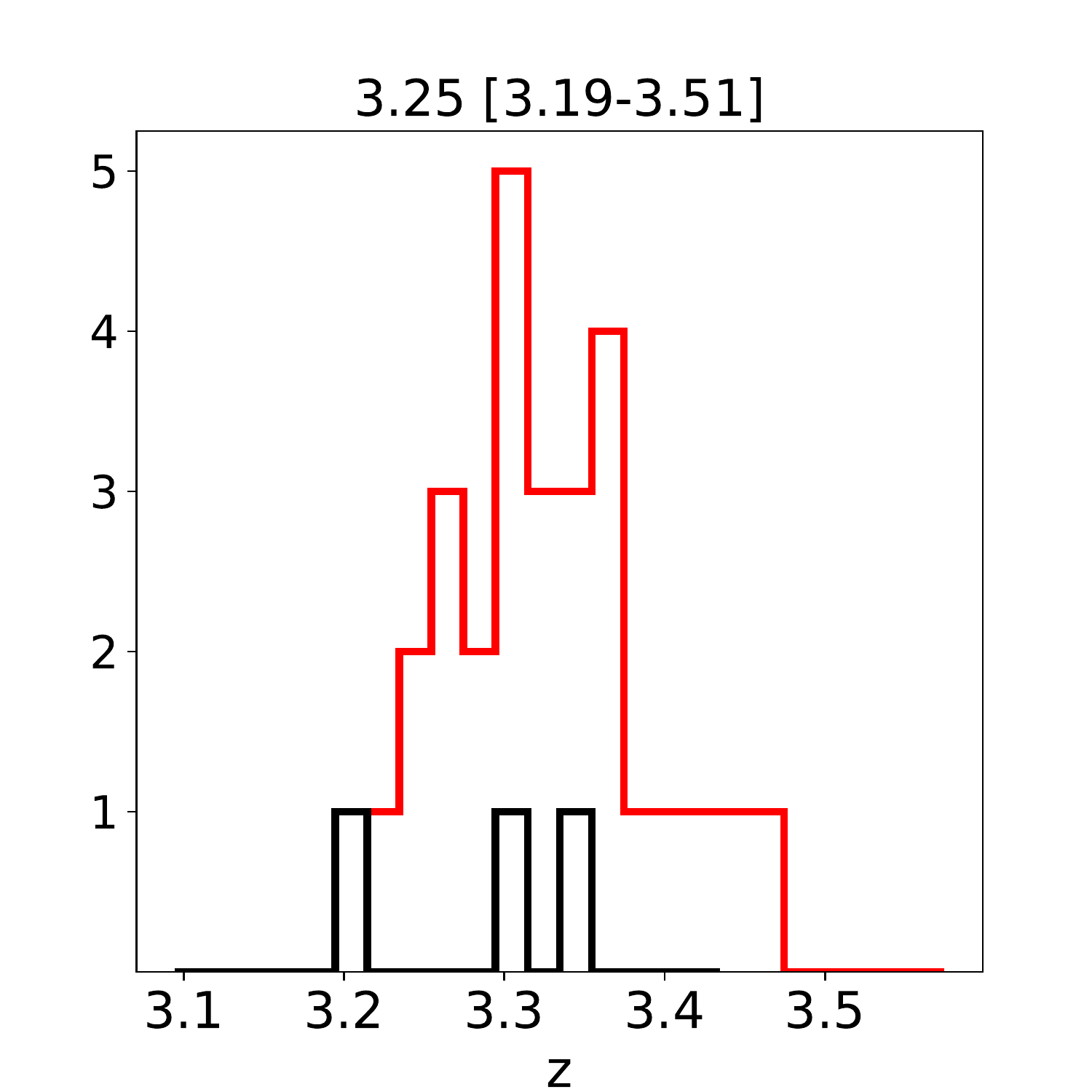}
 \includegraphics[width=5cm]{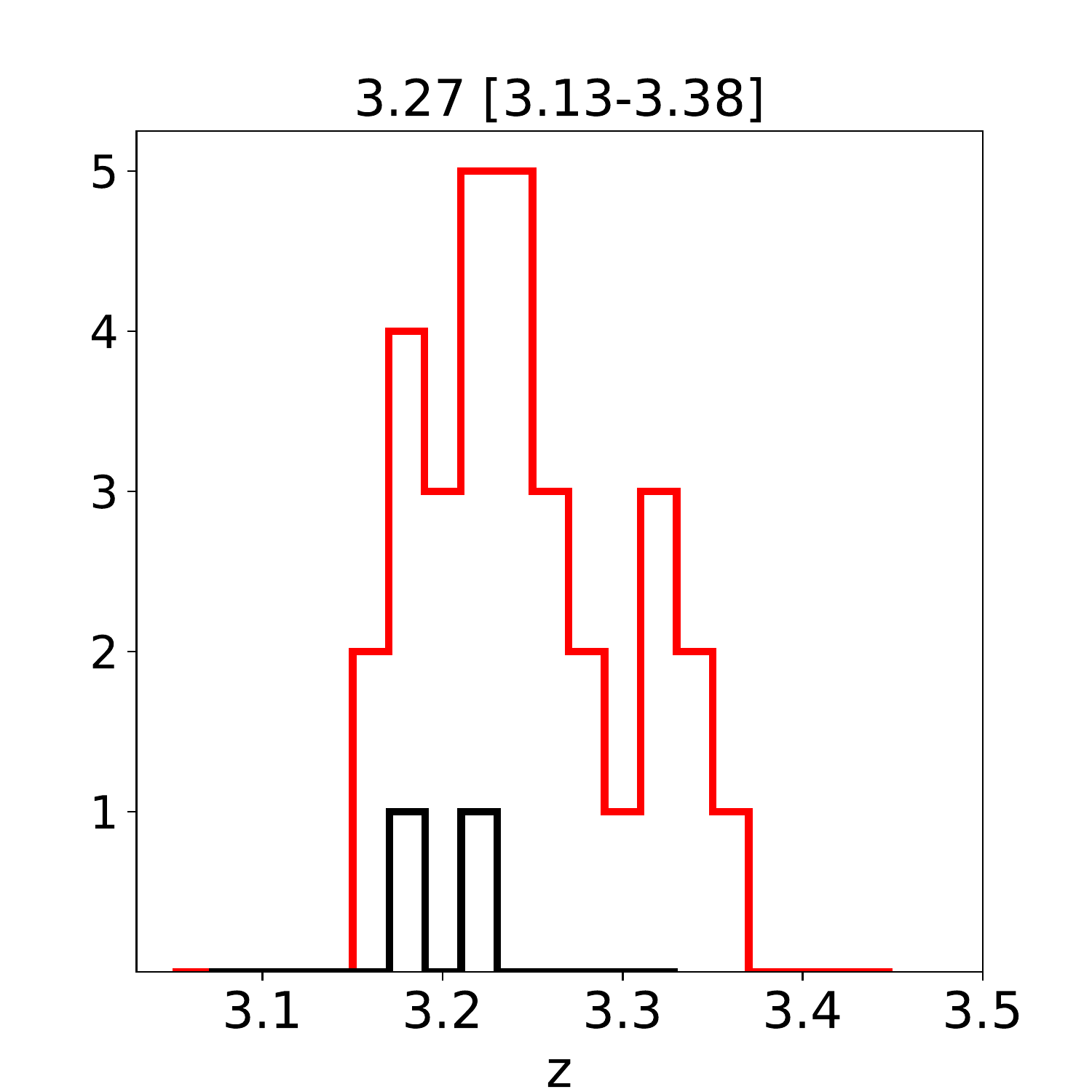}
 \includegraphics[width=5cm]{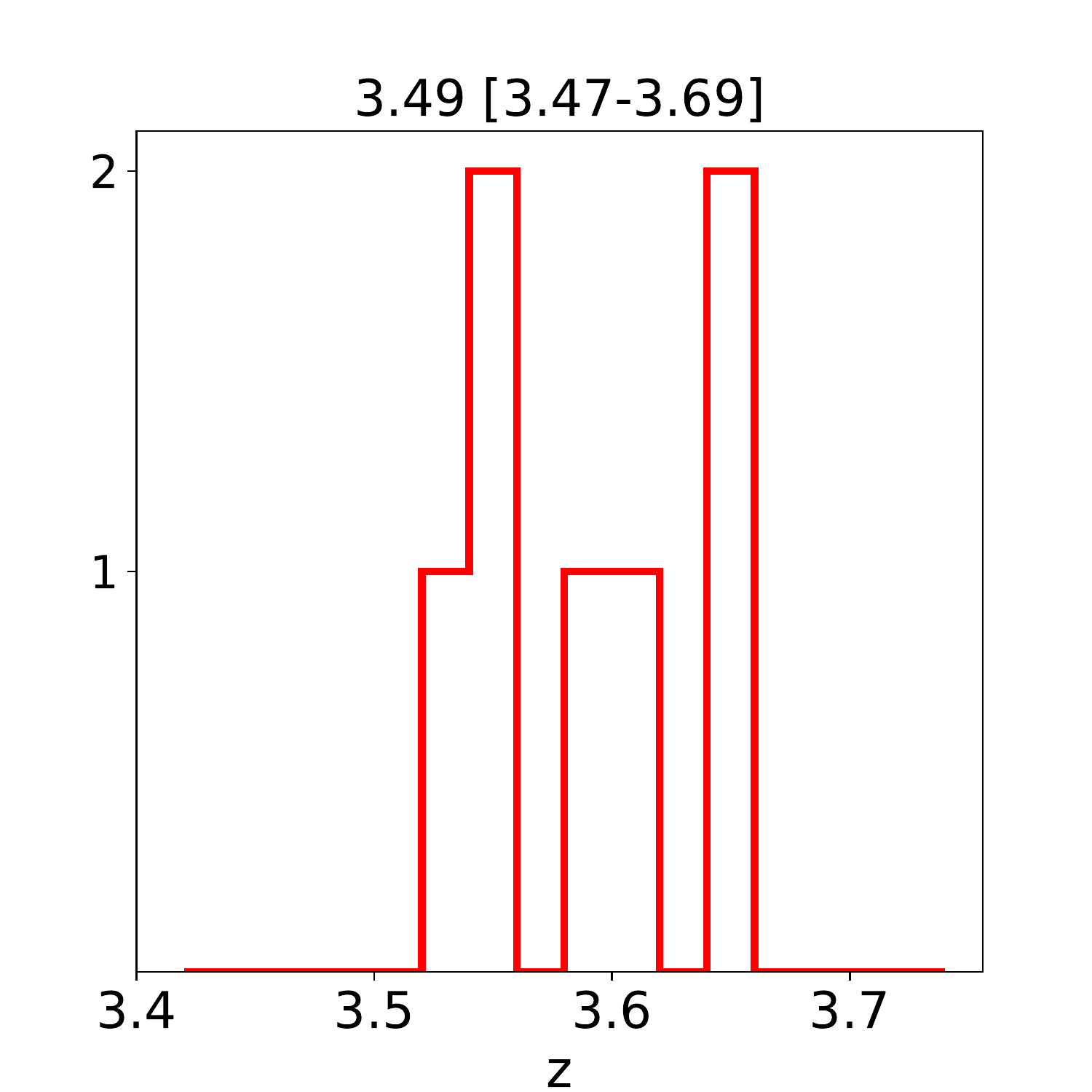}
 \includegraphics[width=5cm]{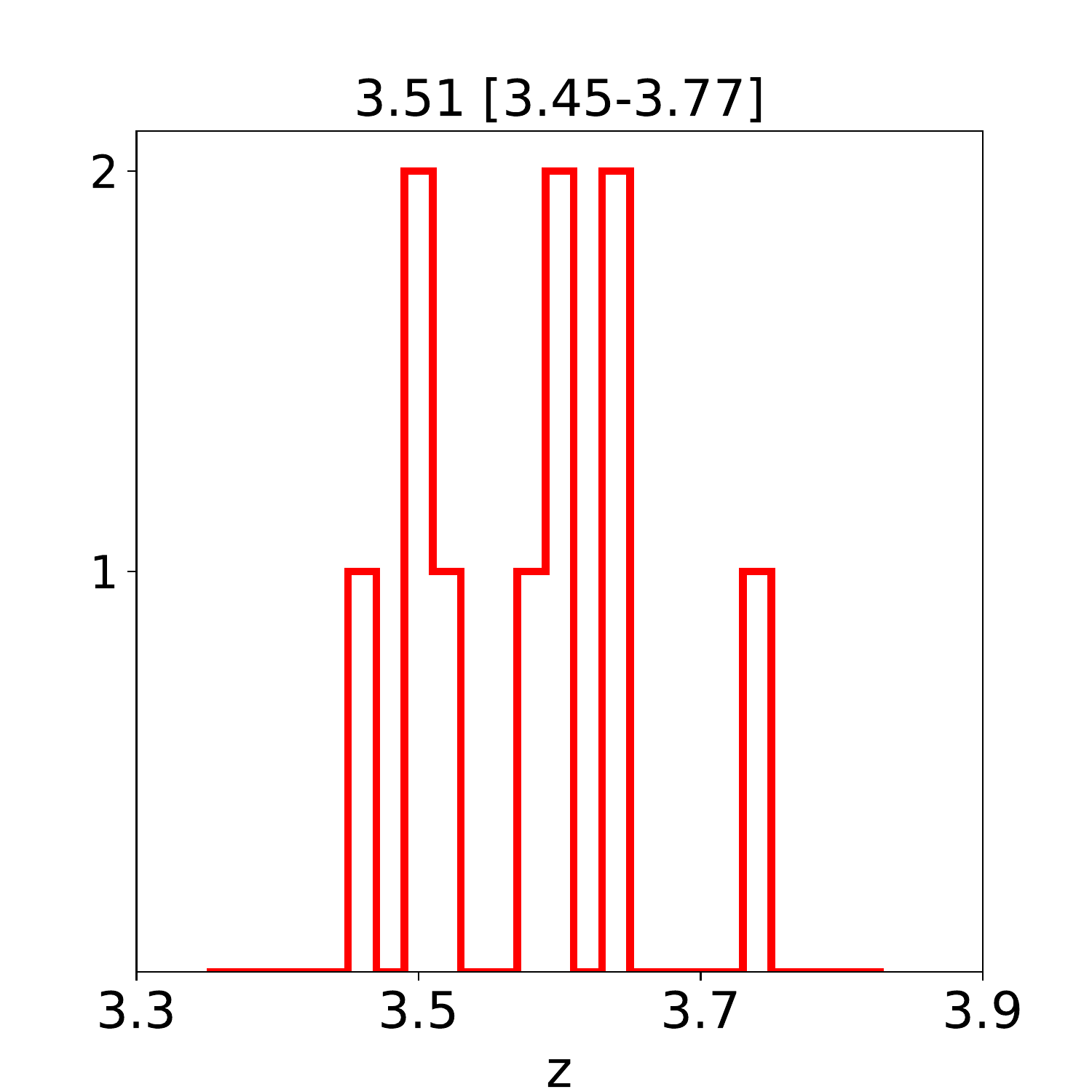}
 \includegraphics[width=5cm]{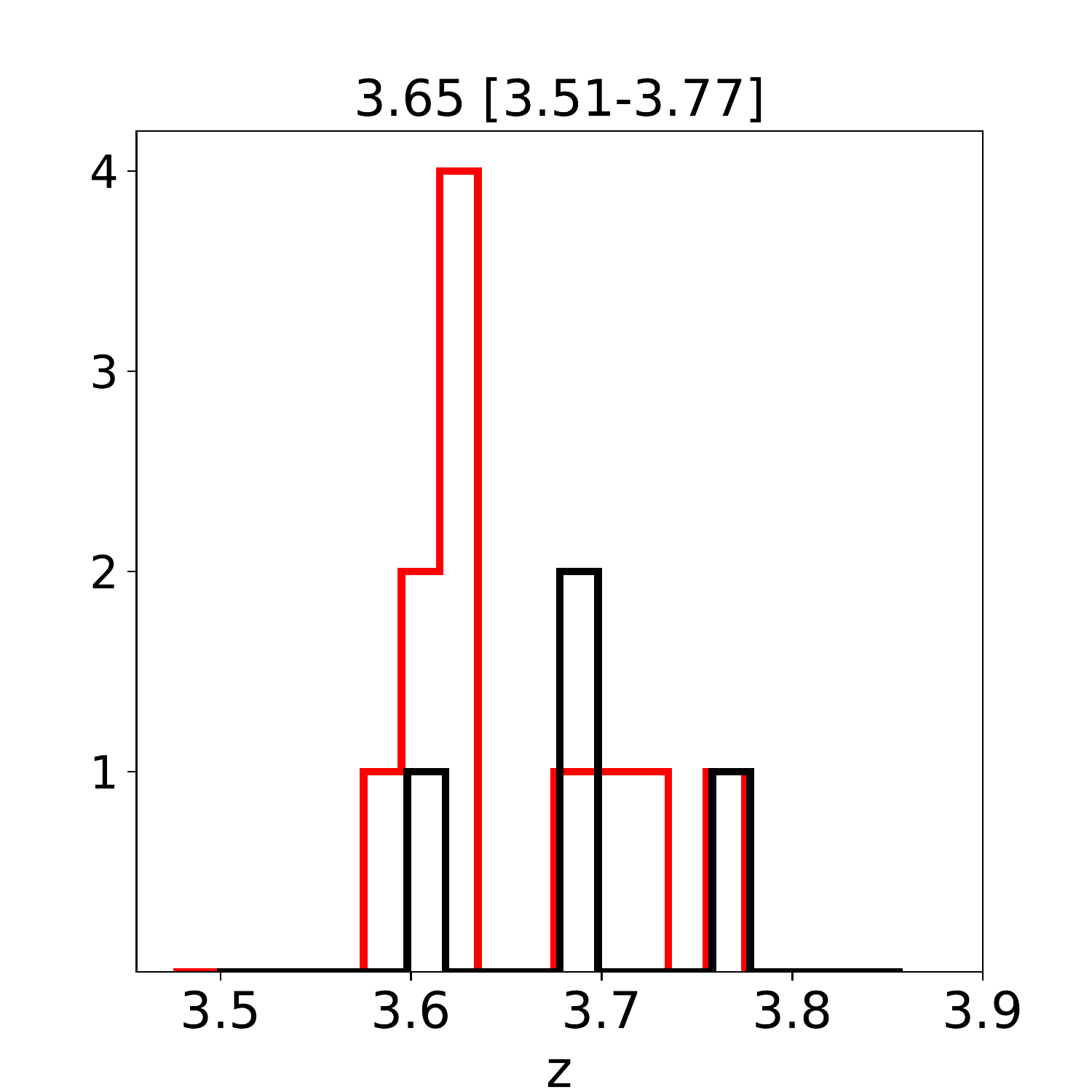}
 \includegraphics[width=5cm]{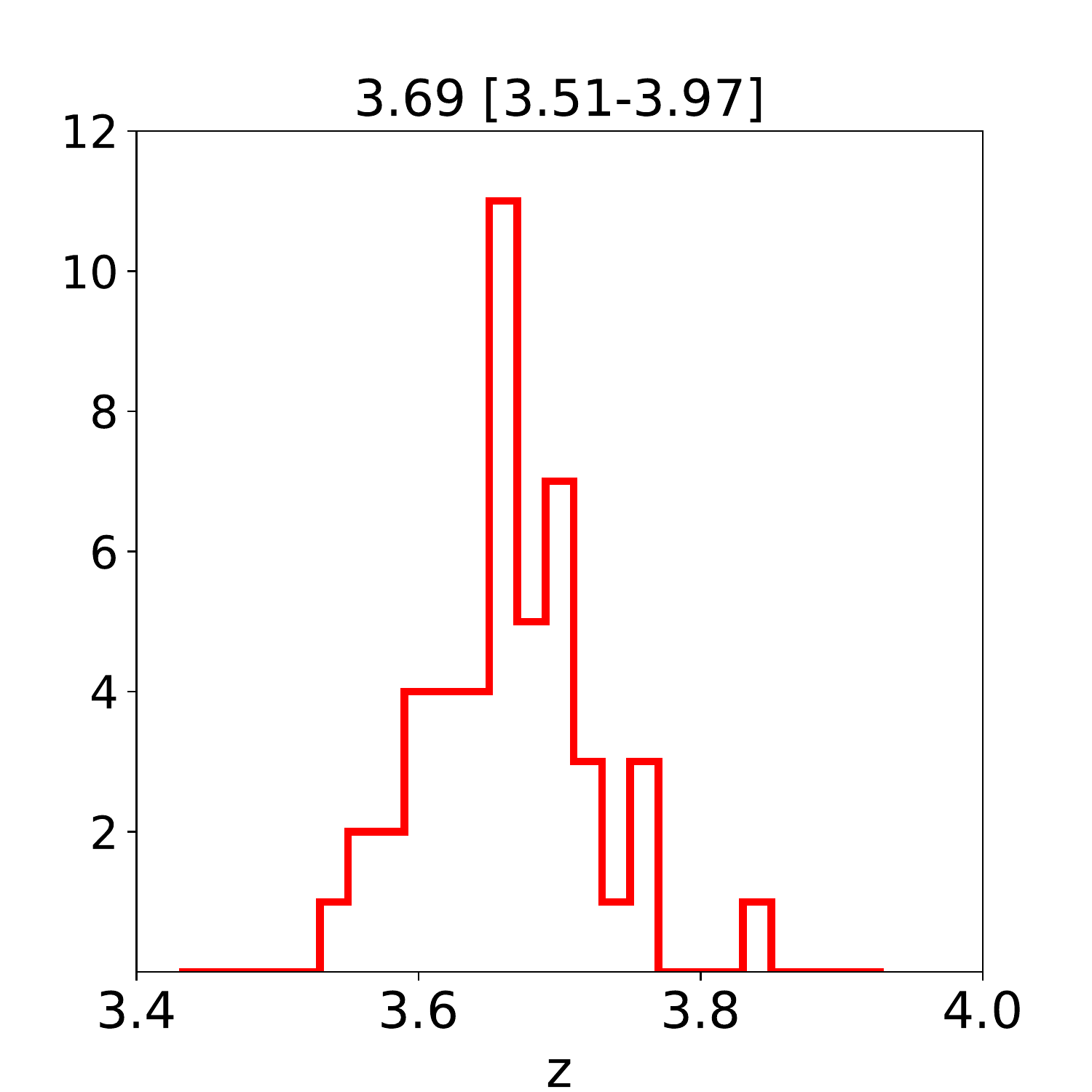}
 \includegraphics[width=5cm]{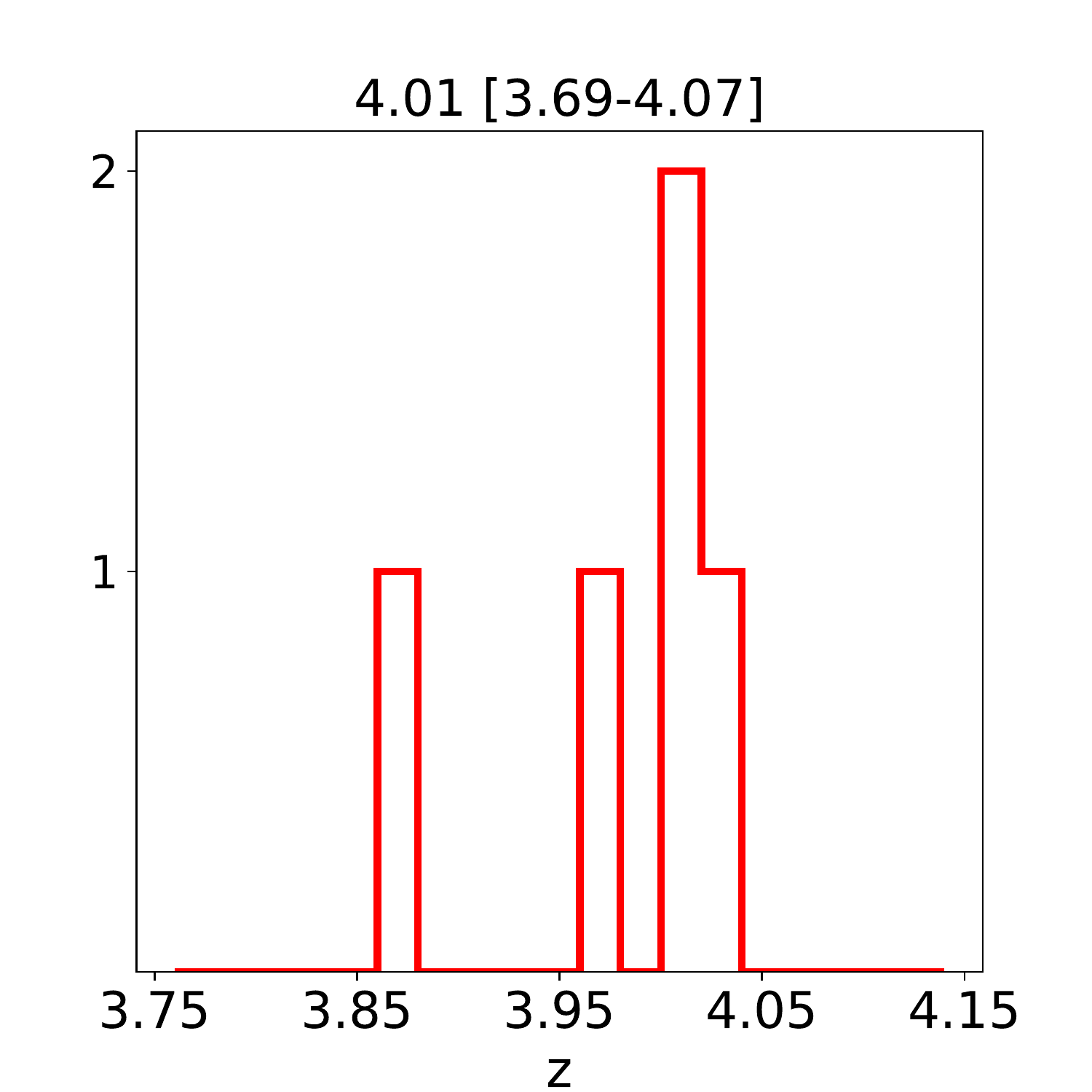}
\caption{Same as Fig. \ref{zdistrCDFS1} for 
%Redshift distribution of the galaxies in 
 the overdensities identified in the UDS. %The color coding is like in Fig. \ref{zdistrCDFS}. 
Some of the overdensities could be composed of more than one density peak, such as the structure at $z\simeq2.33$.
}
\label{zdistrUDS1}%
\end{figure*}

%\begin{figure*}
%\centering
%\includegraphics[width=5cm]{/Volumes/Backup/LUCIA/ChileProjects/ESOvisitingscientist/UDSstructureUPDATEDz365_RA345185833333dec-52355835_allzTH_zdistr.pdf}
 %\includegraphics[width=5cm]{/Volumes/Backup/LUCIA/ChileProjects/ESOvisitingscientist/UDSstructureUPDATEDz369_RA345427483333dec-520225016667_allzTH_zdistr.pdf}
 %\includegraphics[width=5cm]{/Volumes/Backup/LUCIA/ChileProjects/ESOvisitingscientist/UDSstructureUPDATEDz401_RA34329415dec-517225033333_allzTH_zdistr.pdf}
%\caption{Continuation of Fig. \ref{zdistrUDS1}.
%}
%\label{zdistrUDS2}%
%\end{figure*}

%With the scope of estimating the frequency of Ly$\alpha$ emission versus environment, we calculated local densities in CDFS and UDS, and we identified 22 overdensities at $2<z<4$ (Tab. \ref{tab:TabCDFSUDSstructures} and Appendix). 
We provide a qualitative description of the overdensities, leaving a more quantitative analysis for another paper. 
Generally, the overdensities we identify are composed of more than one peak in space and present typical sizes of the order of 1-4 cMpc, comparable with the sizes of the structures detected in the mock catalogs. However, the maximum distance between the members and the highest-density peak is of the order of 3 cMpc at both $z\sim2$ and $z\sim3$ on average. The size in redshift space, expressed as [c $\times$ stdev(z)]/[1+z$_{peak}$], is 4000 km sec$^{-1}$ at both $z\sim2$ and $z\sim3$ on average, even twice larger than the value calculated for mock galaxies. This can indicate that density peaks, overlapping in redshift space, tend to be interpreted as unique structures by our code, more than what was observed in the case of mock galaxies to which we applied photometric redshift uncertainty independent of environment.
%with a scatter of $0.02\times$(1+z). 
Spectroscopic follow-ups of overdensity members are needed to better delineate the redshift-space characteristics of the detected overdensities.

In Appendix C, we also report the Kolmog\'orov-Smirnov (KS) tests performed to evaluate if stellar masses, sSFRs, and rest-frame $U-V$ colors of the members and field galaxies are drawn from the same distribution. We usually find that we can reject the null hypothesis that these three physical properties of the members and field galaxies are drawn from the same distribution at least at 2$\sigma$ (we will highlight in the following the cases in which, instead, we can not reject the null hypothesis). 
 %2sigma per UDS z=3.65 , the others >30 sigma
%mass reject in 5/8 discussed structures, sSFR in 7/8 discussed structures, U-V in 4/8 discussed structures
However, we can not identify a red sequence from the rest-frame $U-V$ color of the members of any of our overdensities. %Also, the difference between the rest-frame $U-V$ color of the members and field galaxies is typically close to zero (see Sect. \ref{stage}). 
%%%%%%%%%%%For the overdensities for which we can reject the null hypothesis that the rest-frame $U-V$ colors of the members and field galaxies are drawn from the same distribution, the difference between the rest-frame $U-V$ color of the members and field galaxies is typically equal to $0.10\pm0.05$. Otherwise, the difference is close to zero ($0.00\pm0.05$ see Sect. \ref{stage}).

According to the value of the highest-density peak and the redshift of our overdensities, we find that they could be progenitor of Fornax-type clusters at $z=0$, which would virialize at $0.2<z<1$. We will highlight in the following the cases in which the total mass estimate is not consistent with the value expected for a Fornax-cluster progenitor at $z\sim2-4$ \citep{Chiang2013}

%In Appendix A, we show the regions occupied by our identified overdensities, together with the location of structures identified by \citet{KangIm2015, Franck2016} in the CDFS. %Usually, a high density region is identified by our code in correspondence of the structures from the literature. However, sometimes the significance of the overdensity is not enough to meet all our criteria, such as the number of members, and the overdensity is not in our list. %, starting with the ones presenting VANDELS Ly$\alpha$ emitters as members.
\begin{table*} %[h]
%\tabletypesize{\scriptsize}
%\rotate
\centering
\caption{Characteristics of the structures identified in the CDFS and in the UDS}
%\tablewidth{0pt}
\label{tab:TabCDFSUDSstructures}
\scalebox{0.78}{
\begin{tabular}{|c|c|c|c|c|c|c|c|c|} 
\hline
%the associated mases are from the FWHM of the radius and the mebers within it
$z_{peak}$ & RA & dec & N & Nspec &($U$-$V$)rest & $V$ rest  & N$_{field}$ &M$_{m}$ \\
&  (J2000)& (J2000) &  &  &      &   AB& &10$^{13}$M$_{\odot}$  \\
(1) & (2) &(3)  & (4)  & (5)  & (6)  &(7) &(8)&(9) \\ 
\hline
CDFS &  & &   &  &  & & &\\ 
\hline 
2.29$^a$ & 53.0629 & -27.7237 & 90 & 12 & 0.70$\pm$0.02 & -19.79$\pm$0.11 & 50 &0.9-0.8 \\
2.30$^a$ & 53.1412 & -27.6871 & 75 & 15 & 0.70$\pm$0.03 & -19.90$\pm$0.16 & 45 &1.0-0.9 \\
2.34 & 53.1421 & -27.8221 & 26 & 2 & 0.61$\pm$0.06 & -19.61$\pm$0.21 & 14 &0.4-0.3 \\
2.55 & 53.1612 & -27.9196 & 11 & 4 & 0.69$\pm$0.10 & -20.12$\pm$0.52 & 2 &0.04-0.03 \\
2.69 & 53.0529 & -27.8779 & 19 & 0 & 0.73$\pm$0.10 & -20.29$\pm$0.30 & & \\
2.80$^a$ & 53.2037 & -27.7746 & 45 & 4 & 0.55$\pm$0.04 & -20.18$\pm$0.15 & 11 & 0.4-0.3 \\
3.17$^a$ & 53.1412 & -27.8612 & 346 & 32 & 0.62$\pm$0.02 & -20.43$\pm$0.05 & &  \\
3.23 & 53.1346 & -27.6954 & 28 & 1 & 0.58$\pm$0.05 & -20.35$\pm$0.16 &  9 &0.4-0.3 \\
3.29 & 53.1229 & -27.7404 & 31 & 2 & 0.66$\pm$0.05 & -20.59$\pm$0.19 & 7&0.3-0.2  \\
3.43 & 53.0137 & -27.7388 & 10 & 6 & 0.62$\pm$0.15 & -21.58$\pm$0.46 & 2&0.05-0.03 \\
3.54 & 53.0985 & -27.8060 & 22 & 4 & 0.68$\pm$0.06 & -20.23$\pm$0.18 & 11 &1.3-1.0 \\
3.55$^a$ & 53.1187 & -27.8596 & 83 & 7 & 0.62$\pm$0.04 & -20.71$\pm$0.12 &  39 &2.4-2.0 \\
3.69 & 53.0712 & -27.6921 & 19 & 3 & 0.63$\pm$0.07 & -20.71$\pm$0.21 & 4 &0.4-0.3 \\
\hline
\hline
UDS &  & &   &  &  & & &\\
\hline
2.33$^a$ & 34.3819 & -5.1273 & 1280 & 165 & 0.74$\pm$0.01 & -19.92$\pm$0.04 & & \\
3.25 & 34.4552 & -5.2023 & 20 & 1 & 0.75$\pm$0.05 & -20.93$\pm$0.16 & 2 &0.3-0.2 \\
3.25 & 34.5203 & -5.1648 & 29 & 3 & 0.65$\pm$0.05 & -20.45$\pm$0.15 &  12 &0.4-0.3 \\
3.27 & 34.2602 & -5.2498 & 31 & 2 & 0.74$\pm$0.05 & -20.88$\pm$0.16 & 7&0.6-0.4 \\
3.49 & 34.4027 & -5.1648 & 7 & 0 & 0.73$\pm$0.13 & -20.50$\pm$0.32 &  &\\
3.51 & 34.3427 & -5.2423 & 10 & 0 & 0.55$\pm$0.08 & -20.40$\pm$0.32 & & \\
3.65 & 34.5186 & -5.2356 & 11 & 4 & 0.68$\pm$0.07 & -20.90$\pm$0.20 & 1 &0.2-0.1 \\
3.69 & 34.5427 & -5.2023 & 48 & 0 & 0.68$\pm$0.04 & -20.88$\pm$0.12 &  &\\
4.01 & 34.3294 & -5.1723 & 5 & 0 & 0.34$\pm$0.09 & -20.44$\pm$0.47 &  &\\
\hline
\hline
\end{tabular}
}
\tablefoot{%The $overdensity$ value is calculated for the volumes corresponding to the cores of the structures. 
(1) Redshift of the highest density peak, (2)(3) position of the highest density peak in RA and dec, (4) total number of structure members, (5) number of spectroscopic redshift among the members of the structures, (6) mean rest-frame $U-V$ colors of the members obtained from the rest-frame absolute magnitudes at 3700 {\AA} and 5500 {\AA}, outputs of the SED fits, (7) mean rest-frame $V$ magnitudes of the members calculated at 5500 {\AA}, also output of the SED fits, (8) number of field galaxies in a volume comparable to the one of the corresponding overdensity, (9) structure mass estimated with Equation \ref{Meq} and assuming a bias factor $b=2$ (left value) and $b=4$ (right value). For the overdensities composed of a large number of multiple peaks and for the overdensities with zero spectroscopic redshifts, we do not report the number of field galaxies or the mass estimates, since the presence of spectroscopic redshifts makes the identified structures more reliable. \\ %The $\sigma_{zspec}$ values are calculated only for the structures with more than one spectroscopic redshift. Their uncertainties correspond to errors of 0.002$\times$(1+z) $\times$2 on the zspec of the galaxies in the spectroscopic input catalog. The M$_{vir}$ value corresponds to the virial mass assuming that the structures are virialized. It is expressed as 3N$\pi$/2G $\times \sigma_{zspec}^2$ R$_{vir}$, where R$_{vir}$ is the Abell radius (172/zref) arcmin. {\bf{WRONG UNITS IN Mvir}}
$^{a}$ the structure could be composed of more than one substructure
%
%$z_{peak}$ & RA & dec & Nspec & N & overdensity & (UV-optical)rest & optical rest & c std(z)/(1+$z_{peak}$) & disp velocity & Mvir200\
%
%
}
\end{table*}
. \\

%{bf{ATT in 3D plot lower grid within +-0.05!!!}}
%{\bf{I think we should say something about every structure}}
%
%
%
%\begin{enumerate}
%\subsubsection{Overdensity at $z=2.29$ in CDFS}
%$\bullet$ pippo
%
$\bullet${\bf{Overdensity at $z=2.29$ in the CDFS}}\\
This is a large overdensity (in space and redshift), probably composed of more than one main peak (Fig. B1). 
As we can see in Fig.\ref{Auno}, a structure detected by \citet{KangIm2015} is included in our overdensity region. % to a dense region visible at $z\sim2.3$. However, the significance of the overdensity is not enough to meet all our criteria (such as the number of members) to be a structure. % of It is not , which however does not meet all our criteria to be listed as an identified overdensity. %{\bf{One of the structures identified by Kang\&Im at a similar redshift in in the same area }}
%Our algorithm estimates 90 members, among which 12 have spectroscopic redshifts. 
One of the main peaks is in the center of the entire overdensity. From this position, filamentary structures depart along the RA direction. %Galaxies with sSFRs much higher than the mean value of the overdensity are located in the outskirts (Appendix B1, D1). 

%The stellar mass, the sSFR, and the rest-frame $U$-$V$ color of the members are consistent with those of the galaxies in the field (Appendix C and D), but the members tend to be fainter in the rest-frame optical magnitude. However, 
%A KS test shows that we can reject the null hypothesis that stellar masses, sSFRs, and rest-frame $U-V$ colors of the members and field galaxies (Fig. C1) are drawn from the same distribution, %(p<D), but 
%and moreover 
%The members of this overdensity are on average 0.4 magnitudes fainter in the rest-frame optical. 
As we can see in Fig. D1, the bluest members with the highest sSFRs have morphologies consistent with disk galaxies. However, the members with the lowest sSFRs can have a variety of morphologies (Sersic index 0<n<5). Even if we can not identify a red sequence from the rest-frame $U-V$ color, there is a tail of field galaxies with $U-V$ colors redder than the members. 
%
%We cannot identify a red sequence from the rest-frame $U-V$ color. 
%The difference between the rest-frame $U-V$ color of the members and field galaxies is zero, but there is a tail of field galaxies with $U-V$ colors redder than the members. 
%We estimated a dispersion velocity over 1600 km sec$^{-1}$ with the gapper method and bootstrapping the errors. 
The mass of the overdensity derived from the matter density is of the order of $9\times 10^{12}$ M$_{\odot}$.\\
%Fornax & 0.2<z<1 & [2.5-4.0] & M<~Mmax
%According to the value of the peak overdensity and its redshift, this structure could be progenitor of a Fornax-type cluster at $z=0$, which would virialize at $0.2<z<1$. The mass we estimate is consistent with the value expected for a Fornax-cluster progenitor at $z\sim2$ \citep{Chiang2013}.\\
%
%
%\subsubsection{Overdensity at $z=2.30$ in CDFS}

$\bullet${\bf{Overdensity at $z=2.30$ in the CDFS}}\\
This overdensity overlaps with the one found in \citet{Salimbeni2009} at $z\sim2.3$ (Fig. \ref{Auno}). It is composed of one main density peak. A secondary peak is located to the north of the main one (Fig. B2).  %According to our algorithm, it contains 75 members, among which 15 have spectroscopic redshifts. 
There are galaxies with sSFRs much higher than the mean value towards the center and also in the outskirts of the overdensity (Fig. B2 and D1).

Stellar masses, sSFRs, and rest-frame $U-V$ colors of the members occupy the same parameter space as field galaxies (Fig. C1 and D1). A KS test shows that we can not reject the hypothesis that the distributions of rest-frame $U-V$ colors are drawn from the same distribution. % and the difference between the mean rest-frame $U-V$ color of overdensity and field galaxies is almost zero. 
The members with spectroscopic redshifts are more massive, brighter in the optical than the other members, and some of them are the reddest members in terms of rest-frame $U-V$ color. %However, we can not see a red sequence of the rest-frame $U-V$ color. 
The members with the highest sSFRs have a variety of morphologies in terms of Sersic index and axis ratio (Fig. D1). 
%
%A few members seem to align is a sort of red sequence if we adopt the method described in \citet{willis13} to calculate the red-sequence UV-optical color. However, they are a small fraction of the members. Also, some of the field galaxies show similar rest-frame UV-optical colors.
%We can not identify a red sequence of the rest-frame $U$-$V$ color. By adopting the gapper method, we estimate a dispersion velocity of about 350 km sec$^{-1}$. Also, 
The mass of the overdensity derived from the matter density is $1\times 10^{13}$ M$_{\odot}$.\\

$\bullet${\bf{Overdensity at $z=2.80$ in the CDFS}}\\
This overdensity is composed of one main peak (Fig. B6) and a tail. %It contains 45 members, among which 4 have spectroscopic redshifts. 
The galaxies composing the main density peak have a large spread in sSFR values and morphologies (Fig. B6 and D2).

Stellar masses, sSFRs, and rest-frame $U-V$ colors of the members expand the same range of values of the galaxies in the field (Fig. C2 and D2). %, but the members tend to be fainter in the rest-frame optical magnitude. 
A KS test shows that we can not reject the null hypothesis that those quantities are drawn from the same distribution.  %(Appendix C). The average value of the rest-frame $U$-$V$ color of the members is $\sim-0.05$ lower than the value in the field.
%We can not identify a red sequence from the rest-frame $U-V$ color and we calculate that the difference between the average rest-frame $U-V$ color of overdensity and field galaxies is -0.05 and 
%The members of the overdensity are also 0.05 magnitudes fainter than the field galaxies in the rest-frame optical.
%
%By adopting the gapper method, we estimate a dispersion velocity of above 1500 km sec$^{-1}$. Also, 
The mass of the overdensity derived from the matter density is about $4\times 10^{12}$ M$_{\odot}$.
%\ Fornax/Virgo & 0.2<z<1/0.7<z<1.6 & [2.5-4.0]/[3.4-5.7] & M<~Mmax
According to the simulations in \citet{Chiang2013},  this structure could be progenitor of a Fornax or Virgo-type cluster at $z=0$, which would not be virialized before $z=1.6$.\\
%
%\subsubsection{Overdensity at $z=3.17$ in CDFS}

$\bullet${\bf{Overdensity at $z=3.17$ in the CDFS}}\\
This overdensity occupies a large area of CDFS at $z\sim3.2$. Our algorithm identifies a region of the space with a high concentration of galaxies distributed around at least three main density peaks (Fig.\ref{zdistrCDFS2} and B7). %Among the 346 members of this overdensity, 34 galaxies have spectroscopic redshifts. %Since the 3 peaks could be close-by overdensities seen overlapping due to photometric redshift uncertainties, we do not comment about their global stage of evolution.  
Two of the Ly$\alpha$-emitting galaxies discussed in the next section are members of this overdensity.

The density peak at dec=-27.76$^{o}$ is mainly composed of galaxies with sSFRs $<6\times 10^{-9}$ yr$^{-1}$, rest-frame $U-V$ colors redder than the field galaxies, and morphologies consistent with that of elliptical galaxies (average Sersic index equal to 2.4; Fig. C2 and D2). Therefore, at that declination the member galaxies are in an evolved state with respect to field galaxies, despite the fact that the stellar mass distribution of the members and field galaxies can be drawn from the same distribution (KS=0.04, p=0.72). %, while the distributions of sSFR and rest-frame $U$-$V$ color . 

\begin{figure*}
 \centering
 \includegraphics[width=9cm]{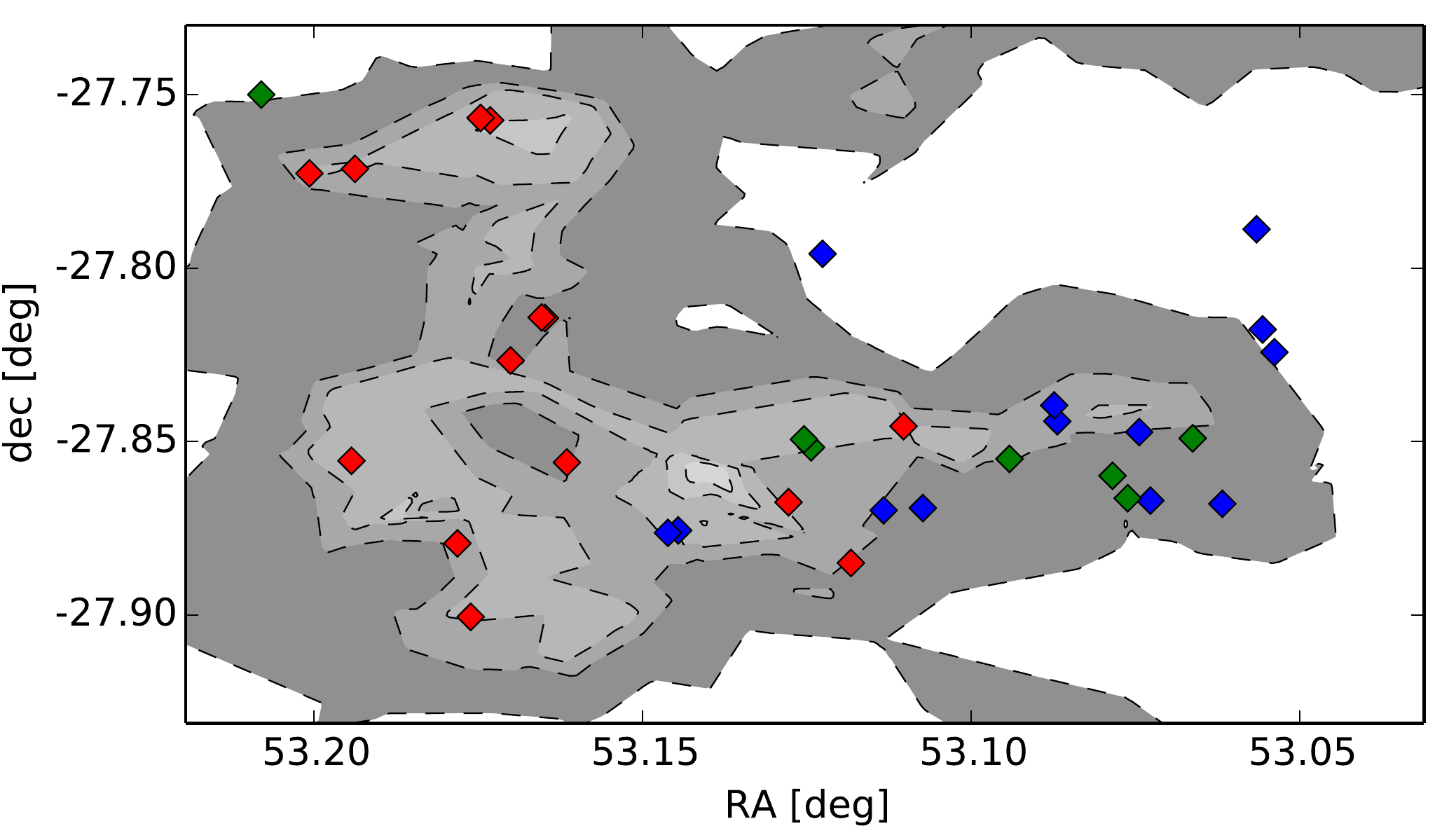}
 \includegraphics[width=9cm]{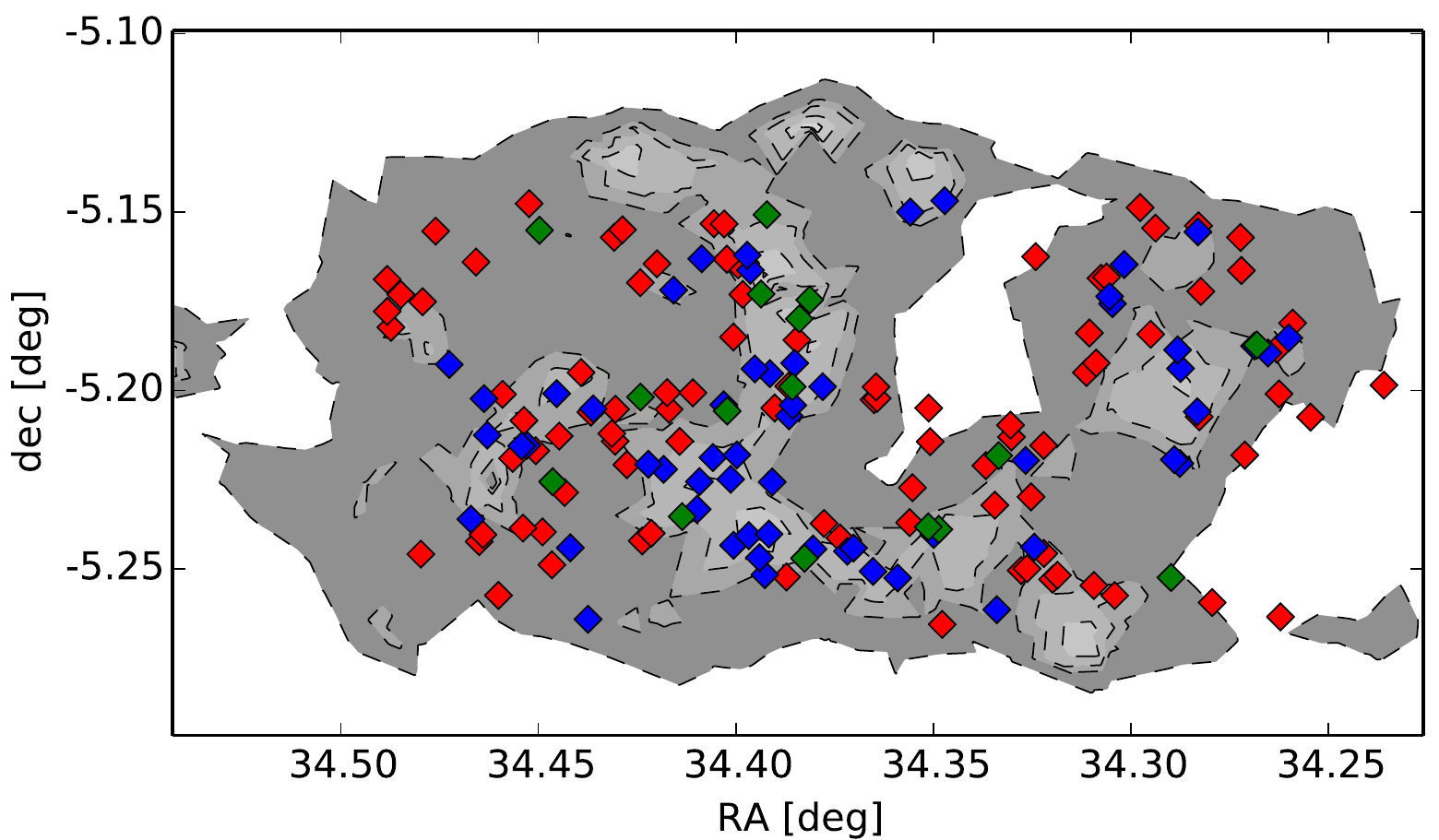}
\caption{Density map of two peculiar overdensities. %showing the structure members with spectroscopic redshifts. 
The structures at $z=3.17$ in the CDFS is shown in the $left ~panel$, the structures at $z=2.33$ in the UDS is shown in the $right ~panel$. Red diamonds correspond to the spectroscopic members within the lowest redshift peaks, blue diamonds to the intermediate redshift peaks, green diamonds to the highest redshift peaks as seen in Fig. \ref{zdistrCDFS2} and Fig. \ref{zdistrUDS1} and explained in the text. }
\label{mapzbins}%
\end{figure*}

The spectroscopic redshift distribution shows three peaks, at $3.0\leq z_{spec}<3.3$, at $3.3\leq z_{spec}<3.6$, and at $3.6\leq z_{spec}<3.9$, respectively composed of 14, 13, and seven sources. As we can see in Fig. \ref{mapzbins}, the lowest redshift peak traces the left side of the overdensity, while the other peaks mainly trace the right side of the overdensity. The highest-redshift peak could be a tail of random alignment. This could indicate that this overdensity is composed of two structures, unlikely to be connected because of the difference in redshift and so unlikely to evolve all together in a cluster at lower redshift.%on uncertain meaning.  
By using the bweight and gapper methods (Sec. 5.1), we estimate the dispersion velocity of the galaxies in the three redshift peaks. %The redshifts of the three peaks are $3.08\pm0.01$, $3.49\pm0.01$, and $3.69\pm0.01$. The dispersion velocities are $3621.4\pm1028.4$, $2938.3\pm663.6$, and $1553.7\pm294.7$ km sec$^{-1}$ if calculated with the bweight method and $1802.2\pm517.1$, $1751.0\pm699.8$, and $1412.8\pm520.8$ km sec$^{-1}$ if calculated with the gapper method.
The redshifts of the three peaks are $3.08\pm0.01$, $3.49\pm0.01$, and $3.69\pm0.01$. The dispersion velocities are $3620\pm1030$, $2940\pm660$, and $1550\pm290$ km sec$^{-1}$ if calculated with the bweight method and %$1800\pm520$, $1750\pm700$, and $1410\pm520$ km sec$^{-1}$ if calculated with the gapper method.
%{\bf{better: 
 $1200\pm290$, $1080\pm420$, and $1090\pm410$ km sec$^{-1}$ if calculated with the gapper method. The difference in these values is due to the different upper limits in velocity applied in the two methods. In an upcoming paper, we will study in detail the galaxies of this overdensity and its fate to lower redshift.\\ %}}

$\bullet${\bf{Overdensity at $z=3.54$ in the CDFS}}\\
This overdensity is composed of two peaks (Fig. B11), one of the two contains galaxies with very low sSFRs relative to the average. %The members tend to be redder and 0.5 magnitudes fainter in the rest-frame optical than field galaxies. 
%Our algorithm finds that the overdensity contains 22 members, among which 4 have spectroscopic redshifts. 
One of the Ly$\alpha$-emitting galaxies discussed in the next section is a member of this overdensity.
%
%A KS test shows that we can reject the null hypothesis that stellar masses, sSFRs, and rest-frame $U-V$ colors of the members and field galaxies are drawn from the same distribution. We can not identify a red sequence from the rest-frame $U-V$ color and The difference between the average rest-frame $U-V$ color of the overdensity and of the field galaxies is 0.12 (Fig. C3 and D3).
%
% By adopting the gapper method, we estimate a dispersion velocity of above 1100 km sec$^{-1}$. Also, 
The mass of the overdensity derived from the matter density is about $1\times 10^{13}$ M$_{\odot}$.\\
%\ Fornax & 0.2<z<1 & [3.5-5.2] & M<Mmax non diventera' cluster?
%The intensity of the peak overdensity and its redshift could indicate that this structure has the characteristics of the progenitor of a Fornax-type cluster at $z=0$, which would not virialize before $z\sim1$. The mass we estimate is consistent with the value expected for a Fornax-cluster progenitor at $z\sim3$ \citep{Chiang2013}.\\

%\subsubsection{Overdensity at $z=3.55$ in CDFS}
$\bullet${\bf{Overdensity at $z=3.55$ in the CDFS}}\\
This overdensity is elongated along the RA direction and it is composed of two main peaks (Fig. B12). 
%Our algorithm estimates it contains 83 members, among which 7 have spectroscopic redshifts. 
One of the Ly$\alpha$-emitting galaxies discussed in the next section is a member of this overdensity. 
The overdensity overlaps with the structure found by \citet{Forrest2017}, detected thanks to the discovery of several [OIII]+H$\beta$-emitting galaxies (Fig. A4). Members with sSFR values larger than the average are located either in the core or in the outskirts of the overdensity (Fig. B12 and D3). 
%
%A KS test shows that we can reject the null hypothesis that the stellar masses, sSFRs, and rest-frame $U-V$ colors of the members and of the field are drawn from the same distribution. We cannot identify a red sequence from the rest-frame $U-V$ color (Fig. C3). The difference between the average rest-frame $U$-$V$ color of the overdensity and of the field galaxies is 0.1.
%The members tend to be redder in the rest-frame $U$-$V$ color than the galaxies in the field. 
%
%By adopting the gapper method, we estimate a dispersion velocity of above 1700 km sec$^{-1}$. 
The mas of the overdensity derived from the matter density is about $2\times 10^{13}$ M$_{\odot}$.\\
% \ Fornax & 0.2<z<1 & [3.5-5.2] & M~Mmax 
%This structure could also be progenitor of a Fornax-type cluster at $z=0$, which would not virialize before $z\sim1$.\\
%%%%dire redshift range, allora Forrest e anche Kang a 3.7
%\subsubsection{Overdensity at $z=3.69$ in CDFS}
%
%This overdensity is composed of one main peak, in which sSFR is higher in the outskirts (Appendix B).
%Our algorithm estimates it contains 19 members, among which 3 have spectroscopic redshifts. 
%
%The stellar mass, the sSFR, and the rest-frame $U$-$V$ color of the members are consistent with those of the galaxies in the field (Appendix C and D). However, the average rest-frame $U$-$V$ color of the members is almost 0.1 magnitude higher than the average of the field galaxies. 
%
%By adopting the gapper method, we estimate a dispersion velocity of about 90 km sec$^{-1}$. Also, The mas of the overdensity derived from the matter density is about 1.5E+13 M$_{\odot}$.
%
%\subsubsection{Overdensity at $z=2.33$ in UDS}

$\bullet${\bf{Overdensity at $z=2.33$ in the UDS}}\\
This overdensity occupies a $6'x15'$ area of the UDS at $z\sim2.3$ (Fig. A5). Our algorithm identified a region of space with a high concentration of galaxies distributed around four main density peaks (Fig. B14). %sAmong the 1280 members of this overdensity, 165 galaxies have spectroscopic redshifts. 
It is not entirely unusual to identify overdensities of this extent. \citet{Balestra2010} also discovered several structures in the central part of CDFS, such as those at $z\simeq2.3$ and $z\simeq2.6$, spatially extended over their entire surveyed area (15 Mpc).

A variety of physical properties characterize the galaxies of the overdensity. %In particular the stellar masses, sSFRs, and rest-frame $U-V$ colors are consistent with those of field galaxies. 
%A KS test shows that we can reject the null hypothesis that the physical parameters of the members are drawn from the same distribution as those for the field galaxies. 
%Since the high-density peaks could be close-by structures seen overlapping due to photometric redshift uncertainties, we do not comment about their global stage of evolution.  
The density peak at dec=-5.18$^{o}$ is mainly composed of galaxies with sSFRs $<3\times 10^{-9}$ yr$^{-1}$ (the average value among all the members), but a variety of morphologies (Fig. B14 and D4). The members with spectroscopic redshifts are the brightest and most massive (Fig. C4) among all. 

The spectroscopic redshift distribution shows three peaks, at $2.00\leq z_{spec}<2.27$, at $2.27\leq z_{spec}<2.37$, and $2.37\leq z_{spec}<2.50$, respectively composed of 92, 57, and 16 sources. As we can see in Fig. \ref{mapzbins}b, the spectroscopic redshifts roughly trace the position of the highest-density regions. In particular, the intermediate and highest redshift peaks trace the overdensity peak at RA = 34.4$^{o}$.  Therefore, the identified overdensity could be composed of a main structure at $z\simeq 2.3$, a filament at lower redshift, and a tail at higher redshift.

By using the bweight and gapper estimators (Sec. 5.1), we estimate the dispersion velocity of the galaxies in the thee main redshift peaks. %The redshifts of the three peaks are $2.188\pm0.004$, $2.305\pm0.003$, and $2.401\pm0.006$. The dispersion velocities are $3705.1\pm188.3$, $1957.7\pm212.5$, and $2339.5\pm479.7$ km sec$^{-1}$ if calculated with the bweight estimator and $3309.4\pm387.8$, $2076.2\pm318.5$, and $1558.4\pm404.4$ km sec$^{-1}$ if calculated with the gapper estimator. 
The redshifts of the three peaks are $2.188\pm0.004$, $2.305\pm0.003$, and $2.401\pm0.006$. The dispersion velocities are $3700\pm190$, $1960\pm210$, and $2340\pm480$ km sec$^{-1}$ if calculated with the bweight estimator,  
%and $3310\pm390$, $2080\pm320$, and $1560\pm400$ km sec$^{-1}$ if calculated with the gapper estimator.
%{\bf{better: 
$2050\pm210$, $1070\pm140$, and $900\pm210$ km sec$^{-1}$ if calculated with the gapper estimator.\\ %}}
%To estimate The mas of the overdensitys contained in these three peaks, we assume the virial theorem and the median distance between the members. We obtain $9\times 10^{18}$, $2\times 10^{18}$, and $3\times 10^{17}$ M$_{\odot}$. These values are upper limits since the peaks are unlikely to be virialized.\\
%
%\subsubsection{Overdensity at $z=3.25$ and RA $=34.45$ in UDS}
%
%This overdensity is composed of one main peak (Appendix B). 
%Our algorithm estimates it contains 20 members, one with spectroscopic redshifts. The members with sSFR more than twice larger than the average value are located in the outskirts of the overdensity, but they have a variety of morphologies. The KS test shows that we can not reject the null hypothesis that stellar mass and sSFR of the members and of the field galaxies are drawn from the same distribution (Appendix C and D). 
%
%We can not identify a red sequence of the rest-frame $U$-$V$ color. However, seven galaxies have a rest-frame $U$-$V$ color of $\sim0.65$ and the average color of the members is 0.1 magnitude redder than the field galaxies. Therefore, for this overdensities we can see a beginning of evolution of its members.
%
%By adopting the gapper method, we estimate a dispersion velocity of above 1500 km sec$^{-1}$. Also, The mas of the overdensity derived from the matter density is about 1E+13 M$_{\odot}$.
%\subsubsection{Overdensity at $z=3.25$ and RA $=34.52$ in UDS}

$\bullet${\bf{Overdensity at $z=3.25$ and RA $=34.52^{o}$ in the UDS}}\\
This overdensity is composed of one main peak (Fig. B16). 
%Our algorithm estimates it contains 29 members, 3 with spectroscopic redshifts. 
The members with sSFRs more than twice above the average are located in the outskirts of the overdensity, and have a variety of morphologies (Fig. B16 and D4). 

The stellar masses and sSFRs are consistent with those of the galaxies in the field. However, a KS test shows that we can not reject the null hypothesis that the distribution of rest-frame $U-V$ colors of the members and field galaxies are the same (Fig. C4 and D4). %We cannot identify a red sequence from the rest-frame $U-V$ color and the average rest-frame $U-V$ colors of the members and field galaxies are consistent.
Two of the Ly$\alpha$-emitting galaxies discussed in the next section are members of this overdensity. Interestingly, their physical properties are consistent with the average properties of the other members. 
%
%By adopting the gapper method, we estimate a dispersion velocity of above 440 km sec$^{-1}$. 
The mass of the overdensity derived from the matter density is $0.4\times 10^{13}$ M$_{\odot}$.\\
%Fornax & 0.2<z<1 & [3.5-5.2] &  M<~Mmax
%This structure could be progenitor of a Fornax-type cluster at $z=0$ and its mass is consistent with that expected for the progenitor at $z\sim3$.\\
%\subsubsection{Overdensity at $z=3.27$ in UDS}

$\bullet${\bf{Overdensity at $z=3.27$ in the UDS}}\\
This overdensity is composed of one main density peak with a tail (Fig. B17) in the RA-dec plain. 
%Our algorithm estimates it contains 31 members, among which two have spectroscopic redshifts. 
One of the Ly$\alpha$-emitting galaxies discussed in the next section is a member of this overdensity. Galaxies with sSFRs lower than the mean value of the overdensity are located either in the peak or in the tail and they tend to have morphologies consistent with disk-like galaxies (Sersic index n<2, Fig. B17 and D5).
%
%The stellar mass and the rest-frame $U$-$V$ color of the members are consistent with those of the galaxies in the field (Appendix C and D). 
%A KS test shows that we can reject the null hypothesis that the stellar masses and rest-frame $U$-$V$ colors of the members and field galaxies are drawn from the same distribution (Fig. C5). The average rest-frame $U$-$V$ color of the members is 0.09 magnitudes redder than that of field galaxies and 
%The members are on average 0.3 magnitudes brighter in the optical than the field galaxies. % We can not identify a red sequence of the rest-frame UV-optical color. 
%
%By adopting the gapper method, we estimate a dispersion velocity of above 400 km sec$^{-1}$. 
The mass of the overdensity derived from the matter density is about $5\times 10^{12}$ M$_{\odot}$.\\

$\bullet${\bf{Overdensity at $z=3.65$ in the UDS}}\\
This overdensity is composed of one main peak (Fig. B20).
%Our algorithm estimates it contains 11 members, among which 4 have spectroscopic redshifts. 
Two of the Ly$\alpha$-emitting galaxies discussed in the next section are members of this overdensity. One of them is located in the outskirts of the structure. Also, the two members with sSFRs more than twice above the average value are located in the outskirts of the overdensity.

%A KS test indicates that we can reject the null hypothesis that the sSFRs and rest-frame $U$-$V$ colors of the members and field galaxies are drawn from the same distribution (Fig. C5). The average rest-frame $U$-$V$ color of the members is 0.1 magnitudes redder than that of the field galaxies and 
%The members are 0.1 o average magnitude brighter in the optical than the field galaxies.
% We cannot identify a red sequence from rest-frame $U-V$ colors redder than those of field galaxies. %However, eight out of 11 members have a similar color of $U-V = 0.7\pm0.1$. 
%
%By adopting the gapper method, we estimate a dispersion velocity of above 500 km sec$^{-1}$. Also, t
The mass of the overdensity derived from the matter density is about $1\times 10^{12}$ M$_{\odot}$.
%Coma & 1.5-2.3 &  [6.8-9.5] & M<Mmax non diventera' cluster?
 The intensity of the peak overdensity and its redshift indicate that this structure could be progenitor of a massive Coma-type cluster at $z=0$, which would virialize at $1.5<z<2.3$. However, the mass we estimate at the current redshift is much lower than the value expected for a Coma-cluster progenitor at $z\sim3.6$ \citep{Chiang2013}. This could indicate that this structure will not be able to actually assemble at $z<3$. We do not have enough supporting evidence to evaluate if the structure will fragment and evolve in several substructures. %before diluting in space.

\subsection{Stage of evolution of the galaxies in the identified overdensities}
\label{stage}

Associations of galaxies at $z>2$ have been identified in simulations \citep{Chiang2013,Muldrew2015} and in observations \citep{Castellano2007,Salimbeni2009,KangIm2015,Franck2016}, and have shown a variety of properties (e.g., size, mass, number of members) and stages of evolution. %First of all, the progenitors of the most massive clusters at $z=0$ can occupy large areas on the sky ($>$10 cMpc on a side)  and can be composed of more than one density peak \citep[][]{Muldrew2015}. Also, they can be already quite massive (more than 10$^{14}$ M$_{\odot}$) and characterized by velocity dispersions (900 km sec$^{-1}$) larger than the values estimated by simulations \citep[][]{KangIm2015,Franck2016}. 
Overdensities could be composed of massive and evolved galaxies, but we could be missing a given galaxy population, depending on the identification method and when our sample only includes a biased type of tracers. 
Moreover, to better identify overdensities and determine their structures, it is desirable to have a high spectroscopic coverage in observing fields larger than 10 cMpc on a side \citep[][]{Muldrew2015}. 

%In this work we focused on the CANDELS catalogs of CDFS and UDS and on the unique spectroscopic dataset provided by VANDELS. We have identified 22 overdensities at $2<z<4$ and described qualitatively their properties in Sect. \ref{structurecdfsuds}. 
Some of the overdensities identified in this work are characterized by properties expected for high-density regions in the local Universe, such as members with low sSFRs in the cores and morphologies consistent with those of elliptical galaxies. %(as described in the previous section)
However, the majority of the overdensities do not show rest-frame $U-V$, $V-J$ colors, and optical magnitudes typical of virialized clusters at $z<1$ \citep{willis13}. 

\begin{figure*}
 \centering
\includegraphics[width=20cm]{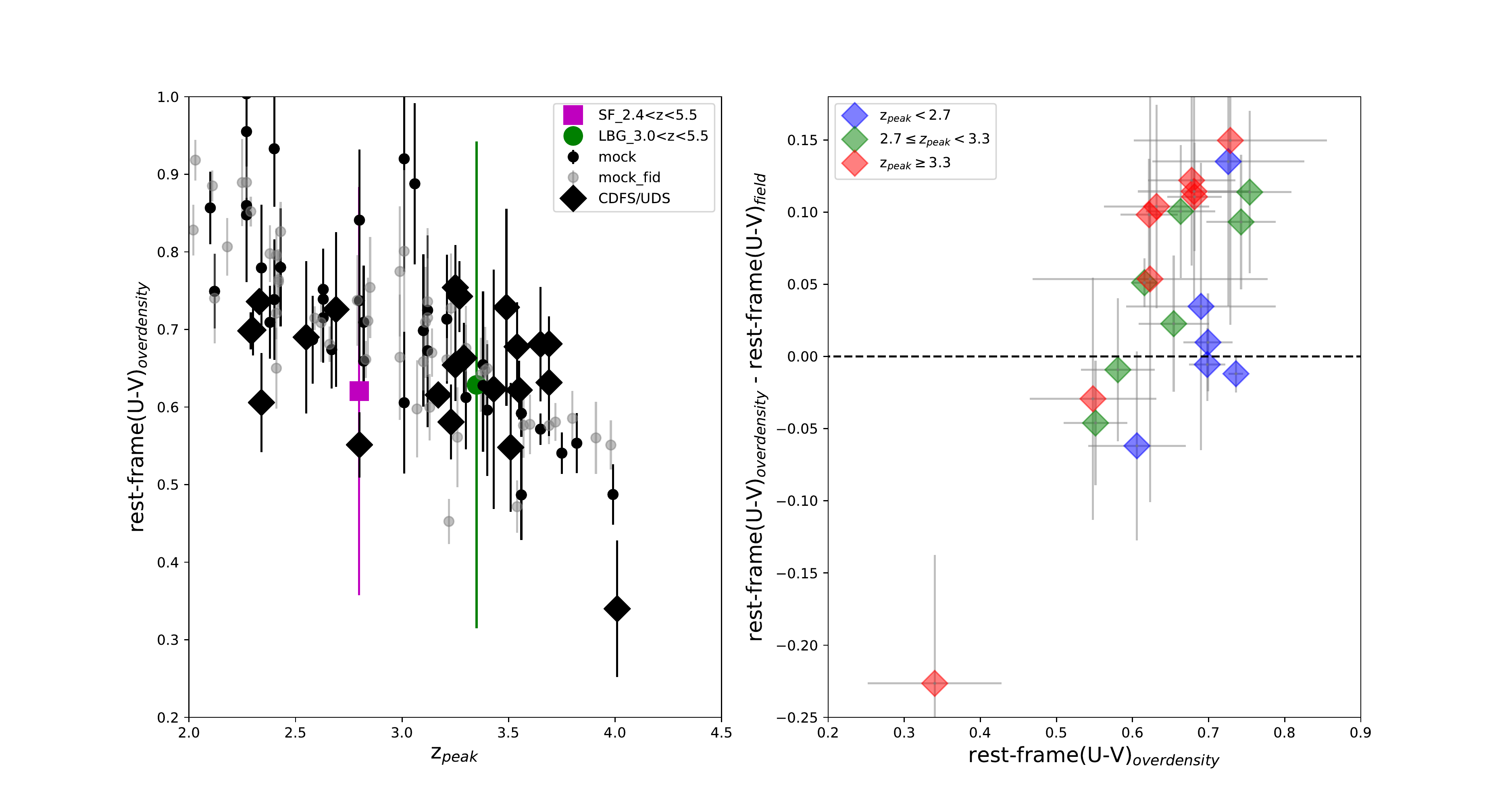}
\caption{$Left ~ Panel$: Mean rest-frame $U-V$ color of the galaxies in the 22 overdensities identified at $2<z<4$ versus redshift of their highest-density peaks (big black diamonds).
. %The structures for which we have at least 2 spectroscopic redshifts are shown as hexagons, for which we have only 1 or even no spectroscopic redshifts with diamonds. For comparison we show as small dots the mean $J-K$ colors versus $K$ magnitudes for the galaxies at the same redshift of the corresponding structures but with density consistent with the field density. 
We, also, show mean color and mean photometric redshift for star-forming galaxies (big magenta squares) and Lyman Break galaxies (big green circles) as classified in \citet{Pentericci2018} and for the structures detected in the mocks, in the fiducial run (small gray circles) and in the run with parameters as in the real data (small black circles).   %not strong above mean
%It is worth noting that the $J$ and $K$ magnitudes come from CFHT WIRCAM filters for CDFS and from UKIRT WFCAM for UDS. The WFCAM $K$ filter is slightly redder than the WIRCAM one, while the $J$ filters occupy the same wavelength range.
$Right ~ Panel$: Difference between the mean rest-frame $U-V$ color of the overdensity members and the mean rest-frame $U-V$ color of field galaxies versus the mean rest-frame $U-V$ color of the overdensity members. The color coding indicates the redshift of the highest-density peak of the identified overdensities, blue for $z<2.7$, green for $2.7\leq z\leq 3.3$, and red for $z>3.3$. The error bars correspond to the standard error of the means. As a reference, the rest-frame $U-V$ color for a representative cluster at $z=0$ is equal to 1.4 and that of field galaxies is 0.5. To derive these values, we assume that a $z=0$ cluster contains a well defined red sequence with a dominance of E and S0 galaxies and that field galaxies are a blue population \citep{Fukugita1995}.
}
\label{figstage}%
\end{figure*}

To investigate the stage of evolution of the galaxies in the overdensities we detected at $2<z<3$ and at $3<z<4$, we show the average rest-frame $U-V$ color (rest-frame $(U-V)_{overdensity}$) as a function of redshift and the difference between the average color of the overdensity members and field galaxies (rest-frame $(U-V)_{overdensity}$ - rest-frame $(U-V)_{field}$) as a function of rest-frame $(U-V)_{overdensity}$ (Fig. \ref{figstage}).
The rest-frame $(U-V)_{overdensity}$ values are broadly consistent with the typical values of star-forming galaxies at $z\sim2.8$ and Lyman break galaxies at $z\sim3.4$ \citep{Pentericci2018}. %{\bf{cut?}}However, for passive galaxies at $z\sim1.3$ the color is much larger, on average equal to 1.7.  
The rest-frame $(U-V)_{overdensity}$ color is independent of redshift for the structures studied here and it is of the order of 0.7 with a large scatter. It is worth noting that the colors of the structures detected in the CDFS and in the UDS are comparable with the values calculated for the overdensities detected in the mocks.%We do not observe a trend of color with redshift {\bf{say better}}. 

The right panel of Fig. \ref{figstage} shows that the typical galaxy members are not systematically redder than field galaxies. The difference between the rest-frame $U-V$ color of the overdensity and its associated field galaxies %_{overdensity}$ - rest-frame $U-V_{field}$ quantity 
is usually much less than 0.2, with a typical uncertainty of 0.06. For half of the overdensities, it is consistent with zero (less than 0.05, with a typical uncertainty of 0.05). There are ten overdensities for which the difference is $0.11\pm0.07$ on average. %rest-frame $U-V_{overdensity}$ - rest-frame $U-V_{field}=0.1-0.15$. 
There is just one overdensity (the one at $z=4.01$ in the UDS without spectroscopic redshifts) for which it is $-0.23\pm0.09$. Therefore, the typical overdensity members are not that different from the field galaxies in terms of rest-frame $U-V$ color. This is reflected in the properties discussed in the previous section as well.

\section{Ly$\alpha$ emitters and environment}
\label{LAEs}

%We build a density grid in bins of 0.1 redshifts and 
%{\bf{non mi e' chiaro se per te i LAE sono solo i vandels o anche le altre categorie
%
%se sono solo i vandels allora quello che fai dopo ha senso senno no...}}
%We study the location of the Ly$\alpha$ emitting galaxies detected in the current VANDELS data with respect to the environment. % and in a few surveys from the literature. %with respect to dense regions. 
%For each 0.1-redshift bin, we estimated the mean and standard deviation values ($\rho_{m}$ and $\sigma_{m}$) of the densities, $\rho$, associated to the galaxies in that redshift bin.  

Useful insights on the properties of galaxies either in dense or in sparse environments could be provided by their Ly$\alpha$ emission. 
The Ly$\alpha$ emission of a star-forming galaxy is related to its recent star formation and it is sensitive to the interstellar (ISM) and circumgalactic medium (CGM) properties, such as HI content and distribution. The confinement or stripping of HI gas due to environmental effects could condition the evolution of the galaxy and also affect the distribution of its Ly$\alpha$ photons. 

\begin{figure*}
 \centering
\includegraphics[width=12cm]{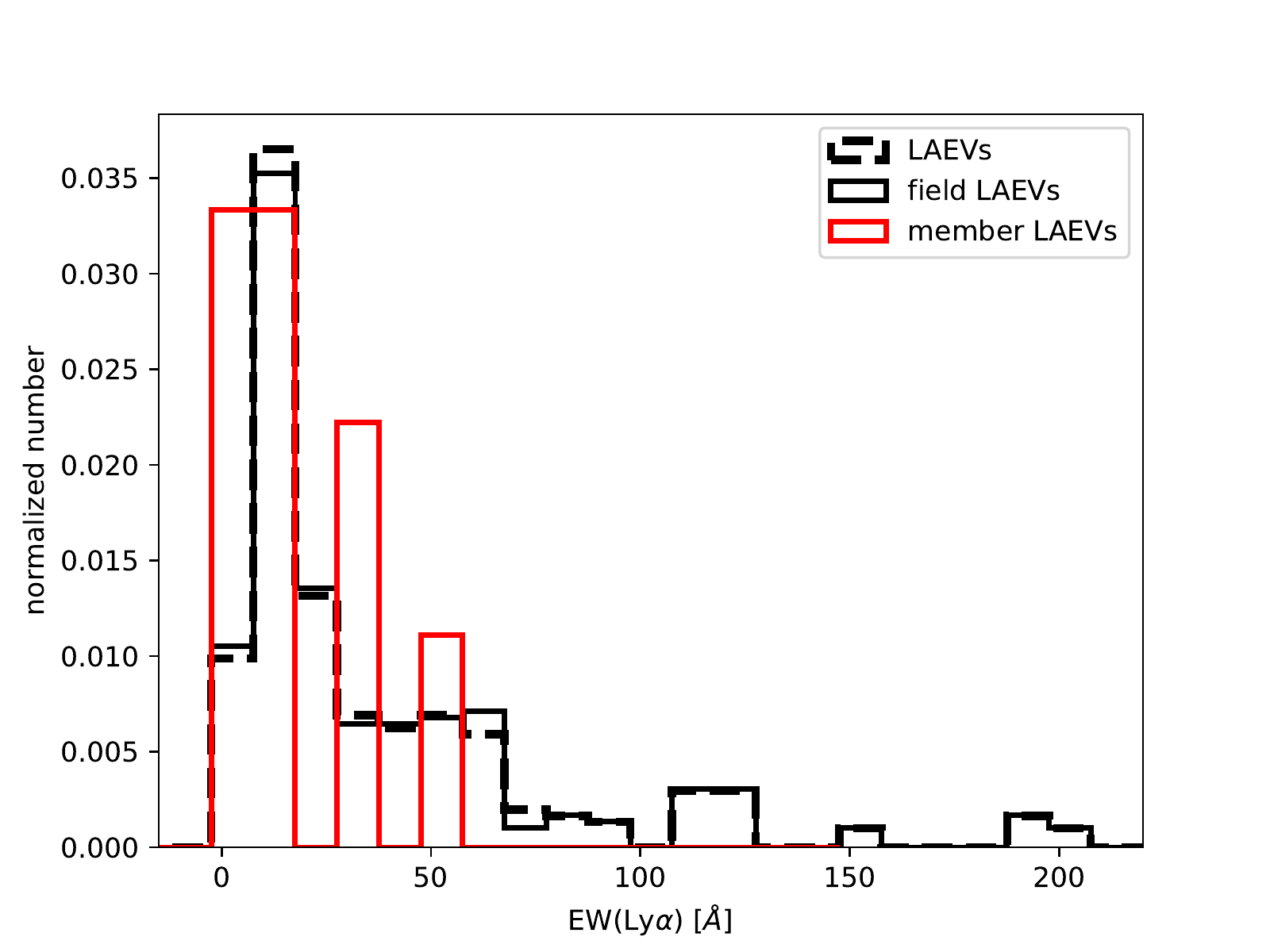}
\caption{Normalized Ly$\alpha$ equivalent width distribution of the Ly$\alpha$-emitting galaxies selected from the VANDELS database (LAEVs). The dashed histogram corresponds to the entire sample of LAEVs. The thin black and red histograms are the distributions for the LAEVs located in the field and members of the detected overdensities (as explained in the text), respectively.
}
\label{EWdistr}%
\end{figure*}

%We select a sample of Ly$\alpha$ emitting galaxies from the current VANDELS archive (LAEVs) In the following section we study the location of an unbiased sample of Ly$\alpha$ emitting galaxies from VANDELS to investigate if their properties can be used as tracers of galaxy properties in overdense regions and if Ly$\alpha$ emitters can be used to trace overdensities \citep{Muldrew2015}. 
%
We select a sample of Ly$\alpha$-emitting galaxies from the current VANDELS archive (LAEVs) and study their location with respect to the environment. As mentioned in Sect. 2, a LAEV is defined as a galaxy with EW(Ly$\alpha) >0$ {\AA} measured in its VANDELS spectrum \citep[e.g.,][]{Pentericci2018, Pentericci2018c, Marchi2019}. 
We estimate the EW(Ly$\alpha$) when the Ly$\alpha$ emission line is detected with a signal-to-noise ratio above 1.5 (to assure a large enough variety of Ly$\alpha$ strengths), assuming an asymmetric Gaussian-profile fit of the emission line above the continuum \citep[see Equation 3 in][]{Guaita2017} and considering the flux density of the continuum between 1300 and 1400 {\AA} rest-frame. The error bar on the EW(Ly$\alpha$) measurement is obtained propagating the error on the integrated flux of Ly$\alpha$ and the rms of the continuum, where the error on the integrated flux is calculated from the equation given by \citet{Lenz1992} for a Gaussian fit. The error bars are larger for the Ly$\alpha$ emission lines detected with signal-to-noise ratio just above 1.5.
In the CDFS (UDS) we consider %51(80)
51 (80) galaxies at $3<z<4$ with $1<$ EW(Ly$\alpha) <200$ {\AA}. Their Ly$\alpha$ equivalent width distribution is shown in Fig. \ref{EWdistr}. %these are the ones with a match in the weighted catalog and that are in the LAElocation file
%
%86(85)  totali
%
Due to their initial selection from drop-out galaxies, they do not all show large values of Ly$\alpha$ equivalent widths, as it is the case of narrow-band selected Ly$\alpha$ emitters \citep[e.g.,][]{Guaita2010}. The median EW(Ly$\alpha$) of the LAEVs in overdense regions is 12 {\AA} and of the LAEVs in the field is 17 {\AA}. 
%
%Also, we consider the Ly$\alpha$ emitters (LAEs) discovered through narrow-band technique by \citet{Zheng2016} (NB466 at $z\sim2.82$ and NB470 at $z\sim2.85$), by the MUSYC collaboration \citep[][at $z\sim3.1$]{Gronwall:2007, Ciradullo2012}, and those discovered in the MUSE-Wide survey and studied in \citet{Herenz2017, Gronke2017}, all in CDFS.
%{\bf{????}}We considered 12 and 24 Ly$\alpha$ absorbers from VANDELS as well. 
%These LAEVs were chosen to be part of the VANDELS masks \citep{Pentericci2018}, as the  %, and so brighter 

Our LAEVs are VANDELS targets \citep{Pentericci2018}, they are galaxies with z$_{spec}$ from VANDELS, and their location was not  preselected in terms of density or environment. % not biased in terms of density of the environment. 
To study the location of galaxies with and without Ly$\alpha$ in emission with respect to environment and to study the relation between Ly$\alpha$ emission and environment, we compare the positions and physical properties of the LAEVs with those of the galaxies without Ly$\alpha$ emission all from VANDELS to avoid sample inhomogeneity (see Sect. 2.1). 

To do this, we define three subsamples of galaxies: (a) all the galaxies in the input catalog (with either photometric or spectroscopic redshift, z$_{photspec}$); (b) the galaxies with spectroscopic redshifts from VANDELS, but without Ly$\alpha$ emission in their spectra (z$_{spec}$V); (c) the LAEVs. Moreover, we define four different density category, and we measure the fraction of z$_{photspec}$, z$_{spec}$V, and LAEVs in each environment with respect to the total number in a certain redshift bin.
%and four density categories. Then, we measure the fraction of z$_{photspec}$, z$_{spec}$V, and LAEVs in each density category with respect to the total number in a certain redshift bin.  

The four density categories are defined as follows. The field (F) category is composed of galaxies with density within 3$\sigma$ of the average local density and in the redshift bin of an identified overdensity; the transition (T) category is composed of galaxies with local density in between the field and the overdensities in the same redshift bin; %We call these density regions as the $transition$ regions (T); 
the overdensity (O) category is composed of galaxies with density values consistent with an overdensity (6$\sigma$ above the average local density), but that are not members of the identified overdensities, in the same redshift bin; and the overdensity-member (M) category is composed of galaxies that are members of the identified overdensity. The OM category is composed of galaxies in the O and M categories. %[{\bf{O and M are summed up in the table, however O is even lower than M}}].

We consider the redshift bins of the identified overdensities and we count the number of z$_{photspec}$, z$_{spec}$V, and LAEVs in each redshift bin. %(z$_{photspec\_zbin}$, z$_{spec}$V$_{zbin}$, and LAEV$_{zbin}$). 
Then we count the number of galaxies per density category in each redshift bin. %(e.g., z$_{spec}$V\_F$_{zbin}$, z$_{spec}$V\_T$_{zbin}$, z$_{spec}$V\_O$_{zbin}$, z$_{spec}$V\_M$_{zbin}$). 
Finally, we measure the fraction of galaxies of a certain density category in a certain redshift bin %(zbin1, ..., zbinN) 
as the ratio of the previously defined numbers. %, \\
%fraction\_zphotzspec\_F$_{zbin1}$=zphotzspec\_F$_{zbin1}$/zphotzspec$_{zbin1}$\\
%fraction\_zphotzspec\_F$_{zbinN}$=zphotzspec\_F$_{zbinN}$/zphotzspec$_{zbinN}$\\
%fraction\_zspecV\_F$_{zbin1}$=zspecV\_F$_{zbin1}$/zspecV$_{zbin1}$\\
%fraction\_zspecV\_F$_{zbinN}$=zspecV\_F$_{zbinN}$/zspecV$_{zbinN}$\\
%fraction\_LAEV\_F$_{zbin1}$=LAEV\_F$_{zbin1}$/LAEV$_{zbin1}$\\
%fraction\_LAEV\_F$_{zbinN}$=LAEV\_F$_{zbinN}$/LAEV$_{zbinN}$
 
In Table \ref{tab:TabFractions}, we show the mean, the mean uncertainty, and the median of the fractions of z$_{photspec}$, z$_{spec}$V, and LAEVs among the redshift bins of all the identified overdensities. The uncertainties on the mean quantities are calculated following the prescription in \citet{Gehrels1986}. In the case of statistically small samples of astrophysical events, they provide a method to calculate the lower and upper values of the confidence levels of a certain number of events \citep[Equation 7 and 14 in][]{Gehrels1986}. We choose the confidence levels corresponding to a 1$\sigma$ limit in the case of Gaussian statistics \citep[Section 1 and Table 1 and 2 in][]{Gehrels1986}.
\begin{table*} %[h]
%\tabletypesize{\scriptsize}
%\rotate
\centering
\caption{Fraction of galaxies versus environment}
%\tablewidth{0pt}
\label{tab:TabFractions}
\scalebox{0.9}{
\begin{tabular}{|c|c|c|c|c|c|c|c|c|} 
\hline
 &    F & T & OM & M &  F & T & OM & M \\
CDFS+UDS  &    \%mean$\pm$error & \%mean$\pm$error & \%mean$\pm$error & \%mean$\pm$error & \%med & \%med & \%med & \%med  \\
\hline
z$_{photspec}$ & 63.1$^{+2.3}_{-2.1}$ & 24.3$^{+1.6}_{-1.3}$ & 12.6$^{+1.2}_{-0.9}$ & 3.1$^{+0.9}_{-0.6}$ & 66 & 23 & 11 & 2\\ %66 & 23 & 10 & 2\\
z$_{spec}$V & 71.4$^{+2.5}_{-2.2}$ & 16.3$^{+1.4}_{-1.1}$ & 12.3$^{+1.2}_{-0.9}$ & 3.7$^{+1.3}_{-0.9}$ & 72 & 16 & 12 & 4 \\ %73 & 16 & 12 & 4 \\
LAEVs & 75.3$^{+2.5}_{-2.2}$ & 16.4$^{+1.7}_{-1.6}$ & 8.3$^{+1.4}_{-1.1}$ & 1.9$^{+1.5}_{-1.2}$ & 76 & 15 & 9 & 0  \\ %75 & 14 & 9 & 0  \\
LAEVs[EW(Ly$\alpha)>20${\AA}]  & 77.4$^{+2.6}_{-2.3}$ & 16.6$^{+1.8}_{-1.6}$ & 6.0$^{+1.7}_{-1.4}$ & 1.7$^{+2.1}_{-1.9}$ & 79 & 15 & 6 & 0 \\ %78 & 14 & 6 & 0 \\
LAEVs[M$_{*}>$medM$_{*}$] & 66.2$^{+2.4}_{-2.1}$ & 18.5$^{+1.9}_{-1.8}$ & 15.3$^{+2.0}_{-1.9}$ & 4.7$^{+2.4}_{-2.3}$ & 67 & 17 & 16 & 0\\ %68 & 18 & 18 & 0\
LAEVs[sSFR$<$medsSFR] & 70.8$^{+2.5}_{-2.2}$ & 17.3$^{+2.1}_{-2.2}$ & 11.9$^{+1.7}_{-1.5}$ & 4.2$^{+2.2}_{-2.1}$ &72 & 18 & 10 & 0 \\ %72 & 20 & 10 & 0 
\hline
\hline
\end{tabular}
} 
\tablefoot{The numbers in the table represent the mean and the median (med) values of the fractions of each galaxy group in the four density categories calculated for the redshift bins of the identified overdensities. The errors on the mean are calculated following the prescription in \citet{Gehrels1986} for statistically small samples, as explained in the text. The galaxy groups are (i) all the galaxies in the input catalog (z$_{photspec}$), (ii) the galaxies with spectroscopic redshift from VANDELS (z$_{spec}$V), (iii) the LAEVs, (iv) the LAEVs with EW(Ly$\alpha)>20${\AA}, (v), the LAEVs with stellar masses larger than the median value of the LAEV sample (M$_{*}>$medM$_{*}$), (vi) the LAEVs with sSFRs lower than the median value of the LAEV sample (sSFR$<$medsSFR). The four density categories are: the field (F), the transition regions (T), the regions with associated density comparable with that of an overdensity (O), the identified structures (M), as explained in the text. The OM group is obtained summing the fractions in the O and M groups. 
The median values of stellar mass (log(M$_{*}$/M$_{\odot}$)) and specific star-formation rate (sSFR) are 9.31 and $3.9\times 10^{-9}$ yr$^{-1}$ respectively. The Ly$\alpha$ equivalent widths span a range of 1-200 {\AA}.
}
\end{table*}

%At $3<z<4$, %the zphotzspec galaxies appear to be equally distributed between field and overdense plus transition regions. 
%Even if 
The fraction of the galaxies of the input catalog %(corresponding to the galaxies with z$_{photspec}$) 
located in the field is 60\%, being the fraction in transition and overdense regions 40\%. 
The galaxies with spectroscopic redshifts coming from VANDELS are 3\% of the total galaxies in the input catalog. 
About 70\% of them happen to be in the field and only 4\% are members of the detected overdensities (Table \ref{tab:TabFractions}). 
 %z$_{spec}$V and of the LAEVs are  About 10\% of the z$_{photspec}$ are located in the overdense regions. 
%
%At $3<z<4$, they happen to be mostly located in the field. 

The LAEVs follow the location of the galaxies with z$_{spec}$V. 
Therefore, our results indicate that, selecting a sample of galaxies unbiased in terms of environment, Ly$\alpha$-emitting galaxies do not predominantly appear to be located inside overdensities. As we can see from the table, less than 2\% of the LAEVs are members of the detected overdensities.
%To better visualize this, we show the mean fractions of z$_{photspec}$, z$_{zpec}$, and LAEVs among the redshift bins of all the identified overdensities in Fig.\ref{fractionFTMO}. 
%
%
%\begin{figure*}
% \centering
%\includegraphics[width=15cm]{figure_withoutxaxis.pdf}
%\caption{Mean fractions of the galaxies with either photometric or spectroscopic redshifts (z$_{photspec}$, circles), with spectroscopic redshifts from VANDELS (z$_spec$V, squares), with spectroscopic redshifts from VANDELS and EW(Ly$\alpha) >0$ {\AA} (LAEVs, stars) in the four density categories, field ($upper ~left$), transition regions ($upper ~right$), overdense regions, but not members of the identified overdensities ($lower ~left$), members of the identified overdensities ($lower ~right$). The fractions are calculated as described in the text. The error bars are calculated following the prescription in \citet{Gehrels1986} for statistically small samples as described in the text.  We also consider three subgroups of VANDELS Ly$\alpha$ emitting galaxies: LAEVs with EW(Ly$\alpha)>20$ {\AA} (red), LAEVs with stellar mass larger than the median value of the entire sample of LAEVs (blue), LAEVs with sSFR larger than the median value of the entire sample of LAEVs (green).  
%}
%\label{fractionFTMO}%
%\end{figure*}

We separate the LAEVs in subgroups, based on EW(Ly$\alpha$), stellar mass, and sSFR (see the following subsection).
About 80\% of the LAEVs with EW(Ly$\alpha)>20$ {\AA} (typical equivalent width cut applied in narrow-band surveys of LAEs). The most massive LAEs (with stellar mass larger than the median value, LAEVs[M$_{*}>$medM$_{*}$]) and the ones with the smallest sSFR (with specific star-formation rate smaller than the median value, LAEVs[sSFR$<$medsSFR]) are located in the field in a lower fraction (66\% and 71\%, respectively). Among the LAEVs, 15\% of the most massive are located in overdense regions.
%The LAEVs in overdense regions tend to have high stellar masses and low sSFRs than the others. %This is also a global behaviour of the VANDELS star-forming galaxies. %, not necessarily Ly$\alpha$ emitters. 
%
%{\bf{questo calcolo era stato suggerito da Gianni, ma da' gli stessi risultati di quello precedente}}It is possible to estimate the fractions of zphotzspec, zspecV, and LAEVs in the entire $3<z<4$ redshift range, without limiting the calculation to the redshift ranges of the detected overdensities. We find that 67\%(22\%)(11\%) of the sources in the entire input catalog, that 76\%(14\%)(10\%) of the sources with a spectroscopic redshift from VANDELS, and that 78\%(15\%)(7\%) of the LAEVs are located in the field(transition)(overdense regions). Among the LAEVs, 81\%(4\%) of the ones with EW(Ly$\alpha)>20$ {\AA} are in the field(overdense regions), while 14\%(70\%) of the high-mass ones are in the field(overdense regions), in agreement with the previous calculation. %{\bf {maybe it is not surprising but it is shown by the fractions}}

\subsection{Physical properties of the Ly$\alpha$-emitting galaxies in the overdense regions}
\label{propLAEs}

\begin{figure*} 
 \centering
\includegraphics[width=16.cm]{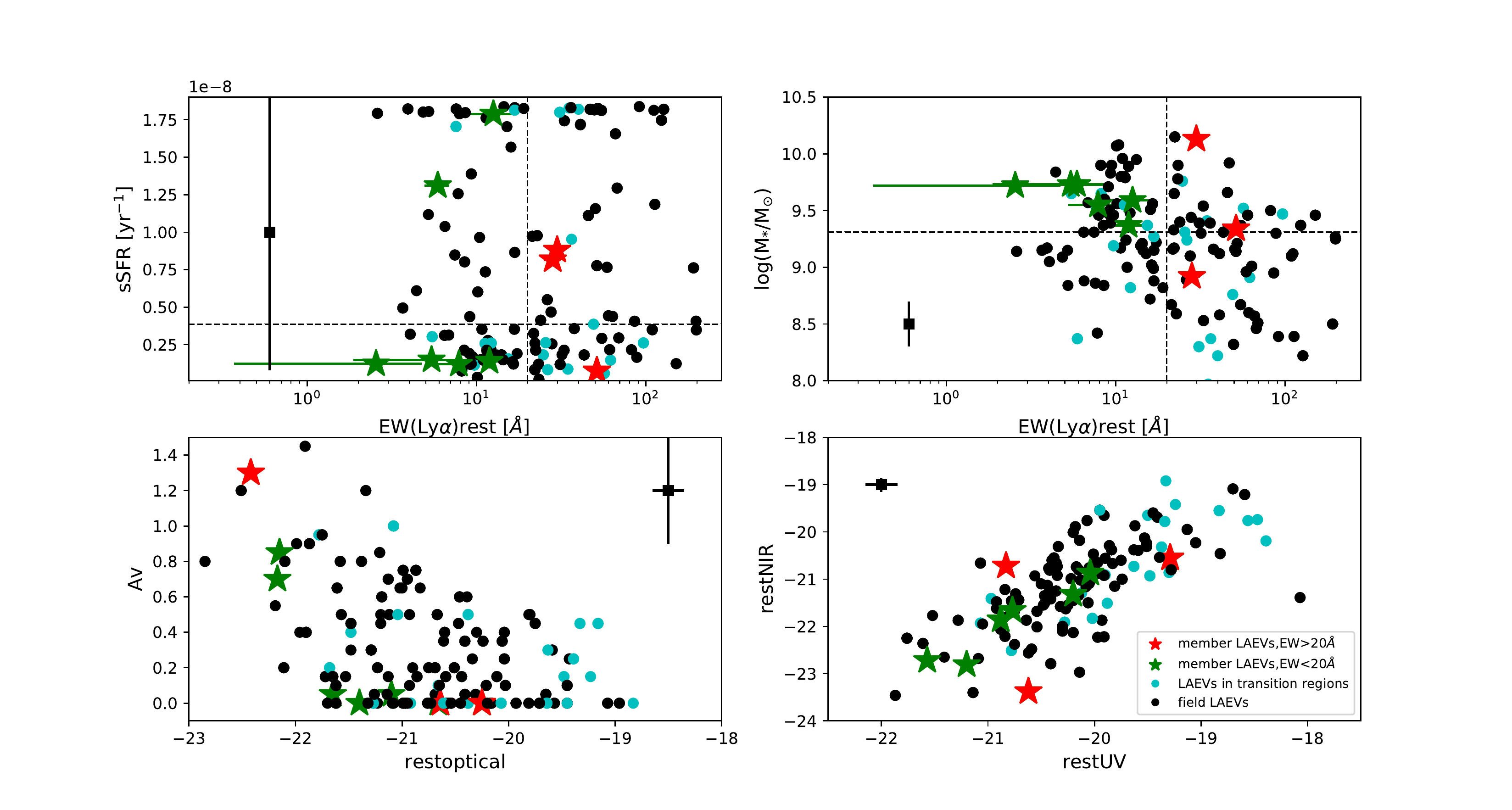}
\includegraphics[width=15.9cm]{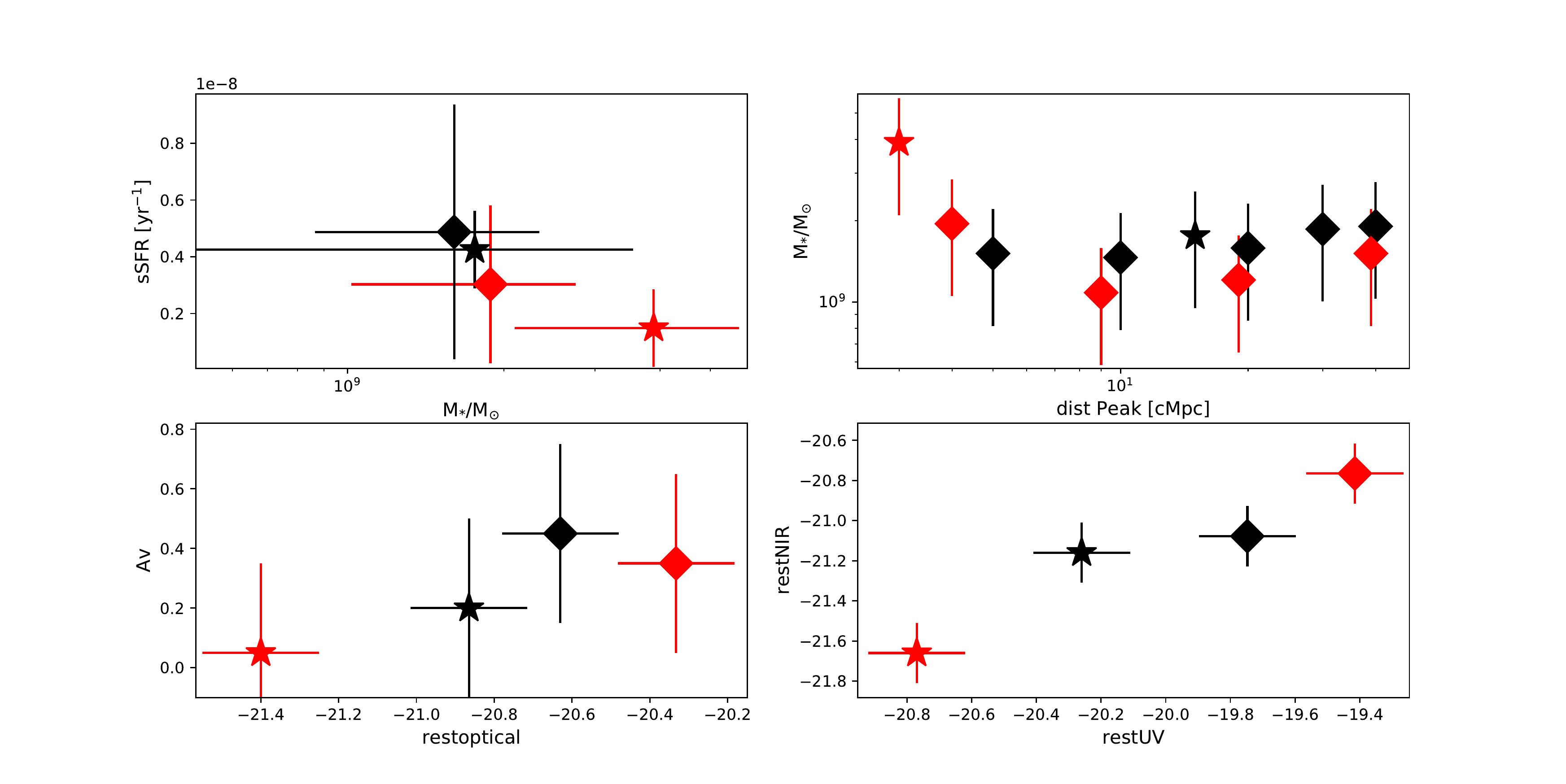} 
\caption{$Upper ~ Panels:$ LAEV physical properties with respect to the environment in the VANDELS fields. The LAEVs located in the identified overdensities and with EW(Ly$\alpha)>20$ {\AA} are shown as red stars, the ones with EW(Ly$\alpha)<20$ {\AA} are shown as green stars, the ones located in transition regions are shown as cyan circles and in the field as black circles. Dashed lines indicate the median values of sSFR, stellar mass, and EW(Ly$\alpha)=20$ {\AA} of the entire LAEV sample. The physical properties we investigate are sSFR versus EW(Ly$\alpha$) ($upper ~left$), stellar mass versus EW(Ly$\alpha$) ($upper ~right$), dust reddening, Av, versus rest-frame optical magnitude ($lower ~left$), rest-frame NIR magnitude versus rest-frame UV magnitude ($lower ~right$). The squares with error bars represent the typical parameter uncertainties coming from the SED fitting  (see the upcoming work on SED fitting parameters). $Lower ~ Panels:$ Median physical properties of the LAEVs and of all the galaxies in the input catalog located in overdensities (red symbols) and in the field (black symbols). %The mean values and the standard deviations of the means are presented as circles for all the galaxies in the input catalog and as open stars for the LAEVs. 
The medians are presented as diamonds for all the galaxies in the input catalog and stars for the LAEVs. 
The error bars on the medians correspond to the uncertainties due to SED fitting.
The physical properties we investigate are sSFR versus stellar mass ($upper ~left$), stellar mass versus distance to the highest-density peak ($upper ~right$) of the identified overdensities, dust reddening, Av, versus rest-frame optical magnitude ($lower ~left$), rest-frame NIR magnitude versus rest-frame UV magnitude ($lower ~right$). %We also show the LAEVs in the detected overdensities as small green and red stars as in the upper panel.
%
% Physical properties of all the galaxies in the input catalog locate in the identified overdensities (red circles), in transition regions (cyan circles), in the field (black circles), and the LAEVs in the identified overdensities (green,red stars). The physical properties we investigate are sSFR versus stellar mass ($upper ~left$), stellar mass versus distance to the highest-density peak ($upper ~right$) of the identified overdensities, dust reddening, Av, versus rest-frame optical magnitude ($lower ~left$), rest-frame NIR magnitude versus rest-frame UV magnitude ($lower ~right$). 
%
}
\label{figpropLAEnonLAE}%
\end{figure*}

In Fig. \ref{figpropLAEnonLAE}, we study the physical properties of the LAEV sample. 
%[to section 2]As described in \citet{McLure2018}, the physical parameters of all the galaxies in the VANDELS database are derived through SED fitting on state of the art multiwavelength catalog (see section). It has been shown \citet[e.g.][]{Santini2015} that stellar mass is almost independent of model assumptions, while SFR and Av are of course more dependent, for instance, on the choice of the star-formation history assumed. Given that we have spectroscopic redshifts and good quality photometry, the typical errors on the rest-frame colours are small - probably at the +/-0.2 mags level.  For the same reasons, the errors on the stellar mass measurements are typically at the level of +/- 0.2 dex. 
%
In comparison to typical narrow-band selected LAEs \citep[e.g.,][]{Gronwall:2007, Hagen2016}, the LAEVs considered here, Lyman break galaxies by selection, have lower sSFRs ($<2\times 10^{-8}$ yr$^{-1}$ versus values up to $10^{-7}$ yr$^{-1}$) and are more massive [$8.5<$ log(M$_{*}$/M$_{\odot})<10$ versus values down to log(M$_{*}$/M$_{\odot})=7$]. 
%
%
%with respect to the environment. 
%We consider the redshift ranges of our detected overdensities and the distance to their highest-density peaks. 
%
However, these values of physical parameters are reasonable and in agreement with models of Ly$\alpha$-emitting galaxies. For instance, \citet{Gurung-Lopez2019b} show the distribution of stellar mass and star-formation rate of Ly$\alpha$-emitting galaxies from a model that incorporates Ly$\alpha$ radiative transfer processes in the interstellar and intergalactic medium \citep{Gurung-Lopez2019}, and that is implemented in the GALFORM semi-analytic model of galaxy formation and evolution \citep{Lacey2016,Baugh2019}. The distribution of stellar masses of Ly$\alpha$-emitting galaxies at $2<z<3$ extends from log(M$_{*}$/M$_{\odot})=7$ to 10, with a peak at log(M$_{*}$/M$_{\odot})=9.0$. According to the SFR distribution, the most probable value of the sSFR would be around $2\times 10^{-9}$ yr$^{-1}$.

The upper panels of the figure show that 70\%(30\%) of the LAEVs with EW(Ly$\alpha)<20$ {\AA} ($>20$ {\AA}) in overdense regions have sSFRs lower than the median value of the entire sample of LAEVs. All the low-EW LAEVs in overdense regions have stellar masses higher than the median value. Also, they tend to be brighter in the rest-frame UV magnitude. %In Tab. \ref{tab:TabPropLAEs}, we list the physical properties of the LAEVs in overdense regions. 
Typically, the LAEVs in overdense regions are more massive and have lower sSFR than the LAEVs in the field. Also, they are brighter in the rest-frame UV, optical, and NIR.

The average EW(Ly$\alpha$) is $17\pm5$ and $35\pm4$ {\AA} for the LAEVs in the identified overdensities and in the field, respectively (see also Fig.\ref{EWdistr}). %, while the median values are 12 and 17 {\AA} respectively. 
However, the KS test shows that we can not reject the null hypothesis that the EWs in the overdensities and in the field are drawn from the same distribution. 
Some of the LAEVs in overdense regions have non-zero dust content (Av$_{SED}$ parameter, despite the large uncertainty we expect for the Av parameter), even though they have high Ly$\alpha$ equivalent widths.

%EWs(Ly$\alpha$) 
We perform a KS test to compare the properties of the LAEVs in the identified overdensities and in the field. According to the test, we can reject the null hypothesis that masses, sSFRs, and rest-frame UV of LAEVs in the overdensities and in the field are drawn from the same distribution at more than 2$\sigma$ ((KS,p)$_{mass}$=(0.40,0.10); (KS,p)$_{sSFR}$=(0.47,0.03); (KS,p)$_{restUV}$=(0.42,0.08)). %Also, the median stellar mass of the LAEVs in overdensities is 10$^{9.6}$ M$_{\odot}$, while that of the LAEVs in the field is 10$^{9.2}$ M$_{\odot}$. 
%The median sSFR is 1.5 $\times 10^{-9}$ and 4.4 $\times 10^{-9}$ yr$^{-1}$ and the median EW(Ly$\alpha$) is 12 and 22 {\AA} for the LAEVs in the overdensities and in the field, respectively.
The average stellar mass of the LAEVs in the identified overdensities is $9.6\pm0.1$ M$_{\odot}$ and that of the LAEVs in the field is $9.25\pm0.04$ M$_{\odot}$ in log scale (the median values are 9.6 and 9.2, respectively).
The average sSFR is $6.0\pm2.1 \times 10^{-9}$ and  $7.7\pm0.7 \times 10^{-9}$ yr$^{-1}$, and the medians are $1.5 \times 10^{-9}$ and $4.2 \times 10^{-9}$ yr$^{-1}$, respectively. The average rest-frame UV is $-20.6\pm0.2$ and $-20.3\pm0.1$ mags, and the medians are -20.8 and -20.3 mags, respectively.

As we can see from the plots in the lower panels of the figure, the LAEVs in overdense regions have stellar mass and sSFR consistent with the galaxies of the input catalog in overdense regions. However, they are brighter in the rest-frame UV and NIR.
%sSFRmember stdev 6.17225041818e-09 1.89953602394e-09
%sSFRfield stdev 7.54335956303e-09 3.82364978422e-10
%
%massmembers stdev 9.55909090909 0.0944212470836
%massfield stdev 9.23098305085 0.0257938449339
%
%EWmembers stdev 19.3416004406 5.48115697406
%EWfield stdev 38.1154055899 2.40799922605
%
%Fmembers stdev 1.22262341443e-17 2.90991827821e-18
%Ffield stdev 2.92544983352e-17 2.98031067955e-18
%
%Lmembers stdev 1.32327293567e+42 3.58299043803e+41
%Lfield stdev 3.19443779983e+42 3.26657917745e+41
%
%The lower panels of Fig. \ref{figpropLAEnonLAE} show the physical properties of all the galaxies in the input catalog. The properties of the LAEVs in overdense regions are comparable to those of the galaxies typically located in overdense regions. 
Even if the LAEVs studied here do not have all the physical properties of typical Ly$\alpha$ emitters, the ones in the overdensities share some  physical properties of the galaxies we find in overdense regions %(lower panels of Fig. \ref{figpropLAEnonLAE}) 
and could provide information about the overdensities.

\subsection{Spectroscopic properties of the Ly$\alpha$-emitting galaxies in the overdense regions}
\label{specLAEs}

The Ly$\alpha$ emission escaping a galaxy is sensitive to the properties of the ISM and of the CGM. Therefore, the information derived from the Ly$\alpha$ of the LAEVs may be used to infer the characteristics of the ISM and of the CGM of the galaxies %(not necessarily Ly$\alpha$ emitters) 
in overdense regions. Moreover, theoretical predictions have shown that bright Ly$\alpha$ emitters reside in, and so could trace, more massive halos \citep[e.g.,][]{Garel2015} on average.
In this subsection, we study the spectroscopic properties of the LAEVs in overdense with respect to field regions, and we investigate if the Ly$\alpha$ emitters can be used to trace the environment characteristics.  

%ere are a few references:
%- Ouchi+03 (fig. 7 in particular: higher clustering for brighter LAEs)
%- Garel+15 (Section 6): brighter LAEs  predicted to reside in more massive haloes on average
%- Khostovan+18 (Fig. 5): brighter LAEs measured to reside in more massive haloes on average

\begin{figure*}
 \centering
 \includegraphics[width=16cm]{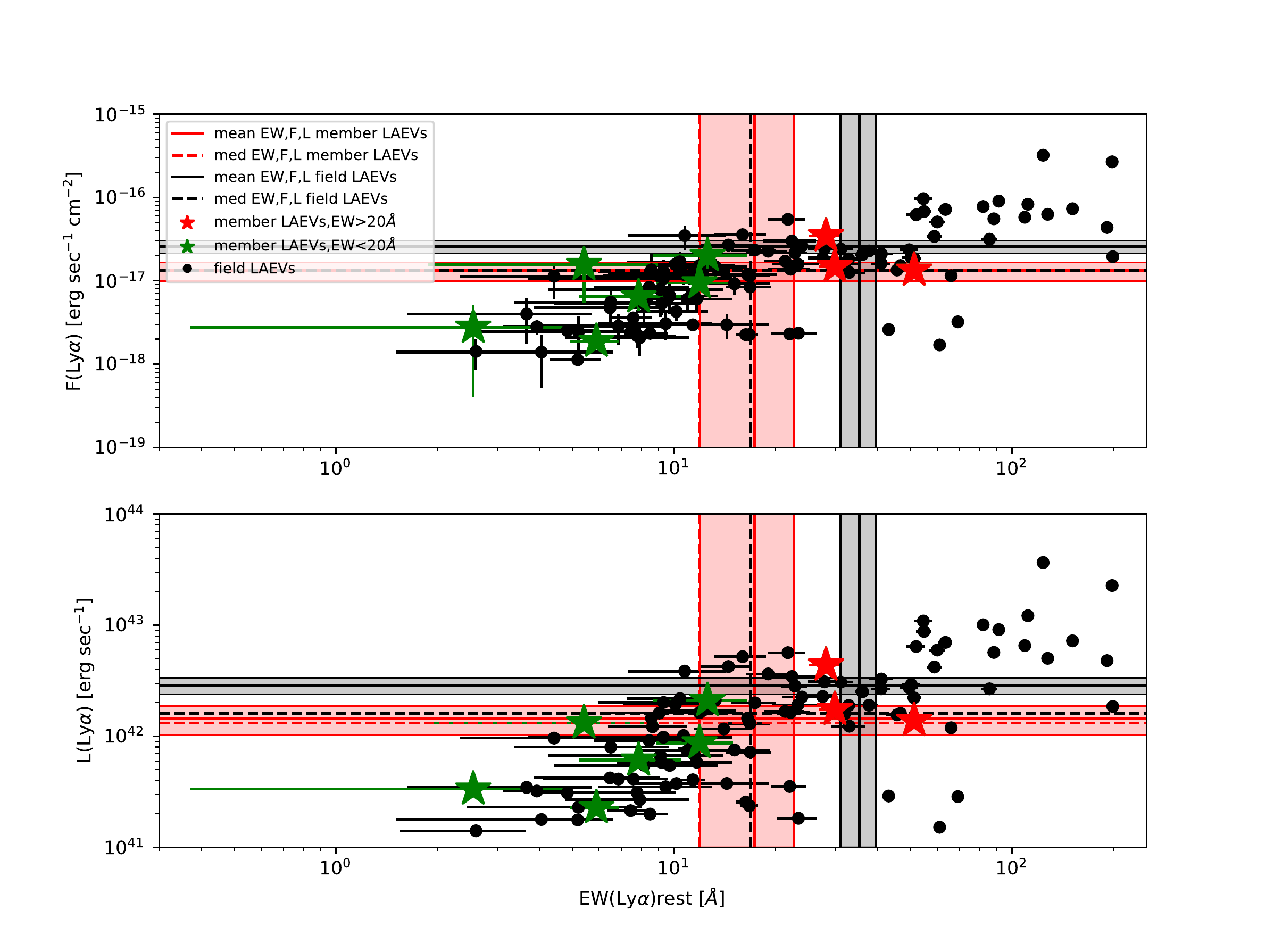} %BetterFlyaLLya_EWLya.pdf} 
\caption{%$Upper ~Panels$: Ly$\alpha$ flux vs restUV. 
$Upper ~ Panel$: Ly$\alpha$ flux vs Ly$\alpha$ equivalent width.  
$Lower ~Panel$: Ly$\alpha$ luminosity vs Ly$\alpha$ equivalent width. Green (red) stars indicate LAEVs in the identified overdensities with EW(Ly$\alpha) <20$ {\AA}  (EW(Ly$\alpha) >20$ {\AA}), black dots indicate the LAEVs in the field. The vertical lines correspond to the mean (solid) and median (dashed) values of the EW(Ly$\alpha$) of LAEVs in overdensities (red) and in the field (black). The horizontal lines correspond to the mean (solid) and median (dashed) values of the F(Ly$\alpha$) and L(Ly$\alpha$) of the LAEVs in the detected overdensities (red) and in the field (black). The shaded areas indicate the standard deviation of the means. The mean and median values of the F(Ly$\alpha$) of the LAEVs in the detected overdensities and the median value of the F(Ly$\alpha$) of the LAEVs in the field overlap.
The mean and median values of the L(Ly$\alpha$) is $1.4\pm0.4 \times 10^{42}$ and $1.3\times 10^{42}$ erg sec$^{-1}$ for the LAEVs in the detected overdensities, while they are $2.9\pm0.5 \times 10^{42}$ and $1.6 \times 10^{42}$ erg sec$^{-1}$ for the LAEVs in the field. The median value of the LAEVs in the identified overdensities with EW(Ly$\alpha)<20$ {\AA}($>20$ {\AA}) is $7.4\times 10^{41}$ erg sec$^{-1}$ ($1.7\times 10^{42}$ erg sec$^{-1}$). 
}
\label{FLLya}%
\end{figure*}

In Fig. \ref{FLLya}, we compare the Ly$\alpha$ flux and luminosity of the LAEVs in the identified overdensities and in the field. 
The median value of the L(Ly$\alpha$) is $1.3\times 10^{42}$ ($1.6 \times 10^{42}$) erg sec$^{-1}$ for the LAEVs in overdense (field) regions. 
%
%Also, the average value of L(Ly$\alpha$) is $1.4\pm0.4 \times 10^{42}$ ($2.9\pm0.5 \times 10^{42}$) erg sec$^{-1}$ for the LAEVs in overdense (field) regions. 
%The range in L(Ly$\alpha$) of the LAEVs in the field is $1\times 10^{41} - 4\times 10^{43}$ erg sec$^{-1}$ and the median value of the LAEVs in the identified overdensities with EW(Ly$\alpha)<20$ {\AA}($>20$ {\AA}) is $7.4\times 10^{41}$ erg sec$^{-1}$ ($1.7\times 10^{42}$ erg sec$^{-1}$). 
The KS test shows that we can not reject the null hypothesis that the Ly$\alpha$ fluxes and luminosities of the LAEVs in overdensities and in the field are drawn from the same distribution ((KS,p)$_{F}$=(0.22,0.78); (KS,p)$_{L}$=(0.26,0.58)).
%Fmembers stdev 1.22262341443e-17 2.90991827821e-18
%Ffield stdev 2.92544983352e-17 2.98031067955e-18
%Lmembers stdev 1.32327293567e+42 3.58299043803e+41
%Lfield stdev 3.19443779983e+42 3.26657917745e+41
%
Therefore, in our sample, we do not observe a trend for which galaxies in overdense regions are significantly brighter in Ly$\alpha$ luminosity than in the field. 
%{\bf{add a comment on the KS test done on physical parameters. KS<p for F(Lya), L(Lya) so we can not reject the null hypothesis that they are drawn from the same distribution INCONCLUSIVE}}

The shape of the Ly$\alpha$ emission line %(presence of a blue bump, separation between blue and red peaks, large scale spatial extension) 
can provide information on the HI column density and kinematics \citep{V2006, Guaita2017} of the emitting galaxy. Due to the interplay of dust and HI gas in the ISM and in the CGM, the Ly$\alpha$ emission line profile can present two peaks, one red shifted and one blue shifted with respect to the systemic redshift. The shift of the main red peak can be related to the HI column density and gas kinematics, such as stellar outflows. Radiative transfer models predict that the intensity of the blue peak and the separation between blue and red peak are also connected with the HI column density \citep{Verhamme2015}. \citet{Guaita2017} showed that the galaxies characterized by low HI column densities (NHI $\sim10^{18}$ atoms cm$^{-2}$) experience stellar-driven outflows with velocities, V$_{exp}$, % {\bf{A me piace scrivere V$_{exp}$ in questo modo}}, 
of the order of a few hundreds of km s$^{-1}$ and present spatially concentrated Ly$\alpha$ emissions. In the galaxies characterized by large HI column densities ($>10^{20}$ atoms cm$^{-2}$), Ly$\alpha$ photons can be efficiently scattered even in the case of static media and can produce extended Ly$\alpha$ halos. However, more complicated geometries of the ISM or of the CGM could also provide interpretations of the NHI values.

Although VANDELS spectra are designed to achieve high signal-to-noise ratio, photospheric stellar absorption lines are intrinsically too weak in individual LAEV spectra to be used as tracers of the systemic redshift. In addition to this, nebular emission lines, such as CIII]1908 and HeII, are not detected in a statistically-meaningful subsample of the Ly$\alpha$-emitting galaxies studied here \citep[e. g.,][]{Marchi2019}. Therefore, we are not able to derive gas kinematics information from the shift of low-ionization absorption lines and systemic redshift for the LAEVs. %, and in particular for those in overdensities. 
We can, however, stack the LAEV spectra centered on the Ly$\alpha$ wavelength to analyse the average shape and spatial extension of the Ly$\alpha$ emission.

\begin{figure*}
 \centering
 \includegraphics[width=20cm]{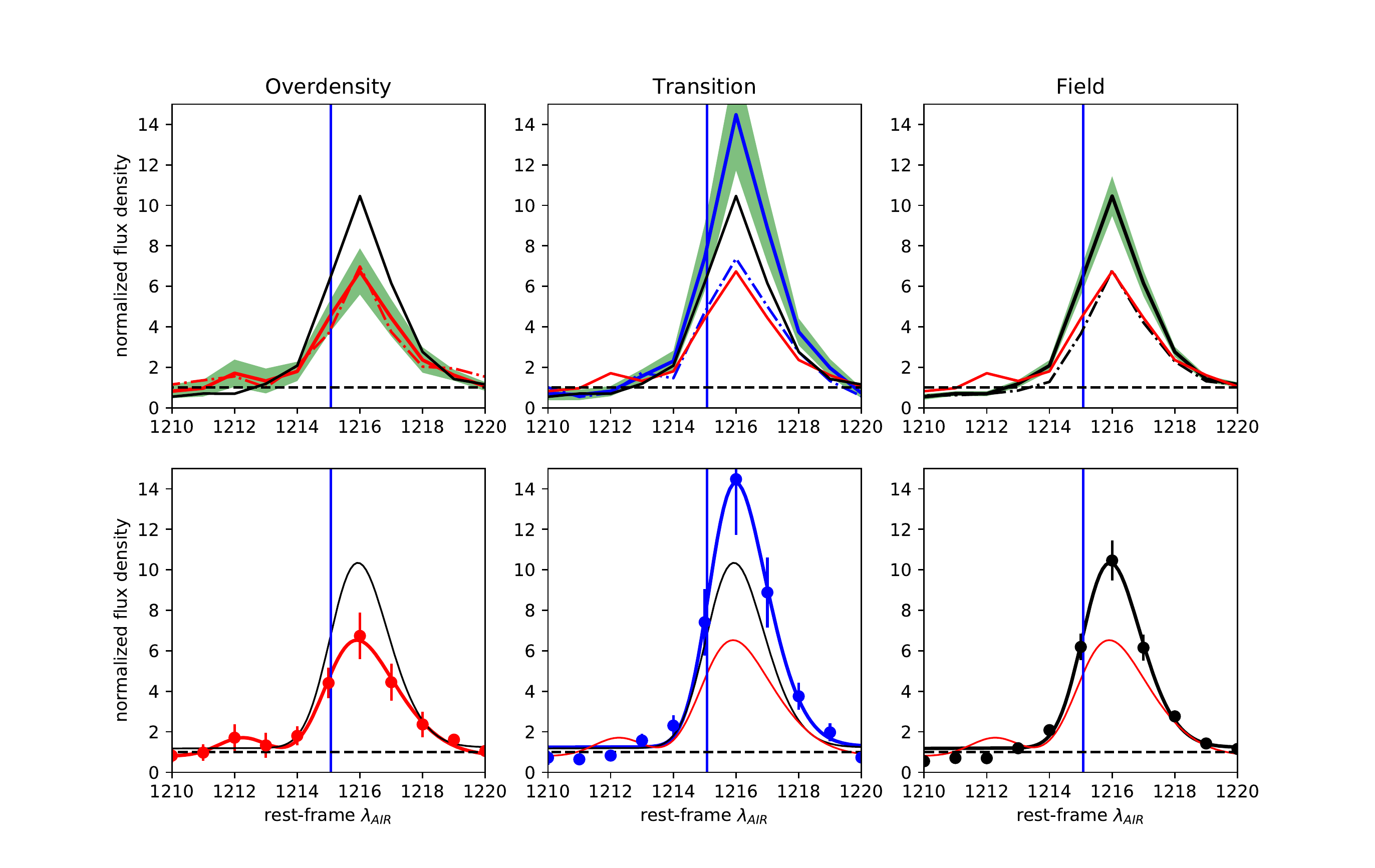}%FigureLyaProfile1Dmeanmed.pdf}   %Lyaprofile_LAEoverdensityfieldtransition_June18_mean.pdf} %/Volumes/Backup/LUCIA/ChileProjects/ESOvisitingscientist/overTallfield_Lyafit_CDFSUDS.png} %CDFSgrid_z282_Zheng2016LAE.png}
\caption{$Upper ~Panels$: Stacks of the Ly$\alpha$ profiles of the LAEVs in the identified overdensities ($left ~ panel$), transition ($middle ~ panel$), and field ($right ~ panel$) regions.   % where the rest-frame wavelengths are calculated from the VANDELS spectroscopic redshifts. 
The green shaded areas indicate the standard deviation of the mean of the fluxes at each wavelength among all the LAEVs in a certain density category. The observed profiles of the LAEVs in overdensity (transition) (field) regions is shown in red (blue) (black) in every panel. The dashed-dotted lines show the median stacks.
$Lower ~Panels$: Asymmetric Gaussian curves, best fits of the stacks (dots with errorbars) in the upper panels. In each panel, red is used for the LAEVs in the identified overdensities, blue for the LAEVs in transition regions, black for the LAEVs in the field. 
The best fits of the asymmetric Gaussian curves provide the following parameters ($\lambda$, FWHM, A) = (1215, 4 {\AA}, 0.86), (1215, 3 {\AA}, 0.95), and (1215, 3 {\AA}, 0.83) for the left, middle, and right stack, respectively. 
}
\label{stackLAE}%
\end{figure*}

In Fig.\ref{stackLAE}, we show the %(mean) 
stack of the normalized 1D spectra of the LAEVs in overdense, transition, and field regions. The stacks are obtained from the rest-frame spectra, aligned to the highest-flux wavelength of the Ly$\alpha$ red peak. %, for the lack of systemic redshifts for the majority of our sources. We fit the stacked Ly$\alpha$ emission lines with asymmetric Gaussian curves and measure the integrated flux (Tab. \ref{table:FLya}). The integrated flux is larger for the field stack and for the transition regions than for the overdensity stack. 
\begin{table*}
\centering
\caption{Ly$\alpha$ Integrated flux from the normalized stacked spectra}  
\label{table:FLya}
\scalebox{1.2}{
\begin{tabular}{|c|c|c|c|}
\hline  
& Overdensity & Transition & Field \\
\hline
red & $16.4^{+2.6}_{-4.2}$& $30.5^{+7.1}_{-5.0}$ & $21.5^{+3.5}_{-1.1}$ \\
blue  & $1.8^{+1.8}_{-1.7}$& --  & -- \\
%\hline
%redRadio & $160^{+55}_{-11}$& -- & -- \\
%blueRadio & $20^{+5}_{-11}$& -- & -- \\
\hline
\end{tabular}
}
\tablefoot{Integrated flux of the normalized stack of Ly$\alpha$ 1D profiles for the nine LAEVs that are members of our identified overdensities, for the 31 LAEVs in transition, and the 100 LAEVs in field regions. `red' refers to the main red peak of the stack, `blue' to the blue peak. %The lower part of the table gives the integrated fluxes of the red and blue peak (redRadio, blueRadio) of the stack of Ly$\alpha$ emitting galaxies detected in the overdense regions traced by the radio galaxies MRC 0052-241, MRC 0943-242 at $z\sim2.9$, MRC 0316-257, and TN J2009-3040 at $z\sim3.1$ \citep{Venemans2007}.
}
\end{table*}
The stack of the galaxies in the identified overdense shows a marginal blue peak, whose integrated flux is eight times smaller than that of the main red peak (see Table \ref{table:FLya}). The blue peak is clearly visible in the spectra of five of the nine sources. % {\bf{it is seen in 5 individually}}. % (Tab. \ref{table:FLya}). 
However, this blue peak is consistent with the continuum level at $\sim1\sigma$. In the lower panels of the figure, we show the best fit asymmetric Gaussian curves \citep[see Equation 3 in][]{Guaita2017} of the three profiles. At the resolution of our data, the levels of asymmetry (A parameter) and the full width half maxima (FWHM parameter) of the three profiles are equivalent.
%The table also shows the integrated fluxes of Ly$\alpha$ peaks of the stack of Ly$\alpha$ emitting galaxies detected in the overdense regions traced by radio galaxies at $z\sim3.1$, which is also 8 times smaller than that of the main red peak.

In principle, the tendency of the galaxies in overdense regions to have low Ly$\alpha$ equivalent widths, high stellar masses, and low sSFRs could indicate that the LAEVs in overdense regions either are forming stars at a lower rate than the LAEVs in the field or they have already experienced intense star-formation phases. In addition to this, the presence of a blue peak may imply that the galaxies in dense environments are characterized by low NHI or %(already consumed into star formation or stripped out by interaction with a common potential well) at least in certain directions. It may also indicate that the galaxies in overdense regions are characterized by 
lower-velocity outflows than in the field, at least in certain directions.

\begin{figure*}
 \centering
\includegraphics[width=8cm]{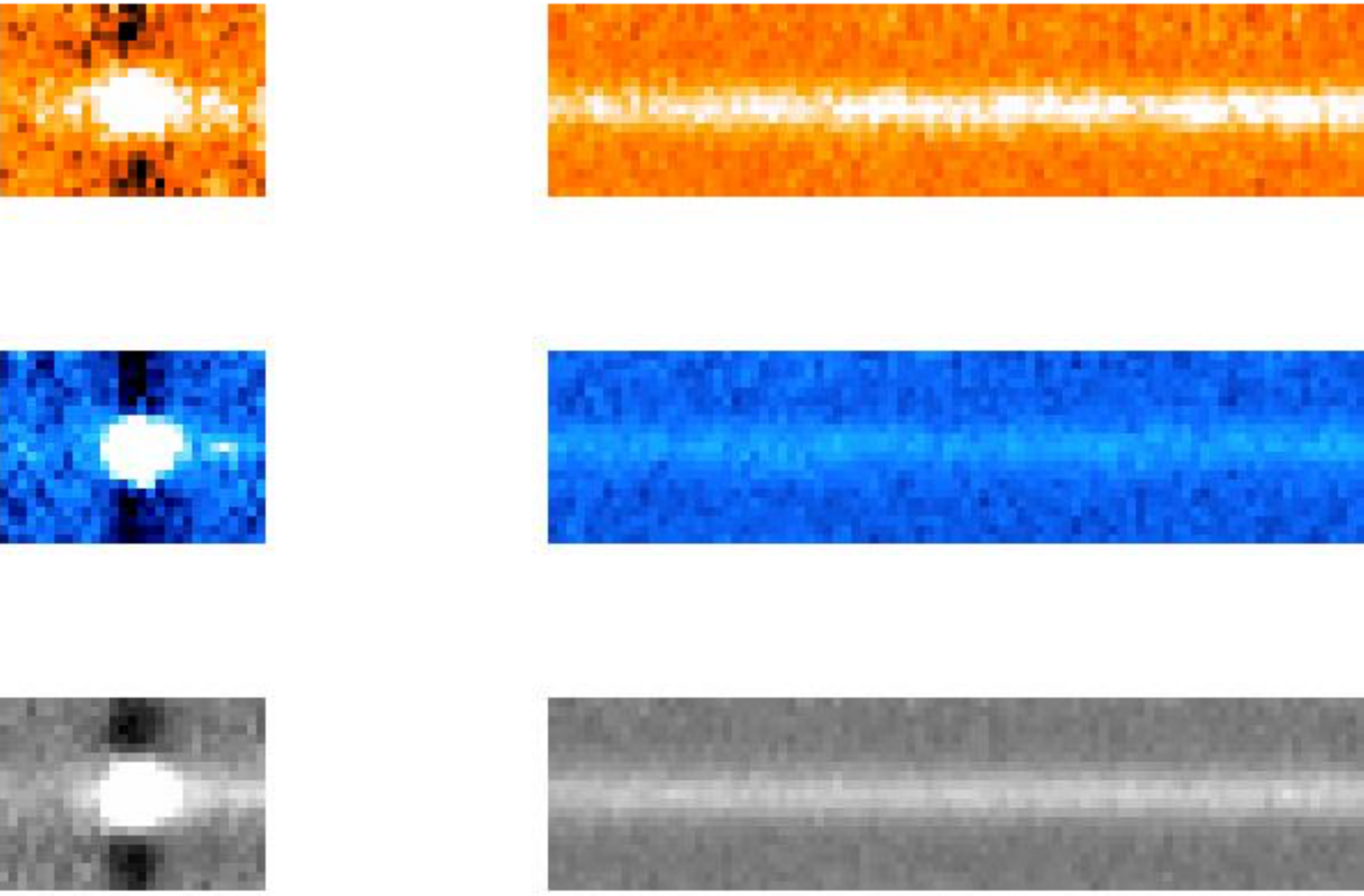} 
\caption{Stacks of the 2D spectra of the LAEVs in the identified overdensities ($upper$), in the transition ($middle$), and field ($lower~ panel$) regions at the wavelength of Ly$\alpha$ ($left ~panels$) and of the UV continuum at 1400 {\AA} ($right ~panels$) rest frame.
}
\label{stackLAEV2D}%
\end{figure*}

\begin{figure*}
 \centering
 \includegraphics[width=20cm]{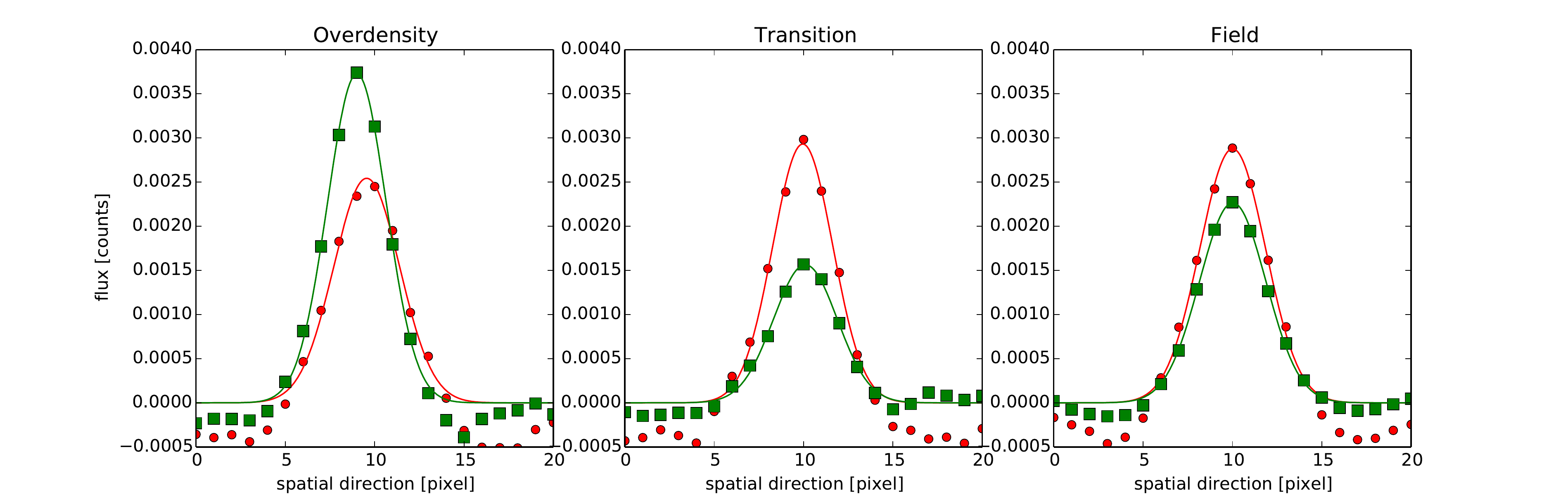} 
\caption{Ly$\alpha$ (red dots) and UV (green squares) profiles in the spatial direction for the LAEVs in the identified overdensities ($left$), in the transition ($middle$) and field ($right ~panel$) regions. The curves represent Gaussian fits to the data.
}
\label{stackLAEV2Dprofile}%
\end{figure*}

%{\bf{stack of LAEs in radio galaxy surroundings.}}
%For comparison, we stack the Ly$\alpha$ profiles of the Ly$\alpha$ emitting galaxies detected in the overdense regions around the radio galaxies MRC 0052-241, MRC 0943-242 at $z\sim2.9$, MRC 0316-257, and TN J2009-3040 at $z\sim3.1$ \citep{Venemans2007}. The resolution of those spectra is originally higher than that of VIMOS. However, the stack shows a component on the blue size of the Ly$\alpha$ emission with integrated flux also 8 time smaller than that of the main red peak.
%\begin{figure*}
 %\centering
 %\includegraphics[width=20cm]{stackLAEmemberwithVenemans.pdf} 
%\caption{Stack of Ly$\alpha$ profiles of Ly$\alpha$ emitting galaxies detected in the overdense regions traced by the radio galaxies MRC 0052-241, MRC 0943-242 at $z\sim2.9$, MRC 0316-257, and TN J2009-3040 at $z\sim3.1$ \citep{Venemans2007}. In the $left ~panel$ we also show the stack of the LAEVs in the field (black) and in the overdense regions (red). In the $left ~panel$ we also show the stack of the Ly$\alpha$ profiles of Ly$\alpha$ emitting galaxies detected in the overdense regions traced by the radio galaxies in yellow and the best 2-Gaussian fit in magenta.
%}
%\label{stackLAEVenemans}%
%\end{figure*}
To investigate if the HI column density could be contributing to the shape of the stacked Ly$\alpha$ profiles, we study the Ly$\alpha$ spatial extension of the LAEVs. Significant NHI should produce Ly$\alpha$ emission which is spatially more extended than the UV continuum \citep{Verhamme2015}. \citet{Guaita2017} estimated the Ly$\alpha$ spatial extension from the 2D spectrum (their Eq. 4 parameter, Ext(Ly$\alpha$-C)) as the difference between the FWHMs of the Gaussian profiles of Ly$\alpha$ and the UV continuum (C) in the spatial direction.
Due to the way sky subtraction is performed in VANDELS (subtracting pairs of dithered exposures), there are bright absorption features on either sides (in the spatial direction) of a bright emission line. In the 2D spectrum stacks, these absorptions increase in signal-to-noise ratio, limiting the ability to measure the wings (in the spatial direction) of an emission line. Since the stack of the LAEVs in the identified overdensities is only composed of nine sources, the signal-to-noise ratio of the absorptions is not that intense and we can see some Ly$\alpha$ line extension. Unfortunately, this is not the case for the stacks in the transition regions and in the field, where the 2D profiles do not provide useful information (see Fig. \ref{stackLAEV2D}).
From the 2D-stack profile (in the spatial direction) of the LAEVs in the identified overdensities (Fig. \ref{stackLAEV2Dprofile}), we estimate a lower limit, Ext(Ly$\alpha$-C) $>0.4$ arcsec ($>3$ physical kpc at $z=3.5$). %Also, we measure an offset of the order of 0.1 arcsec ($\sim0.7$ physical kpc at $z=3.5$) {\bf{CHECK}} between the peak of the Ly$\alpha$ and the UV emissions. 
%Together with the fact that some of 
%Also, the LAEVs in overdense regions have non-negligible dust content (Fig. \ref{figpropLAEnonLAE}) that could indicate that geometrical effects can shape the Ly$\alpha$ emission in overdense regions {\bf{true??}}.
%To test this last point, we make use of the public version of FLaREON \citep{Gurung-Lopez2018}.
%Fixing certain values of dust reddening, expansion velocity, and HI column density, FLaREON predicts the Ly$\alpha$ emission line profile escaping a galaxies with a certain geometry (galaxy surrounded by a thin shell and with a biconical outflow). We assume the dust reddening corresponding to the LAEVs in the overdensities and run FLaREON with low (10 km sec$^{-1}$) and high (200 km sec$^{-1}$) V$_{exp}$, low ($10^{19}$ atoms cm$^{-2}$) and high ($10^{21}$ atoms cm$^{-2}$) HI column densities. The outputs of FLaREON are the Ly$\alpha$ profiles in the wavelength space for each combination of dust reddening, V$_{exp}$, and NHI.
%To account for the presence of a dense environment, FLaREON outputs are combined with the IGM trasmittivity obtained from a state-of-the-art dark matter simulation (PMILLENIUM, Baught et al. in prep.). We consider the median transmittivity of the 50\% most dense environment at $z\sim3$. 
%Q50_Q99 del 50 al 99.9999999999 (mitad en environments mas densos)
%( Mediana de la tranmission)

\subsection{Simulating the Ly$\alpha$ profile of the VANDELS Ly$\alpha$-emitting galaxies}
\label{flareon}

To test if the HI column density and other galaxy properties %geometrical effect 
can produce the Ly$\alpha$ profile we observe for the LAEVs in overdense regions, we make use of the public version of FLaREON \citep{Gurung-Lopez2018}, which is based on LyaRT \citep{Orsi2012}, a radiative transfer Monte Carlo code of Ly$\alpha$ emission. FLaREON predicts the Ly$\alpha$ line profile escaping a galaxy through different outflow configurations. Several outflow geometries are implemented in FLaREON, such as the commonly used thin shell of HI gas \citep{V2006, Gronke2016}, and a galactic wind \citep[Fig. 1 in][]{Gurung-Lopez2019}. Additionally, the escaping line profile depends on the input dust reddening, the V$_{exp}$, and the HI column density.  
The characteristics of the Ly$\alpha$ line profile are strongly coupled with the properties of the outflow. To take into account the diversity of line profiles, we consider different outflow configurations. In particular, we run FLaREON with the following V$_{exp}$, NHI combinations: i) 10 km sec$^{-1}$, $10^{18}$ atoms cm$^{-2}$, ii) 10 km sec$^{-1}$, $10^{21}$ atoms cm$^{-2}$, iii) 200 km sec$^{-1}$, $10^{18}$ atoms cm$^{-2}$, and iv) 200 km sec$^{-1}$, $10^{21}$ atoms cm$^{-2}$. This choice of parameters englobes the lowest and highest values of V$_{exp}$ and NHI, provided by FLaREON. Meanwhile, we fix the dust reddening to the median Av value of the nine LAEVs in overdense regions and of the 100 LAEVs in the field.  
%comment : Falta por casualidad V=200km/s y NHI = 10^19 cm^-2  ?

The Ly$\alpha$ line profile at high redshift is also affected by the presence of neutral hydrogen in the intergalactic medium \citep{Laursen2011}. In \citet{Gurung-Lopez2019b}, the authors develop a method to compute the IGM impact on the Ly$\alpha$ emission in cosmological volumes. Their work is based on the state-of-the-art N-Body simulation P-Millennium \citep{Baugh2018} and the semi-analytic model of galaxy formation and evolution GALFORM \citep{Lacey2016}. The IGM absorption is computed taking into account the local density, velocity, and ionisation state of the IGM. Then, for every position in the simulation, an IGM transmission curve is computed integrating the IGM absorptions along a set of lines of sight. The authors found that, depending on the local density, the IGM transmission changes. In particular, denser regions exhibit higher absorption of the Ly$\alpha$ profile. Therefore, in order to take into account the presence of a dense environment, we convolve the Ly$\alpha$ line profiles produced by FLaREON with the median IGM transmission curve of the top 50\% dense environments at $z=3.0$ (resembling our identified overdensities) computed in the P-Millennium+GALFORM simulation.

\begin{figure*}
 \centering
\includegraphics[width=18cm]{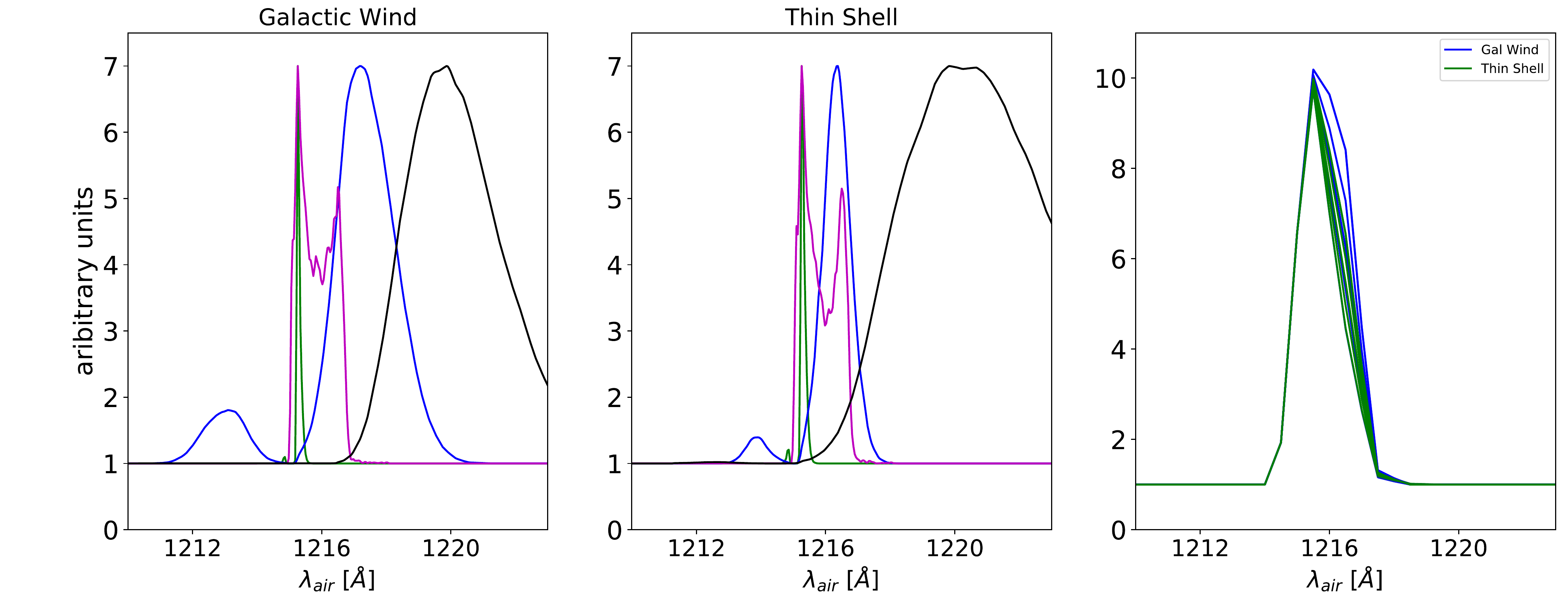}
\caption{Predicted Ly$\alpha$ profiles obtained with the FLaREON model. We present profiles obtained assuming the geometry of a galactic wind ($left$) and of a thin shell ($middle$), the typical dust reddening of the LAEVs in our identified overdensities, and considering the median IGM transmission curve of the top 50\% dense environments at $z=3.0$, as explained in the text. The model curves are arbitrarily normalized to the height of the stacked Ly$\alpha$ profile of LAEVs in our identified overdensities (Fig. \ref{stackLAE}).
%We also show models without including the IGM absorption ($right$). 
%
%{\bf{The black profles in the right panel only show the main red peak.}}
%
The different curves represent the following combinations of parameters, V$_{exp}$=10 km sec$^{-1}$, log(NHI)=18 (green);  V$_{exp}$=10 km sec$^{-1}$, log(NHI)=21.3 (blue); V$_{exp}$=200 km sec$^{-1}$, log(NHI)=18 (magenta); V$_{exp}$=200 km sec$^{-1}$, log(NHI)=21.3 (black) for the galactic wind geometry; 
V$_{exp}$=10 km sec$^{-1}$, log(NHI)=18 (green);  V$_{exp}$=10 km sec$^{-1}$, log(NHI)=21.5 (blue); V$_{exp}$=200 km sec$^{-1}$, log(NHI)=18 (magenta); V$_{exp}$=200 km sec$^{-1}$, log(NHI)=21.5 (black) for the thin shell geometry. 
%The model that performs the best prediction for the Ly$\alpha$ profile we observe in overdense regions is that with V$_{exp}$=10 km sec$^{-1}$ and log(NHI)$\sim$21 for both geometries. We show this last model (normalized to the highest flux of the red peak) with and without including IGM absorption as solid and dashed blue lines, respectively, in the right panel.
%
The $right$ panel shows models obtained assuming V$_{exp}$=160, 170, 180, 190, 200 km sec$^{-1}$ and log(NHI)=19, where the widest profile corresponds to highest velocity for the geometry of a galactic wind (blue) and of a thin shell (green). The curves are arbitrarily normalized to the height of the stacked Ly$\alpha$ profile of LAEVs in the field.
}
\label{model}%
\end{figure*}

In Fig. \ref{model}, we show the Ly$\alpha$ line profiles obtained with FLaREON. Models with V$_{exp}> 150$ km sec$^{-1}$ and NHI$<  10^{20}$ atoms cm$^{-2}$ present a negligible blue peak and a dominant red peak with complicated structures on the red-side wing of the red peak (see for example the magenta curves in the left and middle panels of the figure) that would blend into a broad profile at our resolution. The resulting profile would appear as the one we observe for the stack of the field LAEVs.
However, we want to investigate if there is any reasonable combination of parameters that reproduces a blue peak (in addition to a dominant red peak) as it can be seen in the stack profile of the LAEVs in overdense regions.

Inspecting the predicted line profiles, we find that the profiles with large V$_{exp}$ and high NHI$_{CGM}$ result in a main peak more redshifted and blue-to-red peak separations larger than what we observe in our spectra. Models with low expansion velocities, V$_{exp}$ $ \sim 10$ km sec$^{-1}$, and HI column densities of the order of log(NHI$_{CGM}) \sim 21$
%We find that the biconical outflow geometry with expansion velocity, V$_{exp}$ $=10$ km sec$^{-1}$, and log(NHI$_{CGM}$)=21 
 produce Ly$\alpha$ lines with a visible blue peak, especially after applying the IGM transmission typical of overdense regions. 
As it is shown in Fig. 4 of \citet{Gurung-Lopez2018}, HI column densities larger than $10^{20}$ atoms cm$^{-2}$ could result in blue vs red peak separations even as large as 4 {\AA}, as we observed in our data. In fact, we find that in the case of thin-shell and galactic-wind geometries a blue peak at a separation of about 4 {\AA} from the red peak can be predicted when V$_{exp}$ $\sim10$ km sec$^{-1}$ and log(NHI$_{CGM}) > 21$ \citep[see also Fig 7e in][]{Laursen2011}. 

%as the one we observe in the stack of the LAEVs in overdense regions. However, with the same NHI value and V$_{exp}$ = 200 km sec$^{-1}$ the blue-side flux is reduced significantly. In the case of the thin shell geometry and for log(NHI)=21, the blue peak is also preserved when V$_{exp}$ $\sim10$ km sec$^{-1}$ 
%In Fig. \ref{model}, we show a qualitative comparison with the model and discover that biconical outflow geometry with expansion velocity, V$_{exp}$ $<20$ km sec$^{-1}$, and log(NHI$_{CGM}$)=21 can produce a blue peak of the Ly$\alpha$ line as the one we observe in the stack of the LAEVs in overdense regions. This blue-side flux can be significantly reduced in the case of V$_{exp}$ close to 200 km sec$^{-1}$. In the case of the geometry of a thin shell and for log(NHI)=21, the blue peak is also preserve in the case of a low expansion velocity, V$_{exp}$ $\sim20$ km sec$^{-1}$ \citep[see also Fig 7e in][]{Laursen2011}.
%According to the comparison with FLaREON, the combination of low V$_{exp}$ and large NHI could be the favourable interpretation of our Ly$\alpha$ profile. The low V$_{exp}$ can be related to the fact that the galaxies in overdense regions are forming stars at a low rate, preventing the formation of strong outflows. The NHI could be high because we are studying regions of the Universe which are dense in terms of number density of galaxies and so maybe also dense of HI gas. It is also possible that the HI gas is trapped in the overdensity gravitational potential and stays in the medium surrounding the galaxies \citep{Zheng2016}. In addition to this a large NHI could be characteristic of high mass galaxies which are experiencing elongated phases of star formation, as opposite to bursty phases maybe more typical of low mass LAEVs in the field. 
%
Due to the low statistic, we just perform a qualitative comparison with FLaREON. According to this comparison, the stacked Ly$\alpha$ line profile of the LAEVs in overdense region presented in this work is unlikely to be consistent with a model obtained from a galaxy with large-V$_{exp}$ outflows and low HI column densities. %consistent with that obtained in a galaxy with low-V$_{exp}$ outflows and large HI column densities. 
The low V$_{exp}$, instead, can be related to the fact that the galaxies in overdense regions are currently forming stars at a low rate, preventing the formation of strong or fast outflows. The NHI could be high because we are studying regions of the Universe which are dense in terms of number density of galaxies and so maybe also dense of HI gas. It is also possible that the HI gas is trapped in the overdensity gravitational potential and stays in the medium surrounding the galaxies \citep[e.g.,][]{Zheng2016}. In addition to this, a large NHI could be characteristic of high mass galaxies which are experiencing quiescent star formation, as opposed to bursty phases maybe more typical of low mass LAEVs in the field. In fact, the FLaREON profiles with V$_{exp}> 150$ km sec$^{-1}$ and NHI $<  10^{20}$ atoms cm$^{-2}$   qualitatively reproduce the Ly$\alpha$ stack of the LAEVs in the field.
%
%Despite the low statistics, we have evidence that Ly$\alpha$ emitters, thanks to the information provided by their Ly$\alpha$ emission, can tell us something about the environment in which they and other galaxies live in.
%In this work we have studied XX Ly$\alpha$ emitters and their corresponding Ly$\alpha$ line profile. 
Even if a bigger sample of LAEVs in overdense regions would be desirable, we find evidence that the Ly$\alpha$ line shape is modulated by the surrounding environment. 
%{\bf{stack allora typical behaviour of LAEVs in overdensities}}

%--------------------------------------
\section{Conclusions}
\label{summary}

With the aim of investigating whether Ly$\alpha$-emitting galaxies trace the environment and its properties, we studied local densities in the CDFS and in the UDS. 
We made use of the VANDELS  photometric catalog. The incompleteness of the catalog does not affect in a severe way the estimation of the local densities and the identification of overdense regions in our fields. Photometric redshift uncertainties are the main limit in the study of the environment, as we tested with mocks.

We identified and characterized 22 overdensities at $2<z<4$ (Sect. 5). % that we have qualitatively characterized. 
Some of these overdense regions show one main density peak, but the majority are composed of more than one structure. None of the identified overdensities seems to show properties similar to the virialized structures at $z<2$, but we see that in some of them the members have rest-frame $U-V$ colors redder than the field galaxies. Also, the identified overdensities are typically on the track of the evolution from protoclusters to Fornax-type cluster at $z=0$.

Some of the overdensities occupy a large area of the field and may present more than one dense peak that are unlikely to be connected and to evolve in a unique structure at lower redshift. A test with mocks showed that some of the detected overdensities can be composed by more than one bound structure found in a N-body simulation, even when the positions of the galaxies are not affected by photometric redshift uncertainties. In an upcoming paper, we will study each overdensity and its fate to lower redshift in more detail. However, defining regions with high and low density in the VANDELS field is enough in this work to study the location of the Ly$\alpha$-emitting galaxies with respect to the environment.

We studied the physical and spectroscopic properties of the Ly$\alpha$-emitting galaxies detected in the first two seasons of the VANDELS survey. These LAEVs are VANDELS targets, drop-out galaxies by selection, and their location was not preselected in terms of density or environment. In the CDFS and in the UDS, we considered 51 and 80 galaxies, respectively, at $3<$ z$_{spec}<4$ with $1<$ EW(Ly$\alpha) <200$ {\AA}. 
In comparison to typical narrow-band selected LAEs, the LAEVs have lower sSFRs and are more massive, but have physical properties in agreement with models of Ly$\alpha$-emitting galaxies \citep[e.g.,][]{Gurung-Lopez2019b}. 

Among the selected LAEVs, only nine are members of the detected overdensities and 100 are located in the field.  A KS test showed that we can not reject the null hypothesis that the EW(Ly$\alpha$) (Fig. \ref{EWdistr}) and L(Ly$\alpha$) (Fig. \ref{FLLya}) of the LAEVs in the overdensities and in the field are drawn from the same distribution, but that we can reject the null hypothesis for their stellar masses, sSFRs, and rest-frame UV magnitudes (Fig. \ref{figpropLAEnonLAE}).
We do not typically find the LAEVs in the cores of the most dense regions of the environment. However, the LAEVs in the identified overdensities tend to show higher stellar masses and lower sSFRs with respect to the ones in the field, but are brighter in the rest-frame UV and NIR, and some of them have a clear signature of dust content. Moreover, the LAEVs in overdensities share some physical properties of the galaxies without Ly$\alpha$ in emission in overdense regions.

We stacked the 1D Ly$\alpha$ profiles of the LAEVs in the detected overdensities and in the field (Fig. \ref{stackLAE}). The stack of the LAEVs in overdense regions shows a hint of a blue peak and that the Ly$\alpha$ emission is spatially more extended than the UV continuum.  %that seems to be offset with respect to the UV continuum in the spacial direction. 

We discussed the possible interpretation of the Ly$\alpha$ profiles in terms of the FLaREON model. Geometrical combinations of interstellar media with low expansion velocities and high HI column densities seem favourable to interpret the stack of the Ly$\alpha$ profiles of the LAEVs in our identified overdensities. Models with V$_{exp}> 150$ km sec$^{-1}$ and NHI $<  10^{20}$ atoms cm$^{-2}$, at the resolution of our data, result in a broad red peak that qualitatively reproduce the Ly$\alpha$ stack of the LAEVs in the field. On the other hand, an outflow with low expansion velocity could be related to the low rate at which the galaxies are currently forming stars in overdense regions. %Therefore, the Ly$\alpha$ emitting galaxies can give useful insights on the environment in which they and also other galaxies live in.

\citet{Guaita2017} found that Ly$\alpha$ emitters characterized by expansion velocities of the order of 300 km sec$^{-1}$ and Ly$\alpha$ spatial extension comparable to that of the UV continuum could be experiencing bursty and short phases of star formation which tend to consume the HI gas quickly, preventing an efficient HI scattering and the formation of extended Ly$\alpha$ nebulae. On the other hand, the galaxies with interstellar media consistent with being static and %extended Ly$\alpha$ nebulae related to 
HI column densities of the order $10^{21}$ atoms cm$^{-2}$ could be characterized by longer phases of star formation which consume the HI gas slowly and can be characterized by extended Ly$\alpha$ nebulae. 
%They could also be currently forming stars at a low rate. %by allow an efficient HI scattering of the Ly$\alpha$ photons {\bf {say better}}.

In the overdensities we study here, galaxies in general and LAEVs in particular could be experiencing either long or slow phases of star formation as opposed to short bursty phases due to the effects of environment. %In addition to this, the column density of the HI surrounding the LAEVs in overdense regions could be significant because in overdense regions the HI is already more abundant than in the field and it stays trapped due to the gravitational potential \citep[e.g., ][]{Zheng2016}.
Despite the low statistics, thanks to the information provided by the Ly$\alpha$ emission, Ly$\alpha$ emitters may provide useful insights on the environment in which they reside.

%- Dalle masse del protocluster e masse stellare dei membri potremmo trovare la missing mass.
%-Vedere come si "muovono" le galassie rispetto a quelle del picco (meglio se zspec)
%e capire un po' di dinamica delle strutture. Cosa ci dice l'analisi morphologica sulla dinamica?c'e' AGN nel core?

\begin{acknowledgements}
We thank the ESO staff for their continuous support for the VANDELS survey, particularly the Paranal staff, who helped us to conduct the observations, and the ESO user support group in Garching.
      We thank Manuel Aravena, Jeremy Blazoit, Thibault Garel, Sam Kim, Roberto Gonzalez for useful discussions, Alvaro Orsi and Siddharta Gurung-Lopez for providing radiative transfer models. %, Bram Venemans for providing spectra of LAEs around radio galaxies for comparison.
We thank the ESO scientific visitor program during which part of this project was carried on.
We acknowledge support from CONICYT grants Basal-CATA Basal AFB-170002 (LG, FEB), and Programa de Astronomia FONDO ALMA 2016 31160033 (LG), and the Ministry of Economy, Development, and Tourism's Millennium Science Initiative through grant IC120009, awarded to The Millennium Institute of Astrophysics, MAS (FEB). RA acknowledges support from FONDECYT Regular Grant 1202007.
\end{acknowledgements}

\bibliographystyle{aa}   %% bibliography style file aa.bst from A&A
\bibliography{biblio}

\begin{thebibliography}{103}
\expandafter\ifx\csname natexlab\endcsname\relax\def\natexlab#1{#1}\fi

\bibitem[{{Abell}(1958)}]{Abell1958}
{Abell}, G.~O. 1958, \apjs, 3, 211

\bibitem[{{Andreon} {et~al.}(2014){Andreon}, {Newman}, {Trinchieri},
  {Raichoor}, {Ellis}, \& {Treu}}]{andre14}
{Andreon}, S., {Newman}, A.~B., {Trinchieri}, G., {et~al.} 2014, \aap, 565,
  A120

\bibitem[{{Balestra} {et~al.}(2010){Balestra}, {Mainieri}, {Popesso},
  {Dickinson}, {Nonino}, {Rosati}, {Teimoorinia}, {Vanzella}, {Cristiani},
  {Cesarsky}, {Fosbury}, {Kuntschner}, \& {Rettura}}]{Balestra2010}
{Balestra}, I., {Mainieri}, V., {Popesso}, P., {et~al.} 2010, \aap, 512, A12

\bibitem[{{Baugh} {et~al.}(2018){Baugh}, {Gonzalez-Perez}, {del P Lagos},
  {Lacey}, {Helly}, {Jenkins}, {Frenk}, {Benson}, {Bower}, \&
  {Cole}}]{Baugh2018}
{Baugh}, C.~M., {Gonzalez-Perez}, V., {del P Lagos}, C., {et~al.} 2018, \mnras

\bibitem[{{Baugh} {et~al.}(2019){Baugh}, {Gonzalez-Perez}, {Lagos}, {Lacey},
  {Helly}, {Jenkins}, {Frenk}, {Benson}, {Bower}, \& {Cole}}]{Baugh2019}
{Baugh}, C.~M., {Gonzalez-Perez}, V., {Lagos}, C. d.~P., {et~al.} 2019, \mnras,
  483, 4922

\bibitem[{{Beers} {et~al.}(1990){Beers}, {Flynn}, \& {Gebhardt}}]{Beers1990}
{Beers}, T.~C., {Flynn}, K., \& {Gebhardt}, K. 1990, \aj, 100, 32

\bibitem[{{Bertin} \& {Arnouts}(1996)}]{bertin1996}
{Bertin}, E. \& {Arnouts}, S. 1996, \aaps, 117, 393

\bibitem[{{Bruzual} \& {Charlot}(2003)}]{Bruzual:2003}
{Bruzual}, G. \& {Charlot}, S. 2003, \mnras, 344, 1000

\bibitem[{{Calzetti} {et~al.}(2000){Calzetti}, {Armus}, {Bohlin}, {Kinney},
  {Koornneef}, \& {Storchi-Bergmann}}]{Calzetti2000}
{Calzetti}, D., {Armus}, L., {Bohlin}, R.~C., {et~al.} 2000, \apj, 533, 682

\bibitem[{{Carnall} {et~al.}(2019){Carnall}, {McLure}, {Dunlop}, {Cullen},
  {McLeod}, {Wild}, {Johnson}, {Appleby}, {Dav{\'e}}, {Amorin}, {Bolzonella},
  {Castellano}, {Cimatti}, {Cucciati}, {Gargiulo}, {Garilli}, {Marchi},
  {Pentericci}, {Pozzetti}, {Schreiber}, {Talia}, \& {Zamorani}}]{Carnall2019}
{Carnall}, A.~C., {McLure}, R.~J., {Dunlop}, J.~S., {et~al.} 2019, \mnras, 490,
  417

\bibitem[{{Castellano} {et~al.}(2007){Castellano}, {Salimbeni}, {Trevese},
  {Grazian}, {Pentericci}, {Fiore}, {Fontana}, {Giallongo}, {Santini},
  {Cristiani}, {Nonino}, \& {Vanzella}}]{Castellano2007}
{Castellano}, M., {Salimbeni}, S., {Trevese}, D., {et~al.} 2007, \apj, 671,
  1497

\bibitem[{{Chiang} {et~al.}(2013){Chiang}, {Overzier}, \&
  {Gebhardt}}]{Chiang2013}
{Chiang}, Y.-K., {Overzier}, R., \& {Gebhardt}, K. 2013, \apj, 779, 127

\bibitem[{{Contini} {et~al.}(2016){Contini}, {De Lucia}, {Hatch}, {Borgani}, \&
  {Kang}}]{contini2016}
{Contini}, E., {De Lucia}, G., {Hatch}, N., {Borgani}, S., \& {Kang}, X. 2016,
  \mnras, 456, 1924

\bibitem[{{Cooper} {et~al.}(2012){Cooper}, {Yan}, {Dickinson}, {Juneau},
  {Lotz}, {Newman}, {Papovich}, {Salim}, {Walth}, {Weiner}, \&
  {Willmer}}]{Cooper2012}
{Cooper}, M.~C., {Yan}, R., {Dickinson}, M., {et~al.} 2012, \mnras, 425, 2116

\bibitem[{{Cortese} {et~al.}(2006){Cortese}, {Gavazzi}, {Boselli}, {Franzetti},
  {Kennicutt}, {O'Neil}, \& {Sakai}}]{cortese06}
{Cortese}, L., {Gavazzi}, G., {Boselli}, A., {et~al.} 2006, \aap, 453, 847

\bibitem[{{Cucciati} {et~al.}(2018){Cucciati}, {Lemaux}, {Zamorani}, {Le
  F{\`e}vre}, {Tasca}, {Hathi}, {Lee}, {Bardelli}, {Cassata}, {Garilli}, {Le
  Brun}, {Maccagni}, {Pentericci}, {Thomas}, {Vanzella}, {Zucca}, {Lubin},
  {Amorin}, {Cassar{\`a}}, {Cimatti}, {Talia}, {Vergani}, {Koekemoer}, {Pforr},
  \& {Salvato}}]{Cucciati2018}
{Cucciati}, O., {Lemaux}, B.~C., {Zamorani}, G., {et~al.} 2018, \aap, 619, A49

\bibitem[{{Cucciati} {et~al.}(2014){Cucciati}, {Zamorani}, {Lemaux},
  {Bardelli}, {Cimatti}, {Le F{\`e}vre}, {Cassata}, {Garilli}, {Le Brun},
  {Maccagni}, {Pentericci}, {Tasca}, {Thomas}, {Vanzella}, {Zucca}, {Amorin},
  {Capak}, {Cassar{\`a}}, {Castellano}, {Cuby}, {de la Torre}, {Durkalec},
  {Fontana}, {Giavalisco}, {Grazian}, {Hathi}, {Ilbert}, {Moreau}, {Paltani},
  {Ribeiro}, {Salvato}, {Schaerer}, {Scodeggio}, {Sommariva}, {Talia},
  {Taniguchi}, {Tresse}, {Vergani}, {Wang}, {Charlot}, {Contini}, {Fotopoulou},
  {L{\'o}pez-Sanjuan}, {Mellier}, \& {Scoville}}]{cuccia14}
{Cucciati}, O., {Zamorani}, G., {Lemaux}, B.~C., {et~al.} 2014, \aap, 570, A16

\bibitem[{{Curtis-Lake} {et~al.}(2012){Curtis-Lake}, {McLure}, {Pearce},
  {Dunlop}, {Cirasuolo}, {Stark}, {Almaini}, {Bradshaw}, {Chuter}, {Foucaud},
  \& {Hartley}}]{CurtisLake2012}
{Curtis-Lake}, E., {McLure}, R.~J., {Pearce}, H.~J., {et~al.} 2012, \mnras,
  422, 1425

\bibitem[{{Davis} {et~al.}(1985){Davis}, {Efstathiou}, {Frenk}, \&
  {White}}]{Davis1985}
{Davis}, M., {Efstathiou}, G., {Frenk}, C.~S., \& {White}, S.~D.~M. 1985, \apj,
  292, 371

\bibitem[{{De Lucia} \& {Blaizot}(2007)}]{DeLuciaBlaizot2007}
{De Lucia}, G. \& {Blaizot}, J. 2007, \mnras, 375, 2

\bibitem[{{Durkalec} {et~al.}(2015){Durkalec}, {Le F{\`e}vre}, {Pollo}, {de la
  Torre}, {Cassata}, {Garilli}, {Le Brun}, {Lemaux}, {Maccagni}, {Pentericci},
  {Tasca}, {Thomas}, {Vanzella}, {Zamorani}, {Zucca}, {Amor{\'{\i}}n},
  {Bardelli}, {Cassar{\`a}}, {Castellano}, {Cimatti}, {Cucciati}, {Fontana},
  {Giavalisco}, {Grazian}, {Hathi}, {Ilbert}, {Paltani}, {Ribeiro}, {Schaerer},
  {Scodeggio}, {Sommariva}, {Talia}, {Tresse}, {Vergani}, {Capak}, {Charlot},
  {Contini}, {Cuby}, {Dunlop}, {Fotopoulou}, {Koekemoer}, {L{\'o}pez-Sanjuan},
  {Mellier}, {Pforr}, {Salvato}, {Scoville}, {Taniguchi}, \&
  {Wang}}]{Durkalec2015}
{Durkalec}, A., {Le F{\`e}vre}, O., {Pollo}, A., {et~al.} 2015, \aap, 583, A128

\bibitem[{{Forrest} {et~al.}(2017){Forrest}, {Tran}, {Broussard}, {Allen},
  {Apfel}, {Cowley}, {Glazebrook}, {Kacprzak}, {Labb{\'e}}, {Nanayakkara},
  {Papovich}, {Quadri}, {Spitler}, {Straatman}, \& {Tomczak}}]{Forrest2017}
{Forrest}, B., {Tran}, K.-V.~H., {Broussard}, A., {et~al.} 2017, \apjl, 838,
  L12

\bibitem[{{Franck} \& {McGaugh}(2016)}]{Franck2016}
{Franck}, J.~R. \& {McGaugh}, S.~S. 2016, \apj, 833, 15

\bibitem[{{Fukugita} {et~al.}(1995){Fukugita}, {Shimasaku}, \&
  {Ichikawa}}]{Fukugita1995}
{Fukugita}, M., {Shimasaku}, K., \& {Ichikawa}, T. 1995, \pasp, 107, 945

\bibitem[{{Galametz} {et~al.}(2013){Galametz}, {Grazian}, {Fontana},
  {Ferguson}, {Ashby}, {Barro}, {Castellano}, {Dahlen}, {Donley}, {Faber},
  {Grogin}, {Guo}, {Huang}, {Kocevski}, {Koekemoer}, {Lee}, {McGrath}, {Peth},
  {Willner}, {Almaini}, {Cooper}, {Cooray}, {Conselice}, {Dickinson}, {Dunlop},
  {Fazio}, {Foucaud}, {Gardner}, {Giavalisco}, {Hathi}, {Hartley}, {Koo},
  {Lai}, {de Mello}, {McLure}, {Lucas}, {Paris}, {Pentericci}, {Santini},
  {Simpson}, {Sommariva}, {Targett}, {Weiner}, {Wuyts}, \& {the CANDELS
  Team}}]{Galametz2013}
{Galametz}, A., {Grazian}, A., {Fontana}, A., {et~al.} 2013, \apjs, 206, 10

\bibitem[{{Garel} {et~al.}(2015){Garel}, {Blaizot}, {Guiderdoni},
  {Michel-Dansac}, {Hayes}, \& {Verhamme}}]{Garel2015}
{Garel}, T., {Blaizot}, J., {Guiderdoni}, B., {et~al.} 2015, \mnras, 450, 1279

\bibitem[{{Garilli} {et~al.}(2010){Garilli}, {Fumana}, {Franzetti}, {Paioro},
  {Scodeggio}, {Le F{\`e}vre}, {Paltani}, \& {Scaramella}}]{Garilli2010}
{Garilli}, B., {Fumana}, M., {Franzetti}, P., {et~al.} 2010, \pasp, 122, 827

\bibitem[{{Gehrels}(1986)}]{Gehrels1986}
{Gehrels}, N. 1986, \apj, 303, 336

\bibitem[{{Gladders} \& {Yee}(2000)}]{GladdersYee2000}
{Gladders}, M.~D. \& {Yee}, H.~K.~C. 2000, in Astronomical Society of the
  Pacific Conference Series, Vol. 215, Cosmic Evolution and Galaxy Formation:
  Structure, Interactions, and Feedback, ed. J.~{Franco}, L.~{Terlevich},
  O.~{L{\'o}pez-Cruz}, \& I.~{Aretxaga}, 233

\bibitem[{{Grazian} {et~al.}(2006){Grazian}, {Fontana}, {de Santis}, {Nonino},
  {Salimbeni}, {Giallongo}, {Cristiani}, {Gallozzi}, \&
  {Vanzella}}]{Grazian2006}
{Grazian}, A., {Fontana}, A., {de Santis}, C., {et~al.} 2006, \aap, 449, 951

\bibitem[{{Grogin} {et~al.}(2011){Grogin}, {Kocevski}, {Faber}, {Ferguson},
  {Koekemoer}, {Riess}, {Acquaviva}, {Alexander}, {Almaini}, {Ashby}, {Barden},
  {Bell}, {Bournaud}, {Brown}, {Caputi}, {Casertano}, {Cassata}, {Castellano},
  {Challis}, {Chary}, {Cheung}, {Cirasuolo}, {Conselice}, {Roshan Cooray},
  {Croton}, {Daddi}, {Dahlen}, {Dav{\'e}}, {de Mello}, {Dekel}, {Dickinson},
  {Dolch}, {Donley}, {Dunlop}, {Dutton}, {Elbaz}, {Fazio}, {Filippenko},
  {Finkelstein}, {Fontana}, {Gardner}, {Garnavich}, {Gawiser}, {Giavalisco},
  {Grazian}, {Guo}, {Hathi}, {H{\"a}ussler}, {Hopkins}, {Huang}, {Huang},
  {Jha}, {Kartaltepe}, {Kirshner}, {Koo}, {Lai}, {Lee}, {Li}, {Lotz}, {Lucas},
  {Madau}, {McCarthy}, {McGrath}, {McIntosh}, {McLure}, {Mobasher},
  {Moustakas}, {Mozena}, {Nandra}, {Newman}, {Niemi}, {Noeske}, {Papovich},
  {Pentericci}, {Pope}, {Primack}, {Rajan}, {Ravindranath}, {Reddy}, {Renzini},
  {Rix}, {Robaina}, {Rodney}, {Rosario}, {Rosati}, {Salimbeni}, {Scarlata},
  {Siana}, {Simard}, {Smidt}, {Somerville}, {Spinrad}, {Straughn}, {Strolger},
  {Telford}, {Teplitz}, {Trump}, {van der Wel}, {Villforth}, {Wechsler},
  {Weiner}, {Wiklind}, {Wild}, {Wilson}, {Wuyts}, {Yan}, \& {Yun}}]{Grogin2011}
{Grogin}, N.~A., {Kocevski}, D.~D., {Faber}, S.~M., {et~al.} 2011, \apjs, 197,
  35

\bibitem[{{Gronke} \& {Dijkstra}(2016)}]{Gronke2016}
{Gronke}, M. \& {Dijkstra}, M. 2016, \apj, 826, 14

\bibitem[{{Gronwall} {et~al.}(2007){Gronwall}, {Ciardullo}, {Hickey},
  {Gawiser}, {Feldmeier}, {van Dokkum}, {Urry}, {Herrera}, {Lehmer}, {Infante},
  {Orsi}, {Marchesini}, {Blanc}, {Francke}, {Lira}, \&
  {Treister}}]{Gronwall:2007}
{Gronwall}, C., {Ciardullo}, R., {Hickey}, T., {et~al.} 2007, \apj, 667, 79

\bibitem[{{Guaita} {et~al.}(2010){Guaita}, {Gawiser}, {Padilla}, {Francke},
  {Bond}, {Gronwall}, {Ciardullo}, {Feldmeier}, {Sinawa}, {Blanc}, \&
  {Virani}}]{Guaita2010}
{Guaita}, L., {Gawiser}, E., {Padilla}, N., {et~al.} 2010, \apj, 714, 255

\bibitem[{{Guaita} {et~al.}(2017){Guaita}, {Talia}, {Pentericci}, {Verhamme},
  {Cassata}, {Lemaux}, {Orlitova}, {Ribeiro}, {Schaerer}, {Zamorani},
  {Garilli}, {Le Brun}, {Le F{\`e}vre}, {Maccagni}, {Tasca}, {Thomas},
  {Vanzella}, {Zucca}, {Amorin}, {Bardelli}, {Castellano}, {Grazian}, {Hathi},
  {Koekemoer}, \& {Marchi}}]{Guaita2017}
{Guaita}, L., {Talia}, M., {Pentericci}, L., {et~al.} 2017, \aap, 606, A19

\bibitem[{{Guo} {et~al.}(2013){Guo}, {Ferguson}, {Giavalisco}, {Barro},
  {Willner}, {Ashby}, {Dahlen}, {Donley}, {Faber}, {Fontana}, {Galametz},
  {Grazian}, {Huang}, {Kocevski}, {Koekemoer}, {Koo}, {McGrath}, {Peth},
  {Salvato}, {Wuyts}, {Castellano}, {Cooray}, {Dickinson}, {Dunlop}, {Fazio},
  {Gardner}, {Gawiser}, {Grogin}, {Hathi}, {Hsu}, {Lee}, {Lucas}, {Mobasher},
  {Nandra}, {Newman}, \& {van der Wel}}]{Guo2013}
{Guo}, Y., {Ferguson}, H.~C., {Giavalisco}, M., {et~al.} 2013, \apjs, 207, 24

\bibitem[{{Gurung-L{\'o}pez} {et~al.}(2019{\natexlab{a}}){Gurung-L{\'o}pez},
  {Orsi}, \& {Bonoli}}]{Gurung-Lopez2018}
{Gurung-L{\'o}pez}, S., {Orsi}, {\'A}.~A., \& {Bonoli}, S. 2019{\natexlab{a}},
  \mnras, 490, 733

\bibitem[{{Gurung-L{\'o}pez} {et~al.}(2019{\natexlab{b}}){Gurung-L{\'o}pez},
  {Orsi}, {Bonoli}, {Baugh}, \& {Lacey}}]{Gurung-Lopez2019}
{Gurung-L{\'o}pez}, S., {Orsi}, {\'A}.~A., {Bonoli}, S., {Baugh}, C.~M., \&
  {Lacey}, C.~G. 2019{\natexlab{b}}, \mnras, 486, 1882

\bibitem[{{Gurung-L{\'o}pez} {et~al.}(2020){Gurung-L{\'o}pez}, {Orsi},
  {Bonoli}, {Padilla}, {Lacey}, \& {Baugh}}]{Gurung-Lopez2019b}
{Gurung-L{\'o}pez}, S., {Orsi}, {\'A}.~A., {Bonoli}, S., {et~al.} 2020, \mnras,
  491, 3266

\bibitem[{{Hagen} {et~al.}(2016){Hagen}, {Zeimann}, {Behrens}, {Ciardullo},
  {Grasshorn Gebhardt}, {Gronwall}, {Bridge}, {Fox}, {Schneider}, {Trump},
  {Blanc}, {Chiang}, {Chonis}, {Finkelstein}, {Hill}, {Jogee}, \&
  {Gawiser}}]{Hagen2016}
{Hagen}, A., {Zeimann}, G.~R., {Behrens}, C., {et~al.} 2016, \apj, 817, 79

\bibitem[{{Haines} {et~al.}(2015){Haines}, {Pereira}, {Smith}, {Egami},
  {Babul}, {Finoguenov}, {Ziparo}, {McGee}, {Rawle}, {Okabe}, \&
  {Moran}}]{haines15}
{Haines}, C.~P., {Pereira}, M.~J., {Smith}, G.~P., {et~al.} 2015, \apj, 806,
  101

\bibitem[{{Hatch} {et~al.}(2011){Hatch}, {Kurk}, {Pentericci}, {Venemans},
  {Kuiper}, {Miley}, \& {R{\"o}ttgering}}]{Hatch2011}
{Hatch}, N.~A., {Kurk}, J.~D., {Pentericci}, L., {et~al.} 2011, \mnras, 415,
  2993

\bibitem[{{Hirschmann} {et~al.}(2016){Hirschmann}, {De Lucia}, \&
  {Fontanot}}]{Hirschmann2016}
{Hirschmann}, M., {De Lucia}, G., \& {Fontanot}, F. 2016, \mnras, 461, 1760

\bibitem[{{Kang} \& {Im}(2015)}]{KangIm2015}
{Kang}, E. \& {Im}, M. 2015, Journal of Korean Astronomical Society, 48, 21

\bibitem[{{Kodama} {et~al.}(2007){Kodama}, {Tanaka}, {Kajisawa}, {Kurk},
  {Venemans}, {De Breuck}, {Vernet}, \& {Lidman}}]{Kodama2007}
{Kodama}, T., {Tanaka}, I., {Kajisawa}, M., {et~al.} 2007, \mnras, 377, 1717

\bibitem[{{Koekemoer} {et~al.}(2011){Koekemoer}, {Faber}, {Ferguson}, {Grogin},
  {Kocevski}, {Koo}, {Lai}, {Lotz}, {Lucas}, {McGrath}, {Ogaz}, {Rajan},
  {Riess}, {Rodney}, {Strolger}, {Casertano}, {Castellano}, {Dahlen},
  {Dickinson}, {Dolch}, {Fontana}, {Giavalisco}, {Grazian}, {Guo}, {Hathi},
  {Huang}, {van der Wel}, {Yan}, {Acquaviva}, {Alexander}, {Almaini}, {Ashby},
  {Barden}, {Bell}, {Bournaud}, {Brown}, {Caputi}, {Cassata}, {Challis},
  {Chary}, {Cheung}, {Cirasuolo}, {Conselice}, {Roshan Cooray}, {Croton},
  {Daddi}, {Dav{\'e}}, {de Mello}, {de Ravel}, {Dekel}, {Donley}, {Dunlop},
  {Dutton}, {Elbaz}, {Fazio}, {Filippenko}, {Finkelstein}, {Frazer}, {Gardner},
  {Garnavich}, {Gawiser}, {Gruetzbauch}, {Hartley}, {H{\"a}ussler},
  {Herrington}, {Hopkins}, {Huang}, {Jha}, {Johnson}, {Kartaltepe},
  {Khostovan}, {Kirshner}, {Lani}, {Lee}, {Li}, {Madau}, {McCarthy},
  {McIntosh}, {McLure}, {McPartland}, {Mobasher}, {Moreira}, {Mortlock},
  {Moustakas}, {Mozena}, {Nandra}, {Newman}, {Nielsen}, {Niemi}, {Noeske},
  {Papovich}, {Pentericci}, {Pope}, {Primack}, {Ravindranath}, {Reddy},
  {Renzini}, {Rix}, {Robaina}, {Rosario}, {Rosati}, {Salimbeni}, {Scarlata},
  {Siana}, {Simard}, {Smidt}, {Snyder}, {Somerville}, {Spinrad}, {Straughn},
  {Telford}, {Teplitz}, {Trump}, {Vargas}, {Villforth}, {Wagner}, {Wandro},
  {Wechsler}, {Weiner}, {Wiklind}, {Wild}, {Wilson}, {Wuyts}, \&
  {Yun}}]{Koekemoer2011}
{Koekemoer}, A.~M., {Faber}, S.~M., {Ferguson}, H.~C., {et~al.} 2011, \apjs,
  197, 36

\bibitem[{{Koyama} {et~al.}(2013){Koyama}, {Kodama}, {Tadaki}, {Hayashi},
  {Tanaka}, {Smail}, {Tanaka}, \& {Kurk}}]{Koyama2013}
{Koyama}, Y., {Kodama}, T., {Tadaki}, K.-i., {et~al.} 2013, \mnras, 428, 1551

\bibitem[{{Kriek} {et~al.}(2015){Kriek}, {Shapley}, {Reddy}, {Siana}, {Coil},
  {Mobasher}, {Freeman}, {de Groot}, {Price}, {Sanders}, {Shivaei}, {Brammer},
  {Momcheva}, {Skelton}, {van Dokkum}, {Whitaker}, {Aird}, {Azadi}, {Kassis},
  {Bullock}, {Conroy}, {Dav{\'e}}, {Kere{\v s}}, \& {Krumholz}}]{Kriek2015}
{Kriek}, M., {Shapley}, A.~E., {Reddy}, N.~A., {et~al.} 2015, \apjs, 218, 15

\bibitem[{{Kubo} {et~al.}(2013){Kubo}, {Uchimoto}, {Yamada}, {Kajisawa},
  {Ichikawa}, {Matsuda}, {Akiyama}, {Hayashino}, {Konishi}, {Nishimura},
  {Omata}, {Suzuki}, {Tanaka}, {Yoshikawa}, {Alexander}, {Fazio}, {Huang}, \&
  {Lehmer}}]{Kubo2013}
{Kubo}, M., {Uchimoto}, Y.~K., {Yamada}, T., {et~al.} 2013, \apj, 778, 170

\bibitem[{{Kurk} {et~al.}(2013){Kurk}, {Cimatti}, {Daddi}, {Mignoli},
  {Pozzetti}, {Dickinson}, {Bolzonella}, {Zamorani}, {Cassata}, {Rodighiero},
  {Franceschini}, {Renzini}, {Rosati}, {Halliday}, \& {Berta}}]{Kurk2013}
{Kurk}, J., {Cimatti}, A., {Daddi}, E., {et~al.} 2013, \aap, 549, A63

\bibitem[{{Lacey} {et~al.}(2016){Lacey}, {Baugh}, {Frenk}, {Benson}, {Bower},
  {Cole}, {Gonzalez-Perez}, {Helly}, {Lagos}, \& {Mitchell}}]{Lacey2016}
{Lacey}, C.~G., {Baugh}, C.~M., {Frenk}, C.~S., {et~al.} 2016, \mnras, 462,
  3854

\bibitem[{{Laursen} {et~al.}(2011){Laursen}, {Sommer-Larsen}, \&
  {Razoumov}}]{Laursen2011}
{Laursen}, P., {Sommer-Larsen}, J., \& {Razoumov}, A.~O. 2011, \apj, 728, 52

\bibitem[{{Le F{\`e}vre} {et~al.}(2013){Le F{\`e}vre}, {Cassata}, {Cucciati},
  {Garilli}, {Ilbert}, {Le Brun}, {Maccagni}, {Moreau}, {Scodeggio}, {Tresse},
  {Zamorani}, {Adami}, {Arnouts}, {Bardelli}, {Bolzonella}, {Bondi},
  {Bongiorno}, {Bottini}, {Cappi}, {Charlot}, {Ciliegi}, {Contini}, {de la
  Torre}, {Foucaud}, {Franzetti}, {Gavignaud}, {Guzzo}, {Iovino}, {Lemaux},
  {L{\'o}pez-Sanjuan}, {McCracken}, {Marano}, {Marinoni}, {Mazure}, {Mellier},
  {Merighi}, {Merluzzi}, {Paltani}, {Pell{\`o}}, {Pollo}, {Pozzetti},
  {Scaramella}, {Tasca}, {Vergani}, {Vettolani}, {Zanichelli}, \&
  {Zucca}}]{LeFevre2013}
{Le F{\`e}vre}, O., {Cassata}, P., {Cucciati}, O., {et~al.} 2013, \aap, 559,
  A14

\bibitem[{{Le F{\`e}vre} {et~al.}(2015){Le F{\`e}vre}, {Tasca}, {Cassata},
  {Garilli}, {Le Brun}, {Maccagni}, {Pentericci}, {Thomas}, {Vanzella},
  {Zamorani}, {Zucca}, {Amorin}, {Bardelli}, {Capak}, {Cassar{\`a}},
  {Castellano}, {Cimatti}, {Cuby}, {Cucciati}, {de la Torre}, {Durkalec},
  {Fontana}, {Giavalisco}, {Grazian}, {Hathi}, {Ilbert}, {Lemaux}, {Moreau},
  {Paltani}, {Ribeiro}, {Salvato}, {Schaerer}, {Scodeggio}, {Sommariva},
  {Talia}, {Taniguchi}, {Tresse}, {Vergani}, {Wang}, {Charlot}, {Contini},
  {Fotopoulou}, {L{\'o}pez-Sanjuan}, {Mellier}, \& {Scoville}}]{LeFevre2015}
{Le F{\`e}vre}, O., {Tasca}, L.~A.~M., {Cassata}, P., {et~al.} 2015, \aap, 576,
  A79

\bibitem[{{Lemaux} {et~al.}(2014){Lemaux}, {Cucciati}, {Tasca}, {Le F{\`e}vre},
  {Zamorani}, {Cassata}, {Garilli}, {Le Brun}, {Maccagni}, {Pentericci},
  {Thomas}, {Vanzella}, {Zucca}, {Amor{\'{\i}}n}, {Bardelli}, {Capak},
  {Cassar{\`a}}, {Castellano}, {Cimatti}, {Cuby}, {de la Torre}, {Durkalec},
  {Fontana}, {Giavalisco}, {Grazian}, {Hathi}, {Ilbert}, {Moreau}, {Paltani},
  {Ribeiro}, {Salvato}, {Schaerer}, {Scodeggio}, {Sommariva}, {Talia},
  {Taniguchi}, {Tresse}, {Vergani}, {Wang}, {Charlot}, {Contini}, {Fotopoulou},
  {Gal}, {Kocevski}, {L{\'o}pez-Sanjuan}, {Lubin}, {Mellier}, {Sadibekova}, \&
  {Scoville}}]{Lemaux2014}
{Lemaux}, B.~C., {Cucciati}, O., {Tasca}, L.~A.~M., {et~al.} 2014, \aap, 572,
  A41

\bibitem[{{Lemaux} {et~al.}(2018){Lemaux}, {Le F{\`e}vre}, {Cucciati},
  {Ribeiro}, {Tasca}, {Zamorani}, {Ilbert}, {Thomas}, {Bardelli}, {Cassata},
  {Hathi}, {Pforr}, {Smol{\v c}i{\'c}}, {Delvecchio}, {Novak}, {Berta},
  {McCracken}, {Koekemoer}, {Amor{\'{\i}}n}, {Garilli}, {Maccagni}, {Schaerer},
  \& {Zucca}}]{Lemaux2018}
{Lemaux}, B.~C., {Le F{\`e}vre}, O., {Cucciati}, O., {et~al.} 2018, \aap, 615,
  A77

\bibitem[{{Lenz} \& {Ayres}(1992)}]{Lenz1992}
{Lenz}, D.~D. \& {Ayres}, T.~R. 1992, \pasp, 104, 1104

\bibitem[{{Madau}(1995)}]{Madau:1995}
{Madau}, P. 1995, \apj, 441, 18

\bibitem[{{Maltby} {et~al.}(2016){Maltby}, {Almaini}, {Wild}, {Hatch},
  {Hartley}, {Simpson}, {McLure}, {Dunlop}, {Rowlands}, \&
  {Cirasuolo}}]{Maltby2016}
{Maltby}, D.~T., {Almaini}, O., {Wild}, V., {et~al.} 2016, \mnras, 459, L114

\bibitem[{{Marchi} {et~al.}(2019){Marchi}, {Pentericci}, {Guaita}, {Talia},
  {Castellano}, {Hathi}, {Schaerer}, {Amorin}, {Bolzonella}, {Carnall},
  {Charlot}, {Chevallard}, {Cullen}, {Finkelstein}, {Fontana}, {Fontanot},
  {Garilli}, {Hibon}, {Koekemoer}, {Maccagni}, {McLure}, {Papovich},
  {Pozzetti}, \& {Saxena}}]{Marchi2019}
{Marchi}, F., {Pentericci}, L., {Guaita}, L., {et~al.} 2019, \aap, 631, A19

\bibitem[{{McLure} {et~al.}(2018){McLure}, {Pentericci}, {Cimatti}, {Dunlop},
  {Elbaz}, {Fontana}, {Nandra}, {Amorin}, {Bolzonella}, {Bongiorno}, {Carnall},
  {Castellano}, {Cirasuolo}, {Cucciati}, {Cullen}, {De Barros}, {Finkelstein},
  {Fontanot}, {Franzetti}, {Fumana}, {Gargiulo}, {Garilli}, {Guaita},
  {Hartley}, {Iovino}, {Jarvis}, {Juneau}, {Karman}, {Maccagni}, {Marchi},
  {M{\'a}rmol-Queralt{\'o}}, {Pompei}, {Pozzetti}, {Scodeggio}, {Sommariva},
  {Talia}, {Almaini}, {Balestra}, {Bardelli}, {Bell}, {Bourne}, {Bowler},
  {Brusa}, {Buitrago}, {Caputi}, {Cassata}, {Charlot}, {Citro}, {Cresci},
  {Cristiani}, {Curtis-Lake}, {Dickinson}, {Fazio}, {Ferguson}, {Fiore},
  {Franco}, {Fynbo}, {Galametz}, {Georgakakis}, {Giavalisco}, {Grazian},
  {Hathi}, {Jung}, {Kim}, {Koekemoer}, {Khusanova}, {F{\`e}vre}, {Lotz},
  {Mannucci}, {Maltby}, {Matsuoka}, {McLeod}, {Mendez-Hernandez},
  {Mendez-Abreu}, {Mignoli}, {Moresco}, {Mortlock}, {Nonino}, {Pannella},
  {Papovich}, {Popesso}, {Rosario}, {Salvato}, {Santini}, {Schaerer},
  {Schreiber}, {Stark}, {Tasca}, {Thomas}, {Treu}, {Vanzella}, {Wild},
  {Williams}, {Zamorani}, \& {Zucca}}]{McLure2018}
{McLure}, R.~J., {Pentericci}, L., {Cimatti}, A., {et~al.} 2018, \mnras

\bibitem[{{Miller} {et~al.}(2018){Miller}, {Chapman}, {Aravena}, {Ashby},
  {Hayward}, {Vieira}, {Wei{\ss}}, {Babul}, {B{\'e}thermin}, {Bradford},
  {Brodwin}, {Carlstrom}, {Chen}, {Cunningham}, {De Breuck}, {Gonzalez},
  {Greve}, {Harnett}, {Hezaveh}, {Lacaille}, {Litke}, {Ma}, {Malkan},
  {Marrone}, {Morningstar}, {Murphy}, {Narayanan}, {Pass}, {Perry}, {Phadke},
  {Rennehan}, {Rotermund}, {Simpson}, {Spilker}, {Sreevani}, {Stark},
  {Strandet}, \& {Strom}}]{miller18}
{Miller}, T.~B., {Chapman}, S.~C., {Aravena}, M., {et~al.} 2018, \nat, 556, 469

\bibitem[{{Momcheva} {et~al.}(2016){Momcheva}, {Brammer}, {van Dokkum},
  {Skelton}, {Whitaker}, {Nelson}, {Fumagalli}, {Maseda}, {Leja}, {Franx},
  {Rix}, {Bezanson}, {Da Cunha}, {Dickey}, {F{\"o}rster Schreiber},
  {Illingworth}, {Kriek}, {Labb{\'e}}, {Ulf Lange}, {Lundgren}, {Magee},
  {Marchesini}, {Oesch}, {Pacifici}, {Patel}, {Price}, {Tal}, {Wake}, {van der
  Wel}, \& {Wuyts}}]{Momcheva2016}
{Momcheva}, I.~G., {Brammer}, G.~B., {van Dokkum}, P.~G., {et~al.} 2016, \apjs,
  225, 27

\bibitem[{{Morris} {et~al.}(2015){Morris}, {Kocevski}, {Trump}, {Weiner},
  {Hathi}, {Barro}, {Dahlen}, {Faber}, {Finkelstein}, {Fontana}, {Ferguson},
  {Grogin}, {Gr{\"u}tzbauch}, {Guo}, {Hsu}, {Koekemoer}, {Koo}, {Mobasher},
  {Pforr}, {Salvato}, {Wiklind}, \& {Wuyts}}]{Morris2015}
{Morris}, A.~M., {Kocevski}, D.~D., {Trump}, J.~R., {et~al.} 2015, \aj, 149,
  178

\bibitem[{{Muldrew} {et~al.}(2015{\natexlab{a}}){Muldrew}, {Hatch}, \&
  {Cooke}}]{muldy15}
{Muldrew}, S.~I., {Hatch}, N.~A., \& {Cooke}, E.~A. 2015{\natexlab{a}}, \mnras,
  452, 2528

\bibitem[{{Muldrew} {et~al.}(2015{\natexlab{b}}){Muldrew}, {Hatch}, \&
  {Cooke}}]{Muldrew2015}
{Muldrew}, S.~I., {Hatch}, N.~A., \& {Cooke}, E.~A. 2015{\natexlab{b}}, \mnras,
  452, 2528

\bibitem[{{Newman} {et~al.}(2014){Newman}, {Ellis}, {Andreon}, {Treu},
  {Raichoor}, \& {Trinchieri}}]{Newman2014}
{Newman}, A.~B., {Ellis}, R.~S., {Andreon}, S., {et~al.} 2014, \apj, 788, 51

\bibitem[{{Orsi} {et~al.}(2012){Orsi}, {Lacey}, \& {Baugh}}]{Orsi2012}
{Orsi}, A., {Lacey}, C.~G., \& {Baugh}, C.~M. 2012, \mnras, 425, 87

\bibitem[{{Overzier}(2016)}]{Overzier2016}
{Overzier}, R.~A. 2016, \aapr, 24, 14

\bibitem[{{Pentericci} {et~al.}(2013){Pentericci}, {Castellano}, {Menci},
  {Salimbeni}, {Dahlen}, {Galametz}, {Santini}, {Grazian}, \&
  {Fontana}}]{Pentericci2013}
{Pentericci}, L., {Castellano}, M., {Menci}, N., {et~al.} 2013, \aap, 552, A111

\bibitem[{{Pentericci} {et~al.}(2000){Pentericci}, {Kurk}, {R{\"o}ttgering},
  {Miley}, {van Breugel}, {Carilli}, {Ford}, {Heckman}, {McCarthy}, \&
  {Moorwood}}]{Pentericci2000}
{Pentericci}, L., {Kurk}, J.~D., {R{\"o}ttgering}, H.~J.~A., {et~al.} 2000,
  \aap, 361, L25

\bibitem[{{Pentericci} {et~al.}(2018{\natexlab{a}}){Pentericci}, {McLure},
  {Franzetti}, {Garilli}, \& {the VANDELS team}}]{Pentericci2018c}
{Pentericci}, L., {McLure}, R.~J., {Franzetti}, P., {Garilli}, B., \& {the
  VANDELS team}. 2018{\natexlab{a}}, ArXiv1811.05298

\bibitem[{{Pentericci} {et~al.}(2018{\natexlab{b}}){Pentericci}, {McLure},
  {Garilli}, {Cucciati}, {Franzetti}, {Iovino}, {Amorin}, {Bolzonella},
  {Bongiorno}, {Carnall}, {Castellano}, {Cimatti}, {Cirasuolo}, {Cullen}, {De
  Barros}, {Dunlop}, {Elbaz}, {Finkelstein}, {Fontana}, {Fontanot}, {Fumana},
  {Gargiulo}, {Guaita}, {Hartley}, {Jarvis}, {Juneau}, {Karman}, {Maccagni},
  {Marchi}, {Marmol-Queralto}, {Nandra}, {Pompei}, {Pozzetti}, {Scodeggio},
  {Sommariva}, {Talia}, {Almaini}, {Balestra}, {Bardelli}, {Bell}, {Bourne},
  {Bowler}, {Brusa}, {Buitrago}, {Caputi}, {Cassata}, {Charlot}, {Citro},
  {Cresci}, {Cristiani}, {Curtis-Lake}, {Dickinson}, {Fazio}, {Ferguson},
  {Fiore}, {Franco}, {Fynbo}, {Galametz}, {Georgakakis}, {Giavalisco},
  {Grazian}, {Hathi}, {Jung}, {Kim}, {Koekemoer}, {Khusanova}, {Le F{\`e}vre},
  {Lotz}, {Mannucci}, {Maltby}, {Matsuoka}, {McLeod}, {Mendez-Hernandez},
  {Mendez-Abreu}, {Mignoli}, {Moresco}, {Mortlock}, {Nonino}, {Pannella},
  {Papovich}, {Popesso}, {Rosario}, {Salvato}, {Santini}, {Schaerer},
  {Schreiber}, {Stark}, {Tasca}, {Thomas}, {Treu}, {Vanzella}, {Wild},
  {Williams}, {Zamorani}, \& {Zucca}}]{Pentericci2018}
{Pentericci}, L., {McLure}, R.~J., {Garilli}, B., {et~al.} 2018{\natexlab{b}},
  \aap, 616, A174

\bibitem[{{Press} \& {Davis}(1982)}]{PressDavis1982}
{Press}, W.~H. \& {Davis}, M. 1982, \apj, 259, 449

\bibitem[{{Rhoads} {et~al.}(2009){Rhoads}, {Malhotra}, {Pirzkal}, {Dickinson},
  {Cohen}, {Grogin}, {Hathi}, {Xu}, {Ferreras}, {Gronwall}, {Koekemoer},
  {K{\"u}mmel}, {Meurer}, {Panagia}, {Pasquali}, {Ryan}, {Straughn}, {Walsh},
  {Windhorst}, \& {Yan}}]{Rhoads2009}
{Rhoads}, J.~E., {Malhotra}, S., {Pirzkal}, N., {et~al.} 2009, \apj, 697, 942

\bibitem[{{Salimbeni} {et~al.}(2009){Salimbeni}, {Castellano}, {Pentericci},
  {Trevese}, {Fiore}, {Grazian}, {Fontana}, {Giallongo}, {Boutsia},
  {Cristiani}, {de Santis}, {Gallozzi}, {Menci}, {Nonino}, {Paris}, {Santini},
  \& {Vanzella}}]{Salimbeni2009}
{Salimbeni}, S., {Castellano}, M., {Pentericci}, L., {et~al.} 2009, \aap, 501,
  865

\bibitem[{{Salimbeni} {et~al.}(2008){Salimbeni}, {Giallongo}, {Menci},
  {Castellano}, {Fontana}, {Grazian}, {Pentericci}, {Trevese}, {Cristiani},
  {Nonino}, \& {Vanzella}}]{Salimbeni2008}
{Salimbeni}, S., {Giallongo}, E., {Menci}, N., {et~al.} 2008, \aap, 477, 763

\bibitem[{{Santini} {et~al.}(2015){Santini}, {Ferguson}, {Fontana}, {Mobasher},
  {Barro}, {Castellano}, {Finkelstein}, {Grazian}, {Hsu}, {Lee}, {Lee},
  {Pforr}, {Salvato}, {Wiklind}, {Wuyts}, {Almaini}, {Cooper}, {Galametz},
  {Weiner}, {Amorin}, {Boutsia}, {Conselice}, {Dahlen}, {Dickinson},
  {Giavalisco}, {Grogin}, {Guo}, {Hathi}, {Kocevski}, {Koekemoer},
  {Kurczynski}, {Merlin}, {Mortlock}, {Newman}, {Paris}, {Pentericci},
  {Simons}, \& {Willner}}]{Santini2015}
{Santini}, P., {Ferguson}, H.~C., {Fontana}, A., {et~al.} 2015, \apj, 801, 97

\bibitem[{{Shimakawa} {et~al.}(2015){Shimakawa}, {Kodama}, {Tadaki}, {Hayashi},
  {Koyama}, \& {Tanaka}}]{Shimakawa2015}
{Shimakawa}, R., {Kodama}, T., {Tadaki}, K.-i., {et~al.} 2015, \mnras, 448, 666

\bibitem[{{Springel} {et~al.}(2005){Springel}, {White}, {Jenkins}, {Frenk},
  {Yoshida}, {Gao}, {Navarro}, {Thacker}, {Croton}, {Helly}, {Peacock}, {Cole},
  {Thomas}, {Couchman}, {Evrard}, {Colberg}, \& {Pearce}}]{Springel2005}
{Springel}, V., {White}, S.~D.~M., {Jenkins}, A., {et~al.} 2005, \nat, 435, 629

\bibitem[{{Steidel} {et~al.}(1998){Steidel}, {Adelberger}, {Dickinson},
  {Giavalisco}, {Pettini}, \& {Kellogg}}]{Steidel1998}
{Steidel}, C.~C., {Adelberger}, K.~L., {Dickinson}, M., {et~al.} 1998, \apj,
  492, 428

\bibitem[{{Steidel} {et~al.}(2005){Steidel}, {Adelberger}, {Shapley}, {Erb},
  {Reddy}, \& {Pettini}}]{Steidel2005}
{Steidel}, C.~C., {Adelberger}, K.~L., {Shapley}, A.~E., {et~al.} 2005, \apj,
  626, 44

\bibitem[{{Stott} {et~al.}(2012){Stott}, {Hickox}, {Edge}, {Collins}, {Hilton},
  {Harrison}, {Romer}, {Rooney}, {Kay}, {Miller}, {Sahl{\'e}n}, {Lloyd-Davies},
  {Mehrtens}, {Hoyle}, {Liddle}, {Viana}, {McCarthy}, {Schaye}, \&
  {Booth}}]{Stott2012}
{Stott}, J.~P., {Hickox}, R.~C., {Edge}, A.~C., {et~al.} 2012, \mnras, 422,
  2213

\bibitem[{{Straughn} {et~al.}(2009){Straughn}, {Pirzkal}, {Meurer}, {Cohen},
  {Windhorst}, {Malhotra}, {Rhoads}, {Gardner}, {Hathi}, {Jansen}, {Grogin},
  {Panagia}, {di Serego Alighieri}, {Gronwall}, {Walsh}, {Pasquali}, \&
  {Xu}}]{Straughn2009}
{Straughn}, A.~N., {Pirzkal}, N., {Meurer}, G.~R., {et~al.} 2009, \aj, 138,
  1022

\bibitem[{{Strazzullo} {et~al.}(2016){Strazzullo}, {Daddi}, {Gobat},
  {Valentino}, {Pannella}, {Dickinson}, {Renzini}, {Brammer}, {Onodera},
  {Finoguenov}, {Cimatti}, {Carollo}, \& {Arimoto}}]{Strazzullo2016}
{Strazzullo}, V., {Daddi}, E., {Gobat}, R., {et~al.} 2016, \apjl, 833, L20

\bibitem[{{Strazzullo} {et~al.}(2013){Strazzullo}, {Gobat}, {Daddi}, {Onodera},
  {Carollo}, {Dickinson}, {Renzini}, {Arimoto}, {Cimatti}, {Finoguenov}, \&
  {Chary}}]{Strazzullo2013}
{Strazzullo}, V., {Gobat}, R., {Daddi}, E., {et~al.} 2013, \apj, 772, 118

\bibitem[{{Trevese} {et~al.}(2007){Trevese}, {Castellano}, {Fontana}, \&
  {Giallongo}}]{Trevese2007}
{Trevese}, D., {Castellano}, M., {Fontana}, A., \& {Giallongo}, E. 2007, \aap,
  463, 853

\bibitem[{{Trump} {et~al.}(2013){Trump}, {Konidaris}, {Barro}, {Koo},
  {Kocevski}, {Juneau}, {Weiner}, {Faber}, {McLean}, {Yan},
  {P{\'e}rez-Gonz{\'a}lez}, \& {Villar}}]{Trump2013}
{Trump}, J.~R., {Konidaris}, N.~P., {Barro}, G., {et~al.} 2013, \apjl, 763, L6

\bibitem[{{van der Wel} {et~al.}(2012){van der Wel}, {Bell}, {H{\"a}ussler},
  {McGrath}, {Chang}, {Guo}, {McIntosh}, {Rix}, {Barden}, {Cheung}, {Faber},
  {Ferguson}, {Galametz}, {Grogin}, {Hartley}, {Kartaltepe}, {Kocevski},
  {Koekemoer}, {Lotz}, {Mozena}, {Peth}, \& {Peng}}]{vanderWel2012}
{van der Wel}, A., {Bell}, E.~F., {H{\"a}ussler}, B., {et~al.} 2012, \apjs,
  203, 24

\bibitem[{{Vanzella} {et~al.}(2008){Vanzella}, {Cristiani}, {Dickinson},
  {Giavalisco}, {Kuntschner}, {Haase}, {Nonino}, {Rosati}, {Cesarsky},
  {Ferguson}, {Fosbury}, {Grazian}, {Moustakas}, {Rettura}, {Popesso},
  {Renzini}, {Stern}, \& {GOODS Team}}]{Vanzella2008}
{Vanzella}, E., {Cristiani}, S., {Dickinson}, M., {et~al.} 2008, \aap, 478, 83

\bibitem[{{Vanzella} {et~al.}(2009){Vanzella}, {Giavalisco}, {Dickinson},
  {Cristiani}, {Nonino}, {Kuntschner}, {Popesso}, {Rosati}, {Renzini}, {Stern},
  {Cesarsky}, {Ferguson}, \& {Fosbury}}]{Vanzella2009}
{Vanzella}, E., {Giavalisco}, M., {Dickinson}, M., {et~al.} 2009, \apj, 695,
  1163

\bibitem[{{Venemans} {et~al.}(2007){Venemans}, {R{\"o}ttgering}, {Miley}, {van
  Breugel}, {de Breuck}, {Kurk}, {Pentericci}, {Stanford}, {Overzier}, {Croft},
  \& {Ford}}]{Venemans2007}
{Venemans}, B.~P., {R{\"o}ttgering}, H.~J.~A., {Miley}, G.~K., {et~al.} 2007,
  \aap, 461, 823

\bibitem[{{Verhamme} {et~al.}(2015){Verhamme}, {Orlitov{\'a}}, {Schaerer}, \&
  {Hayes}}]{Verhamme2015}
{Verhamme}, A., {Orlitov{\'a}}, I., {Schaerer}, D., \& {Hayes}, M. 2015, \aap,
  578, A7

\bibitem[{{Verhamme} {et~al.}(2017){Verhamme}, {Orlitov{\'a}}, {Schaerer},
  {Izotov}, {Worseck}, {Thuan}, \& {Guseva}}]{Verhamme2017}
{Verhamme}, A., {Orlitov{\'a}}, I., {Schaerer}, D., {et~al.} 2017, \aap, 597,
  A13

\bibitem[{{Verhamme} {et~al.}(2006){Verhamme}, {Schaerer}, \&
  {Maselli}}]{V2006}
{Verhamme}, A., {Schaerer}, D., \& {Maselli}, A. 2006, \aap, 460, 397

\bibitem[{{Wainer} \& {Thissen}(1976)}]{WainerThissen1976}
{Wainer}, H. \& {Thissen}, D. 1976, in {Psychometrika}, Vol.~41, 9

\bibitem[{{Willis} {et~al.}(2013){Willis}, {Clerc}, {Bremer}, {Pierre},
  {Adami}, {Ilbert}, {Maughan}, {Maurogordato}, {Pacaud}, {Valtchanov},
  {Chiappetti}, {Thanjavur}, {Gwyn}, {Stanway}, \& {Winkworth}}]{willis13}
{Willis}, J.~P., {Clerc}, N., {Bremer}, M.~N., {et~al.} 2013, \mnras, 430, 134

\bibitem[{{Wuyts} {et~al.}(2008){Wuyts}, {Labb{\'e}}, {F{\"o}rster Schreiber},
  {Franx}, {Rudnick}, {Brammer}, \& {van Dokkum}}]{Wuyts2008}
{Wuyts}, S., {Labb{\'e}}, I., {F{\"o}rster Schreiber}, N.~M., {et~al.} 2008,
  \apj, 682, 985

\bibitem[{{Wuyts} {et~al.}(2009){Wuyts}, {van Dokkum}, {Franx}, {F{\"o}rster
  Schreiber}, {Illingworth}, {Labb{\'e}}, \& {Rudnick}}]{Wuyts2009}
{Wuyts}, S., {van Dokkum}, P.~G., {Franx}, M., {et~al.} 2009, \apj, 706, 885

\bibitem[{{Zabludoff} {et~al.}(1993){Zabludoff}, {Geller}, {Huchra}, \&
  {Vogeley}}]{Zabludoff1993}
{Zabludoff}, A.~I., {Geller}, M.~J., {Huchra}, J.~P., \& {Vogeley}, M.~S. 1993,
  \aj, 106, 1273

\bibitem[{{Zheng} {et~al.}(2016){Zheng}, {Malhotra}, {Rhoads}, {Finkelstein},
  {Wang}, {Jiang}, \& {Cai}}]{Zheng2016}
{Zheng}, Z.-Y., {Malhotra}, S., {Rhoads}, J.~E., {et~al.} 2016, \apjs, 226, 23

\bibitem[{{Zirm} {et~al.}(2012){Zirm}, {Toft}, \& {Tanaka}}]{Zirm2012}
{Zirm}, A.~W., {Toft}, S., \& {Tanaka}, M. 2012, \apj, 744, 181

\bibitem[{{Zoldan} {et~al.}(2017){Zoldan}, {De Lucia}, {Xie}, {Fontanot}, \&
  {Hirschmann}}]{Zoldan2017}
{Zoldan}, A., {De Lucia}, G., {Xie}, L., {Fontanot}, F., \& {Hirschmann}, M.
  2017, \mnras, 465, 2236

\end{thebibliography}

%\extrafloats{100} 
%\Online
\onecolumn

\begin{appendix} %First appendix
\section{Location of the overdensities identified in the CDFS and in the UDS}
\label{appendix1}
We present here the position of the identified overdensities in space. We also show the position of the structures detected in the literature in the same areas.

%weighted_CDFS_grid_forfigure.py
\begin{figure*}[h!]
 \centering
\includegraphics[width=8.4cm]{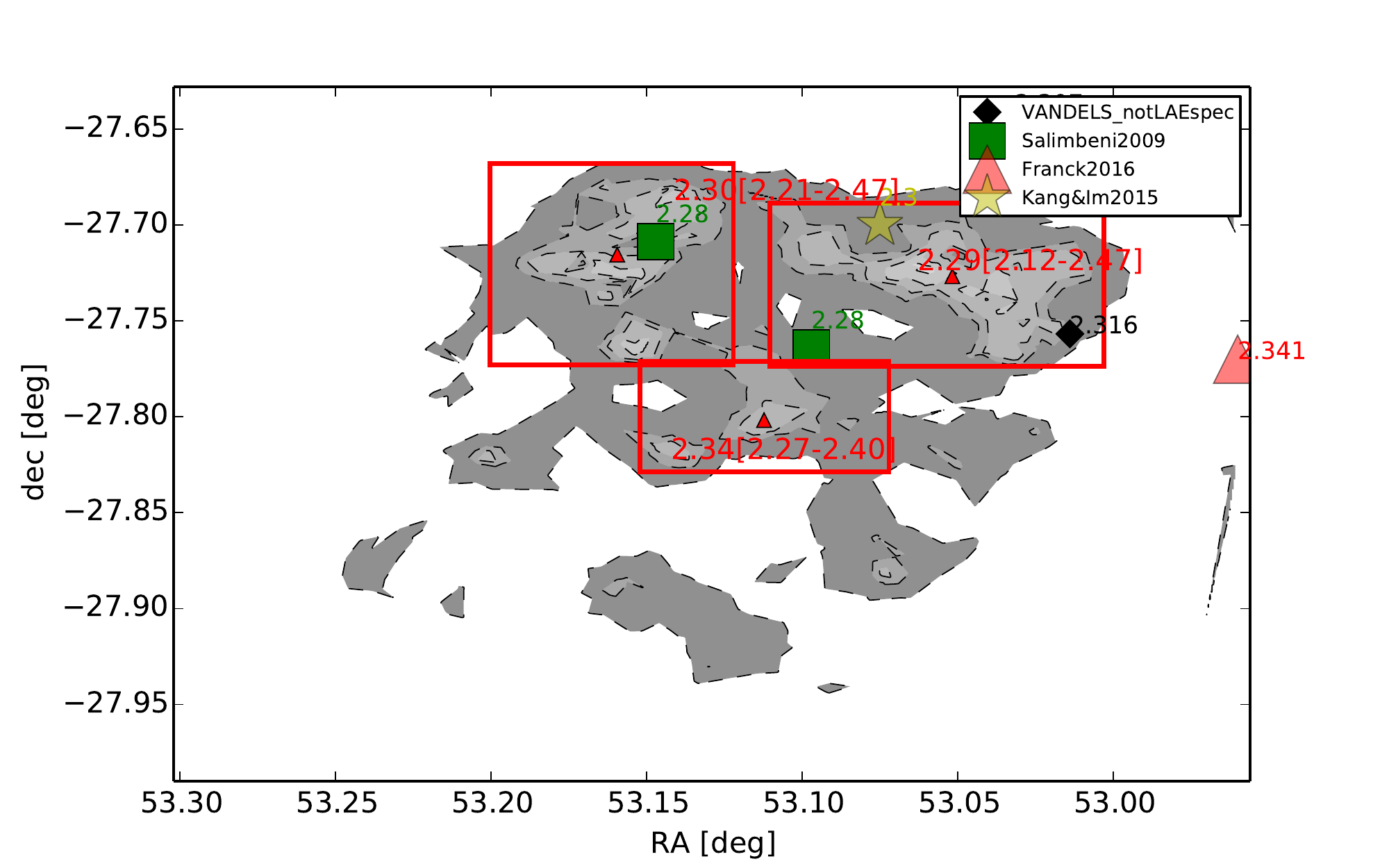}
\includegraphics[width=8.4cm]{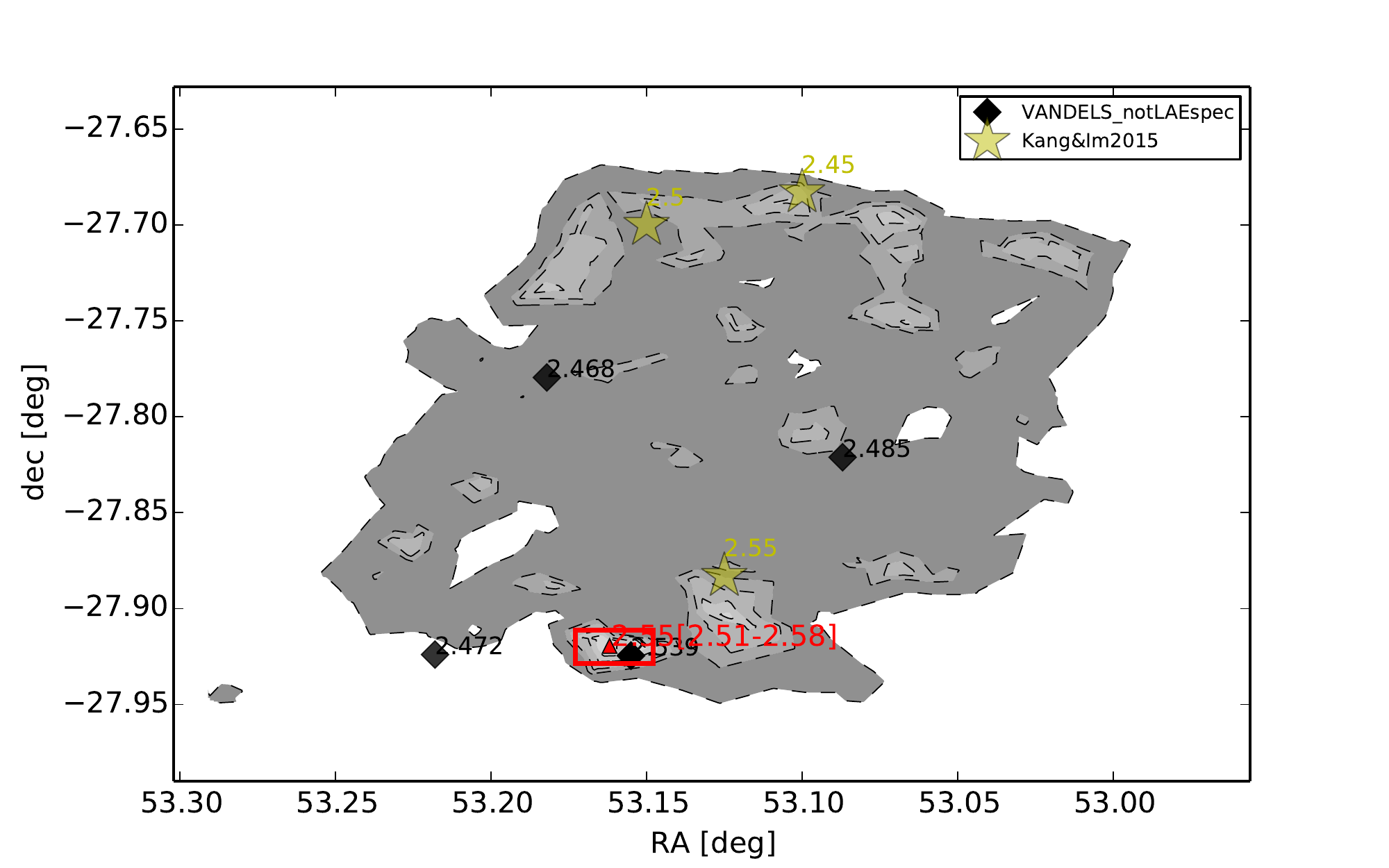}
\caption{Density levels for galaxies at $2.25<z<2.35$ ($left$) and $2.45<z<2.55$ ($right$) in the CDFS. The contours represent 3, 2, 1, 0.5$\sigma$ over the mean value of the local densities associated to the galaxies in the indicated redshift bin. Spectroscopic redshifts from VANDELS (VANDELS\_notLAEspec) %z$_{spec}$V) 
are indicated as black diamonds, VANDELS LAEs are indicated as red stars. For them, we also indicate the spectroscopic redshifts. Red rectangles represent the location of the identified overdensities. The center of these structures is shown with red triangles. The redshift of the highest-density peak and the redshift range of the structures is written in red. We also show the position of the center of structures detected in the literature in the same field and in the same redshift ranges, \citet{Salimbeni2009} (green squares), \citet{Franck2016} (big red triangles), \citet{KangIm2015} (yellow stars). We write the redshifts of the literature-structure cores in green, red, and yellow, respectively.} % LAE only in the bin of $deltaz=0.1$}
\label{Auno}%
\end{figure*}

\begin{figure*}[h!]
 \centering
\includegraphics[width=8.4cm]{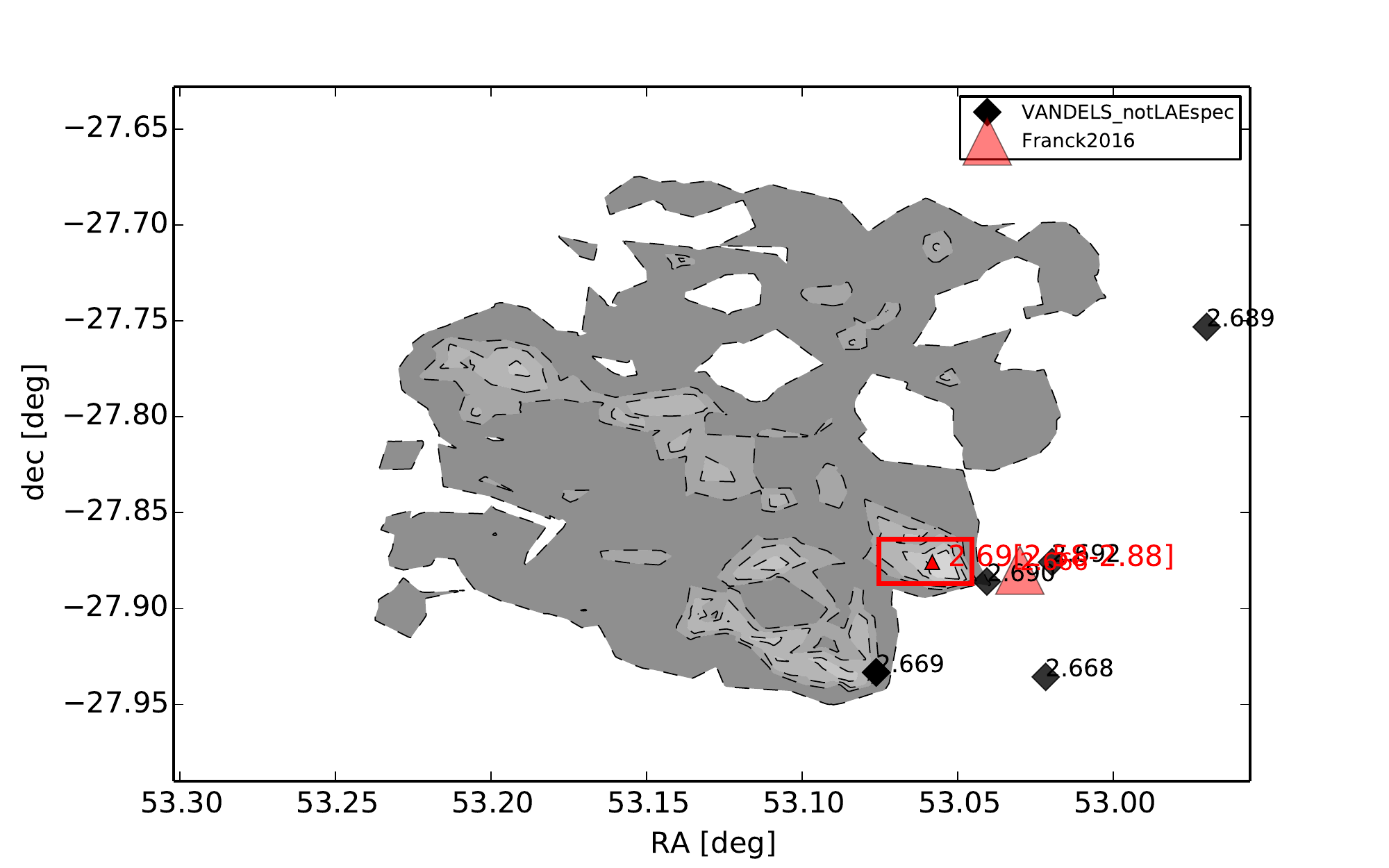}
\includegraphics[width=8.4cm]{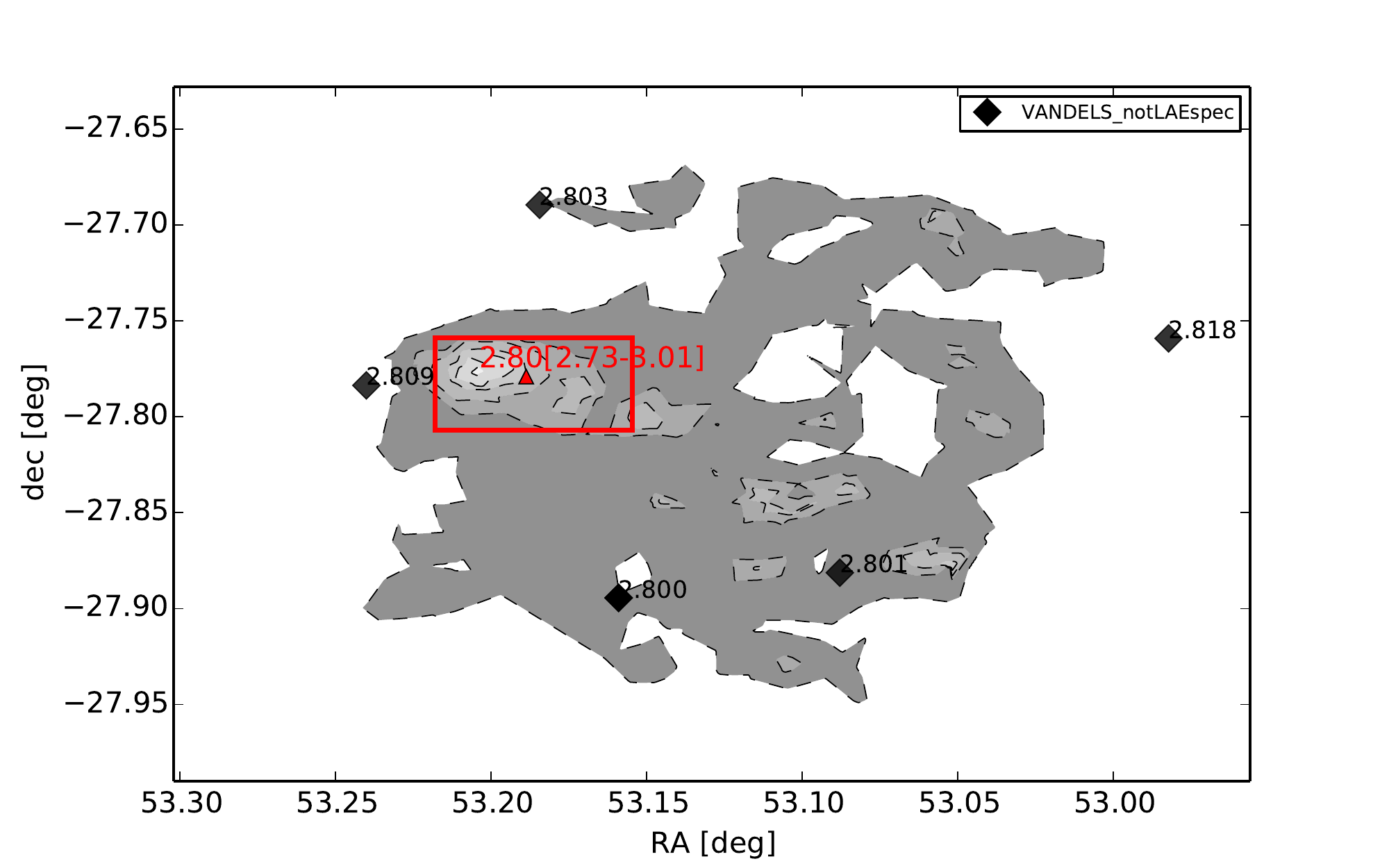}
\caption{Density levels for galaxies at $2.65<z<2.75$ ($left$) and $2.75<z<2.85$ ($right$) in the CDFS.}
\label{Adue}%
\end{figure*}

\begin{figure}
 \centering
\includegraphics[width=8.4cm]{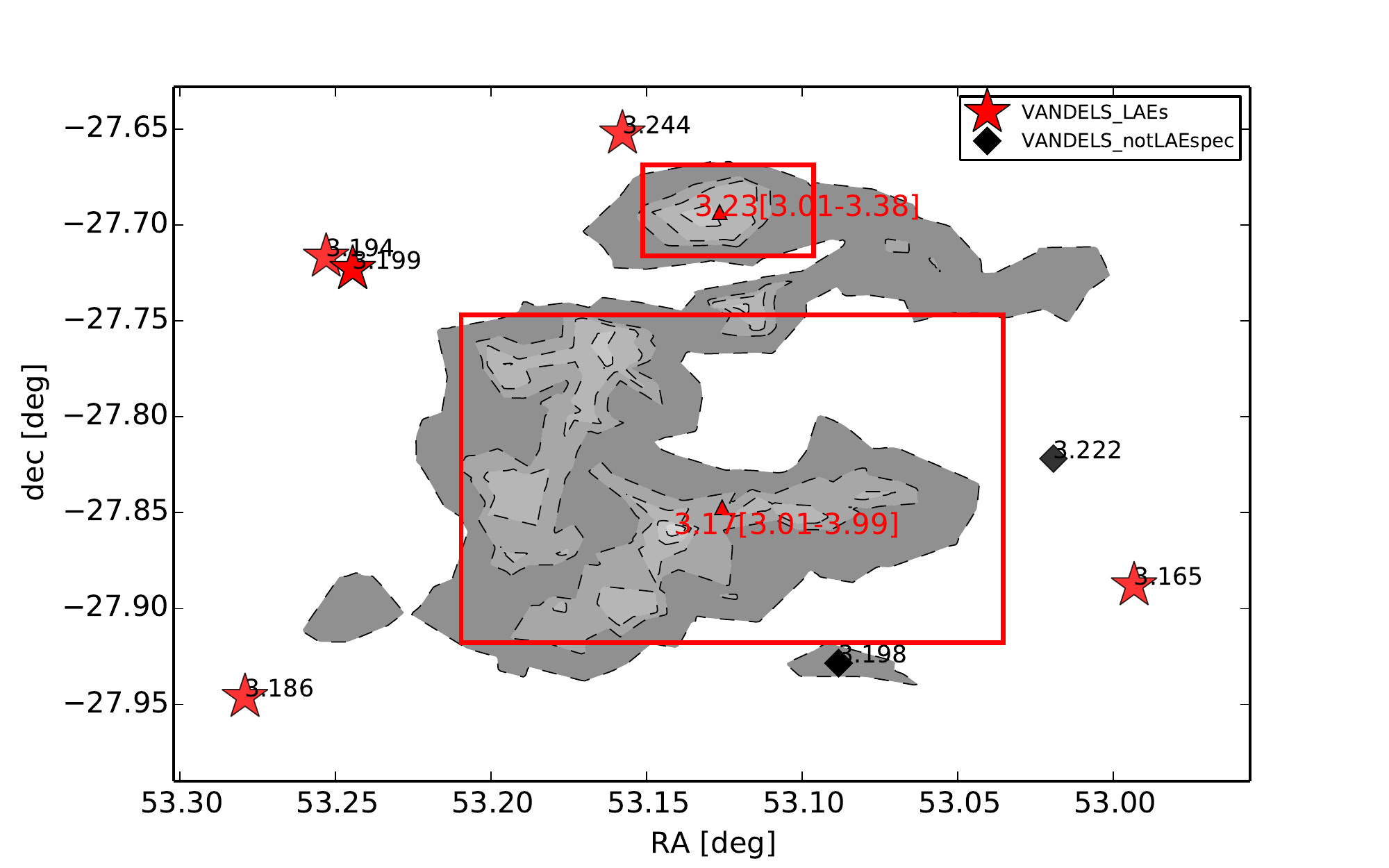}
\includegraphics[width=8.4cm]{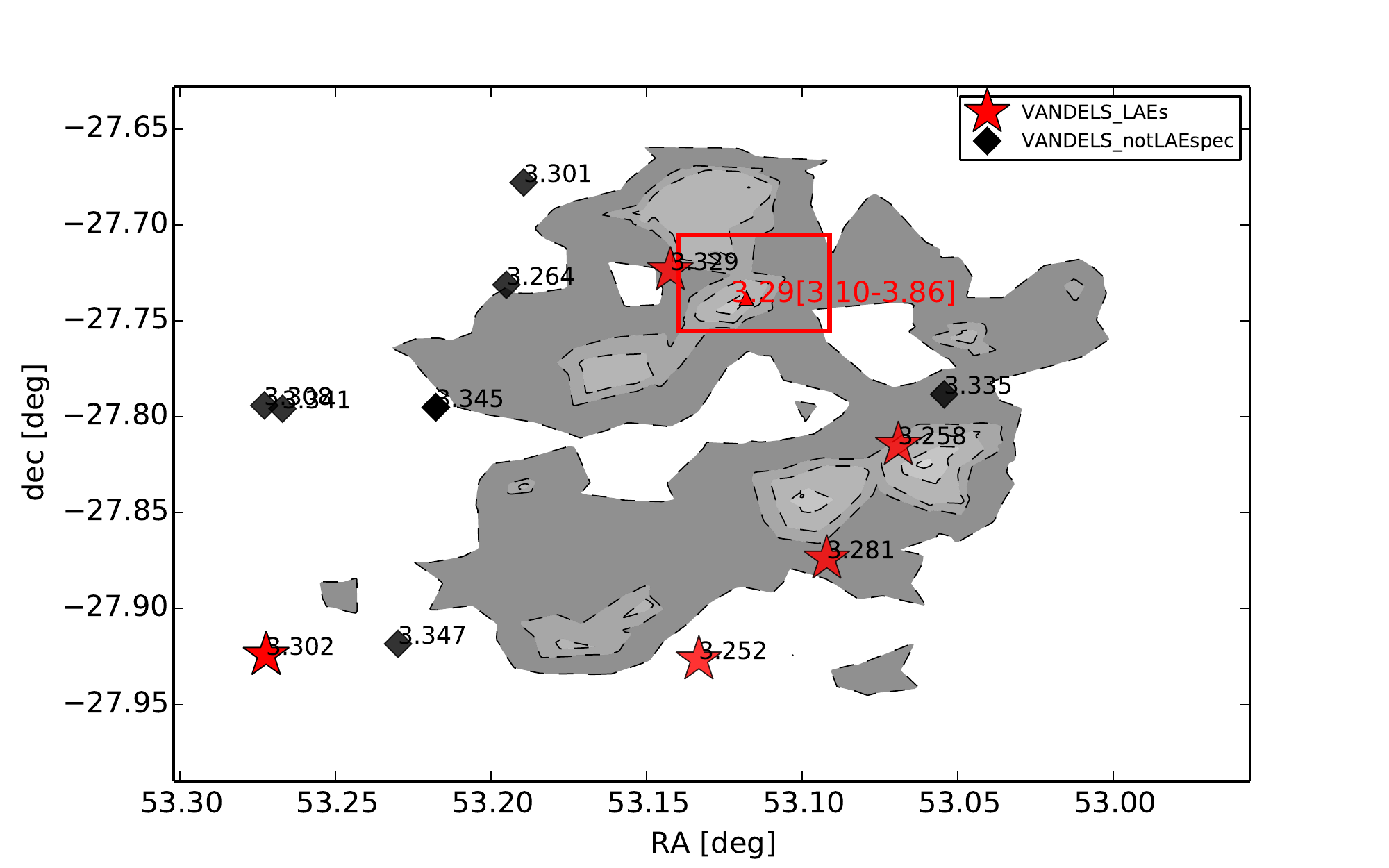}
\caption{Density levels for galaxies at $3.15<z<3.25$ ($left$) and $3.25<z<3.35$ ($right$) in the CDFS.}
\label{Atre}%
\end{figure}

\begin{figure}
 \centering
\includegraphics[width=8.4cm]{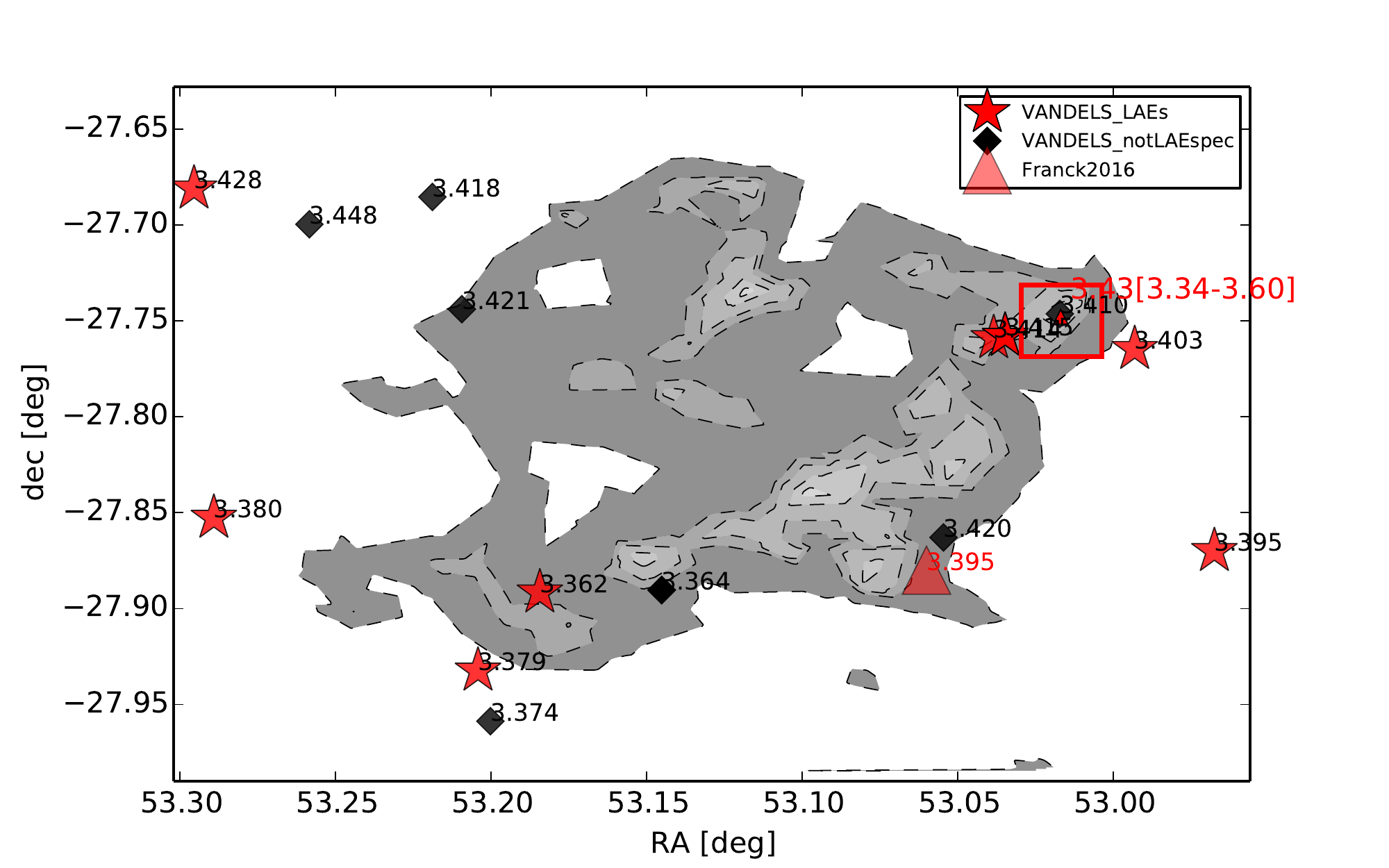}
\includegraphics[width=8.4cm]{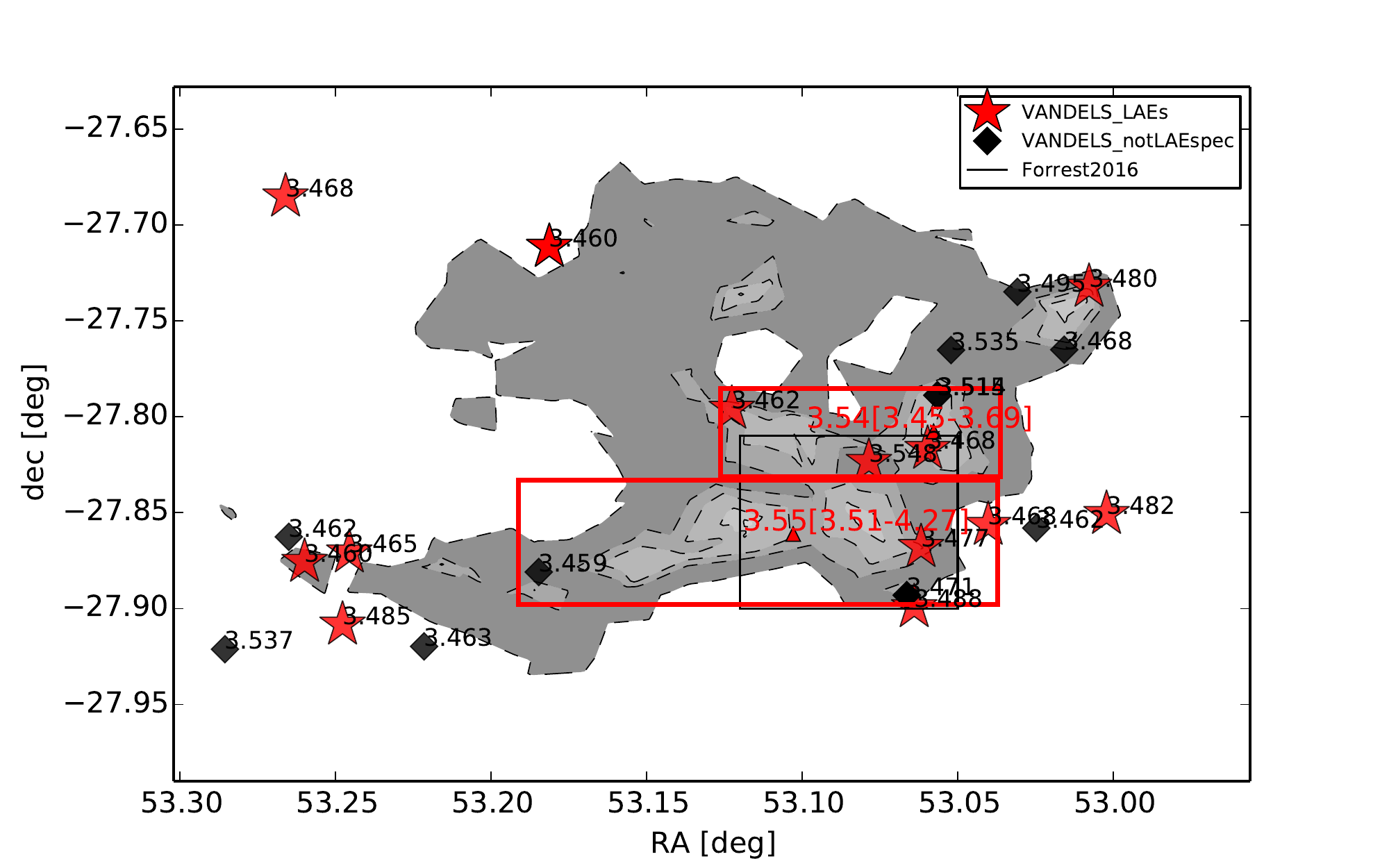}
\caption{Density levels for galaxies at $3.35<z<3.45$ ($left$) and $3.45<z<3.55$ ($right$) in the CDFS.}
\label{Aquattro}%
\end{figure}

\begin{figure}
 \centering
\includegraphics[width=8.4cm]{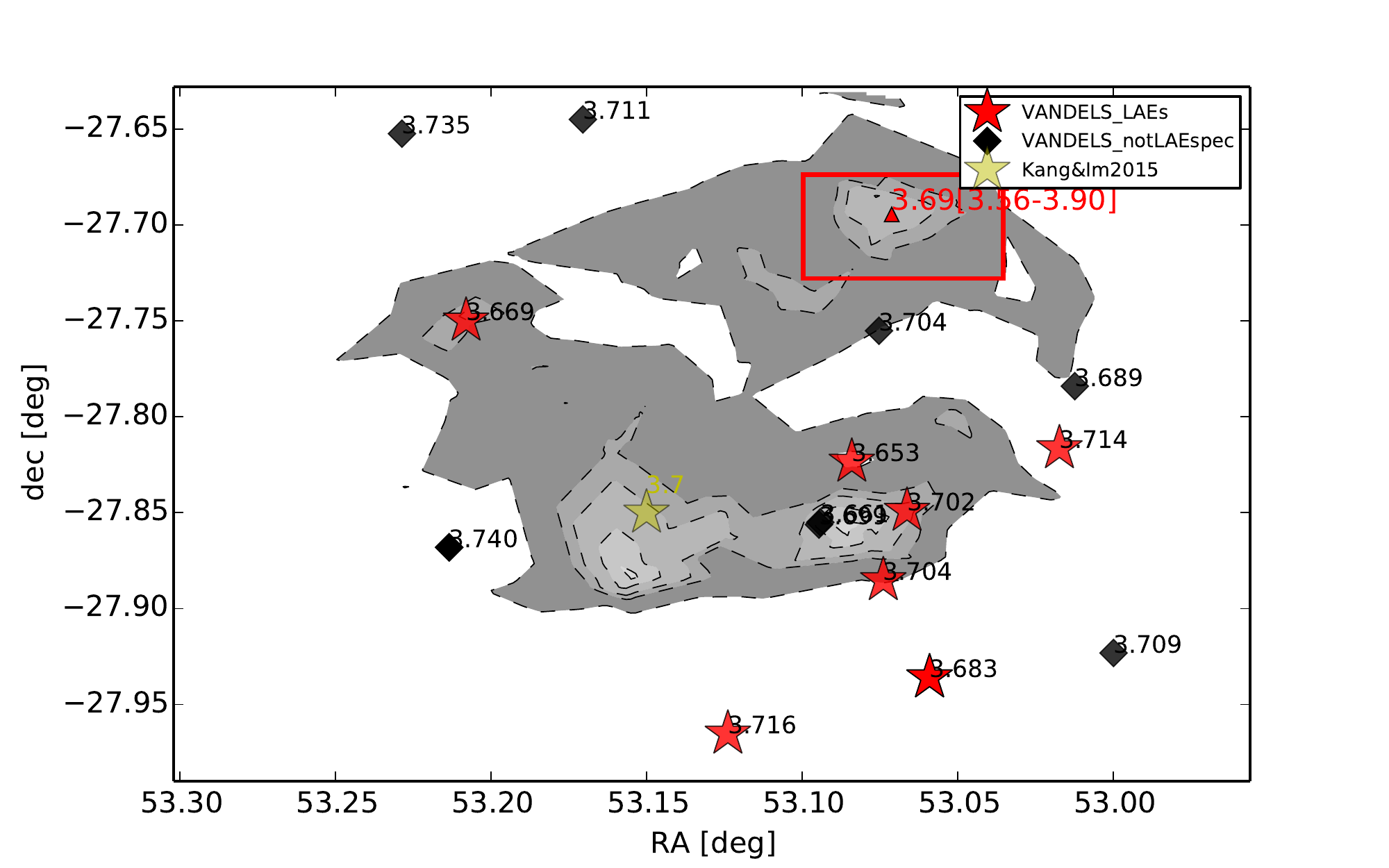}
\includegraphics[width=8.4cm]{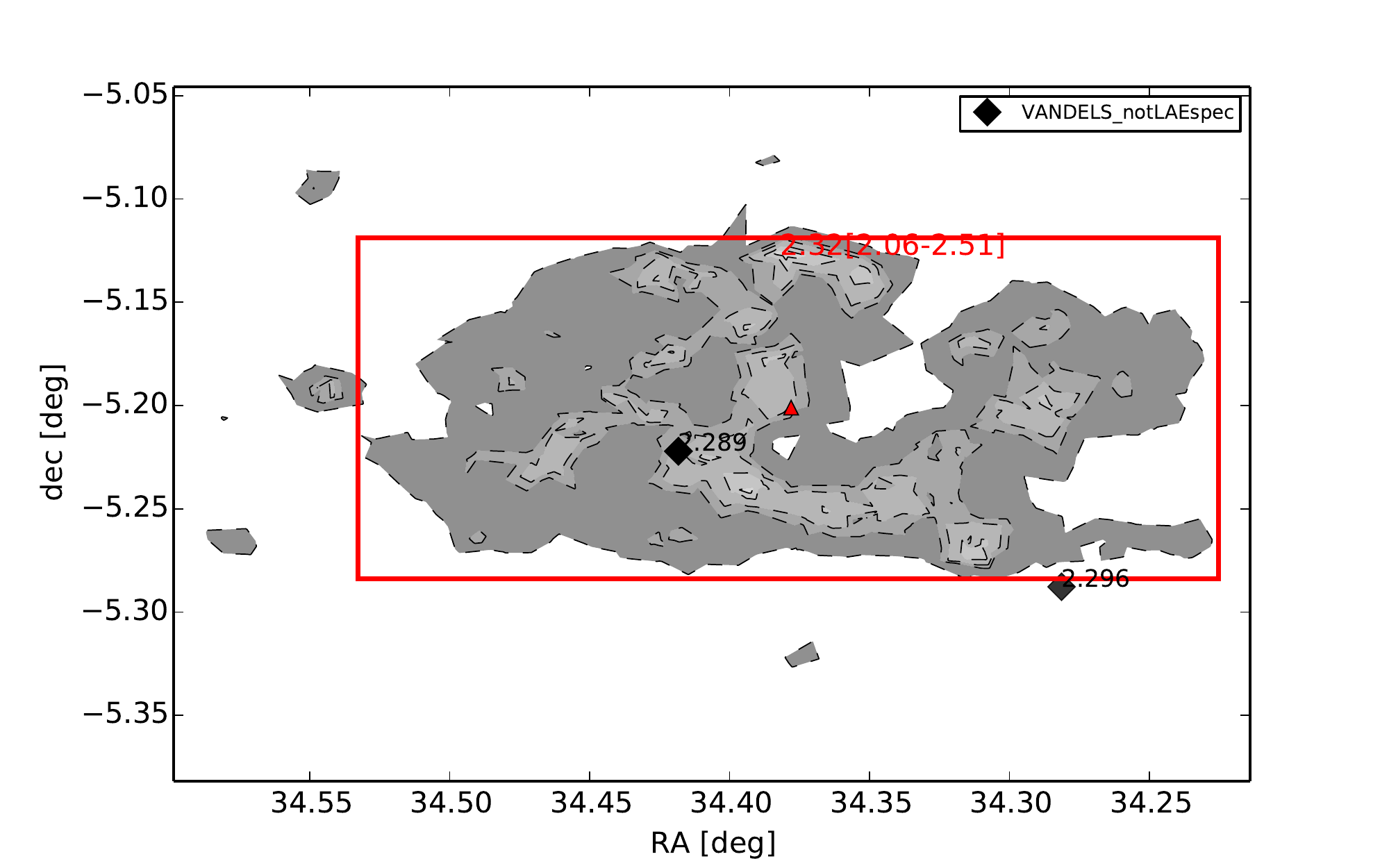}
\caption{Density levels for galaxies at $3.65<z<3.75$ ($left$) in the CDFS and $2.25<z<2.35$ ($right$) in the UDS.}
\label{Acinque}%
\end{figure}

\begin{figure}
 \centering
\includegraphics[width=8.4cm]{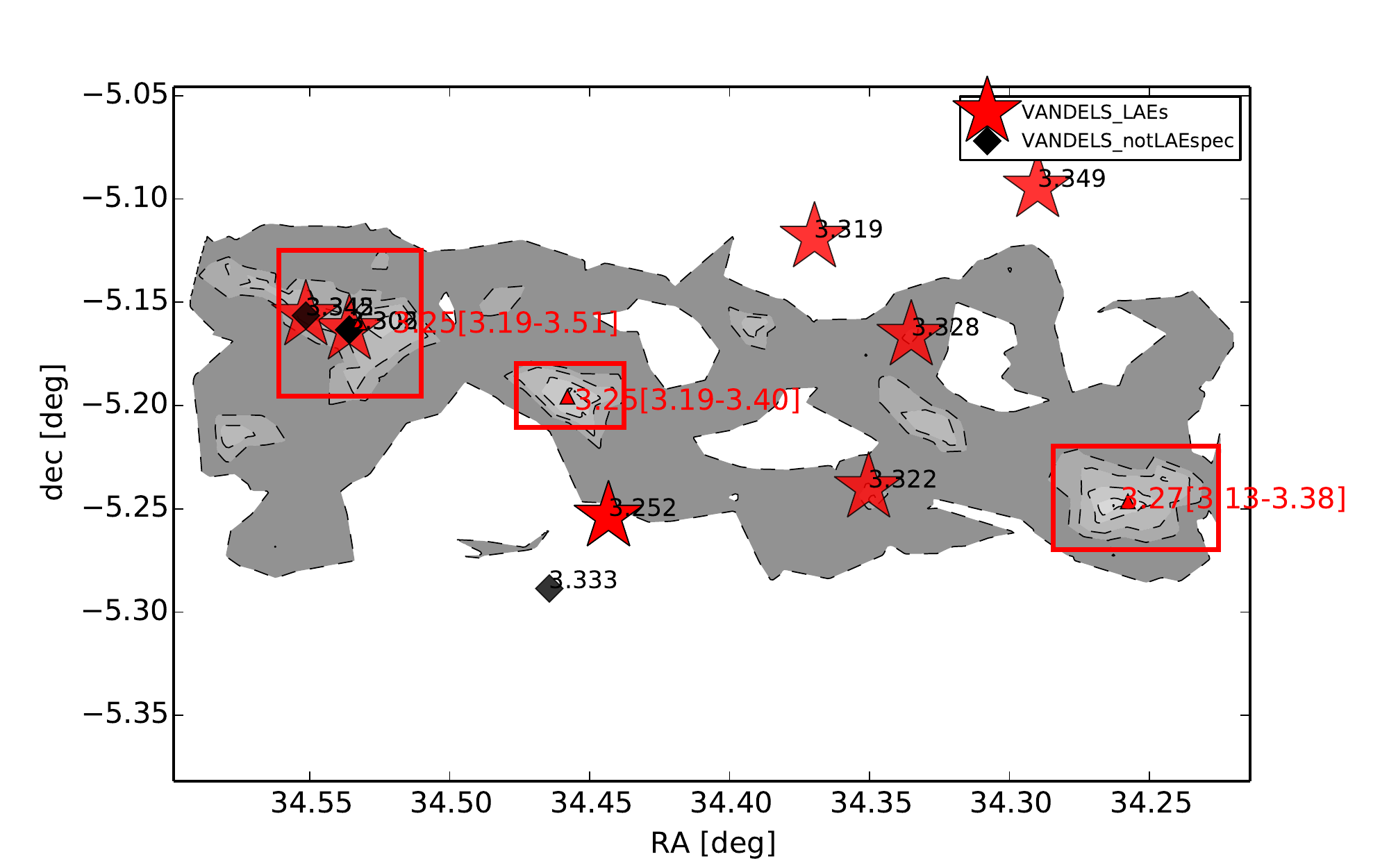} %CDFSstructureUPDATEDwithcomparison_z33.pdf}
\includegraphics[width=8.4cm]{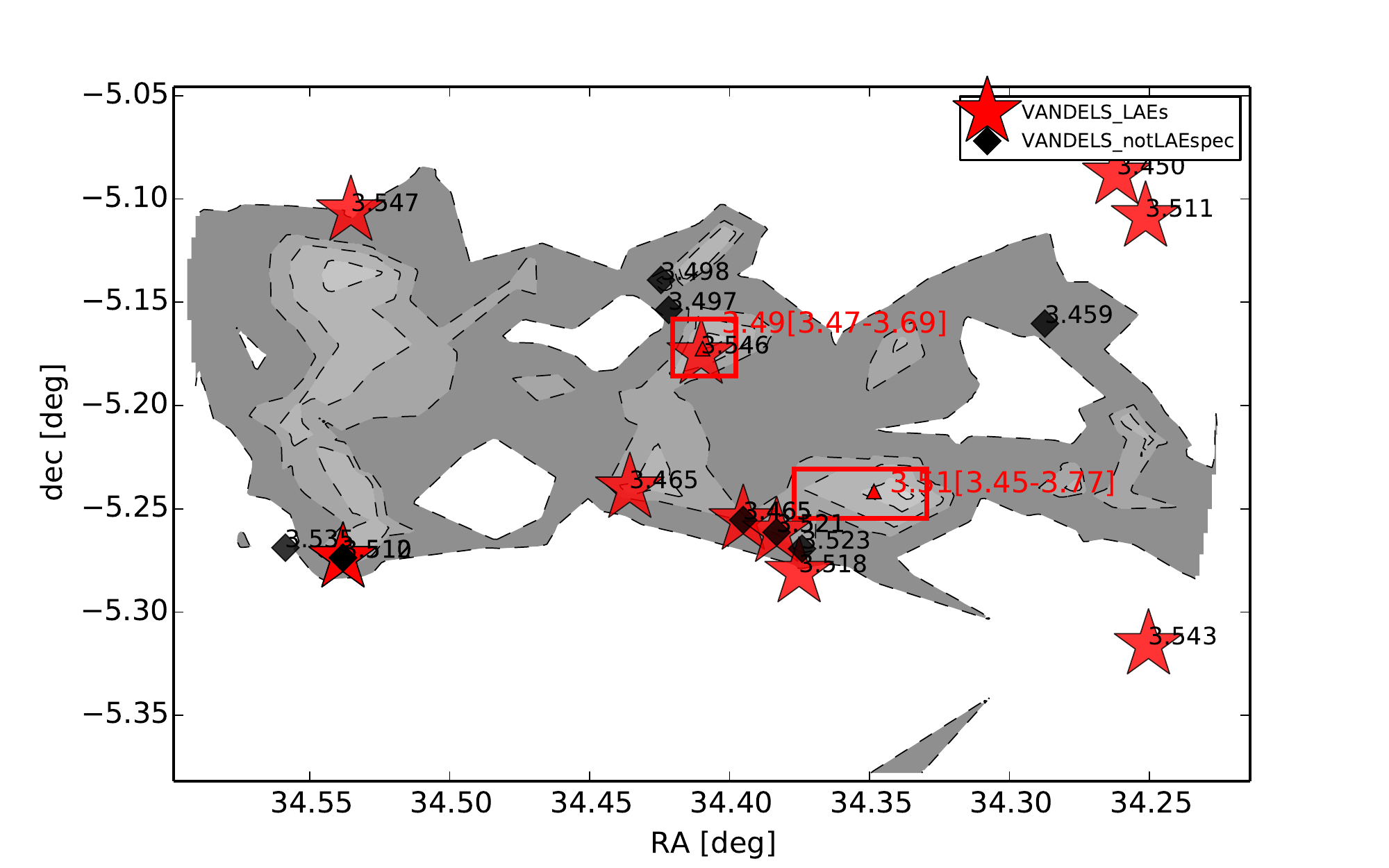}
\caption{Density levels for galaxies at $3.25<z<3.35$ ($left$) and $3.45<z<3.55$ ($right$) in the UDS.}
\label{Asei}%
\end{figure}

\begin{figure}
 \centering
\includegraphics[width=8.4cm]{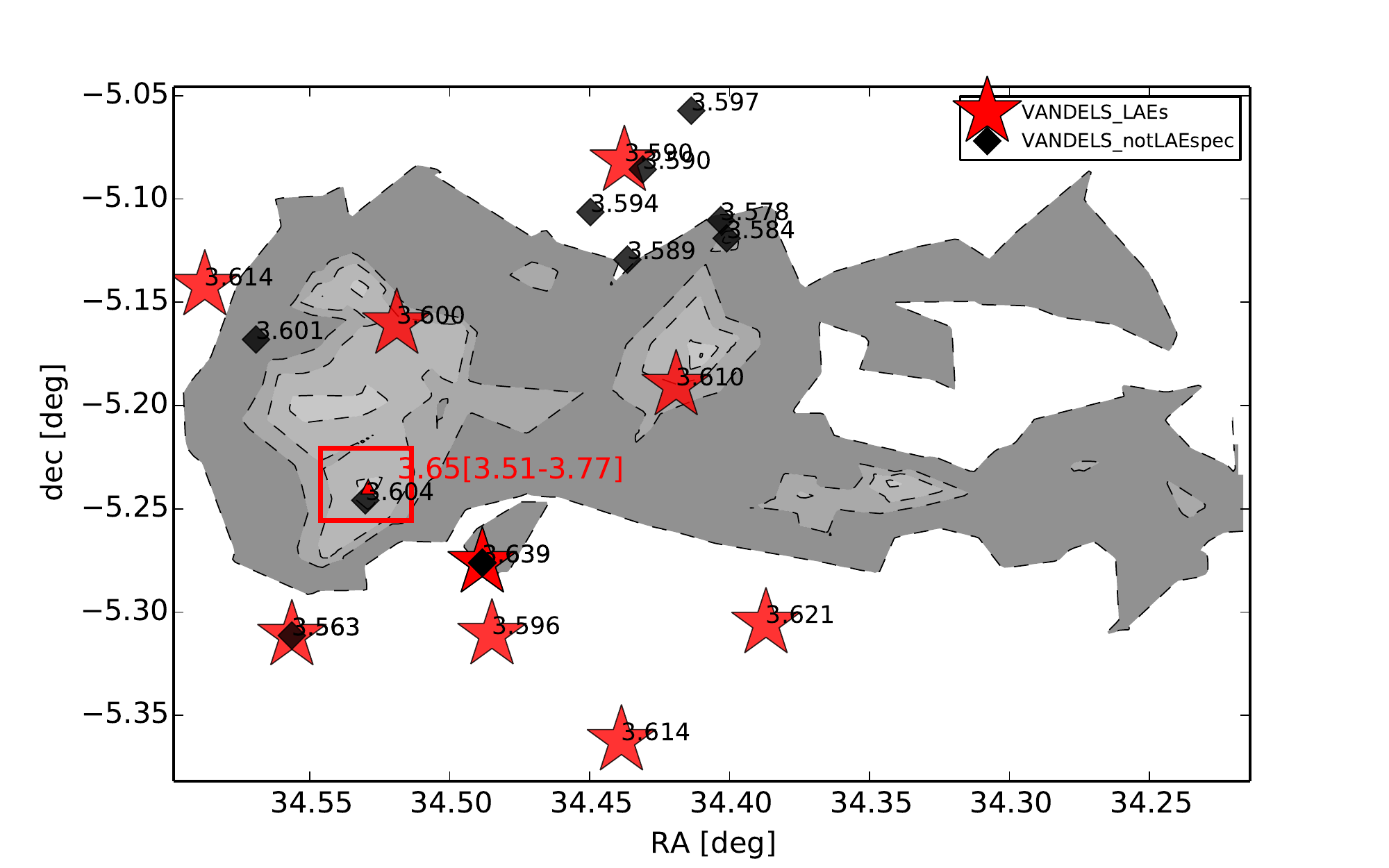} %CDFSstructureUPDATEDwithcomparison_z36.pdf}
\includegraphics[width=8.4cm]{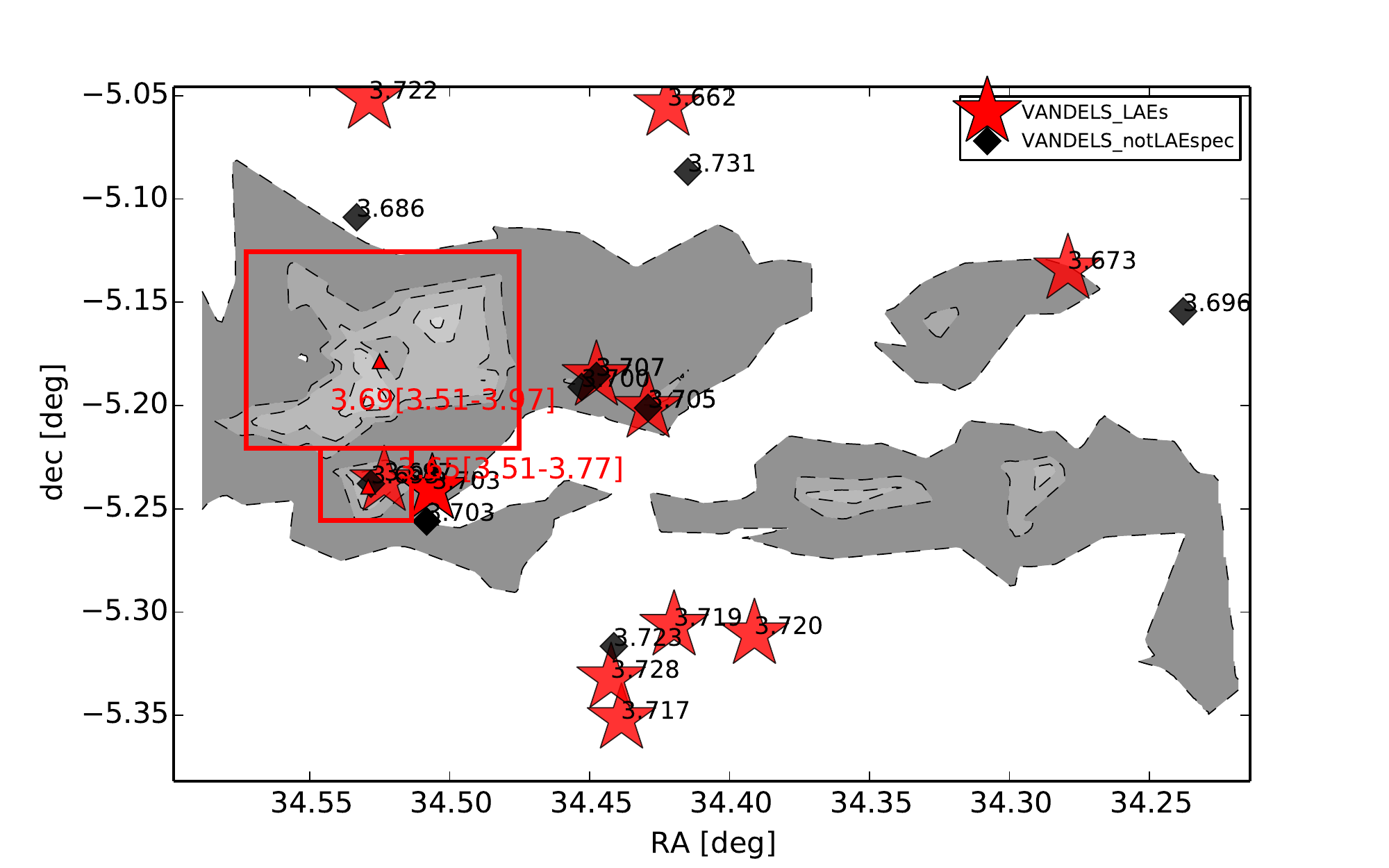}
\caption{Density levels for galaxies at $3.55<z<3.65$ ($left$) and $3.65<z<3.75$ ($right$) in the UDS.}
\label{Asette}%
\end{figure}

%\end{appendix}
%\end{document}
%~ \\
%~ \\

\clearpage

\section{3D visualization of the overdensities identified in the CDFS and in the UDS}
\label{appendix2}

In this section, we show the 3D visualization of the identified overdensities. It is worth noting that the redshift direction can comprise much larger physical sizes than the RA and declination. This is a result of the way our code detects overdensities. In fact, to calculate local densities, we choose an initial cell size in the redshift direction proportional to the photometric redshift uncertainty, the largest uncertainty we have. The extension of each cell is characterized by a parallelepiped volume, elongated towards the redshift direction (as described in Sect. 3). Within the detected overdensities, there exist dense cores, compact also in the redshift direction, that could evolve in protoclusters.

\begin{figure*}[h!]
 \centering
\includegraphics[width=9cm]{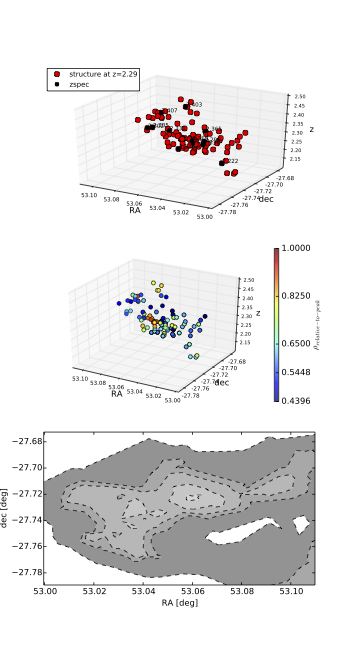}
\includegraphics[width=9cm]{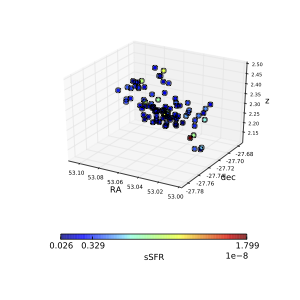}
\caption{3D visualization of the overdensity identified at $z=2.29$ in the CDFS. In the $first~ panel$, we show the overdensity members in the RA,dec,redshift space in red. The members with spectroscopic redshifts either from VANDELS or from the literature are shown in black with the black numbers indicating the $z_{spec}$. In the $second~ panel$, the color coding is the density associated to each member with respect to density associated to the highest-density peak. The $third ~panel$ shows the density map in 2D. The contours represent 3, 2, 1, 0.5$\sigma$ over the mean value of the local densities associated to the galaxies in the redshift range, $z_{peak}-0.05<z<z_{peak}+0.05$, where $z_{peak}$ is the redshift of the highest-density peak and it is shown for reference. %In the $right~ panel$, the color coding is the sSFR. 
The $fourth ~panel$ is as the first one, but the color coding of the symbols refers to the sSFR. The horizontal color-bar label shows the minimum, the mean, and the maximum value of sSFR.}
\label{Bcdfs229}%
\end{figure*}

\twocolumn

\begin{figure}
 %\centering [width=9.5cm]
%\resizebox{10cm}{!}
\includegraphics[width=8cm]{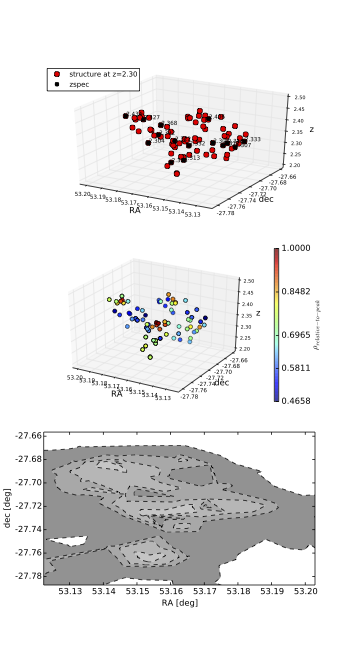}
\includegraphics[width=8cm]{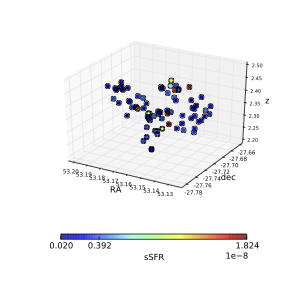}
\caption{The same as Fig. \ref{Bcdfs229} for the overdensity detected at $z=2.30$ in the CDFS.}
\label{Bcdfs230}%
\end{figure}

\begin{figure}
 \centering
\includegraphics[width=8cm]{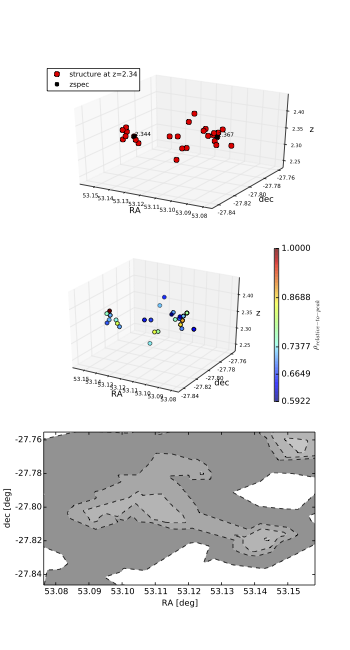}
\includegraphics[width=8cm]{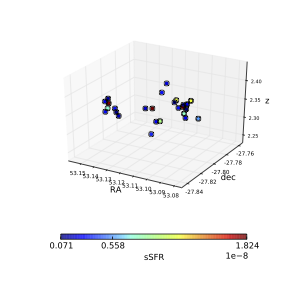}
\caption{The same as Fig. \ref{Bcdfs229} for the overdensity detected at $z=2.34$ in the CDFS.}
\label{Bcdfs234}%
\end{figure}

\begin{figure}
 \centering
\includegraphics[width=8cm]{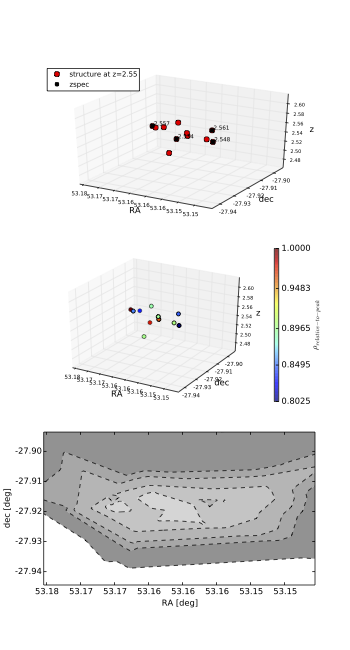}
\includegraphics[width=8cm]{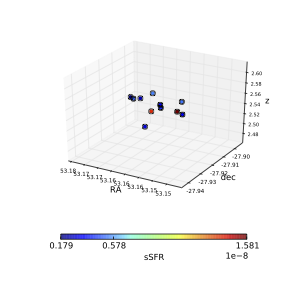}
\caption{The same as Fig. \ref{Bcdfs229} for the overdensity detected at $z=2.55$ in the CDFS.}
\label{Bcdfs255}%
\end{figure}

\begin{figure}
 \centering
\includegraphics[width=8cm]{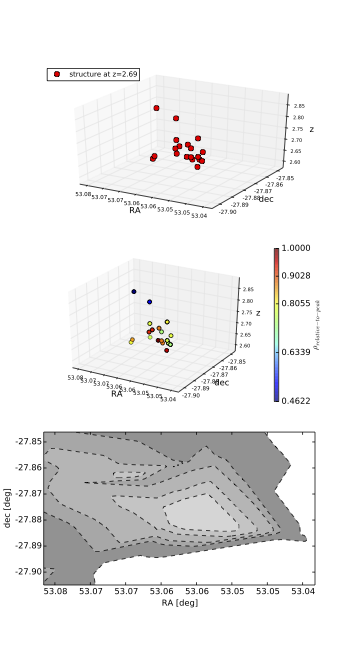}
\includegraphics[width=8cm]{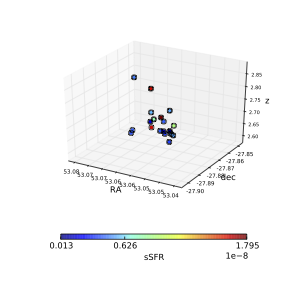}
\caption{The same as Fig. \ref{Bcdfs229} for the overdensity detected at $z=2.69$ in the CDFS.}
\label{Bcdfs269}%
\end{figure}

\begin{figure}
 \centering
\includegraphics[width=8cm]{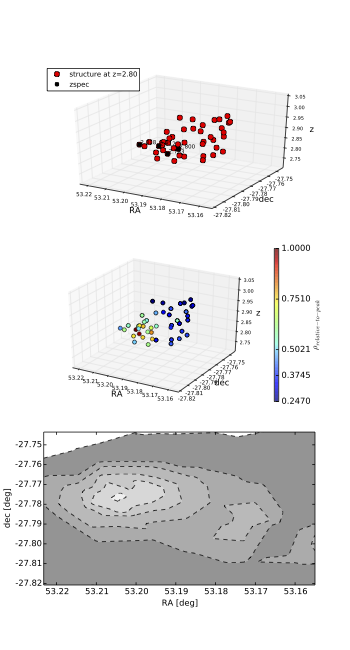}
\includegraphics[width=8cm]{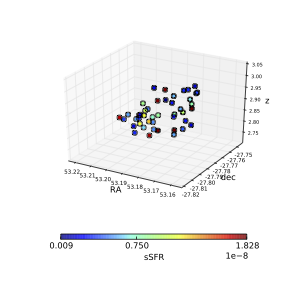}
\caption{The same as Fig. \ref{Bcdfs229} for the overdensity detected at $z=2.80$ in the CDFS.}
\label{Bcdfs280}%
\end{figure}

\begin{figure}
 \centering
\includegraphics[width=8cm]{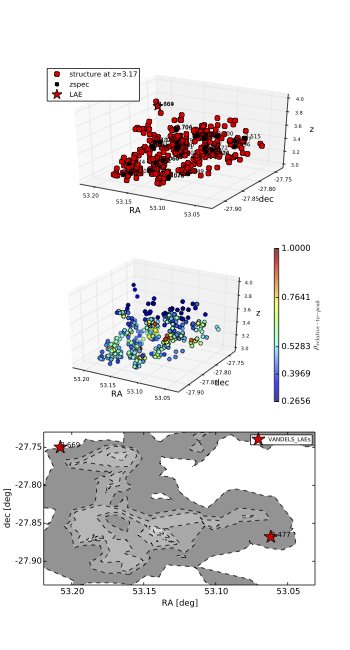}
\includegraphics[width=8cm]{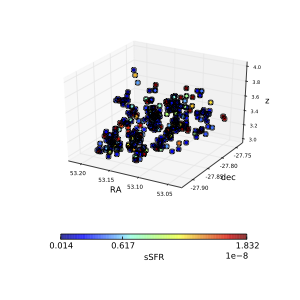}
\caption{The same as Fig. \ref{Bcdfs229} for the overdensity detected at $z=3.17$ in the CDFS.}
\label{Bcdfs317}%
\end{figure}

\begin{figure}
 \centering
\includegraphics[width=8cm]{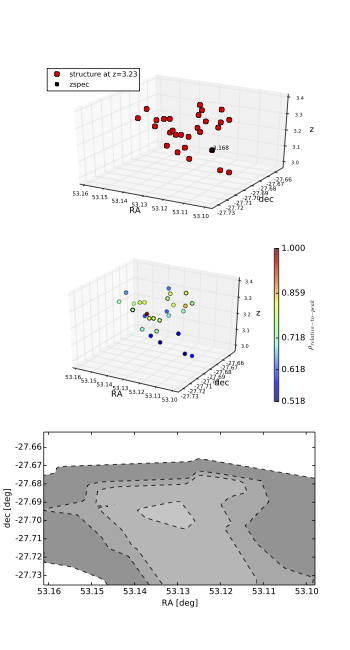}
\includegraphics[width=8cm]{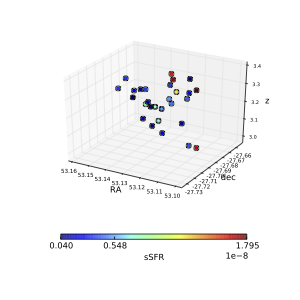}
\caption{The same as Fig. \ref{Bcdfs229} for the overdensity detected at $z=3.23$ in the CDFS.}
\label{Bcdfs323}%
\end{figure}

\begin{figure}
 \centering
\includegraphics[width=8cm]{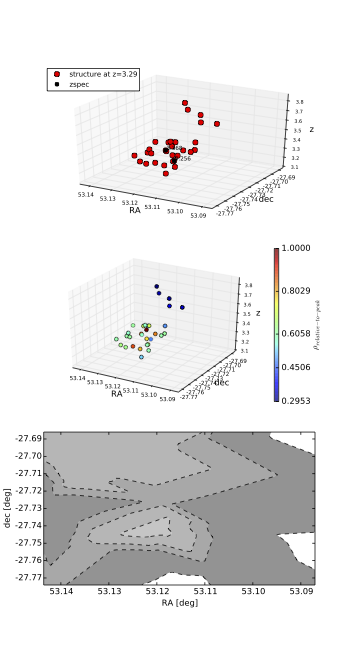}
\includegraphics[width=8cm]{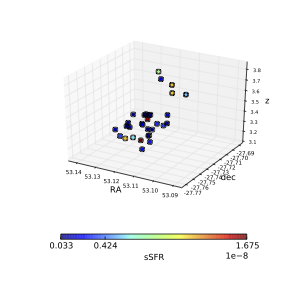}
\caption{The same as Fig. \ref{Bcdfs229} for the overdensity detected at $z=3.29$ in the CDFS.}
\label{Bcdfs329}%
\end{figure}

\begin{figure}
 \centering
\includegraphics[width=8cm]{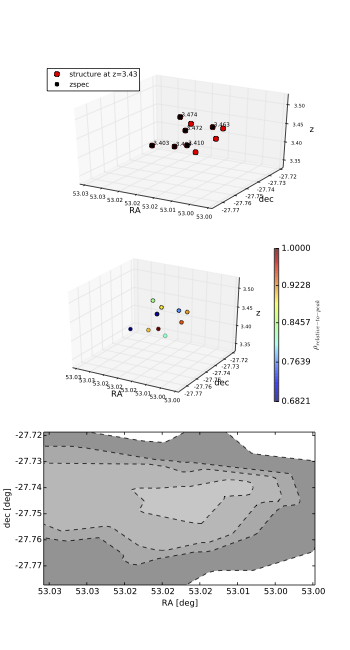}
\includegraphics[width=8cm]{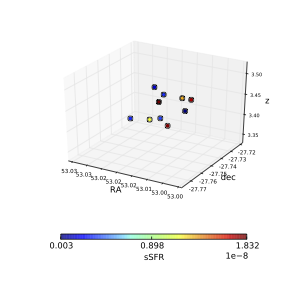}
\caption{The same as Fig. \ref{Bcdfs229} for the overdensity detected at $z=3.43$ in the CDFS.}
\label{Bcdfs343}%
\end{figure}

\begin{figure}
 \centering
\includegraphics[width=8cm]{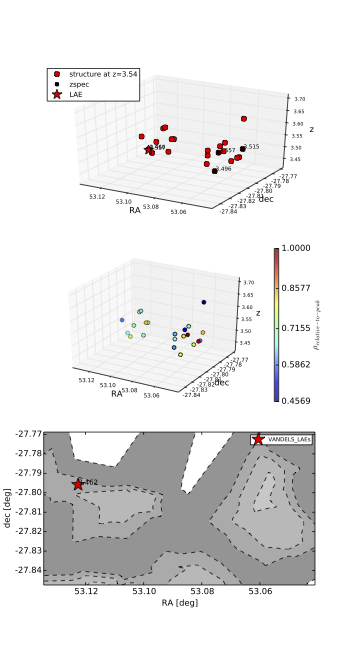}
\includegraphics[width=8cm]{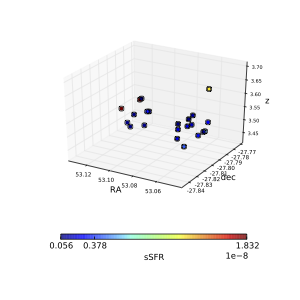}
\caption{The same as Fig. \ref{Bcdfs229} for the overdensity detected at $z=3.54$ in the CDFS.}
\label{Bcdfs354}%
\end{figure}

\begin{figure}
 \centering
\includegraphics[width=8cm]{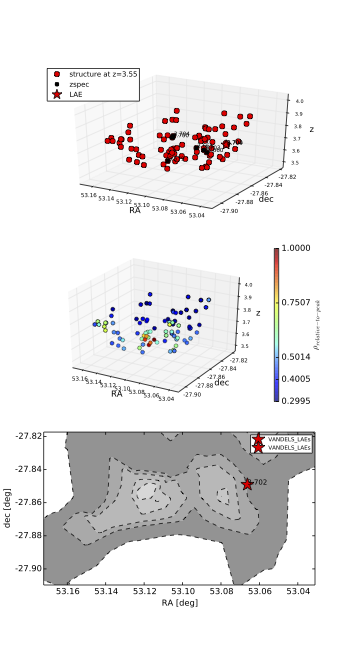}
\includegraphics[width=8cm]{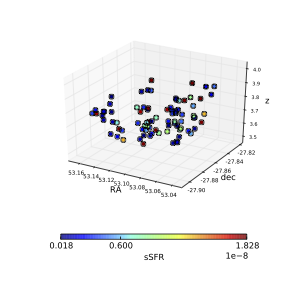}
\caption{The same as Fig. \ref{Bcdfs229} for the overdensity detected at $z=3.55$ in the CDFS.}
\label{Bcdfs355}%
\end{figure}

\begin{figure}
 \centering
\includegraphics[width=8cm]{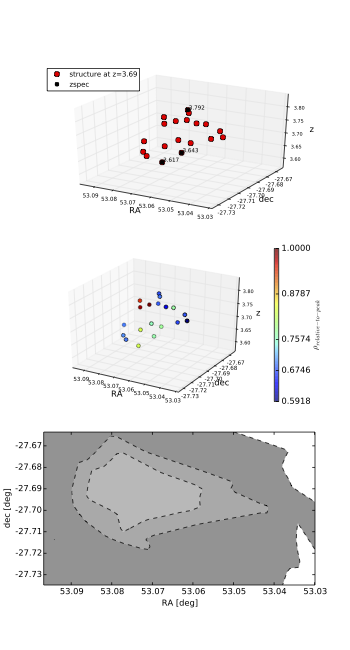}
\includegraphics[width=8cm]{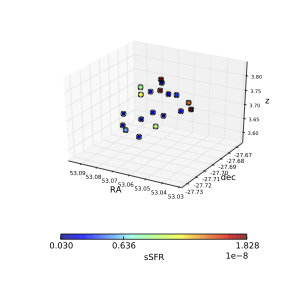}
\caption{The same as Fig. \ref{Bcdfs229} for the overdensity detected at $z=3.69$ in the CDFS.}
\label{Bcdfs369}%
\end{figure}

\clearpage

\begin{figure}
 \centering
\includegraphics[width=8cm]{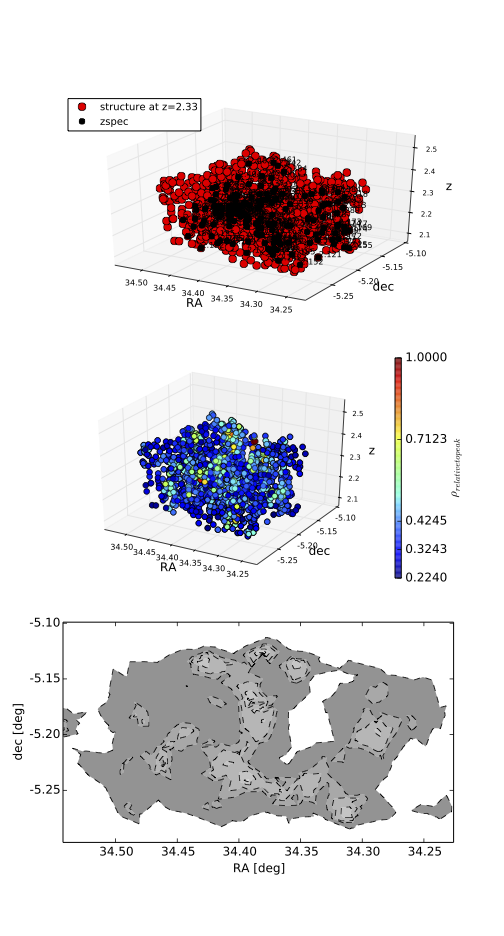}
\includegraphics[width=8cm]{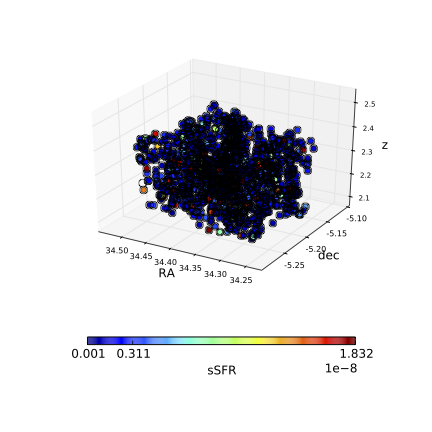}
\caption{The same as Fig. \ref{Bcdfs229} for the overdensity detected at $z=2.33$ in the UDS.}
\label{Buds233}%
\end{figure}

\begin{figure}
 \centering
\includegraphics[width=8cm]{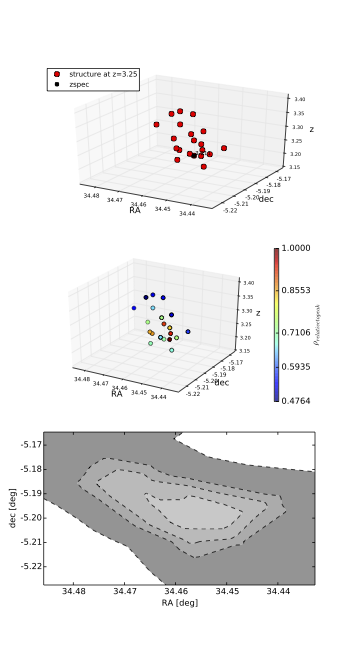}
\includegraphics[width=8cm]{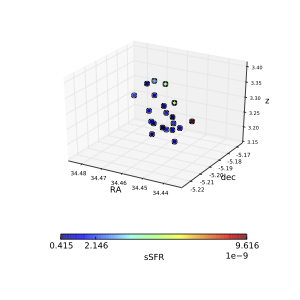}
\caption{The same as Fig. \ref{Bcdfs229} for the overdensity detected at $z=3.25$ in the UDS.}
\label{Buds3253445}%
\end{figure}

\begin{figure}
 \centering
\includegraphics[width=8cm]{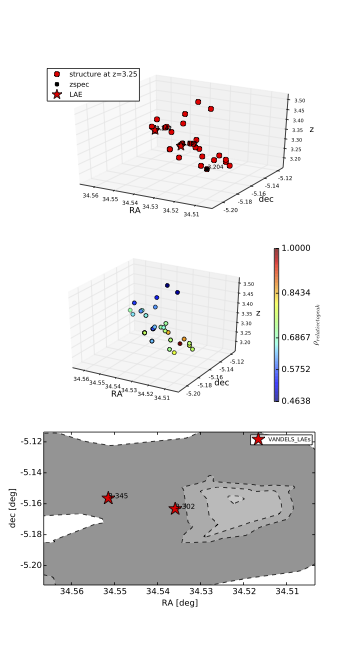}
\includegraphics[width=8cm]{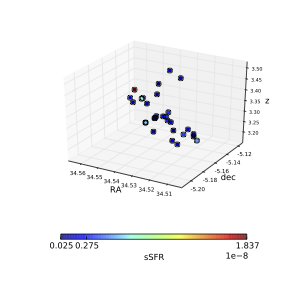}
\caption{The same as Fig. \ref{Bcdfs229} for the overdensity detected at $z=3.25$ in the UDS.}
\label{Buds3253453}%
\end{figure}

%\clearpage

\begin{figure}
 \centering
\includegraphics[width=8cm]{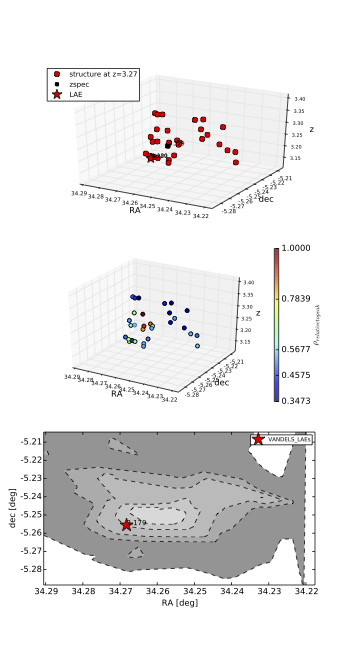}
\includegraphics[width=8cm]{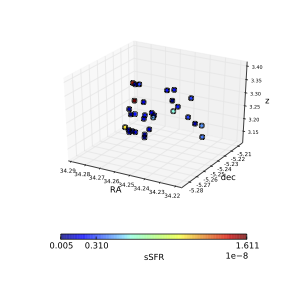}
\caption{The same as Fig. \ref{Bcdfs229} for the overdensity detected at $z=3.27$ in the UDS.}
\label{Buds327}%
\end{figure}

\begin{figure}
 \centering
\includegraphics[width=8cm]{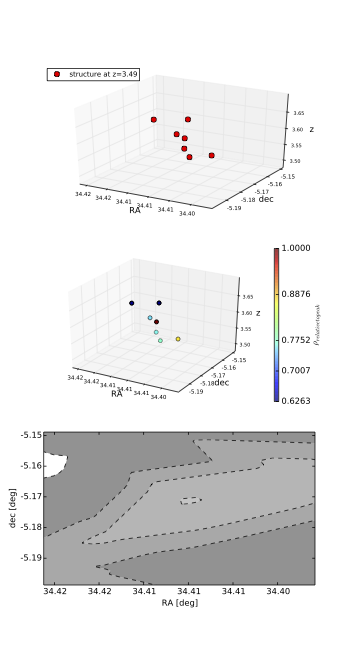}
\includegraphics[width=8cm]{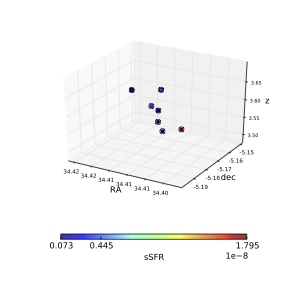}
\caption{The same as Fig. \ref{Bcdfs229} for the overdensity detected at $z=3.49$ in the UDS.}
\label{Buds349}%
\end{figure}

\begin{figure}
 \centering
\includegraphics[width=8cm]{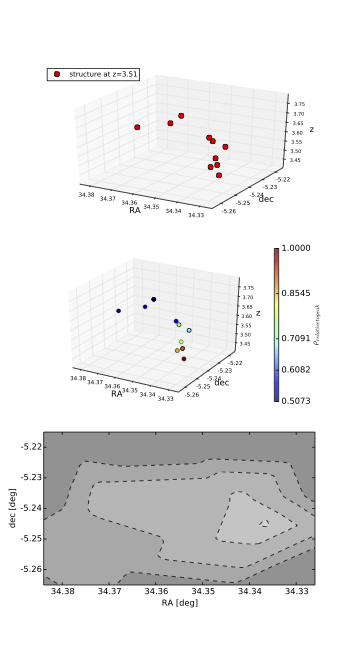}
\includegraphics[width=8cm]{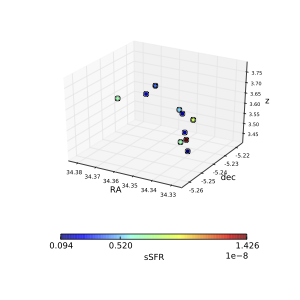}
\caption{The same as Fig. \ref{Bcdfs229} for the overdensity detected at $z=3.51$ in the UDS.}
\label{Buds351}%
\end{figure}

\clearpage

\begin{figure}
 \centering
\includegraphics[width=8cm]{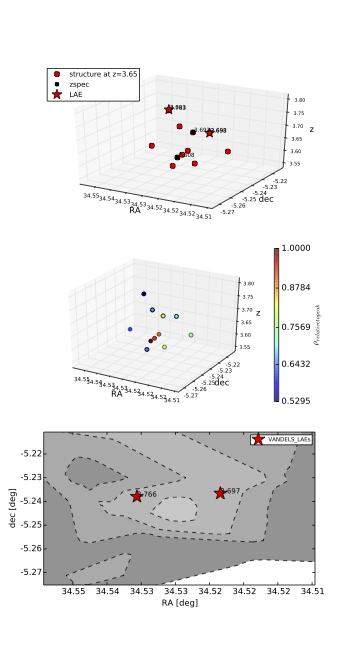}
\includegraphics[width=8cm]{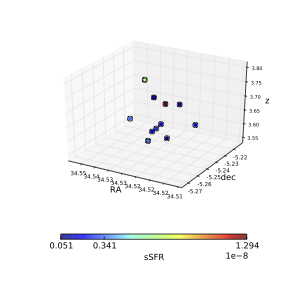}
\caption{The same as Fig. \ref{Bcdfs229} for the overdensity detected at $z=3.65$ in the UDS.}
\label{Buds365}%
\end{figure}

\begin{figure}
 \centering
\includegraphics[width=8cm]{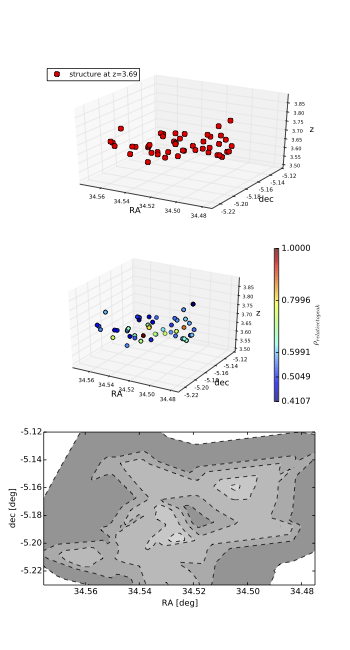}
\includegraphics[width=8cm]{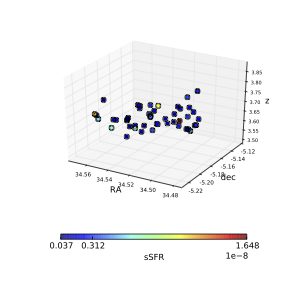}
\caption{The same as Fig. \ref{Bcdfs229} for the overdensity detected at $z=3.69$ in the UDS.}
\label{Buds369}%
\end{figure}

\begin{figure}
 \centering
\includegraphics[width=8cm]{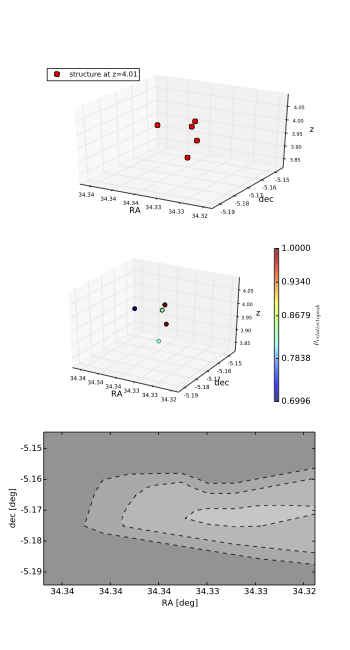}
\includegraphics[width=8cm]{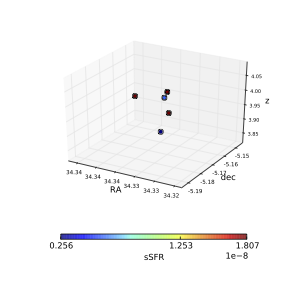}
\caption{The same as Fig. \ref{Bcdfs229} for the overdensity detected at $z=4.01$ in the UDS.}
\label{Buds401}%
\end{figure}

\clearpage 
\onecolumn

%ONLY EXAMPLE
\section{Color, magnitude, stellar mass, and specific star-formation rate versus environment}
\label{appendix3}

For the overdensities discussed in Sect. 5.2, we present here stellar mass ($upper ~left$), sSFR ($upper ~right$), rest-frame $U-V$ color ($lower ~right$) versus the density to highest-density peak, and rest-frame $U-V$ color versus rest-frame $V$ magnitude for the overdensity members (big red dots) and field galaxies (small black dots). The overdensity members with spectroscopic redshifts either from VANDELS or from the literature are shown as big black dots. The stars indicate the properties of the LAEVs in the overdensities described in Sect. 7. We show KS statistics for the distributions of stellar masses, sSFRs, and rest-frame $U-V$ colors of the overdensity members and field galaxies as KS and p parameters. The red horizontal lines in the lower right panels indicate the position of the red sequence and its uncertainty according to the definition of \citet{willis13}. 
%{\bf{the other plots are attached}}

\begin{figure*}[h!]
\includegraphics[width=10cm]{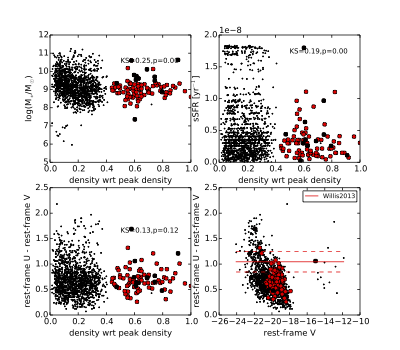}
\includegraphics[width=10cm]{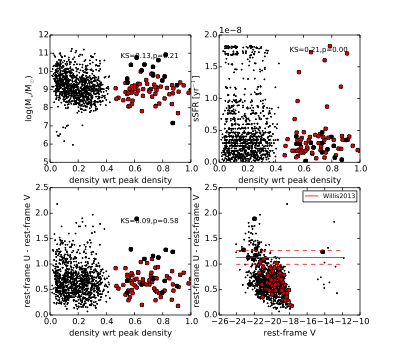}
\caption{As described in the text of this section for the overdensities detected at $z=2.29$ ($left ~ panels$) and  $z=2.30$ ($right ~ panels$) in the CDFS.
}
\label{uno}%
\end{figure*}

\begin{figure}
\includegraphics[width=10cm]{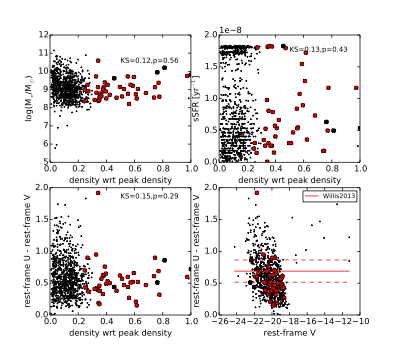}
\includegraphics[width=10cm]{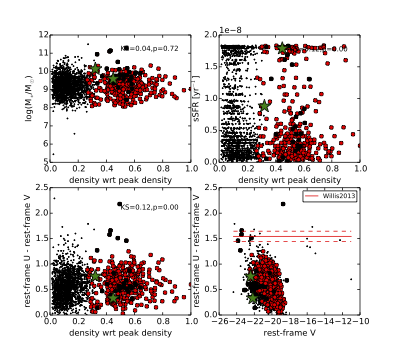}
\caption{As described in the text of this section for the overdensities detected at $z=2.80$ ($left ~ panels$) and  $z=3.17$ ($right ~ panels$) in the CDFS.
}
\label{due}%
\end{figure}

\begin{figure}
\includegraphics[width=10cm]{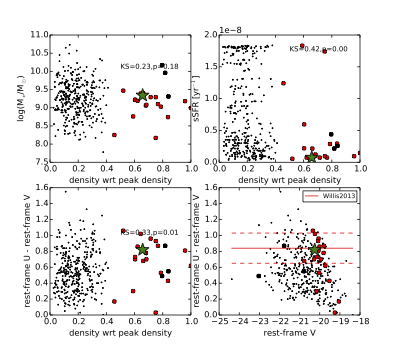}
\includegraphics[width=10cm]{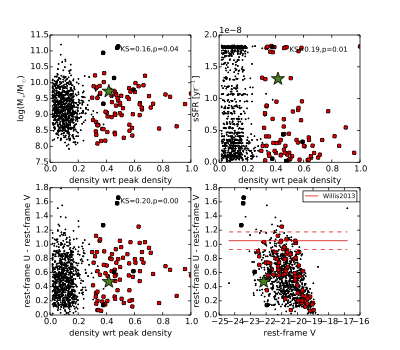}
\caption{As described in the text of this section for the overdensities detected at $z=3.54$ ($left ~ panels$) and  $z=3.55$ ($right ~ panels$) in the CDFS.
}
\label{tre}%
\end{figure}

\begin{figure}
\includegraphics[width=10cm]{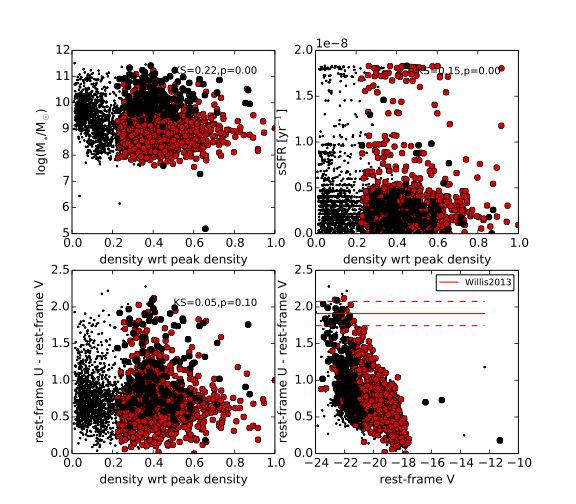}
\includegraphics[width=10cm]{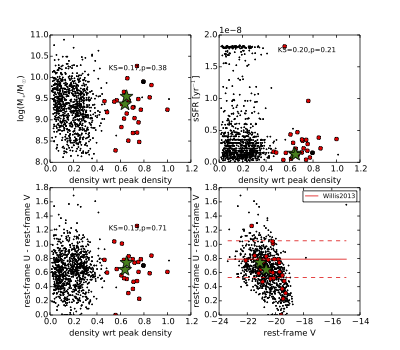}
\caption{As described in the text of this section for the overdensities detected at $z=2.33$ ($left ~ panels$) and  $z=3.25$ ($right ~ panels$) in the UDS.
}
\label{quattro}%
\end{figure}

\begin{figure}
\includegraphics[width=10cm]{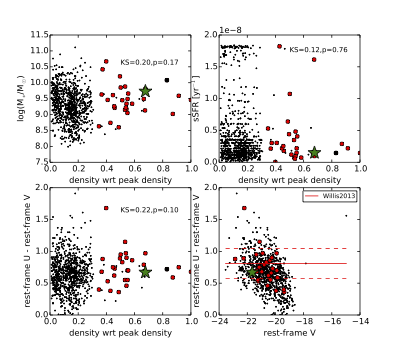}
\includegraphics[width=10cm]{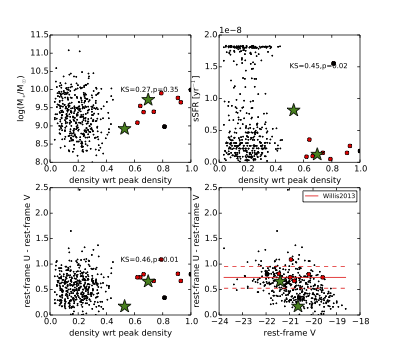}
\caption{As described in the text of this section for the overdensities detected at $z=3.27$ ($left ~ panels$) and  $z=3.65$ ($right ~ panels$) in the UDS.
}
\label{cinque}%
\end{figure}

%\clearpage

\section{Color-color diagram versus environment and galaxy shape}
\label{appendix4}

For the overdensities discussed in Sect. 5.2, we present here rest-frame $U-V$ colors as a function of rest-frame $V-J$ colors for the overdensity members (red dots) and field galaxies (small black dots) ($upper ~left$), for the overdensity members and color coded according to the sSFR ($upper ~ right$), for the overdensity members and color coded according to the Sersic index obtained with GALFIT ($lower ~left$), for the overdensity members and color coded according to the GALFIT fit axis ratio ($lower ~right$). The morphological parameters come from the analysis described in \citet{vanderWel2012}. The ticks in the vertical color bars indicate the maximum, the average, and the minimum value of labelled quantity. Generally, the members with higher rest-frame $U-V$ colors are characterized by sSFR values lower than the mean value among the members. 
%{\bf{the other plots are attached}}

\begin{figure*}[h!]
% \centering
\includegraphics[width=10cm]{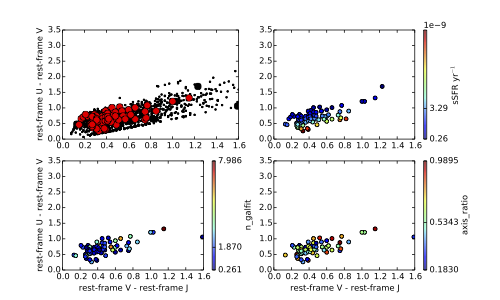}
\includegraphics[width=10cm]{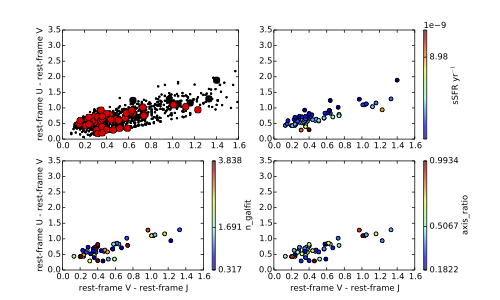}
\caption{As described in the text of this section for the overdensities detected at $z=2.29$ ($left ~ panels$) and $z=2.30$ ($right ~ panels$) in the CDFS.}
\label{Duno}%
\end{figure*}

\begin{figure*}[h!]
% \centering
\includegraphics[width=10cm]{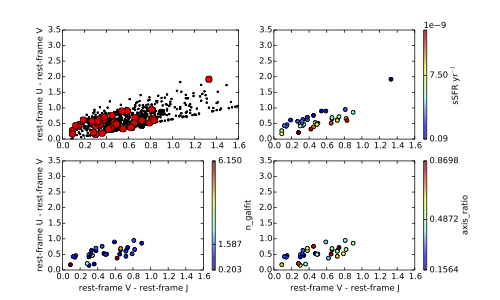}
\includegraphics[width=10cm]{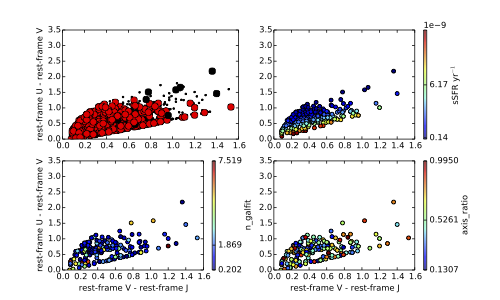}
\caption{As described in the text of this section for the overdensities detected at $z=2.80$ ($left ~ panels$) and $z=3.17$ ($right ~ panels$) in the CDFS.}
\label{Ddue}%
\end{figure*}

\begin{figure*}[h!]
\includegraphics[width=10cm]{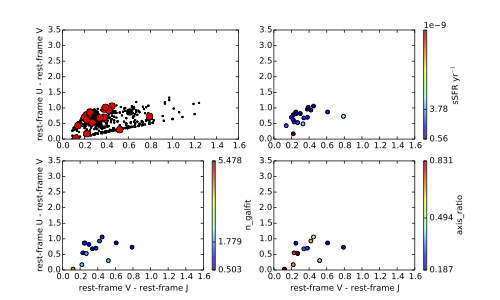}
\includegraphics[width=10cm]{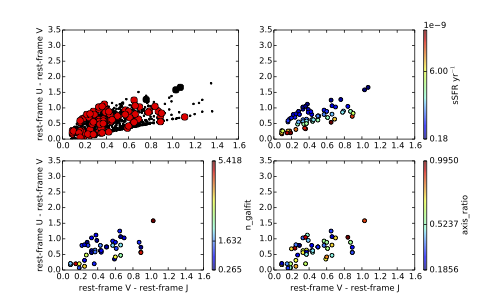}
\caption{As described in the text of this section for the overdensities detected at $z=3.54$ ($left ~ panels$) and $z=3.55$ ($right ~ panels$) in the CDFS.}
\label{Dtre}%
\end{figure*}

\begin{figure*}[h!]
\includegraphics[width=10cm]{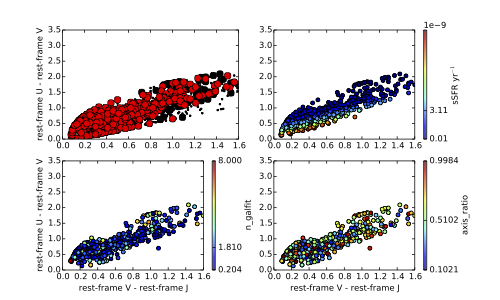}
\includegraphics[width=10cm]{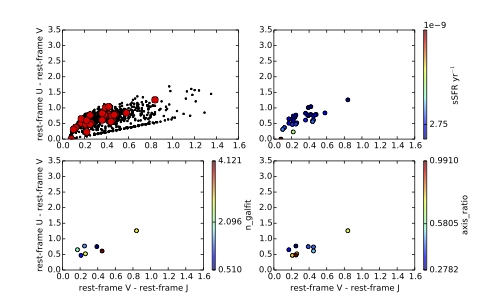}
\caption{As described in the text of this section for the overdensities detected at $z=2.33$ ($left ~ panels$) and $z=3.25$ ($right ~ panels$) in the UDS.}
\label{Dquattro}%
\end{figure*}

%do not compile
\begin{figure*}
\includegraphics[width=10cm]{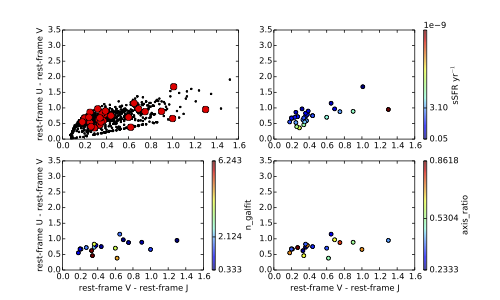}
\includegraphics[width=10cm]{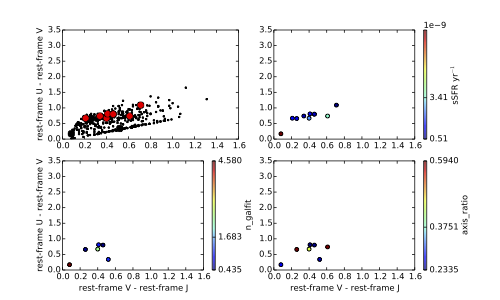}
\caption{As described in the text of this section for the overdensities detected at $z=3.27$ ($left ~ panels$) and $z=3.65$ ($right ~ panels$) in the UDS.}
\label{Dcinque}%
\end{figure*}

\end{appendix}
\end{document}